\documentclass[11pt]{article}
\usepackage{stackengine}
\stackMath

\usepackage{pifont}

\usepackage{dsfont}

\usepackage{amsmath,amssymb,graphicx}
\usepackage{hyperref}
\usepackage[nosort]{cite}

\usepackage{caption}
\usepackage{slashed}
\usepackage[makeroom]{cancel}

\DeclareFontFamily{U}{matha}{\hyphenchar\font45}
\DeclareFontShape{U}{matha}{m}{n}{
      <5> <6> <7> <8> <9> <10> gen * matha
      <10.95> matha10 <12> <14.4> <17.28> <20.74> <24.88> matha12
      }{}
\DeclareSymbolFont{matha}{U}{matha}{m}{n}

\DeclareMathSymbol{\oleft}{2}{matha}{"68}
\DeclareMathSymbol{\oright}{2}{matha}{"69}

\usepackage{multicol,xcolor,longtable}

\usepackage[T1]{fontenc}
\usepackage{hyphenat}
\usepackage{amsfonts}
\usepackage{mathrsfs}
\usepackage{mathdots}
\definecolor{darkred}{rgb}{0.65,0.15,0}
\definecolor{newgreen}{rgb}{0.2,0.62,0.14}
\hypersetup{pdfborder={0 0 0},colorlinks=true,urlcolor=blue,citecolor=blue,linkcolor=darkred,linktocpage=true}

\usepackage{array}
\newcolumntype{P}[1]{>{\centering\arraybackslash}p{#1}}
\usepackage{lscape}
\usepackage{multirow}

\usepackage{empheq}

\newcommand{\twelve}{\scalebox{0.5}{$12$}}
\newcommand{\ten}{{\scalebox{0.5}{$10$}}}

\newcommand{\ALT}{\textrm{\large{$\wedge$}}}
\newcommand{\SYM}{\textrm{\large{$\vee$}}}
\newcommand{\adjhat}{\widehat{\rm adj}}
\newcommand{\adjhhat}{\stackon[-11.5pt]{\stackon[-10.5pt]{\rm adj }{\widehat{\phantom{adj}}}}{\widehat{\phantom{adj}}}}

\newcommand{\be}{\begin{equation}}
\newcommand{\ee}{\end{equation}}
\newcommand{\bea}{\setlength\arraycolsep{2pt} \begin{eqnarray}}
\newcommand{\eea}{\end{eqnarray}}

\newcommand{\eq}[1]{\eqref{#1}}

\newcommand{\ord}[1]{{\scriptscriptstyle (#1)}}
\newcommand{\dgr}[1]{{\scalebox{0.5}{$(#1)$}}}

\newcommand{\nn}{\nonumber}

\newcommand{\lin}{\text{\tiny{(lin.)}}}

\newcommand{\ints}{\mathds{Z}}
\newcommand{\reals}{\mathds{R}}

\newcommand{\cM}{\mathcal{M}}
\newcommand{\cV}{\mathcal{V}}
\newcommand{\cE}{\mathcal{E}}
\newcommand{\cT}{\mathcal{T}}
\newcommand{\cF}{\mathcal{F}}
\newcommand{\cL}{\mathcal{L}}

\newcommand{\lb}{\left[}
\newcommand{\rb}{\right]}

\newcommand{\mf}[1]{{\mathfrak{#1}}}

\newcommand{\eprint}[1]{{\href{http://arxiv.org/abs/#1}{[\texttt{#1}]}}}
\newcommand{\eprintN}[1]{{\href{http://arxiv.org/abs/#1}{[\texttt{#1 [hep-th]}]}}}

\newcommand{\eprintGR}[1]{{\href{http://arxiv.org/abs/#1}{[\texttt{#1 [math.GR]}]}}}
\newcommand{\eprintRT}[1]{{\href{http://arxiv.org/abs/#1}{[\texttt{#1 [math.RT]}]}}}
\newcommand{\doi}[2]{{\hypersetup{urlcolor=darkred}\href{http://dx.doi.org/#2}{#1}\hypersetup{urlcolor=blue}}}

\newcommand{\CR}{\nonumber \\}
\newcommand{\w}[1]{\\[0.#1cm]}
\newcommand{\cU}{\,\mathcal{U}}
\newcommand{\cJ}{\mathcal{J}}
\newcommand{\cN}{\mathcal{N}}

\newcommand{\Fs}[1]{{}^{\scalebox{0.6}{$(\frac{#1}{2})$}}{}\hspace{-0.6mm}{\cF}}
\newcommand{\MLthree}{{M_{\scalebox{0.5}{$3$}}}}
\newcommand{\NLthree}{{N_{\scalebox{0.5}{$3$}}}}

\newcommand{\ta}{{\tilde\alpha}}
\newcommand{\tb}{{\tilde\beta}}
\newcommand{\tg}{{\tilde\gamma}}
\newcommand{\td}{{\tilde\delta}}

\newcommand{\wa}{{\widehat\alpha}}
\newcommand{\wb}{{\widehat\beta}}
\newcommand{\wg}{{\widehat\gamma}}

\newcommand{\wwa}{\stackunder[-3.8pt]{\stackunder[-4.3pt]{ \scriptstyle \widehat{\phantom{\alpha}}}{\scriptstyle \widehat{\phantom{\alpha}}}}{ \scriptstyle \alpha}}
\newcommand{\wwb}{\stackunder[-3.8pt]{\stackunder[-4.3pt]{ \scriptstyle \widehat{\phantom{\alpha}}}{\scriptstyle \widehat{\phantom{\alpha}}}}{ \scriptstyle \beta}}
\newcommand{\wwg}{\stackunder[-3.8pt]{\stackunder[-4.3pt]{ \scriptstyle \widehat{\phantom{\alpha}}}{\scriptstyle \widehat{\phantom{\alpha}}}}{ \scriptstyle \gamma}}

\newcommand{\wwatext}{\stackon[-6pt]{\stackon[-5.2pt]{   \alpha }{ \widehat{\phantom{\alpha}}}}{  \widehat{\phantom{\alpha}}}}

\newcommand{\tI}{{\tilde{I}}}
\newcommand{\tJ}{{\tilde{J}}}
\newcommand{\tK}{{\tilde{K}}}

\newcommand{\scal}[1]{\bigl ({#1} \bigr )}
\newcommand{\lsharp}{\text{\raisebox{-0.45ex}{\Large\guilsinglleft}}}
\newcommand{\rsharp}{\text{\raisebox{-0.45ex}{\Large\guilsinglright}}}

\newcommand{\tL}{{\widetilde\Lambda}}
\newcommand{\hL}{{\widehat\Lambda}}
\newcommand{\tXi}{{\widetilde\Xi}}
\newcommand{\hXi}{{\widehat\Xi}}

\makeatletter

\@addtoreset{equation}{section}
\makeatother

\begin{document}

\begin{flushright} CPHT-RR017.032021\\MI-TH-21 \end{flushright} 
 \vspace{8mm}

\begin{center}

{\LARGE \bf \sc  a master exceptional field theory}\\[5mm]

\vspace{6mm}
\normalsize
{\large  Guillaume Bossard${}^{1}$, Axel Kleinschmidt${}^{2,3}$ and Ergin Sezgin${}^4$}

\vspace{10mm}
${}^1${\it Centre de Physique Th\'eorique, CNRS,  Institut Polytechnique de Paris\\
91128 Palaiseau cedex, France}
\vskip 1 em
${}^2${\it Max-Planck-Institut f\"{u}r Gravitationsphysik (Albert-Einstein-Institut)\\
Am M\"{u}hlenberg 1, DE-14476 Potsdam, Germany}
\vskip 1 em
${}^3${\it International Solvay Institutes\\
ULB-Campus Plaine CP231, BE-1050 Brussels, Belgium}
\vskip 1 em
${}^4${\it Mitchell Institute for Fundamental Physics and Astronomy\\ Texas A\&M University
College Station, TX 77843, USA}

\vspace{20mm}

\hrule

\vspace{5mm}

 \begin{tabular}{p{14cm}}

We construct a pseudo-Lagrangian that is invariant under rigid $E_{11}$ and transforms as a density under $E_{11}$ generalised diffeomorphisms. The gauge-invariance requires the use of a section condition studied in previous work on $E_{11}$ exceptional field theory and the inclusion of constrained fields that transform in an indecomposable $E_{11}$-representation together with the $E_{11}$ coset fields. We show that, in combination with gauge-invariant and $E_{11}$-invariant duality equations, this pseudo-Lagrangian reduces to the bosonic sector of non-linear eleven-dimensional supergravity for one choice of solution to the section condition. For another choice, we reobtain the $E_8$ exceptional field theory and conjecture that our pseudo-Lagrangian and duality equations produce all exceptional field theories with maximal supersymmetry in any dimension. We also describe how the theory entails non-linear equations for higher dual fields, including the dual graviton in eleven dimensions. Furthermore, we speculate on the relation to the $E_{10}$ sigma model.
\end{tabular}

\vspace{6mm}
\hrule
\end{center}

\thispagestyle{empty}

\newpage

\setcounter{page}{1}
\pagenumbering{roman}

\setcounter{tocdepth}{2}
\tableofcontents


\section{Introduction}
\setcounter{page}{1}
\pagenumbering{arabic}

It has been proposed by West~\cite{West:2001as,West:2003fc,Tumanov:2016abm} that it should be possible to write (the bosonic part of) $D=11$ supergravity in a way that utilises the infinite-dimensional Kac--Moody symmetry $E_{11}$.\footnote{See~\cite{Julia,Julia2} for early discussions of Kac--Moody symmetries in supergravity and~\cite{Damour:2002cu} for a different proposal involving~$E_{10}$. See also Appendix~\ref{app:E10} for a discussion on the relation of our model to $E_{10}$.} The proposed construction involves a non-linear realisation of $E_{11} / K(E_{11})$ where $K(E_{11})\subset E_{11}$ denotes a subgroup that generalises the eleven-dimensional Lorentz group $SO(1,10)$ and will be defined in more detail below.\footnote{
It is (a spin cover of) this group that is relevant for the fermionic fields as has been discussed for $E_{10}$ in~\cite{Damour:2005zs,deBuyl:2005sch,Damour:2006xu} and for $E_{11}$ in~\cite{Kleinschmidt:2006tm,Kleinschmidt:2007zd,Steele:2010tk,Bossard:2019ksx}. This subgroup is called $I_c(E_{11})$ in~\cite{West:2001as,West:2003fc,Tumanov:2016abm}.} 
The fields of the non-linear realisation depend on space-time coordinates $z^M$ that also transform under $E_{11}$ in an infinite-dimensional highest weight representation~\cite{West:2003fc} that we shall call $R(\Lambda_1)$ in this paper. The Maurer--Cartan derivatives $\partial_M \cV \cV^{-1} $ of the $E_{11} / K(E_{11})$ coset field $\cV$ are invariant under rigid right multiplication of $\cV$ by $E_{11}$ and only transform under the $K(E_{11})$ subgroup.\footnote{This transformation includes an inhomogeneous connection piece.}
In~\cite{Tumanov:2016abm} a set of first-order `modulo equations' were proposed for the coset field $\cV$. These were constructed out of the Maurer--Cartan derivatives that transform into each other under $K(E_{11})$ and are generated from the matter duality equation $F_4= \star F_7$. The terminology of modulo equation means a (first-order) equation that is not gauge-invariant (for gauge parameters depending on eleven coordinates) but only holds up to certain gauge transformations that can be eliminated by passing to a higher-order equation~\cite{Tumanov:2016abm,Tumanov:2017whf,Glennon:2020qpt}. For the graviton the gauge-invariant equations are second order, and for more complicated fields gauge-invariance requires differential equations of arbitrarily high order~\cite{Tumanov:2016abm,Tumanov:2017whf,Glennon:2020qpt}. The gauge transformations and the intrinsic multiplet structure of the whole $K(E_{11})$-multiplet of first-order modulo equations are not known to the best of our knowledge. Another interesting feature of this proposal is that it involves infinite dualisations of the physical fields~\cite{Riccioni:2006az}, a bit in the spirit of unfolding of equations of motion~\cite{Vasiliev:2005zu,Bekaert:2006py,Boulanger:2015mka}. A similar infinite dualisation appears for the $E_9$ symmetry of $D=11$ supergravity reduced to two space-time dimensions~\cite{Nicolai:1987kz} and also plays a r\^ole in the $E_{10}$ proposal via the gradient conjecture~\cite{Damour:2002cu}.

In a different, but not unrelated, strand of research, field theories with extended space-time symmetries and exceptional symmetry groups have been constructed. These so-called exceptional field theories (ExFT)~\cite{Berman:2010is,Hohm:2013vpa,Hohm:2013uia,Hohm:2014fxa,Hohm:2015xna,Abzalov:2015ega,Baguet:2015xha,Musaev:2015ces,Berman:2015rcc,Bossard:2021jix} possess fields in a non-linear realisation of $E_n$ (for $n\leq 9$) and these fields depend on extended (internal) coordinates $Y^M$ involving representations of $E_n$, see also~\cite{Hull:2007zu,Coimbra:2011nw,Coimbra:2012af}. Exceptional field theories have proved to be very powerful tools in analysing Kaluza--Klein reductions in supergravity \cite{Hull:2007zu,Godazgar:2013dma,Lee:2014mla,Baron:2014bya,Ciceri:2014wya,Baguet:2015sma,Varela:2015ywx,Kruger:2016agp,Cassani:2016ncu,Malek:2018zcz,Hohm:2019bba,Malek:2019eaz,Berman:2020tqn,Guarino:2020flh}. Their (pseudo-)actions are uniquely fixed by the symmetries of the ExFT and these symmetries importantly include a generalised gauge symmetry called the generalised Lie derivative or generalised diffeomorphisms~\cite{Berman:2011cg,Berman:2012vc,Aldazabal:2013via}.\footnote{This generalised similar structures in double field theory and generalised geometry~\cite{Siegel:1993th,Hitchin:2004ut,Gualtieri:2003dx,Hull:2009mi,Hull:2009zb,Hohm:2010pp}.} Closure of this gauge symmetry and consistency of the whole procedure crucially depends on the {\em section constraint} (stated in~\eqref{eq:SC}) that restricts the dependence of all objects on the extended coordinates. Choosing a particular solution to this section constraint renders ExFT fully equivalent to unreduced $D=11$ supergravity or type IIB supergravity~\cite{Blair:2013gqa}. However, while the formulation has $E_n$ symmetry, picking a solution to the section constraint breaks the $E_n$ symmetry and one is left with the known symmetries of the supergravity theory in question. Besides the generalised diffeomorphism invariance and the section constraint, another crucial feature of ExFT is the occurrence of {\em constrained fields} that go beyond the $K(E_n)/ E_n$ coset fields and the usual supergravity tensor hierarchy~\cite{deWit:2008ta} but are central to the invariance of the theory \cite{Hohm:2013uia,Hohm:2014fxa}. The fields are constrained in the sense of the section constraint and do not carry additional degrees of freedom.

In an effort to define an exceptional field theory for $E_{11}$
we have recently proposed a non-linear set of first-order duality equations~\cite{Bossard:2019ksx} that can be written as
\begin{align}
\label{eq:DEintro}
 \cM_{IJ} F^I = \Omega_{IJ} F^J\,,
\end{align}
where $F^I$ denotes an infinite collection of non-linear field strengths that transform under $E_{11}$ in a representation that is defined by its tensor hierarchy algebra~\cite{Bossard:2017wxl}. This representation is neither highest nor lowest weight but can be shown to carry a symplectic form that we write as $\Omega_{IJ}$ and that generalises at the same time the usual Levi--Civita symbol that appears in duality equations and the symplectic form familiar from electric-magnetic duality relations in $D=4$. The generalised metric $\cM_{IJ}$ on the left-hand side is the one constructed from the $E_{11} / K(E_{11})$ coset acting in the representation of the field strengths  and a non-degenerate $K(E_{11})$-invariant bilinear form $\eta_{IJ}$.\footnote{Since we are working in a metric formulation the local $K(E_{11})$ invariance is automatic and all equations are $E_{11}$-covariant. We note, however, that in order to properly define $\cM_{IJ}$ we have to construct it from a vielbein in a parabolic $K(E_{11})$ gauge for the coset $E_{11} / K(E_{11})$. This will be discussed in more detail in Section~\ref{sec:GL11}.} The existence of $\eta_{IJ}$ is a key assumption in our construction and we shall summarise evidence for it in Section~\ref{sec:THA}.
The definition of the field strengths $F^I$ crucially involves an infinite set of constrained fields that go beyond the $E_{11}$ coset fields.  These fields are necessary from the construction of the tensor hierarchy algebra and sit in an {\em indecomposable} representation with the $E_{11}$ coset fields~\cite{Bossard:2017wxl}.\footnote{The need for additional fields can be seen most directly when considering linearised gauge-invariance for the dual graviton equation where the trace of spin connection needs to be included~\cite{Bossard:2017wxl}. Moreover, the latter transforms indecomposably together with the dual graviton potential under the Poincar\'e algebra \cite{West:2002jj,Boulanger:2008nd} and indecomposable representations of the Poincar\'e algebra also  occur for unfolded formulations of gauge fields~\cite{Boulanger:2015mka}.} A similar feature was also observed in the context of $E_9$ ExFT~\cite{Bossard:2018utw}. We emphasise that the $E_{11}$-covariance of $F^I$ and the mere existence proof of the representation labelled by $I$ depends on the tensor hierarchy algebra $\cT(\mf{e}_{11})$ introduced in~\cite{Bossard:2017wxl}. Besides the use of the tensor hierarchy algebra our approach differs from West's original $E_{11}$ proposal in other aspects, such as the section constraint, as discussed in more detail in~\cite{Bossard:2017wxl,Bossard:2019ksx}.

We showed in~\cite{Bossard:2019ksx} that the duality equation~\eqref{eq:DEintro} is invariant under $E_{11}$ generalised diffeomorphisms if an appropriate section constraint is obeyed. The argument in~\cite{Bossard:2019ksx} depended on a certain $E_{11}$ group-theoretic identity, which we refer to as the \textit{master identity}. This was checked partially in the reference and in this paper we give a modified version of this identity and provide strong evidence for its validity. 

The duality equation~\eqref{eq:DEintro} by itself is not sufficient to fully determine the dynamics of $E_{11}$ ExFT since one also requires equations of motion for the constrained fields, as explained in~\cite{Bossard:2019ksx}. In the present paper we shall provide these equations, thereby completing the construction of the $E_{11}$ exceptional field theory.

We shall arrive at these equations of motion by constructing a pseudo-Lagrangian whose variation provides all equations for the constrained fields as well as a projection (mediated by the constrained fields) of the duality equations~\eqref{eq:DEintro}. This situation is completely analogous to what happens for ExFTs in other dimensions as is the structure of the pseudo-Lagrangian \cite{Hohm:2013uia,Bossard:2021jix}. The pseudo-Lagrangian takes the schematic form
\begin{align}
\label{eq:Lagin}
\mathcal{L}= \mathcal{L}_{\text{pot}_1} + \mathcal{L}_{\text{pot}_2}+ \mathcal{L}_{\text{kin}} + \mathcal{L}_{\text{top}}
\,.
\end{align}
The terms occurring in this pseudo-Lagrangian are summarised in \eqref{AllTermsL}. For the purposes of this introduction we shall only describe these individual pieces qualitatively.
\begin{itemize}
\item
The (first) potential term $\mathcal{L}_{\text{pot}_1}$ is the standard one that appears in all ExFTs and takes a universal form that is given for example in~\cite{Cederwall:2017fjm}. It depends only on the $E_{11}$ coset fields.
\item 
The (second) potential term $\mathcal{L}_{\text{pot}_2}$ generalises a similar term for $E_8$~\cite{Hohm:2014fxa} and $E_9$ ExFT~\cite{Bossard:2018utw} and is related to the non-closure of the algebra of generalised Lie derivatives in the absence of ancillary parameters. It was generalised to any simply laced finite-dimensional group $G$ in~\cite{Cederwall:2019bai}.  For Kac--Moody groups $\mathcal{L}_{\text{pot}_2}$ depends on both the $E_{11}$ coset fields and the constrained field transforming indecomposably with the ${\mf e}_{11}$ current.
\item
The kinetic term $\mathcal{L}_{\text{kin}}$ generalises the usual field strength squared terms and involves both the $E_{11}$ coset fields and the constrained fields.\footnote{Strictly speaking, the sign of the kinetic term is the opposite of the usual sign. In Section~\ref{sec:altL}, we give an alternative form of the pseudo-Lagrangian where we combine the terms differently to bring out the standard sign.}
\item 
The topological term $\mathcal{L}_{\text{top}}$ generalises the topological term of other ExFTs that does not depend on the external metric. It depends on the $E_{11}$ coset fields only through the $\mf{e}_{11}$ current, without the explicit appearance of the generalised metric $\cM$. It is defined as a rigid $E_{11}$-invariant completion of the total derivative of a constrained field transforming in an indecomposable representation together with the ${\mf e}_{11}$ current, as does the topological term in $E_9$ ExFT~\cite{Bossard:2021jix}.
\end{itemize}
Each of the individual terms is invariant under rigid $E_{11}$ but only a specific combination of the four terms is invariant under $E_{11}$ generalised diffeomorphisms. In $E_n$ ExFT, the invariant terms are invariant by themselves under $E_n$ generalised diffeomorphisms and connected by external diffeomorphisms. Here, all these diffeomorphisms are subsumed in $E_{11}$ generalised diffeomorphisms that therefore fix everything.
All objects in this pseudo-Lagrangian depend on generalised space-time coordinates $z^M$ in the $R(\Lambda_1)$ representation of $E_{11}$ and the construction crucially requires the associated section constraint. 
However, when decomposing $E_{11}$ into $GL(D)\times E_{11-D}$ to make contact with exceptional field theory in $D$ dimensions, the actual kinetic, topological  and potential terms will generically get contributions from all the different parts of~\eqref{eq:Lagin}. In that sense the naming of the terms in~\eqref{eq:Lagin} is  somewhat arbitrary and chosen because of structural similarities with those in $E_n$ ExFT.

The construction of the pseudo-Lagrangian~\eqref{eq:Lagin} will be one of the central results of this paper. Its gauge-invariance will depend on several new $E_{11}$ identities that have not been known to the best of our knowledge. We can prove many of them and provide supporting partial checks that cover complete $E_{11}$ representations for the others.

The pseudo-Lagrangian~\eqref{eq:Lagin} and the duality equation~\eqref{eq:DEintro} are fairly formal objects since $E_{11}$ is an infinite-dimensional algebra whose exact structure is not known. The infinite-dimensionality in particular means that one has to be sure that the pseudo-Lagrangian~\eqref{eq:Lagin} is well-defined as it involves infinite, potentially ill-defined sums. We shall address this issue in {\em level decomposition} of $E_{11}$ where a finite-dimensional subgroup of $E_{11}$ is used as an organising principle~\cite{Damour:2002cu,West:2002jj,Nicolai:2003fw,Kleinschmidt:2003jf}. By employing an associated `semi-flat' formulation and partial gauge-fixing of the local $K(E_{11})$-invariance, we can show that the pseudo-Lagrangian becomes a well-defined object. It is important to stress that our results do not rely on a truncated level decomposition of the pseudo-Lagrangian, but hold to {\it all} levels.

To underline this point, we analyse in detail two cases. In the first, the finite-dimensional subgroup is $GL(11)\subset E_{11}$, corresponding to diffeomorphisms in eleven dimensions. The pseudo-Lagrangian~\eqref{eq:Lagin} in that case will be shown to describe 
eleven-dimensional supergravity and in particular its Euler--Lagrange equations include the non-linear Einstein equation. Of course, this also requires choosing the corresponding solution to the $E_{11}$ section constraint. The pseudo-Lagrangian~\eqref{eq:Lagin} can moreover be used to obtain an infinite class of Lagrangians that describe the infinite set of dual fields in the theory. We shall exhibit in particular the non-linear Lagrangian for the dual graviton and the three-form potential gradient dual.\footnote{Gradient duals generalise the $(D{-}2)$-form dual to scalar fields to arbitrary $p$-form potentials. $E_{11}$ exceptional field theory as the $E$ theory of \cite{Riccioni:2006az}  includes an infinite tower of successive gradient duals to the three-form, its dual six-form and the dual graviton field.}

The second case we analyse is for the finite-dimensional subgroup $GL(3)\times E_8$, associated to $E_8$ exceptional field theory. We will show in this case that the pseudo-Lagrangian reproduces the Lagrangian derived in~\cite{Hohm:2014fxa}. Because the individual pseudo-Lagrangians in~\eqref{eq:Lagin} are not invariant under generalised diffeomorphisms, the reconstruction of the covariant derivative and field strengths requires to recombine all contributions. 

In general we expect the same to be true for any Levi subgroup $L_D$ associated to the fundamental weight $\Lambda_D$ in the convention of Figure~\ref{fig:e11dynk}, with\footnote{We here restrict to the standard $GL(D){\times} E_{11-D}$ subgroups that are obtained by deleting node $D$ from the Dynkin diagram. Choosing an $E_{11}$ conjugate of such a subgroup can lead to theories with multiple time directions~\cite{Keurentjes:2004bv}. Time-like T- und U-dualities and their effect on the signature of space-time have been investigated in~\cite{Hull:1998br,Hull:1998ym}, also in the context of exceptional field theory \cite{Malek:2013sp,Blair:2016xnn}.} 
\begin{align}
\label{eq:Levis}
L_D &= GL(D) \times E_{11-D} \, , \quad \text{for } 3\le D\le 8\; , \quad &\text{$E_{11-D}$ ExFT} \; , \CR
L_9 &= GL(10) \times SL(2)\; ,  \quad &\text{type IIB} \; ,  \CR
L_{10} &= GL(1) \times Spin_{+\!}(10,10)\; , \quad &\text{double field theory}  \; , \CR
L_{11} &= GL(11) \; ,  \quad &\text{D=11 supergravity}\; . 
\end{align}
Choosing a Levi subgroup of this type singles out a $GL(1)\subset GL(D)$ factor that can be used to define a $\ints$-grading on $E_{11}$ that we shall refer to as the `level'.
For all $L_D$, the dynamical fields appear at level $k\in\ints$ in the range $0\le k \le  -(\Lambda_1,\Lambda_D)$ in terms of the canonically normalised inner product between weights. For instance, for $L_{11}$ one has $-(\Lambda_1,\Lambda_{11})=\tfrac32$ and thus only levels $k=0,1$ appear and they correspond to the usual propagating fields, namely the metric and the three-form. The other propagating fields are dual to the dynamical fields with a duality equation $\cE_k=0$ between fields of level $k$ and fields of level $-2(\Lambda_1,\Lambda_D)-k$. Solving partly the section constraint in the $L_D$ level decomposition, the pseudo-Lagrangian decomposes as
\be \label{LinallD}
\cL = \cL_{L_D}   -\tfrac14 \hspace{-5mm} \sum_{k>-(\Lambda_1,\Lambda_D)} |\cE_k|^2 \; , 
\ee
up to total derivative terms,
such that the corresponding Euler--Lagrange equations subject to the duality equation $\cE_k=0$ are equivalent to the Euler--Lagrange equations of the \mbox{(pseudo-)} Lagrangian $\cL_{L_D}$. Although $\cL$ depends on the infinitely many fields of the $E_{11} / K(E_{11})$ coset, $\cL_{L_D}$ only depends on the fields of level $ k \le  -(\Lambda_1,\Lambda_D)$, which are the standard fields in the corresponding theory. For $GL(3)\times E_8$ we compute that $\cL_{L_D}$ is the exceptional field theory Lagrangian~\cite{Hohm:2014fxa} and we expect that $\cL_{L_D}$ is the exceptional field theory Lagrangian for odd $D$ between $2$ and $8$, and the pseudo-Lagrangian for even $D$. Similarly $\cL_{L_9}$ is expected to be the type IIB pseudo-Lagrangian and $\cL_{L_{10}}$ the double field theory pseudo-Lagrangian  \cite{Hohm:2011dv}.

In the list~\eqref{eq:Levis}, we have not included cases where the Levi subgroup is infinite-dimensional. One of those cases is $L_2=SL(2) \times E_9$ whose ExFT version has been constructed recently~\cite{Bossard:2021jix} and has served as an inspiration for the present paper. The other Kac--Moody group is $L_1=GL(1)\times E_{10}$ and we find that the same decomposition \eqref{LinallD} of our model applies to $D=1$ for $E_{10}$, although the content of the field strength representation is more conjectural. We describe this in Appendix~\ref{app:E10} and also discuss a possible relation to the $E_{10}$ sigma model that arises in the analysis of the cosmological billiard~\cite{Damour:2002cu}. The proposed relation constrains the sigma model conserved charge to lie in the $E_{10}$-orbit of its positive Borel subalgebra.\footnote{We did not include  $D=9$ supergravity~\cite{Bergshoeff:1995as} in the list \eqref{eq:Levis}, because it is associated to a non-maximal parabolic with Levi $L_{9,11} = GL(9)\times GL(2)$. But we expect the same result with $k> - ( \Lambda_1 , \Lambda_9 {+} \Lambda_{11})$ in \eqref{LinallD}. Another case that is not included in the list is that of massive type IIA which requires a background that does not satisfy the section constraint, or alternatively, a deformation of the gauge structure~\cite{Ciceri:2016dmd}. Massive IIA in the context of $E_{11}$ was also discussed in~\cite{Schnakenburg:2002xx,Tumanov:2016dxc}.}
 
\medskip

The structure of the article is as follows. We first introduce the necessary group-theoretic facts and notation for $E_{11}$, its irreducible representations and the indecomposable representation extending the adjoint $\mf{e}_{11}$ that is part of the tensor hierarchy algebra in Section~\ref{sec:e11}. This section also contains the definition of the $E_{11}$ coset and constrained fields as well as their transformations under generalised diffeomorphisms. In Section~\ref{sec:dyn}, we review the duality equation~\eqref{eq:DEintro} in more detail and construct the pseudo-Lagrangian~\eqref{eq:Lagin}. We verify that the pseudo-Lagrangian is consistent with the duality equation~\eqref{eq:DEintro} in Section~\ref{sec:EOMchi}, postponing the derivation of the equation for the constrained fields to Section~\ref{sec:EOM}. In Section~\ref{sec:GI}, we prove gauge invariance of the pseudo-Lagrangian under generalised diffeomorphisms. In order to analyse the pseudo-Lagrangian for a given solution of the section condition one has to choose a level decomposition and the necessary steps for performing such an analysis are given in Section~\ref{sec:SF}. 
In Section~\ref{sec:GL11}, we then study the pseudo-Lagrangian in level decomposition under $GL(11)$ and show that it gives exactly non-linear $D=11$ supergravity. Section~\ref{HigherLevelSection} is devoted to studying the consequences of our model for the higher level fields and how they relate to dual formulations of the theory. In Section~\ref{sec:E8}, we perform the level decomposition for $GL(3)\times E_8$ and reproduce $E_8$ ExFT. Appendix \ref{app:ids} contains the many proofs and supporting evidence for the group-theoretic identities that are used in the construction of the theory. 
In particular, in Table~\ref{tab:ids}, we summarise these identities and  recall in the conclusions the main assumptions stated above, namely  the master identity and the existence of $\eta_{IJ}$.
In Appendix~\ref{app:ext}, we formalise some aspects of indecomposable representations in the language of Lie algebra cohomology. Appendices \ref{app:GL11} and \ref{app:E8} collect details on the $GL(11)$ and $GL(3)\times E_8$ level decompositions of the various fields and tensors. Appendix~\ref{app:E10} contains details of the $GL(1)\times E_{10}$ decomposition and remarks on the relation to the $E_{10}$ sigma model. 

Since this is a rather long paper, readers primarily interested in seeing how eleven-dimensional supergravity emerges from the proposed master exceptional field theory may focus on Sections~\ref{sec:e11}, \ref{sec:dyn}, \ref{sec:SF} and~\ref{sec:GL11}.


\section{Preliminaries}
\label{sec:e11}

In this section, we introduce the basic group-theoretic building blocks, fields and transformations laws that will be essential for constructing the pseudo-Lagrangian of $E_{11}$ exceptional field theory. Throughout this section several $E_{11}$ objects will be introduced with specific index conventions. These will be summarised at the end in section~\ref{sec:notsum} in table form for the convenience of the reader. The derivation of the group theoretical identities is relegated to Appendix~\ref{app:ids}.

\subsection{\texorpdfstring{Building blocks from $E_{11}$ and its tensor hierarchy algebra}{Building blocks from E11 and its tensor hierarchy algebra}}
\label{sec:THA}

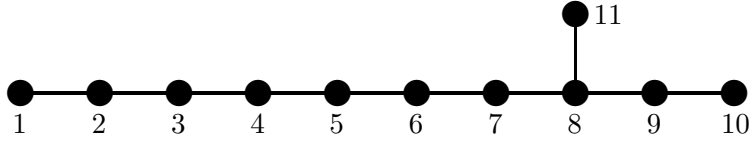
\begin{figure}[t!]
\centering
\begin{picture}(300,50)
\thicklines
\multiput(10,10)(30,0){10}{\circle*{10}}
\put(10,10){\line(1,0){270}}
\put(220,40){\circle*{10}}
\put(220,10){\line(0,1){30}}
\put(7,-5){$1$}
\put(37,-5){$2$}
\put(67,-5){$3$}
\put(97,-5){$4$}
\put(127,-5){$5$}
\put(157,-5){$6$}
\put(187,-5){$7$}
\put(217,-5){$8$}
\put(247,-5){$9$}
\put(275,-5){$10$}
\put(227,36){$11$}
\end{picture}
\caption{\label{fig:e11dynk}{\sl Dynkin diagram of $E_{11}$ with labelling of nodes used in the text.}}
\end{figure}

The Lie algebra $\mf{e}_{11}$ is a Kac--Moody algebra defined from its Dynkin diagram depicted in Figure~\ref{fig:e11dynk} and we consider its split real form, see~\cite{Kac}. This means in particular that the subalgebra corresponding to nodes $1$ to $10$ of the diagram is the $\mf{sl}(11)$ (over $\reals$) that can be extended to $\mf{gl}(11)$ by taking the Cartan generator associated with node $11$.

For the development of the general formalism we shall denote the generators of $E_{11}$ as $t^\alpha$ with commutation relations
\begin{align}
\label{eq:e11}
\bigl[  t^\alpha, t^\beta \bigr] = f^{\alpha\beta}{}_\gamma t^\gamma\,.
\end{align}
There is a non-degenerate $E_{11}$-invariant bilinear form on $\mf{e}_{11}$ that we shall denote by $\kappa^{\alpha\beta}$ and that can be used to raise and lower indices on $E_{11}$-tensors such as $f^{\alpha\beta}{}_\gamma$. As usual $f^{\alpha\beta\gamma}$ is totally antisymmetric.

One can define a `temporal involution' on $\mf{e}_{11}$ that generalises the `minus-transpose' operation on matrices~\cite{Kac,Englert:2003py}. We shall denote the fixed-point algebra of this involution by $K(\mf{e}_{11})\subset \mf{e}_{11}$. Its intersection with $\mf{gl}(11)$ discussed above is $\mf{so}(1,10)$ and therefore $K(\mf{e}_{11})$ should be thought of as an infinite generalisation of the Lorentz algebra.\footnote{The split real Lie algebra $\mf{e}_{11}$ has a `maximal compact' subalgebra that is fixed by the standard Chevalley involution and whose intersection with $\mf{gl}(11)$ is $\mf{so}(11)$ rather than $\mf{so}(1,10)$. We write $K(\mf{e}_{11})$ for the subalgebra fixed by the temporal involution. For example, for the split real $\mf{e}_7$, the corresponding subalgebras are $\mf{su}(8)$ as the maximal compact and $\mf{su}^*(8)$ as the fixed point algebra of a temporal involution.} We shall denote the corresponding group by $K(E_{11})$ and it is known that it has a two-fold cover $\widetilde{K(E_{11})}$~\cite{Harring:2019}, generalising the spin groups and that has finite-dimensional spinor representations~\cite{Kleinschmidt:2006tm,Bossard:2019ksx}. We shall not consider fermions in this paper.

Associated with the Lie algebra $\mf{e}_{11}$ is a tensor hierarchy super-algebra $\cT(\mf{e}_{11})$ that was defined in~\cite{Bossard:2017wxl}, see also Appendix~\ref{app:ids} and~\cite{Cederwall:2021ymp}. The concept of tensor hierarchy super-algebra was first introduced by Palmkvist in~\cite{Palmkvist:2013vya}, see also~\cite{Cremmer:1997ct,Cremmer:1998px,HenryLabordere:2002dk,Henneaux:2010ys} for related ideas. It is a super-algebra with a $\ints$-grading that is consistent with the Grassmann $\ints_2$-grading

\begin{align}
\label{eq:Tdec}
\cT(\mf{e}_{11}) = \bigoplus_{p\in\ints} \cT_p
\end{align}
and $\mf{e}_{11}$ is the maximal simple subalgebra of $\cT_0$. The $E_{11}$ modules $\cT_p$ play a prominent r\^ole in the construction of the theory and we will now discuss some of their properties. 

Let us first introduce further notation.
We use the symbol $\oleft$ to represent an indecomposable representation sum  $M_1 \oleft M_2$, meaning that $M_1 \subset M_1 \oleft M_2$ is a proper submodule while $M_2 \cong M_1 \backslash  (M_1 \oleft M_2) $ is only a quotient module, but not a submodule. Throughout the paper we shall use the notation that $R(\lambda)$ is the irreducible highest weight module of $\mf{e}_{11}$ with highest weight $\lambda = \sum_{i=1}^{11} k_i \Lambda_i$, with $\Lambda_i$ the $i$th fundamental weight in the numbering convention of Figure~\ref{fig:e11dynk}. We shall  use instead $L(\lambda)$ for a completely reducible \textit{bounded weight module} of  $\mf{e}_{11}$ with highest weight $\lambda$, which is generally a specific countable direct sum of irreducible highest weight modules $R(\lambda^\prime)$ with weights $\lambda^\prime \le \lambda$ including $R(\lambda)$.\footnote{With $\lambda^\prime\le \lambda$ we mean that $\lambda-\lambda^\prime$ is a non-negative linear combination of simple roots. In addition, $\lambda^\prime < \lambda$ denotes the stronger two conditions $\lambda^\prime \le \lambda$ and $\lambda^\prime \neq \lambda$.} 
As we do not know the full structure of the individual $\cT_p$ as $E_{11}$-representations, we shall sometimes just list the first low-lying irreducible representations and use the notation $L(\lambda)$ to include all the lower weight modules that are not known.\footnote{
These can be ordered by the {\em height} of the highest weight $\lambda$, where height refers to the sum of the coefficients of $\lambda$ in a simple root basis.}

In particular $\cT_0, \cT_{\pm 1}$ and $\cT_{\pm 2}$ play a special  r\^ole in the construction of the theory. The algebra $\cT_0$ includes ${\mf e}_{11}$ as a proper subalgebra and decomposes into $E_{11}$ representations as the direct sum~\cite{Bossard:2017wxl}
\be
\cT_0 = \adjhat \oplus
D_0\ ,
\label{conjectured}
\ee
where  $\adjhat$ is an indecomposable representation that is built on the adjoint of $\mf{e}_{11}$ and that we write as
\be 
\label{adjhatDef} 
\adjhat = {\mf{e}_{11} \oleft  L(\Lambda_2})\ ,
\ee
whereas $D_0$ is completely reducible. 
In~\eqref{adjhatDef}, $L(\Lambda_2)=R(\Lambda_2)\oplus\ldots $ is a direct sum of highest weight modules of $E_{11}$ that form an indecomposable Lie algebra extension of $\mf{e}_{11}$~\cite{Bossard:2017wxl}. Which representations occur exactly is fixed by the tensor hierarchy algebra. To the extent that it has been analysed, the only known representation in $L(\Lambda_2)$ at present is the irreducible module $R(\Lambda_2)$. The reader is invited to think of $L(\Lambda_2)$ as just $R(\Lambda_2)$ everywhere in this paper. Our results do not depend on the exact knowledge of additional representations in $L(\Lambda_2)$.\footnote{The next potential candidate that we identified and that we could not rule out on cohomological grounds is $R(2\Lambda_{10})$.}
From the investigations in~\cite{Bossard:2017wxl} it is known that the completely reducible $D_0$ is a sum of highest weight modules and that its first summand is given by $R(\Lambda_{10})$.

The $\mf{e}_{11}$ representation $\cT_1$ is a bounded weight module of highest weight $\Lambda_1$ given by 
\begin{align}
\label{eq:T1exp}     
\cT_1 =  R(\Lambda_1) \oplus R(\Lambda_1 + \Lambda_{10}) \oplus R(\Lambda_{11}) \oplus R(\Lambda_1+2\Lambda_3)  \oplus  \ldots\,,
\end{align}
where we recognise the irreducible module that was already mentioned in the introduction for the space-time coordinates~\cite{West:2003fc}. In the theory we shall only consider the generators $P^M$ in $R(\Lambda_1) \subset \cT_1$.

Since $\cT$ is a super-algebra, the level $\cT_2$ is obtained by anticommutators $\{ P^M, P^N\}$ and thus is contained in the symmetric product of $\cT_1$ with itself. It is a bounded weight module of highest weight $\Lambda_{10}$ with 
\begin{align}
\label{eq:THA2}
    \cT_2 =  L(\Lambda_{10}) \oplus D_2 \,,
\end{align}
where 
\be
L(\Lambda_{10}) =  R(\Lambda_1) \SYM R(\Lambda_1) \ominus R(2\Lambda_1)\ ,
\label{eq:L10def} 
\ee
and $D_2$ is a fully reducible module that can be ignored in our construction. Due to the defining properties of the tensor hierarchy algebra $\cT$,  $R(2\Lambda_1)$ is not part of $\cT_2$. 
Since $R(\Lambda_1)$ is the representation of the derivatives~\cite{West:2003fc}, $L(\Lambda_{10})$ is recognised as the symmetric section condition~\cite{Bossard:2017wxl}. This will be discussed in more detail in Section~\ref{sec:coord}. 

The construction of the theory as defined in \cite{Bossard:2019ksx} relies on the assumption that $D_0= L(\Lambda_{10}) \oplus D'_0$, where $D'_0$ denotes a direct sum of further possible highest weight $E_{11}$ representations. 
In Appendix~\ref{app:MID}, we prove $L(\Lambda_{10})\subset \cT_2$ and provide evidence for the conjecture that $L(\Lambda_{10})\subset \cT_0$ in Appendices~\ref{Casimir} and \ref{app:e8tha}. Although this would make the algebraic construction of the theory from the tensor hierarchy algebra more uniform, we do not require this assumption in this paper. 

We now discuss in more detail the indecomposable representation $\adjhat$ in~\eqref{adjhatDef} and introduce notation for its basis elements to be used throughout the paper. We write the generators in $\adjhat$ as 
\begin{align}
t^{\wa}=(t^\alpha, t^\ta)\,,
\end{align}
 which besides the $E_{11}$ generators $t^\alpha$, include the basis elements $t^{\ta}$ in $L(\Lambda_2)$ that satisfy
\begin{align}
\label{eq:T0CR}
\lb t^\alpha, t^\ta \rb = -T^{\alpha \ta}{}_\tb t^\tb - K^{\alpha \ta}{}_\beta t^\beta\,,
\end{align}
where $T^{\alpha \ta}{}_\tb$ are representation matrices of $\mf{e}_{11}$ and $K^{\alpha\ta}{}_{\tb}$ is a Lie-algebra cocycle. It represents the fact that the $t^\ta$ are in an indecomposable representation with the adjoint of $\mf{e}_{11}$ since the action of $\mf{e}_{11}$ on $t^\ta$ gives back $\mf{e}_{11}$ generators. We give some more information on Lie algebra cocycles in Appendix~\ref{app:ext}. 

The action of $E_{11}$ on the generators $t^\ta$ leads via a Jacobi identity in $\cT_0$ to the two identities
\begin{subequations}
\begin{align}
\label{eq:Ttrm}
T^{\alpha \ta}{}_\tg T^{\beta\tg}{}_\tb -  T^{\beta \ta}{}_\tg T^{\alpha\tg}{}_\tb &= f^{\alpha\beta}{}\gamma T^{\gamma \ta}{}_\tb\,,\\
\label{eq:Ktrm}
K^{\alpha \ta}{}_\gamma f^{\beta\gamma}{}_\delta- K^{\beta \ta}{}_\gamma f^{\alpha \gamma}{}_\delta - T^{\alpha \ta}{}_\tg K^{\beta\tg}{}_\delta +  T^{\beta \ta}{}_\tg K^{\alpha\tg}{}_\delta  &= -f^{\alpha\beta}{}_\gamma K^{\gamma \ta}{}_\delta\,.
\end{align}
\end{subequations}
The first identity expresses the fact that the $T^{\alpha\ta}{}_\tb$ are representation matrices of $\mf{e}_{11}$ and thus transform as (invariant) tensors under $E_{11}$. The second identity is the cocycle identity and shows that $K^{\alpha \ta}{}_\beta$ is {\em not} a tensor under $E_{11}$ when $\ta$ is viewed as an index of the representation $L(\Lambda_2)$. The cocycle is defined modulo a redefinition of the generator $t^{\ta} = t^{\ta} + K^{\ta}{}_{\beta} t^\beta$ such that $K^{\alpha \ta}{}_\beta$ and $K^{\prime \alpha \ta}{}_\beta $ satisfying 
\be
K^{\alpha \ta}{}_\beta = K^{\prime \alpha \ta}{}_\beta + T^{\alpha \ta}{}_\tb K^\tb{}_\beta + f^{\alpha\gamma}{}_\beta K^\ta{}_\gamma \;  
\ee
are equivalent cocycles. We shall encounter several $E_{11}$ invariant tensors with one index $\wa$ in the representation $\adjhat$, as for example the structure constants $f^{\wa\wb}{}_{\wg}$ for two generators in $\adjhat$. Because of the indecomposable representation, their components written with $\wa = ( \alpha , \tilde{\alpha})$ are not $E_{11}$-invariant tensors for the indices $\alpha$ in the adjoint and $\tilde{\alpha}$ in  $L(\Lambda_2)$, like $f^{\alpha\tilde{\beta}}{}_\gamma=-K^{\alpha\tilde{\beta}}{}_\gamma$ for example. It will be convenient to define the failure to transform under $E_{11}$ according to the direct sum representation  $\mathfrak{e}_{11} \oplus L(\Lambda_2)$ as a `non-covariant' variation $\Delta^\alpha$. The notation we employ for an object $O$ is 
\begin{align}
\label{eq:DNC}
\Delta^\alpha O = \text{(transformation of $O$ under $t^\alpha$) - (na\"ive transformation suggested by the indices).}
\end{align}
Equation~\eqref{eq:Ktrm} can be written in this notation as
\begin{align}
\label{eq:Dcoc}
\Delta^\alpha K^{\beta\ta}{}_\delta =  - K^{\alpha \ta}{}_\gamma f^{\beta\gamma}{}_\delta -T^{\beta \ta}{}_\tg K^{\alpha\tg}{}_\delta \,,
\end{align}
while~\eqref{eq:Ttrm} is $\Delta^\alpha T^{\beta\ta}{}_\tb =0$. The na\"ive transformation is simply given by the action of the corresponding representation matrix; for the adjoint, the matrix representation of $t^\alpha$ is  $T^{\alpha \beta}{}_{\gamma} = -f^{\alpha\beta}{}_\gamma$.

It is important to note that due to the indecomposable structure of $\adjhat$ there is no invariant tensor $\kappa^{\wa\wb}$ that can be used for raising and lowering indices on $\adjhat$. The dual module is instead $\adjhat^* = \mf{e}_{11}^* \oright \overline{L(\Lambda_2)}\subset \cT_{-2}$.

\medskip

We shall denote the generators of $L(\Lambda_{10})\subset \mathcal{T}_2$ collectively as $P^\Lambda$. They transform under $E_{11}$ with representation matrices as
\begin{align}
\lb t^\alpha , P^\Lambda \rb = - T^{\alpha \Lambda}{}_\Sigma P^\Sigma\,,
\end{align}
and there is no non-covariance associated with this action of $E_{11}$, thus $\Delta^\alpha T^{\beta \Lambda}{}_\Sigma =0$.

\medskip

The first negative level $\cT_{-1}$ of the tensor hierarchy algebra $\cT(\mf{e}_{11})$ was identified in~\cite{Bossard:2017wxl} to be the one relevant for defining field strengths. For $n\le9$,  $\cT_{-1}({\mf e}_n)$ is the so-called embedding tensor representation~\cite{deWit:2005hv,deWit:2008ta} in supergravity in $11-n$ dimensions~\cite{Palmkvist:2013vya}. $\cT_{-1}$ is therefore the natural representation for the generalised fluxes in exceptional geometry \cite{Blair:2014zba,Chatzistavrakidis:2019seu,Fernandez-Melgarejo:2018yxq}. We shall label the generators of $\cT_{-1}$ by $t_I$ and they transform as
\begin{align}
\lb t^\alpha , t_I \rb  =  T^{\alpha J}{}_I t_J
\end{align}
under $E_{11}$ with proper tensors $T^{\alpha I}{}_J$. As generators of the Lie super-algebra $\cT(\mf{e}_{11})$, the $t_I$ are $\ints_2$-odd. 
In~\cite{Bossard:2017wxl} it was shown that $\cT_{-1}$ admits a non-degenerate antisymmetric tensor $\Omega_{IJ}$ that we shall refer to as the symplectic form.    The tensor $\Omega_{IJ}$ is invariant under $E_{11}$ (and $\cT_0$) and is the restriction of the $\mathds{Z}_2$-symmetric invariant bilinear form of degree $p=-2$ that implies $\cT_{-2-p} = \cT_p^{\; *}$.

As a representation of $E_{11}$, the space $\cT_{-1}$ is neither a highest nor lowest weight representation, but rather is expected to decompose as a vector space sum as
\begin{align}
\label{eq:Tminus1}
\cT_{-1}=   R_{-1} \oplus  \bigoplus_{\lambda} R(\lambda)  \oplus  \bigoplus_{\lambda} \overline{R(\lambda) }  \ .
\end{align}
where $R_{-1}$ is the piece that does not admit a highest or lowest weight and $R(\lambda)$ and $\overline{R(\lambda)}$ denote a (possibly infinite) sequence of highest and lowest weight representations, respectively. $R_{-1}$ is the quotient of $\cT_{-1}$ by the maximal proper submodule given by the sum of highest and lowest weight representations. 

Our construction crucially requires the existence of a non-degenerate $K(E_{11})$-invariant bilinear form $\eta_{IJ}$, for instance for writing $\cM_{IJ}$ in the duality equation~\eqref{eq:DEintro}. 
If $\cT_{-1}$ were completely reducible, in which case $R_{-1}$ would be an $E_{11}$-submodule of $\cT_{-1}$, the existence of a $K(E_{11})$-invariant bilinear form $\eta_{IJ}$ on $\cT_{-1}$ would be guaranteed. However, there are also examples where such a form exists without the requirement of complete reducibility.\footnote{By analysing examples in the branching of $\cT_{-1}$ under $\mf{e}_9$ or $\mf{e}_{10}$ (see \eqref{E10Dind}), we have checked that it is possible to have indecomposable representations of $\mf{e}_n$ within~\eqref{eq:Tminus1} that still admit a $K(\mf{e}_n)$-invariant bilinear form.} In the following we shall assume that $\eta_{IJ}$ exists. Our investigations to date have not unveiled any submodule $R(\lambda)$ so that $\cT_{-1}$ might be irreducible, but at present~\eqref{eq:Tminus1} is an assumption and we do not have proof of the existence of $\eta_{IJ}$ whose existence is crucial for our model.  
We have evidence for the existence of $\eta_{IJ}$ from considering subalgebras of $\mf{e}_{11}$. In~\cite{Bossard:2017wxl} we computed  the explicit components  of $\eta_{IJ}$ in $GL(11)$ level decomposition that are $K(E_{11})$-invariant for the first seven $\mf{gl}(11)$ levels. Considering the $\mf{e}_9$ subalgebra of $\mf{e}_{11}$ as in~\cite{Bossard:2021jix} establishes the existence of $\eta_{IJ}$ on an infinite-dimensional submodule, including the first three $\mf{e}_9$ levels.

Assuming $\eta_{IJ}$ exists, one has the identity  
\begin{align}
\Omega_{IJ^\prime}\eta^{J^\prime \! K} \Omega_{KI^\prime} \eta^{I^\prime\! J} =\delta_I^J\,,
\end{align}
which allows for the definition of a twisted self-duality equation. This identity was also checked in level decomposition.

\subsection{Space-time, fields, currents and field strengths}
\label{sec:coord}

In this section, we introduce the generalised space-time of $E_{11}$ exceptional field theory, define the fields of the theory and a set of currents and field strengths. 

\subsubsection{Space-time and section constraint}

In order to define the fields and the pseudo-Lagrangian, we need to introduce a generalised space-time on which they live. As West~\cite{West:2003fc} we take this to be given by coordinates $z^M$ transforming in the highest weight representation $R(\Lambda_1)$ of $E_{11}$. This is also the natural coordinate representation obtained by extrapolating the pattern of coordinate representations for other exceptional field theories~\cite{Coimbra:2012af,Berman:2012vc} or equivalently the vector fields of maximal supergravity theories~\cite{deWit:2008ta}. The representation also occurs in the tensor hierarchy algebra component $\cT_1$ as described in \eqref{eq:T1exp}~\cite{Bossard:2017wxl}. We recall that $\cT_1$ is completely reducible, and the index $M$ will always refer to the first direct summand $R(\Lambda_1)$ only.

The action of $E_{11}$ in the $R(\Lambda_1)$ representation is through invariant tensors $T^{\alpha M}{}_N$ that satisfy
\begin{align}
\label{eq:Rlamf}
T^{\alpha M}{}_P T^{\beta P}{}_N -T^{\beta M}{}_P T^{\alpha P}{}_N =  f^{\alpha\beta}{}_\gamma T^{\gamma M}{}_N\,.
\end{align}

Partial derivatives with respect to the coordinates $z^M$ will be written as $\partial_M$.
Either by considering a complex of functions from the tensor hierarchy algebra or by using the general form of the section constraint of other exceptional field theories one is led to requiring that
\begin{align}
\label{eq:SC}
\kappa_{\alpha\beta} T^{\alpha P}{}_M T^{\beta Q}{}_N \partial_P \otimes \partial_Q = -\frac12 \partial_M\otimes \partial_N + \partial_N \otimes \partial_M
\end{align}
when the partial derivatives act on any pair of fields or parameters in the theory. The equation above is the section constraint of $E_{11}$ exceptional field theory and follows very naturally as a Jacobi identity when considering the tensor hierarchy algebra $\cT(\mf{e}_{11})$~\cite{Bossard:2017wxl}. It will play a central r\^ole in our construction. Some of the group-theoretic identities we write will only be valid `on section' and we shall make this manifest by contracting the corresponding indices with dummy derivatives as in \eqref{eq:SC}.

In terms of representation theory, the section constraint~\eqref{eq:SC} picks out specific pieces of the symmetric ($\SYM$) and antisymmetric $(\ALT$) second powers of $R(\Lambda_1)$. As $E_{11}$ representations these are given by~\cite{Bossard:2017wxl,Bossard:2019ksx}\footnote{Tensor products of highest weight representations of $E_{11}$ are completely reducible~\cite{Kac} but yield an infinite sum of highest weight modules. These can be partially ordered by the {\em height} of the highest weight $\Lambda$, where height refers to the sum of the coefficients of $\Lambda$ in a simple root basis. When writing the result of the tensor product, we only show the first few $E_{11}$ representations in this order. Branching to subalgebras $\mf{gl}(D)\oplus \mf{e}_{11-D}$ introduces a {\em level} that shall be used to organise the corresponding representations. This will be explained in more detail starting from Section~\ref{sec:GL11}. For many computations the SimpLie software~\cite{Bergshoeff:2007qi} was very useful.}
\begin{subequations}
\label{eq:L1tens}
\begin{align}
R(\Lambda_1)  \SYM R(\Lambda_1) &= R(2\Lambda_1)  \oplus \overbrace{\big( R(\Lambda_{10})\oplus R(\Lambda_2 +\Lambda_{10}) \oplus \ldots\big)}^{L(\Lambda_{10})}\,,\\
R(\Lambda_1) \ALT R(\Lambda_1) &= R(\Lambda_2) \oplus \underbrace{\big(R(\Lambda_4)\oplus\ldots\big)}_{L(\Lambda_4)}\,.
\end{align}
\end{subequations}
The section constraint~\eqref{eq:SC} is the statement that the components along the representations $L(\Lambda_{10})\oplus L(\Lambda_4)$ have to vanish. Note that although $L(\Lambda_{10})$ and $ L(\Lambda_4)$ are infinitely reducible, they should intuitively be thought of as being `small' compare to  $R(2\Lambda_1)$ and $R(\Lambda_2)$. For example one can find two linearly independent $X_M$ and $Y_M$ such that $X_M Y_N$ vanishes in $L(\Lambda_{10})\oplus L(\Lambda_4)$, but one cannot find non-trivial vectors such that $X_M Y_N$ vanishes in  $R(2\Lambda_1)  $ and its vanishing in $R(\Lambda_2)$ implies that they are linearly dependent. 

The action of the algebra $\mathcal{T}_0$ in the representation $\mathcal{T}_1$ \eqref{eq:T0CR} moreover implies that $T^{\Lambda M}{}_N=0$ and $T^{\ta M}{}_N$ satisfies the following identity\footnote{Note that it makes sense to restrict the indices to $R(\Lambda_1)$ because $\mathfrak{e}_{11}$ acts on $R(\Lambda_1)$. Therefore if all the uncontracted $\mathcal{T}_1$ indices are in $R(\Lambda_1)$, all the contracted $\mathcal{T}_1$ indices are also in  $R(\Lambda_1)$.}
\begin{align}
\label{eq:THA01}
T^{\ta M}{}_P T^{\beta P}{}_N - T^{\beta M}{}_P T^{\ta P}{}_N = T^{\beta \ta}{}_\tb T^{\tb M}{}_N + K^{\beta \ta}{}_\alpha T^{\alpha M}{}_N
\end{align}
that will be used when checking gauge-invariance of the pseudo-Lagrangian.

\subsubsection{Fields}

The fields of exceptional field theory will come from $\mf{e}_{11}^* \oright \overline{L(\Lambda_2)}\oplus \overline{L(\Lambda_{10})} \subset \cT_{-2}$, the dual representation of $\cT_0\supset \mf{e}_{11}$. The fields of the theory are locally functions on the module $\overline{R(\Lambda_1)}$, and the generalised space-times is locally isomorphic to the module $\overline{R(\Lambda_1)}$. They include the representative $\cV(z)\in E_{11}$ of the coset $E_{11} / K(E_{11})$, which is associated with a non-linear realisation of $E_{11}$. It transforms under rigid $E_{11}$ and local $K(E_{11})$ as
\begin{align} \label{Vrightleft} 
\cV(z) \to k(z) \cV(g^{-1}z) g\,,
\end{align}
where the rigid $g\in E_{11}$ acts on the coordinates $z^M$ according to the $R(\Lambda_1)$ representation.\footnote{In~\cite{West:2003fc}, this is expressed by saying that one considers a non-linear realisation of the semi-direct product $E_{11} \ltimes R(\Lambda_1)$.}
One can consider $\cV(z)$ in any representation of $E_{11}$ and we shall typically suppress the space-time dependence in the equations. For example, writing $\cV$ in the $R(\Lambda_1)$ representation leads to $\cV^A{}_M$, where $A$ is a flat tangent space index transforming under $K(E_{11})$ and $\cV$ is to be thought of as the vielbein on the generalised space-time with local coordinates $z^M$.\footnote{We stress that this is only a local concept and, moreover, the generalised space-time dependence is always restricted by the section constraint~\eqref{eq:SC}.}
We also note that at the level of the coordinates $z^M$ there is no meaning to the usual distinction between `external' and `internal' coordinates. This distinction arises only when considering a level decomposition $GL(D) \times E_{11-D} \subset E_{11}$ of the type we consider starting from Section~\ref{sec:GL11}.

Besides the vielbein $\cV$ we will make ample use of the `generalised metric'
\begin{align}
\cM= \cV^\dagger \eta \cV\,,
\end{align}
 where $\eta$ is a $K(E_{11})$-invariant symmetric tensor on the representation in which one wants to evaluate $\cM$. For instance, the $R(\Lambda_1)$ representation possesses such an invariant tensor $\eta_{MN}$ (as do all highest or lowest weight representations~\cite{Kac}) and this leads to the generalised metric $\cM_{MN}$ that is symmetric in its indices and transform only under rigid $E_{11}$ but is inert under local $K(E_{11})$. As we have explained above in Section~\ref{sec:THA}, we assume a similar $K(E_{11})$-invariant metric $\eta_{IJ}$ also exists on the field strength representation $\cT_{-1}$ and we can define $\cM_{IJ}$. We should stress that the generalised metrics $\cM$ and $\cM_{IJ}$ are at the moment formal objects and plagued with infinite sums and questions of being well-defined. We shall explain in Section~\ref{sec:SF} how to define properly the theory using the vielbein formulation in a (parabolic) gauge for the coset $E_{11} / K(E_{11})$. The use of $\cM_{MN}$ is nonetheless very useful to simplify the equations, and $E_{11}$ representation theory ensures that all the identities formally derived using $\cM_{MN}$ imply well-defined identities in a  parabolic gauge. All the expressions we shall write in the following are formal in the same sense but can be made meaningful in level decomposition.\footnote{If one restricts all elements to the so-called minimal Kac--Moody group, defined in Section~\ref{sec:subtle}, the expressions are meaningful as they stand.  We shall argue in Section~\ref{sec:subtle} that, for physical reasons, we need to work with a completion of the minimal group and explain how the model remains well-defined.}

Starting from the generalised metric $\cM$ function of the fields in the coset $E_{11} / K(E_{11})$,  we define the current
\begin{align}
\label{eq:curdef}
\cM^{-1} \partial_M \cM = J_{M\alpha} t^\alpha 
 \quad\quad \Longrightarrow\quad\quad
\cM^{PN}\partial_M \cM_{NQ} =  J_{M\alpha} T^{\alpha P}{}_Q\,,
\end{align}
that takes values in the $\mf{e}_{11}$ Lie algebra. We have also given its expression in the fundamental representation $R(\Lambda_1)$. Since the generalised metric $\cM_{MN}$ is symmetric in its indices we note that the definition implies the identity 
\begin{align}
\label{eq:Jherm}
J_{M\alpha} T^{\alpha P}{}_N  \cM^{NQ}  = J_{M\alpha}  T^{\alpha Q}{}_N  \cM^{NP} \,.
\end{align}
The current also satisfies the usual Maurer--Cartan equation
\begin{align}
\label{eq:MC}
2 \partial_{[M} J_{N] \alpha} + f^{\beta\gamma}{}_\alpha J_{M\beta} J_{N\gamma}=0
\end{align}
Since the current is valued in $\mf{e}_{11}$ we can use the Killing--Cartan metric $\kappa_{\alpha\beta}$ to raise and lower the adjoint index and $J_M{}^\alpha$ are the components of an element in the co-adjoint representation $\mf{e}_{11}^*$. 

Besides the fields in the coset  $E_{11} / K(E_{11})$, $E_{11}$ exceptional field theory also includes additional constrained fields. A constrained (generalised) one-form $w_M$ is a field with an index $M$ in $\overline{R(\Lambda_1)}$ that satisfies the section constraint with any derivative of any field $\Phi$ or any other constrained one-form \cite{Hohm:2013uia,Hohm:2014fxa}, i.e.
\begin{align}
\label{eq:SCconstrained}
\kappa_{\alpha\beta} T^{\alpha P}{}_M T^{\beta Q}{}_N \, w_P \, \partial_Q  \Phi = -\frac12 w_M\,  \partial_N \Phi +w_N \,  \partial_M \Phi\; . 
\end{align}
Examples of constrained (generalised) one-forms are $J_{M\alpha}$ and
$\partial_M \Phi$ for any function $\Phi$ since the form index is carried by an explicit derivative. 
Together with the current $J_M{}^\alpha$, the constrained field $\chi_M{}^\ta$ parametrises a constrained one-form in the $\cT_0$ co-adjoint representation $\cT_{-2} = \cT_0^*$
\be 
dz^M ( J_M{}^\alpha \bar t_\alpha + \chi_M{}^\ta \bar t_{\ta}  ) \in \mf{e}_{11} \oright \overline{L(\Lambda_2)} \subset \cT_{-2} \; .  
\label{JTHA} 
\ee
The constrained field $\chi_M{}^\ta$ does not belong to a representation of $\mf{e}_{11}$ by itself, but the components 
\be \label{Jhat}
J_M{}^\wa\equiv (J_M{}^\alpha, \chi_M{}^\ta)
\ee
transform together in the representation $\overline{R(\Lambda_1)} \otimes \adjhat$.\footnote{Note that the field components are in the dual representation of the fields themselves.} It will be convenient to think of $\chi_M{}^\ta$ as transforming in the representation $\overline{R(\Lambda_1)} \otimes L(\Lambda_2)$ with the non-covariant transformation
\be 
\Delta^\alpha \chi_M{}^{\tilde{\beta}} = K^{\alpha\tilde{\beta}}{}_\gamma J_M{}^\gamma \; .  
\label{eq:Dchi} 
\ee

As will be justified in what follows, we also need to introduce two more constrained fields. The first one, $\zeta_M{}^\Lambda$ in the representation $\overline{R(\Lambda_1)} \otimes L(\Lambda_{10})$, was already assumed as part of $\mathcal{T}_0$ in~\cite{Bossard:2019ksx}, while $\zeta_M{}^\tL$ in $\overline{R(\Lambda_1)} \otimes L(\Lambda_{4})$ was overlooked. We shall find that they both are necessary for the duality equation to be gauge invariant.\footnote{The representation $L(\Lambda_4)$ is beyond the level truncation that has been considered in the literature, but the all-level considerations in Appendix~\ref{app:MID} show that it is needed.}
So we have 
\begin{align}
\label{eq:CF}
\chi_M{}^\ta &\quad \text{constrained fields in  $L(\Lambda_2)$ such that } (dz^M J_M{}^{\alpha},dz^M \chi_M{}^{\tilde{\alpha}} ) \in \adjhat \  ,
\nn\\
\zeta_M{}^\Lambda &\quad \text{constrained fields in $L(\Lambda_{10}) = R(\Lambda_1) \SYM R(\Lambda_1) \ominus R(2\Lambda_1)$}\ ,
\\
\zeta_M{}^\tL & \quad \text{constrained fields in $L(\Lambda_4) = R(\Lambda_1) \wedge R(\Lambda_1) \ominus R(\Lambda_2) $}\ ,\nn
\end{align}
where by `constrained field in $L(\lambda)$' we mean fields in $\overline{R(\Lambda_1)} \otimes L(\lambda)$ with a constrained index in $\overline{R(\Lambda_1)}$.
We will use the combined notation $\zeta_M{}^\hL$ for the constrained field 
\be 
\zeta_M{}^\hL= (\zeta_M{}^\Lambda , \zeta_M{}^{\tL}) \; , 
\ee
which is valued in the section constraint representation $L(\Lambda_{10}) \oplus L(\Lambda_4)$, and can schematically be thought of as $\zeta_{M}{}^{P;Q}$ with $\zeta_M{}^{P;Q} \partial_P \otimes \partial_Q = 0 $. The semi-colon here is used to denote a tensor product $R(\Lambda_1)\otimes R(\Lambda_1)$.

\subsubsection{Field strengths}

As explained in~\cite{Bossard:2017wxl}, the tensor hierarchy algebra implies the existence of an $E_{11}$-invariant tensor defining a map from $\cT_1 \otimes \cT_{-2} \rightarrow \cT_{-1}$ which enters in the definition of a field strength in $\cT_{-1}$. In general there is a nilpotent differential operator acting on fields in $ \Phi(z) \in \cT_p$ to give fields in $\cT_{p+1}$ as
\be 
\label{dadP} 
d \Phi(z) = ({\rm ad} P^M) (\partial_M \Phi(z)  )
\ee
where $ {\rm ad}(P^M)$ is the graded commutator with $P^M$ and nilpotency $d^2=0$ follows from the tensor hierarchy algebra~\cite{Bossard:2017wxl}.
This differential derived from the tensor hierarchy algebra provides a projection of the current components $J_M{}^\wa$~\eqref{Jhat} to the field strength representation $\cT_{-1}$
\be 
[ P^M ,   J_M{}^\wa  \bar t_\wa  ]
=C^{I M}{}_\wa\, J_M{}^\wa t_I\; . 
\label{oldF}
\ee
As we shall see in the next section when discussing the duality equation,  the appropriate field strength requires additional terms and is defined as
\bea
F^I &=& C^{I M}{}_\wa\, J_M{}^\wa +  \widetilde{C}^{IM}{}_{\hL}\, \zeta_M{}^\hL
\nn\\
&=&  C^{I M}{}_\alpha\, J_M{}^\alpha + C^{IM}{}_{\ta}\, \chi_M{}^\ta +  \widetilde{C}^{IM}{}_\Lambda\, \zeta_M{}^\Lambda + \widetilde{C}^{IM}{}_\tL\,  \zeta_M{}^\tL\ .
\label{eq:FStemp}
\eea
This includes an additional dependence on the constrained fields $(\zeta_M{}^\Lambda, \zeta_M{}^\tL)$. Compared to~\cite{Bossard:2019ksx}, the definition of the field strength differ in two regards. First, as mentioned above, the constrained field $\zeta_M{}^\tL$ in $L(\Lambda_4)$ were absent in~\cite{Bossard:2019ksx}. Secondly, the constrained field $\zeta_M{}^\Lambda$ in $L(\Lambda_{10})$ here appears with the tensor $\widetilde{C}^{IM}{}_\Lambda$ that is defined from $\cT$ as the conjugate of the commutator $[\cT_{2},\cT_{-1}]$ (see~\eqref{NewCtilde}) while in~\cite{Bossard:2019ksx} we used the commutator $[\cT_1,\cT_{-2}]$ similar to~\eqref{oldF}. We have evidence that these two definitions might be equivalent, but this will not be essential for the construction of the theory with the definition~\eqref{eq:FStemp}.  

In the following we will drop the tildes on $\widetilde{C}^{IM}{}_\Lambda$ and $\widetilde{C}^{IM}{}_\tL$ for ease of notation, thus writing
\begin{align}
\label{eq:FS}
F^I = C^{I M}{}_\wa\, J_M{}^\wa +  C^{IM}{}_{\hL}\, \zeta_M{}^\hL\,.
\end{align}
The tensor $C^{IM}{}_\hL$ is not defined from the tensor hierarchy algebra $\mathcal{T}$,   but will be defined in Section~\ref{sec:DE} to ensure gauge-invariance of the duality equation and is discussed in detail in Appendix~\ref{app:MID}.

Turning to the structure constants above that do arise in the  $\cT(\mf{e}_{11})$ algebra,
since the index $\wa$ of the indecomposable representation is downstairs, we are dealing with the dual representation $\mf{e}_{11}^* \oright \overline{R(\Lambda_2)}$ and thus the component $C^{IM}{}_\ta$ is an $E_{11}$-tensor while $C^{IM}{}_\alpha$ is not. Its non-covariant variation is given by
\begin{align}
\label{eq:DeltaC}
\Delta^\alpha C^{IM}{}_\beta &= - K^{\alpha \ta}{}_\beta C^{IM}{}_\ta\,.
\end{align}
The field strength~\eqref{eq:FS} transforms covariantly under rigid $E_{11}$ in the $\cT_{-1}$ representation, using \eqref{eq:Dchi}.

\medskip

The tensors defined above satisfy a number of important identities when their generalised space-time indices are on section. In particular, $\cT = \oplus_p \cT_p$ defines a graded complex of functions satisfying the section with a nilpotent exterior differential defined by \eqref{dadP}. The first identity comes from $d^2 \xi = 0 $ when acting on a gauge parameter $\xi\in \overline{R(\Lambda_1)}\subset \cT_{-3} = \cT_1^{\; *}$ \begin{align}
\label{eq:THA1}
C^{IM}{}_{\widehat\alpha} T^{\widehat\alpha N}{}_P \,\partial_M \partial_N  =0 \; . 
\end{align}
We have written dummy partial derivatives to emphasise the fact that the identity only holds when the indices $M$ and $N$ are projected onto the irreducible representation $R(2\Lambda_1)$ associated to the section constraint~\eqref{eq:SC} on a \textit{symmetric} tensor. It is important that here there is no contribution from $\overline{D}_0$ since $\{ P^M, \overline{P}_N\} = T^{\wa M}{}_N \bar{t}_{\wa}  \subset \adjhat^*\subset \cT_{-2}$, as can be checked by $E_{11}$ representation theory since the bounding weight of $D_0$ is $\Lambda_{10}$.\footnote{These standard methods rely on analysing the weight diagrams of highest weight modules and this is the only methods we are aware of for general indefinite Kac--Moody algebras. For the case of affine algebras, one could also use the Virasoro algebra that can be constructed in the universal enveloping algebra and make its coset form act on tensor products~\cite{Feingold:1981,Feingold:1983,Feingold:1985,Kac:1987}. A similar structure for $E_{11}$ or other indefinite Kac--Moody algebras is not known. 
} 

The same derivation from $d^2 \Phi =0 $ for a field $\Phi \in \cT_{-2}$ gives
\begin{align}
\label{eq:sympart}
\Omega_{IJ} C^{IM}{}_{\wa}  C^{JN}{}_{\wb}  \,\partial_M \partial_N=0 \; . 
\end{align}

It is proved in Appendix \ref{app:Moth} that this generalises to the following  identities when no symmetry is assumed for the two derivatives:
\begin{subequations}
\begin{align}
\label{eq:ID4}
 \Omega_{IJ} C^{IM}{}_{\alpha} C^{JN}{}_{\beta} \,\partial_M\otimes \partial_N &=-2 \Pi_\ta{}^{MN} K_{(\alpha}{}^\ta{}_{\beta)}\,\partial_M\otimes \partial_N \, ,\\
 \label{eq:ID2}
\Omega_{IJ} C^{IM}{}_{\tilde{\alpha}} C^{JN}{}_{\beta}  \partial_M \otimes \partial_N  & = -   \Pi_\tb{}^{MN} T_\beta{}^{\tb}{}_\ta  \partial_M \otimes \partial_N \; , \\
\label{eq:ID3}
\Omega_{IJ} C^{IM}{}_\ta C^{JN}{}_\tb \,\partial_M\otimes \partial_N&=0 \; ,
\end{align}
\end{subequations}
where $\Pi_\ta{}^{MN}$ is the intertwiner for the injection $R(\Lambda_2) \subset R(\Lambda_1) \ALT R(\Lambda_1)$. 
The left-hand side of~\eqref{eq:ID4} does not need to be symmetrised in $\alpha$ and $\beta$ since it is automatically antisymmetric in $M$ and $N$.

One checks by standard methods for $E_{11}$ tensor products~\cite{Kac} that\footnote{Using the notation $A^{MN} \partial_M \otimes \partial_N = 0 $ means that the projection of $MN$ in $A^{MN}$ to the representations $R(2\Lambda_1)$ and $R(\Lambda_2)$ vanishes, see~\eqref{eq:L1tens}. In particular, if the solution to the section constraint is realised by a linear subspace and the indices $M$ and $N$ belong to this linear subspace, the corresponding components of $A^{MN}$ vanish.}
\begin{subequations}
\begin{align}
\label{eq:ID7}
\Omega_{IJ} C^{IM}{}_\ta C^{JN}{}_\hL \,\partial_M\otimes \partial_N&=0 \,,\\
\label{eq:ID8}
\Omega_{IJ} C^{IM}{}_\hXi C^{JN}{}_\hL \,\partial_M\otimes \partial_N &=0 \,.
\end{align}
\end{subequations}

\subsection{Generalised diffeomorphisms}
\label{sec:gendiff}

Besides the rigid $E_{11}$ transformations of the fields we shall also require local gauge transformations provided by the generalised diffeomorphisms. These were defined in~\cite{Bossard:2017wxl,Bossard:2019ksx} following the general pattern of any generalised Lie derivative for arbitrary groups and representations~\cite{Bossard:2017aae} on a vector $V^M$ in $R(\Lambda_1)$ by
\begin{align}
\mathcal{L}_\xi V^M = \xi^N \partial_N V^M - \kappa_{\alpha\beta} T^{\alpha P}{}_Q \partial_P \xi^Q T^{\beta M}{}_N V^N +w  \, \partial_N \xi^N V^M
\end{align}
The subscript $\xi^M$ represents the gauge parameter in the $R(\Lambda_1)$ representation of $E_{11}$. The vector representation $R(\Lambda_1)$ of $E_{11}$ implies that the gauge parameter $\xi^M$ has weight $w=-\tfrac12$~\cite{Bossard:2017aae}. 

Closure of the algebra of generalised diffeomorphisms only holds when the section constraint~\eqref{eq:SC} is fulfilled~\cite{Berman:2012vc} and only modulo ancillary transformations that arise for the first time for $E_8$~\cite{Hohm:2014fxa,Cederwall:2015ica,Cederwall:2017fjm}. Computing the closure of the algebra here leads to the relation
\begin{align}
\label{eq:GAfail}
\lb \mathcal{L}_{\xi_1} , \mathcal{L}_{\xi_2} \rb V^M -\mathcal{L}_{[\xi_1,\xi_2]} V^M =  \frac12 \big(f_{\alpha\beta\gamma} T^{\gamma P}{}_R T^{\beta Q}{}_S - 2 \delta^{(P}_{[R} T_{\alpha}{}^{Q)}_{\ }{}_{S]}^{\ } \big) \xi_1^R \partial_P \partial_Q \xi_2^S T^{\alpha M}{}_N V^N - (\xi_1{\leftrightarrow}\xi_2) \,,
\end{align}
where the section condition was already used and $[\xi_1,\xi_2]=\frac12 \mathcal{L}_{\xi_1} \xi_2 - \frac12 \mathcal{L}_{\xi_2} \xi_1$. In order to absorb this failure of closure one requires ancillary transformations.  Extrapolating the structure of the ancillary transformations from $E_8$ and $E_9$~\cite{Hohm:2014fxa,Bossard:2017aae,Bossard:2018utw} one expects that there is a first gauge parameter in the representation $\overline{R(\Lambda_1)}\otimes R(\Lambda_3)$, i.e., with the index structure $\Sigma_M{}^{\hspace{-0.3mm}N_1\hspace{-0.5mm} N_2\hspace{-0.3mm}  N_3}$, where the upper $R(\Lambda_1)$ indices are completely antisymmetric and the lower index is section constrained. One needs in fact a more general ancillary gauge parameter $\Sigma_M{}^\tI$ with an index $\tI$ that labels the completely reducible bounded weight  representation
\begin{align}
\label{eq:tildeI}
L(\Lambda_3) = R(\Lambda_1)\otimes L(\Lambda_2)  \ominus R(\Lambda_1 + \Lambda_2) = R(\Lambda_3) \oplus R(\Lambda_1 + \Lambda_{10}) \oplus \ldots
\end{align}
We also impose the tracelessness condition
\begin{align}
\label{eq:sigmatr}
\Pi_{\tI}{}^{M \ta} \Sigma_M{}^\tI = 0\,,
\end{align}
where $\Pi_{\tI}{}^{M \ta}$ is an $E_{11}$ intertwiner tensor that projects out the $R(\Lambda_1 + \Lambda_2)$ from the tensor product 
\begin{align}
R(\Lambda_1)\otimes L(\Lambda_2) =  R(\Lambda_1 + \Lambda_2) \oplus R(\Lambda_3) \oplus R(\Lambda_1+\Lambda_{10}) \oplus R(\Lambda_1+\Lambda_4) \oplus R(\Lambda_1+2\Lambda_3) \oplus \ldots
\end{align}
to the representation $L(\Lambda_3) $ in~\eqref{eq:tildeI}. There must be an extension of the theory for which the scalar fields $\cV(z)$ are defined in a bigger coset associated to (the conjugate $\mf{e}_{11} \oleft \overline{L(\Lambda_2)}$ under the Cartan involution  of)  $\adjhat \subset \cT_0$ and the constraint \eqref{eq:sigmatr} is relaxed since it can be contracted with a generator $\bar T_{\ta}{}^M{}_N$ in $\overline{L(\Lambda_2)}$. This is important in the construction of the supersymmetric theory including the fermions \cite{Bossard:2019ksx}, but we shall not attempt to define it in this paper.\footnote{For $E_9$ this is equivalent to the formulation extending $E_9$ by the generator $L_{-1}$ and the associated field $\tilde{\rho}$ in (bosonic) exceptional field theory \cite{Bossard:2018utw,Bossard:2021jix}. At the level of supergravity, the field $\tilde{\rho}$  transforms non-trivially under supersymmetry \cite{Nicolai:1988jb,Nicolai:2004nv} and  $\tilde{\rho}$ is necessary for $\widetilde{K(E_{9})}$ to commute with the supersymmetry transformations.}

Here, we take $\Pi_\tI{}^{M\ta}$ to be the direct sum of the canonically normalised projectors to the individual representations in~\eqref{eq:tildeI}. We define its conjugate tensor by 
\be
C^{\tI}{}_{M\ta}= \eta^{\tI\tJ} \eta_{MN} \eta_{\ta\tb} \Pi_\tJ{}^{N\tb}\ , 
\ee
where, due to the reducibility of the $\tI$ (and potentially $\ta$) indices there is some freedom in the normalisation of the $K(E_{11})$-invariant tensors $\eta^{\tI\tJ}$ (and $\eta_{\ta\tb}$). We will fix this freedom below.  We prove in Appendix  \ref{app:ids} that the tensor $C^{\tI}{}_{M\ta}$ is part of a larger structure constant $C^\tI{}_{M\wa} = (C^\tI{}_{M\alpha},C^{\tI}{}_{M\ta})$ in $L(\Lambda_3) \otimes \overline{R(\Lambda_1)} \otimes \adjhat^*$, whose component $C^\tI{}_{M\alpha}$ transforms non-covariantly as
\begin{align}
\label{eq:DeltaC2}
\Delta^\alpha C^\tI{}_{M\beta} &= - K^{\alpha \ta}{}_\beta C^\tI{}_{M \ta }\, .
\end{align}
We use this tensor to define an extended generalised Lie derivative
\begin{align}
\label{eq:extGL}
\mathcal{L}_{(\xi,\Sigma)} V^M = \mathcal{L}_\xi V^M  - \eta_{\tI\tJ}\eta^{PQ} \eta^{\alpha\beta} C^{\tJ}{}_{Q \alpha} \Sigma_P{}^\tI T_\beta{}^{M}{}_N V^N\,.
\end{align}
Even though only the non-tensorial part $C^\tJ{}_{Q\alpha}$ appears in the extension of the generalised Lie derivative, this expression is $E_{11}$-covariant due to~\eqref{eq:sigmatr}. One computes that the algebra of generalised diffeomorphisms \eqref{eq:extGL}
\begin{align}
\label{eq:GA2}
\lb \mathcal{L}_{(\xi_1,\Sigma_1)} , \mathcal{L}_{(\xi_2,\Sigma_2)} \rb V^M =\mathcal{L}_{([\xi_1,\xi_2],\Sigma_{12})} V^M \,,
\end{align}
indeed closes with the parameter
\begin{align}
\label{eq:sigma12}
\Sigma_{12\, M}{}^\tI =  \mathcal{L}_{\xi_1} \Sigma_{2\, M}{}^\tI + \frac12 C^{\tI}{}_{N\wa} T^{\wa P}{}_Q \xi_1^N \partial_M\partial_P \xi_2^Q + \frac12 T_\beta{}^{\tI}{}_\tJ \Sigma_{1 M}{}^{\tJ}  \eta_{\tK\tilde{L}}\eta^{PQ} \eta^{\beta\gamma} C^{\tK}{}_{P \gamma} \Sigma_{2 Q}{}^{\tilde{L}} - (1{\leftrightarrow}2)  \,.
\end{align}
The Lie derivative of  the weightless constrained parameter $\Sigma_M{}^\tI$ is
\begin{align}
\mathcal{L}_\xi \Sigma_M{}^\tI &= \xi^N \partial_N \Sigma_M{}^\tI - T^{\alpha P}{}_Q T_\alpha{}^\tI{}_\tJ \partial_P\xi^Q \Sigma_M{}^\tJ + T^{\alpha P}{}_Q T_\alpha{}^N{}_M \partial_P\xi^Q \Sigma_N{}^\tI\nn\\
&= \xi^N \partial_N \Sigma_M{}^\tI - T^{\alpha P}{}_Q T_\alpha{}^\tI{}_\tJ \partial_P\xi^Q \Sigma_M{}^\tJ  + \partial_M\xi^Q \Sigma_Q{}^\tI - \frac12 \partial_N\xi^N \Sigma_M{}^\tI\ ,
\end{align}
where we have written the second line by using the section constraint~\eqref{eq:SC} to make it manifest that the result is again section constrained on the index $M$. Moreover, one checks by $E_{11}$-invariance of $\Pi_\tI{}^{M\ta}$ that $\Pi_\tI{}^{M\ta} \mathcal{L}_\xi \Sigma_M{}^\tI=0$, i.e., the Lie derivative preserves the tracelessness condition~\eqref{eq:sigmatr}.
To obtain the last term in \eqref{eq:sigma12} and show that it satisfies \eqref{eq:sigmatr}, we use that $\Sigma_{1\, M}{}^{\tI}$ and $\Sigma_{2\, M}{}^{\tI}$ do and the identity 
\begin{align}
\label{eq:ID9}
C^\tI{}_{Q\wa} T^{\wa P}{}_R  \eta^{QM} \eta^{RN} \partial_M \otimes \partial_N =  0 \; , \end{align}
which follows from the property that neither $R(2\Lambda_1)$ not $R(\Lambda_2)$ are contained in $R(\Lambda_1) \otimes L(\Lambda_3)$. The structure on the upper indices is that of the tensor product $L(\Lambda_3)\otimes R(\Lambda_1)\subset R(\Lambda_1)\otimes R(\Lambda_2)\otimes R(\Lambda_1)$ and we need to check that this does not contain the representations $R(2\Lambda_1)$ and $R(\Lambda_2)$ that are non-trivial when the section constraint is fulfilled, see~\eqref{eq:L1tens}. 

To show that the failure~\eqref{eq:GAfail} of the generalised Lie derivative to form a closed gauge algebra is indeed reabsorbed by the ancillary transformation of parameter $ \frac12 C^{\tI}{}_{N\wa} T^{\wa P}{}_Q \xi_1^N \partial_M\partial_P \xi_2^Q$ in \eqref{eq:sigma12}, we use the identity  
\begin{align}
\label{eq:ID5}
C^{\tI}{}_{P \wb} T^{\wb (M}{}_Q  \eta^{N)R} \eta_{\tI\tJ} \eta^{\alpha\gamma}  C^{\tJ}{}_{R \gamma}\,  \partial_M \otimes  \partial_N = \Bigl( f^{\alpha\beta}{}_\gamma T^{\gamma (M}{}_P T_\beta{}^{N)}{}_Q -2 \delta^{(M}_{[P} T^{\alpha N)}{}_{Q]} \Bigr) \partial_M\otimes   \partial_N\,,
\end{align}
which we demonstrate in Appendix~\ref{app:ids}. To moreover show that this second term in $\Sigma_{12}$ satisfies the traceless condition~\eqref{eq:sigmatr}, we use the additional identity
\begin{align}
\label{eq:ID6}
\eta^{\ta\tilde{\gamma}} \eta_{IJ} \eta_{NP} C^{I P}{}_{\tilde{\gamma}}   C^{JM}{}_{\wb} = \Pi_{\tI}{}^{M\ta} C^{\tI}{}_{N\wb}\; ,
\end{align}
where we recall that $\wb=(\beta,\tb)$, which implies,  together with \eqref{eq:THA1}, that $\Sigma_{12}$ indeed satisfies the tracelessness condition. Identity~\eqref{eq:ID6} links the representation $L(\Lambda_3)$ with index $\tI$ to the representation $\cT_{-1}$ with index $I$. 
It is proved in Appendix~\ref{Sweds} and fixes the freedom in the definition of $\eta_{\tI\tJ}$.

As we shall see the tensor $C^{\tI}{}_{P \wb}$ also plays a r\^ole in the construction of the pseudo-Lagrangian of $E_{11}$ exceptional field theory as it enters what was called the second potential term in~\eqref{eq:Lagin}. This is not surprising since also for $E_8$ the gauge algebra only closes when an additional ancillary gauge transformation $\Sigma_M$ is introduced and also the potential acquires a new term compared to $E_n$ with $n\leq 7$~\cite{Hohm:2014fxa}. The same identity also appears in the definition of the potential for $E_9$ \cite{Bossard:2018utw}. More generally is was shown in \cite{Cederwall:2019bai} that the identity \eqref{eq:ID5} appears in the closure of the algebra and the definition of the potential for all  simply laced  finite-dimensional Lie groups $G$ and any highest weight coordinate module $R(\lambda)$. 

\medskip

Having discussed the general Lie derivative and its closure we shall now describe how the fields of $E_{11}$ exceptional field theory transform under generalised diffeomorphisms. In this we shall focus on the parameter $\xi^M$ only.\footnote{The gauge transformations of parameter $\Sigma_M{}^{\tI}$ for the generalised metric $\cM$ and the current $J_M{}^\alpha$ follow from the definition, but the gauge transformations of the constrained fields  $(\chi_M{}^\wa, \zeta_M{}^{\widehat{\Lambda}})$  are not straightforward to get. It is expected that the closure of the algebra of generalised diffeomorphisms on the constrained fields requires introducing additional gauge transformations with two constrained indices and under which all the covariant objects $\cM$, $\cE^I$ (where $\cE^I=0$ is the duality equation for $F^I$) are strictly invariant.}
We write the generalised Lie derivative of any object $O$ as
\begin{align} \label{GeneLie}
\mathcal{L}_\xi O = \xi^M \partial_M O - T_{\alpha}{}^N{}_P \partial_N \xi^P t^\alpha O + w \partial_M \xi^M O\,,
\end{align}
where $t^\alpha O$ is the action of the $E_{11}$ generator $t^\alpha$ on $O$ and $w$ represents the weight of the object.

The generalised diffeomorphism will be written as $\delta_\xi O$ and need not coincide with the generalised Lie derivative as there can be non-covariant terms when additional derivatives or constrained indices are involved. We shall write this non-covariant gauge transformation as
\begin{align}
\label{eq:defDxi}
\Delta_\xi = \delta_\xi - \mathcal{L}_\xi\,.
\end{align}
The non-covariant gauge transformation defined above should not be confused with the `non-covariant' rigid $\mf{e}_{11}$ transformation $\Delta^\alpha$ in \eqref{eq:Dchi} associated to the indecomposable structure. In particular $\mathcal{L}_\xi \chi_M{}^{\tilde{\alpha}}$ takes into account the indecomposable representation and includes the term $-T_{\alpha}{}^N{}_P \partial_N \xi^P \Delta^\alpha$ according to \eqref{GeneLie}.

For instance, the generalised metric transforms under generalised diffeomorphisms as always in exceptional field theory
\begin{align}
\label{eq:GTM}
\delta_\xi \cM = \mathcal{L}_\xi \cM = \xi^M  \partial_M \cM + \kappa_{\alpha\beta} T^{\alpha M}{}_N \partial_M\xi^N (\cM t^\beta +t^{\beta \dagger} \cM)
\end{align}
and this coincides with its Lie derivative as it is a fully covariant object. Therefore, $\Delta_\xi \cM =0$ in any $E_{11}$ representation admitting a $K(E_{11})$-invariant bilinear form.

By contrast, the gauge transformation of the current $J_{M}{}^\alpha$ follows from~\eqref{eq:GTM} and using its definition~\eqref{eq:curdef} as~\cite[Eq.~(3.18)]{Bossard:2019ksx}
\begin{align}
\label{eq:dxiJ}
\delta_\xi J_M{}^\alpha &= \xi^N \partial_N J_M{}^\alpha + T_\beta{}^N{}_P \partial_N \xi^P  f^{\beta\alpha}{}_\gamma J_M{}^\gamma +T_\beta{}^P{}_Q \partial_P \xi^Q T^\beta{}^N{}_M J_N{}^\alpha +\frac12 \partial_N \xi^N J_M{}^\alpha  \nn
\\
&\hspace{10mm} +T^{\alpha N}{}_P \left( \partial_M\partial_N \xi^P + \cM_{NQ} \cM^{PR} \partial_M\partial_R\xi^Q\right)\ .
\end{align}
The first line coincides with the Lie derivative of a weight $w=1/2$ object (since the derivative index is downstairs) but the second line is a non-covariant gauge transformation
\begin{align}
\label{eq:NCJ}
\Delta_\xi J_{M}{}^\alpha =  T^{\alpha N}{}_P \left( \partial_M\partial_N \xi^P + \cM_{NQ} \cM^{PR} \partial_M\partial_R\xi^Q\right) \,,
\end{align}
proportional to second derivatives of the parameter $\xi$ that appear in all exceptional field theories.\footnote{This is similar to the fact that Christoffel symbols transform non-covariantly under diffeomorphism whereas the metric is covariant.}

\medskip

The gauge transformation of the constrained fields was established in~\cite[Eq.~(3.20)]{Bossard:2019ksx}. Since the constrained fields are in some sense completions of the currents according to~\eqref{eq:FS}, their transformation is very similar. However, they have an additional non-covariant piece due to the existence of additional invariant tensors. Their transformation under generalised diffeomorphisms is\footnote{The invariant tensors $T^{\ta N}{}_P$ and $T^{\Lambda N}{}_P$ appearing in the equations below correspond to the action of $\cT_0$ in the tensor hierarchy algebra on its level one representation $\cT_{+1}$~\cite{Bossard:2019ksx}.}
\begin{subequations}
\begin{align}
\label{eq:chiGT}
\delta_\xi \chi_M{}^\ta &=  \xi^N \partial_N \chi_M{}^{\tilde\alpha} -T_\alpha{}^N{}_P \partial_N \xi^P \Big(T^{\alpha\tilde\alpha}{}_{\tilde\beta} \chi_M{}^\tb  + K^{\alpha\tilde\alpha}{}_{\beta} J_M{}^\beta - T^{\alpha Q}{}_M \chi_{Q}{}^\ta \Bigr)+ \frac12 \partial_N \xi^N \chi_M{}^\ta
\nn\\
&\quad\quad + T^{\tilde\alpha N}{}_P \left( \partial_M\partial_N \xi^P + \cM_{NQ} \cM^{PR} \partial_M\partial_R\xi^Q\right)+ {\Pi^{\tilde\alpha}{}_{ Q P }} \cM^{NQ} \partial_M \partial_N \xi^P \,,\\
\delta_\xi \zeta_M{}^\hL &=   \xi^N \partial_N \zeta_M{}^\hL -T_\alpha{}^N{}_P \partial_N \xi^P \Big(T^{\alpha\hL}{}_{\widehat\Sigma} \zeta_M{}^{\widehat\Sigma}  - T^{\alpha Q}{}_M \zeta_{Q}{}^\hL \Big)+ \frac12 \partial_N \xi^N \zeta_M{}^\hL
\nn\\
&\quad\quad 
+ {\Pi^{\hL}{}_{ Q P }} \cM^{NQ} \partial_M \partial_N \xi^P \,,
\end{align}
\end{subequations}
where compared to~\cite{Bossard:2019ksx} we have used the section constraint~\eqref{eq:SC}.
Moreover, we have additionally used the fact that there is no tensor $T^{\hL N}{}_P$ as can be checked by $E_{11}$ representation theory. This simplifies the gauge transformation of $\zeta_M{}^\hL$ compared to that of $\chi_M{}^\ta$.
It is important to note that the algebra of diffeomorphisms generated by $(\xi^M, \Sigma_M{}^{\tilde{I}})$ does not close on the constrained fields $\chi^\ta$ and $\zeta^\hL$, but involves instead higher gauge transformations starting with a parameter involving two constrained indices. In particular, in the commutation rules one finds terms involving  $\partial_P \xi^R \partial_M \partial_N \xi^Q + \partial_N \xi^Q \partial_M \partial_P \xi^R$ which cannot be interpreted as a $\Sigma_M{}^{\tilde{I}}$ variation. The precise form of these higher gauge transformations remains to be determined. Nonetheless, they are not needed to demonstrate the gauge invariance of the dynamics under generalised diffeomorphisms with parameter $\xi^M$ only.

Turning to the $\xi$-transformations in \eq{eq:chiGT}, the first lines of these transformations are again the Lie derivatives so that we can read off the non-covariant gauge transformation as
\begin{subequations}
\label{eq:DxiCS}
\begin{align}
\label{eq:Dxichi}
\Delta_\xi \chi_M{}^\ta &= T^{\tilde\alpha N}{}_P \left( \partial_M\partial_N \xi^P + \cM_{NQ} \cM^{PR} \partial_M\partial_R\xi^Q\right)+ {\Pi^{\tilde\alpha}{}_{ Q P }} \cM^{NQ} \partial_M \partial_N \xi^P \,,\\
\label{eq:Dxizeta}
\Delta_\xi \zeta_M{}^\hL &=
 \Pi^\hL{}_{ Q P } \cM^{NQ} \partial_M \partial_N \xi^P \,.
\end{align}
\end{subequations}
The extra $E_{11}$-invariant tensors that occur in these transformations are $\Pi^\ta{}_{MN}$ and $\Pi^\hL{}_{MN}=(\Pi^\Lambda{}_{MN}, \Pi^\tL{}_{MN})$. They correspond to the projections of the antisymmetric and symmetric product of two $R(\Lambda_1)$ representations, see~\eqref{eq:L1tens}, onto the corresponding representations $R(\Lambda_2) \subset L(\Lambda_2)$, $L(\Lambda_4)$ and $L(\Lambda_{10})$, respectively. Thus they have the symmetry properties
\begin{align}
\label{eq:Pitensors}
\Pi^\ta{}_{MN} = \Pi^\ta{}_{[MN]}\,,\quad\quad
\Pi^\tL{}_{MN} = \Pi^\tL{}_{[MN]}\,,\quad\quad
\Pi^\Lambda{}_{MN} = \Pi^\Lambda{}_{(MN)}\,.
\end{align}

Using the generalised metric in the various representations, we can also define dual tensors to the projector $\Pi^\ta{}_{MN} $ by
\begin{align}
\label{eq:dualPi}
\Pi_\ta{}^{MN} = \cM_{\ta\tb} \cM^{MP} \cM^{NQ} \Pi^\tb{}_{PQ} = \eta_{\ta\tb} \eta^{MP} \eta^{NQ} \Pi^\tb{}_{PQ}
\end{align}
where $\eta^{MN}$ and $\eta^{\ta\tb}$ are uniquely defined for the irreducible representation $R(\Lambda_1)$ and $R(\Lambda_2)$, respectively. We also note that we have the following identities involving the $E_{11}$-invariant projectors  
\begin{align}
\label{eq:ID1}
\Pi_\tb{}^{MN}  T^{\alpha\tb}{}_\ta = -2 \Pi_\ta{}^{P[M}T^{\alpha N]}{}_P \,,\quad
\Pi^\ta{}_{MN}  T^{\alpha\tb}{}_\ta = -2 \Pi^\tb{}_{P[M}T^{\alpha P}{}_{N]} \,.
\end{align}

\begin{table}[th]
\centering
\begin{tabular}{c|c|c}
Index & $E_{11}$ representation & Object \\
\hline\hline &&
\\[-2mm]
$\wa = \bigg\{\begin{array}{l} \alpha \\ \ta \end{array}$ & $\adjhat\,
  \bigg\{\begin{array}{l} \mf{e}_{11} \text{(adjoint)} \\ L(\Lambda_2)\end{array}
  $    &    $ J_M{}^\wa = \bigg\{ 
  \begin{array}{cl} J_M{}^\alpha & \text{(current~\eqref{eq:curdef})}\\ \chi_M{}^\ta & \text{(constr. field~\eqref{eq:CF})} \end{array}$
  \\[5mm]\hline&&
  \\[-3mm]
 $ \hL = \bigg\{\begin{array}{l} \Lambda\\ \tL\end{array}$ 
  &  $ \bigg\{ \begin{array}{l} L(\Lambda_{10})= R(\Lambda_1) \SYM R(\Lambda_1) \ominus R(2\Lambda_1) 
  \\  L(\Lambda_4)=R(\Lambda_1) \wedge R(\Lambda_1) \ominus R(\Lambda_2) \end{array} \,$ 
 &  $ \zeta_M{}^\hL = \bigg\{ \begin{array}{cl} \zeta_M{}^\Lambda & \text{(constr. field~\eqref{eq:CF})}\\ \zeta_M{}^\tL & \text{(constr. field~\eqref{eq:CF})} \end{array}$
 \\[5mm]\hline&&
  \\[-3mm]
 $M$ & $R(\Lambda_1)$ & $\partial_M \quad\text{(space-time derivative \eqref{eq:SC})}$\\[2mm]\hline&&
 \\[-3mm]
 $I$ & $\cT_{-1}$ \eqref{eq:Tminus1} & $F^I \quad \text{(field strength \eqref{eq:FS})}$\\[2mm]\hline&&
 \\[-3mm]
 $\tI$ & $L(\Lambda_3)=  R(\Lambda_1)\otimes L(\Lambda_2) \ominus R(\Lambda_1{+}\Lambda_2)$
 & $\Sigma_M{}^\tI \quad\text{(anc. parameter~\eqref{eq:sigmatr})}$ 
\end{tabular}
\caption{\label{tab:sum1}{\sl Summary of the $E_{11}$-representations that occur (up to conjugation) in the construction of $E_{11}$ exceptional field theory with some associated objects and where they were defined.}}
\end{table}

\begin{table}[h!]
\centering
\begin{tabular}{c|c|c}
Object & occurs where & $E_{11}$ tensor?  \\\hline\hline&&\\[-2mm]
$f^{\alpha \beta}{}_\gamma$ & $\mf{e}_{11}$ structure constant~\eqref{eq:e11} & yes\\[2mm]\hline&&
\\[-3mm]
$K^{\alpha \ta}{}_\beta$ & cocycle for indecomposable representation~\eqref{eq:T0CR} & no \eqref{eq:Dcoc}\\[2mm]\hline&&
\\[-3mm]
$\left\{\begin{array}{c}T^{\alpha M}{}_N\\ T^{\ta M}{}_N\end{array}\right.$ & action of generators of $\adjhat$ on coord. module $R(\Lambda_1)$ \eqref{eq:THA01} & $\left\{\begin{array}{c}\text{yes} \\ \text{no} \end{array}\right.$\\[2mm]\hline&&
\\[-3mm]
$\left\{\begin{array}{c}C^{IM}{}_\alpha\\ C^{IM}{}_\ta \\ C^{IM}{}_\Lambda\\C^{IM}{}_\tL \end{array}\right.$ & definition of field strength \eqref{eq:FS} & $\left\{\begin{array}{c} \text{no \eqref{eq:DeltaC}}\\ \text{yes} \\ \text{yes}\\\text{yes} \end{array}\right.$\\[2mm]\hline&&
\\[-3mm]
$\Omega_{IJ}$ & symplectic form on $\cT_{-1}$ & yes\\[2mm]\hline&&
\\[-3mm]
$\left\{\begin{array}{c} C^\tI{}_{M\alpha} \\ C^{\tI}{}_{M\ta}\end{array}\right.$ & anc. gauge transformation \eqref{eq:extGL}, \eqref{eq:ID5} & $\left\{\begin{array}{c}\text{no}\\ \text{yes}\end{array}\right.$ \\[4mm]\hline&&
\\[-3mm]
$\Pi^\ta{}_{MN}$ & intertwiner $R(\Lambda_1)\ALT R(\Lambda_1)\to R(\Lambda_2) $ \eqref{eq:L1tens}, \eqref{eq:Pitensors}& yes\\[2mm]\hline&&
\\[-3mm]
$\Pi^\Lambda{}_{MN}$ & intertwiner $R(\Lambda_1)\SYM R(\Lambda_1)\to L(\Lambda_{10})$ \eqref{eq:L1tens}, \eqref{eq:Pitensors}& yes\\[2mm]\hline&&
\\[-3mm]
$\Pi^\tL{}_{MN}$ & intertwiner $R(\Lambda_1)\ALT R(\Lambda_1)\to L(\Lambda_4) $ \eqref{eq:L1tens}, \eqref{eq:Pitensors}& yes
\end{tabular}
\caption{\label{tab:sum2}{\sl Summary of some  $E_{11}$ objects  that occur (up to conjugation) in the construction of $E_{11}$ exceptional field theory and whether they are covariant under rigid $E_{11}$ for the fully reducible representation.}}
\end{table}

\subsection{Summary of notation}
\label{sec:notsum}

Since we have introduced a fair number of $E_{11}$ objects, we briefly summarise the most relevant notation for the convenience of the reader in Tables~\ref{tab:sum1} and~\ref{tab:sum2}. The identities for the various tensors mentioned in this section are 
summarised in Appendix~\ref{app:ids}. Here we call tensors the invariant tensors in the fully reducible modules only, so the component of the tensors involving the indecomposable representation $\adjhat$ or its conjugate are not called $E_{11}$ tensors when they are not invariant tensors for the fully reducible representation $\mf{e}_{11} \oplus L(\Lambda_2)$ or its conjugate. 


\section{Dynamics}
\label{sec:dyn}

Equipped with the preliminaries regarding all fields and their rigid and gauge transformations, we construct the pseudo-Lagrangian of $E_{11}$ exceptional field theory in this section. It will be a pseudo-Lagrangian in the sense that its Euler--Lagrange equations have to be supplemented by duality equations. These duality equations were already constructed in our last work~\cite{Bossard:2019ksx} and we review them first to set the scene. 

\subsection{Duality equation}
\label{sec:DE}

The duality equation was already given in~\eqref{eq:DEintro} in the introduction and we repeat it here for convenience:
\begin{align}
\label{eq:DE}
\cM_{IJ} F^J = \Omega_{IJ} F^J\,,
\end{align}
where the field strength $F^I$ with values in $\cT_{-1}$ was defined in~\eqref{eq:FS}, the symplectic form $\Omega_{IJ}$ at the end of Section~\ref{sec:THA} and $\cM_{IJ} = (\cV^\dagger \eta \cV)_{IJ}$ is obtained by using $\eta_{IJ}$ defined also in Section~\ref{sec:THA} and evaluating the $E_{11}$ vielbein $\cV$ in the representation $\cT_{-1}$. Its inverse is $\cM^{IJ}$. Note that we have $\Omega_{IJ^\prime} \cM^{J^\prime K} \Omega_{KL} \cM^{LJ} =\delta_I^J$.

Following~\cite{Bossard:2019ksx}, where the gauge invariance of \eq{eq:DE}, albeit for the field strength $F^{I}_{\scalebox{0.6}{0}}$ in \eqref{oldF}, was studied, we 
now compute the transformation of the field strength $F^I$ to be
\begin{align}
\delta_\xi F^I &=    \xi^M \partial_M F^I -  T_\alpha{}^N{}_M \partial_N \xi^M T^\alpha{}^I{}_J F^J +\frac12 \partial_M\xi^M F^I
\nn\\
&\quad  + \Bigl(\bigl( C^{IM}{}_ \alpha T^{\alpha R}{}_Q+ C^{IM}{}_{ \tilde\alpha} T^{\tilde\alpha R}{}_Q \bigr)  \cM^{QN}\cM_{RP} \CR
& \hspace{40mm} + C^{IM}{}_{\tilde\alpha} \Pi^{\tilde\alpha}{}_{QP}\cM^{QN}+ C^{IM}{}_\hL \Pi^\hL{}_{QP}\cM^{QN} \Bigr)   \partial_M \partial_N \xi^P \,,
\label{deltaF}
\end{align}
such that the non-covariant gauge transformation is
\begin{align}
\Delta_\xi F^I &= \Bigl(\bigl( C^{IM}{}_ \alpha T^{\alpha R}{}_Q+ C^{IM}{}_{ \tilde\alpha} T^{\tilde\alpha R}{}_Q \bigr)  \cM^{QN}\cM_{RP} \CR
& \hspace{40mm} + C^{IM}{}_{\tilde\alpha} \Pi^{\tilde\alpha}{}_{QP}\cM^{QN}+ C^{IM}{}_\hL \Pi^\hL{}_{QP}\cM^{QN}\Bigr)   \partial_M \partial_N \xi^P \,.
\label{nablaF}
\end{align}
As explained in~\cite{Bossard:2019ksx}, there is no gauge covariant field strength unlike for other $E_n$ ExFTs. Note that the $E_7$ field strengths of the vectors are covariant under internal diffeomorphisms but not under external diffeormorphisms, and only their duality equation is covariant under both~\cite{Hohm:2013uia}. 

The gauge-invariance of the duality equation~\eqref{eq:DE} requires the non-trivial equality of  the non-covariant gauge transformations of its two sides
\begin{align}
\label{eq:DxiFI}
\cM_{IJ} \Delta_\xi F^J = \Omega_{IJ} \Delta_\xi F^J
\end{align}
and this equality is equivalent to the conjectural tensorial identity
\begin{align}
\label{eq:MIDnew}
&&  \cM_{IJ} C^{JM}{}_{\widehat\alpha} T^{\widehat\alpha Q}{}_R \cM_{QP} \cM^{RN}&=\Omega_{IJ}C^{JN}{}_{\ta} \Pi^{\ta}{}_{QP}  \cM^{QM} + \Omega_{IJ} C^{JN}{}_\hL \Pi^\hL{}_{QP} \cM^{QM}\,,
\nn\\
 &\Leftrightarrow&  \Omega_{IJ} C^{JM}{}_{\widehat\alpha} T^{\widehat\alpha Q}{}_R \cM_{QP} \cM^{RN}&=\cM_{IJ}C^{JN}{}_{\ta} \Pi^{\ta}{}_{QP}  \cM^{QM} +\cM_{IJ} C^{JN}{}_\hL \Pi^\hL{}_{QP} \cM^{QM} \,.
\end{align}
In the above equation,  we recall that $C^{JN}{}_\ta$ and $\Pi^\ta{}_{QP} $ are $E_{11}$ invariant tensors and that $T^{\Lambda N}{}_P=0$. 
These equations can also be written without explicit reference to the scalar matrix $\cM$, since they are tensorial, as follows
 \be 
 \Omega_{IJ} C^{J M}{}_{\wa} T^{\wa N}{}_Q =  \overline{C}_{I Q}{}^{\tilde{\beta}} \Pi_{\tilde{\beta}}{}^{MN} + \overline{ C}_{I Q}{}^{\hL} \Pi_{\hL}{}^{MN}  \ ,
 \label{NewMaster}  
 \ee
where the conjugate tensors are defined as 
\begin{subequations}
\begin{align}
\label{CL2}  \overline{C}_{IP}{}^{\tilde{\alpha}} & \equiv \eta_{IJ} \eta_{PQ} \eta^{\tilde{\alpha}\tilde{\beta}} C^{JQ}{}_{\tilde{\beta}} = \cM_{IJ} \cM_{PQ} \cM^{\tilde{\alpha}\tilde{\beta}} C^{JQ}{}_{\tilde{\beta}} \; , \\
\label{CL10} \overline{C}_{IP}{}^{\hL} &\equiv \eta_{IJ} \eta_{PQ} \eta^{\hL\hXi} C^{JQ}{}_{\hXi} = \cM_{IJ} \cM_{PQ} \cM^{\hL\hXi} C^{JQ}{}_{\hXi} \; .
 \end{align}
 \end{subequations}
We shall refer to the conjectured relation \eq{NewMaster}  as the \emph{master identity} because of its central r\^ole for gauge invariance of our model. Equations \eq{deltaF}, \eq{nablaF} and \eq{eq:MIDnew}, differ from those given  in \cite{Bossard:2019ksx} by extending the sum over the index $\Lambda$ to one over $\hL$ and by the property that $\Pi_{\tilde{\beta}}{}^{MN}$ is the intertwiner in the irreducible representation $R(\Lambda_2)$ only. 
In Appendix \ref{app:MID} we prove that, disregarding~\eqref{CL2}, the identity~\eqref{NewMaster} is guaranteed to be satisfied for some invariant tensors $\overline{C}_{IQ}{}^{\hL}$ and  $\overline{C}_{IQ}{}^{\tilde{\beta}}$.
The non-trivial claim that we are making with~\eqref{NewMaster} is that 
$\overline{C}_{IQ}{}^{\tilde{\beta}}$ can be  identified with $C^{IM}{}_{\tilde{\alpha}}$ according to equation \eqref{CL2}.  
We establish  this identification in Appendix~\ref{app:MID} for the irreducible component of $\cT_{-1}$ that contains the supergravity duality equations.
A sufficient, but not necessary, condition for the proof of the master identity would therefore be the complete reducibility of $\cT_{-1}$. If $\cT_{-1}$ turned out to be indecomposable instead, one may have to modify the definition of $\overline{C}_{IP}{}^{\tilde{\alpha}}$ for $I$ valued in the corresponding invariant subspace for~\eqref{NewMaster} to hold. 
The validity of the master identity beyond what we prove in Appendix~\ref{app:MID} remains one of the key assumptions of our construction.

Independent of the proof of the master identity, we note that ${C}^{IP}{}_{\Lambda}$ in  \eqref{CL10} might be structure constants of the tensor hierarchy algebra. We provide some evidence  that $L(\Lambda_{10}) \subset \cT_0$ in  Appendix \ref{app:MID}, a statement that is further supported when we construct a $\cT_0$-homomorphism from $L(\Lambda_{10})\subset \cT_2$ to $\cT_0$  in Appendix \ref{Casimir}. 
We prove moreover in~\eqref{asforTHA} that the tensor $C^{IM}{}_{\Lambda}$ that we use in~\eqref{NewMaster} satisfies 
\be  
\Omega_{IJ} C{}^{I M}{}_{\alpha}  C^{J N}{}_{\Lambda}   \partial_M \partial_N =  0 \; ,
\ee
just as it would if $C^{IM}{}_{\Lambda}$ was identified with the structure constant of the tensor hierarchy algebra, similarly as in \eqref{eq:sympart}. 

Turning to ${C}^{I Q}{}_{\tL}$, we prove in Appendix \ref{app:MID} that they cannot be structure constants of $\cT(\mf{e}_{11})$, and in particular that  
 \be  
 \Omega_{IJ} C{}^{I M}{}_{\alpha}  C^{J N}{}_{\tL}   \partial_M \partial_N \ne   0 \;  . 
 \ee
Thus, even though the tensor hierarchy algebra $\cT(\mf{e}_{11})$ is a very useful and comprehensive tool for encoding the structure of $E_{11}$ exceptional field theory, it does not provide all the ingredients needed. We shall comment further on this in the conclusions. 

\subsection{\texorpdfstring{The pseudo-Lagrangian of $E_{11}$ exceptional field theory}{The pseudo-Lagrangian of E11 exceptional field theory}}
\label{sec:Lag}

In this section, we define the individual pieces of the schematic pseudo-Lagrangian
\begin{align}
\label{eq:Lag}
\mathcal{L}= \mathcal{L}_{\text{pot}_1} + \mathcal{L}_{\text{pot}_2} +  \mathcal{L}_{\text{kin}} + \mathcal{L}_{\text{top}} 
\end{align}
stated in~\eqref{eq:Lagin} of the introduction. We shall only show the rigid $E_{11}$-invariance of the individual pieces in this section but already put the correct relative coefficients for the combination to be invariant under generalised diffeomorphisms. This invariance will be checked in detail in Section~\ref{sec:GI}.

\subsubsection*{The first potential term}

All exceptional field theories possess a contribution to the potential term that takes the form
\begin{align}
\label{eq:Lpot1}
\mathcal{L}_{\text{pot}_1} =  -\frac14 \kappa_{\alpha\beta} \cM^{MN} J_M{}^\alpha J_N{}^\beta + \frac12 J_{M \alpha} T^{\beta M}{}_P \cM^{PQ} T^{\alpha N}{}_Q J_{N\beta}\,.
\end{align}
It only uses the $E_{11}$ current~\eqref{eq:curdef} and depends solely on the $E_{11}$ fields, but not on any of the constrained fields. It is manifestly $E_{11}$-invariant as it is only constructed from $E_{11}$ tensors. We have taken the canonical coefficients for this generic term~\cite{Cederwall:2017fjm}.\footnote{This potential term is more commonly written as $  \frac{1}{4 c}  \cM^{MN} \partial_M \cM^{PQ} \partial_N \cM_{PQ} - \frac{1}{2} \cM^{MP} \partial_M \cM^{NQ} \partial_N \cM_{PQ} $ for finite-dimensional groups $E_n$  \cite{Berman:2010is}, with $c$ defined such that $T_{\alpha}{}^P{}_Q T_\beta{}^Q{}_P = c \, \kappa_{\alpha\beta}$. However $c$ diverges for Kac--Moody groups.}  They are canonical in that they produce the Ricci scalar with unit coefficient, see Section~\ref{sec:GL11}.

\subsubsection*{The second potential term}

The second potential term is the generalisation of a similar term in the $E_8$ exceptional field theory~\cite{Hohm:2014fxa}, that has been generalised to any finite-dimensional simply laced Lie group in~\cite{Cederwall:2019bai}. Its presence is closely tied to the fact that there are ancillary gauge transformations whose parameter we called $\Sigma_M{}^\tI$ in Section~\ref{sec:gendiff}, with an associated tensor $C^\tI{}_{M\wa}$, see~\eqref{eq:ID5}. Using this we can write the expression
\begin{align}
\label{eq:Lpot2}
\mathcal{L}_{\text{pot}_2} = - \frac12 \cM_{\tI\tJ} C^{\tI}{}_{P\widehat\alpha} C^{\tJ}{}_{Q\widehat\beta} \cM^{QM} \cM^{PN} J_M{}^{\widehat\alpha} J_N{}^{\widehat\beta}\,,
\end{align}
where we recall that $\tI$ labels the reducible representation $L(\Lambda_3)$ defined in~\eqref{eq:tildeI}. Since $\wa$-indices range over the $\adjhat$ part of $\cT_0$ only,
this potential term depends only of the $E_{11}$ fields via the current $J_M{}^\alpha$ and of the constrained fields $\chi_M{}^\ta$ but not on of the other constrained fields $\zeta_M{}^\hL$. 
The second potential term is manifestly $E_{11}$-invariant. The tensor $C^{\tI}{}_{P\widehat\alpha}$ was already conjectured to exist in~\cite{Bossard:2017wxl} in the analysis of the linearised equations. We prove its existence and the relevant algebraic identities it satisfies in Appendix~\ref{Sweds}. 

The $E_{11}$-representation $L(\Lambda_3)\supset R(\Lambda_3)$ is a bounded weight module of highest weight $\Lambda_3$. When this representation is branched to $GL(3)\times E_8 \subset E_{11}$, its first element in level decomposition is a singlet and the first component of the tensor $C^{\tI}{}_{P \alpha}$ is the $E_8$ Cartan--Killing form, such that \eqref{eq:Lpot2} produces the new term that appears in the $E_8$ ExFT compared to $E_n$ ExFT with $n\leq 7$. We shall see this correspondence in much more detail when we consider the $E_8$ level decomposition in Section~\ref{sec:E8}.

\subsubsection*{Kinetic term}

The kinetic term resembles a generalisation of the covariant field strength-squared terms in other ExFTs. It uses the $E_{11}$ tensors appearing in the covariant field strengths $F^I$ defined in~\eqref{eq:FS}, but combines them in a slightly different way.

The kinetic term is given by
\begin{align}
\label{eq:Lkin}
\mathcal{L}_{\text{kin}} = \frac14 \cM_{IJ} C^{IM}{}_{\!\wa} C^{JN}{}_{\!\wb} J_M{}^\wa J_N{}^\wb - \frac12 \cM_{IJ} C^{IM}{}_{\!\wa} C^{JN}{}_{\!\hL} J_M{}^\wa \zeta_N{}^\hL -\frac14 \cM_{IJ} C^{IM}{}_{\!\hL} C^{JN}{}_{\!\hXi} \zeta_M{}^\hL \zeta_N{}^\hXi \,,
\end{align}
where we have fixed the overall coefficients knowing the result of the gauge invariance of the full pseudo-Lagrangian that will be studied in Section~\ref{sec:GI}.  The kinetic term depends on the constrained fields $\chi_M{}^\ta, \zeta_M{}^\Lambda$ and $\zeta_M{}^\tL$. 

Note that one might have expected to get the opposite sign for the first term in \eqref{eq:Lkin} such that one would have obtained the expected kinetic term $ \widetilde{\mathcal L}_{\text{kin}}=- \frac14 \cM_{IJ} F^I F^J$ in the democratic formulation of the theory, in which all fields and their duals appear at the same time. In fact  we shall see in Section \ref{sec:altL} below that there exists an alternative decomposition of the pseudo-Lagrangian in which the kinetic term is $ \widetilde{\mathcal L}_{\text{kin}}$, but the potential term instead does not take the expected form. This is a consequence of the fact that there is no natural split of the $E_{11}$ pseudo-Lagrangian into a kinetic term and a potential term given that the coordinates do not split into internal and external ones.

\subsubsection*{Topological term}

The topological term in exceptional field theory for finite-dimensional groups is the term that neither depends on the external nor the internal metric. There is no such term for $E_{11}$ exceptional field theory, but we call the topological term a term in the pseudo-Lagrangian that does not depend explicitly on the generalised metric $\cM_{MN}$, but only implicitly through the current $J_M{}^\alpha$. We shall first explain how to obtain such a term that is invariant under rigid $E_{11}$ transformations.

For this we take inspiration from $E_9$ exceptional field theory~\cite{Bossard:2021jix} and look for an $E_{11}$-invariant completion of the derivative of the constrained field $\chi_M{}^\ta$, i.e., we start from 
\begin{align}
\Pi_\ta{}^{MN} \partial_M \chi_N{}^\ta\,.
\end{align}
Recall that $\Pi_\ta{}^{MN}$ is antisymmetric in $M$ and $N$, see~\eqref{eq:Pitensors}, and that only the component in $R(\Lambda_2) \subset L(\Lambda_2)$ is non-vanishing because of the section constraint. 
The term above is not $E_{11}$-invariant because $\chi_M{}^\ta$ transforms indecomposably under $E_{11}$ according to~\eqref{eq:Dchi}. Since $\Pi_\ta{}^{MN}$ is an $E_{11}$-tensor, we only have to parametrise the completion of $\partial_{[M} \chi_{N]}{}^\ta$. Making an ansatz in terms of the tensors at our disposal and requiring it to be $E_{11}$-covariant leads uniquely to the combination
\begin{align}
\label{eq:Theta}
\Theta_{MN}{}^\ta = 2\partial_{[M} \chi_{N]}{}^\ta  + J_{[M}{}^{\alpha} T_{\alpha}{}^{\ta}{}_\tb \chi_{N]}{}^\tb + J_{M}{}^\alpha K_{[\alpha}{}^{\ta}{}_{\beta]} J_{N}{}^\beta \,,
\end{align}
where adjoint $E_{11}$ indices are raised and lowered freely using the Killing--Cartan form $\kappa_{\alpha\beta}$.
Let us briefly check that this is indeed $E_{11}$-covariant by computing its non-covariant transformation
\begin{align} \label{NonManifestE11} 
\Delta^\gamma \Theta_{MN}{}^\ta =  2 \partial_{[M} J_{N]}{}^\beta K^{\gamma\ta}{}_\beta + J_{[M}{}^\alpha T_{\alpha}{}^{ \ta}{}_\tb K^{\gamma \tb}{}_\beta J_{N]}{}^\beta - J_{M}{}^\alpha \big( K^{\gamma \tb}{}_{[\beta} T_{\alpha]}{}^\ta{}_\tb  - K^{\gamma \ta}{}_\delta f_{\alpha\beta}{}^\delta  \big) J_{N}{}^\beta =0
\end{align}
using~\eqref{eq:Dchi}, \eqref{eq:Dcoc} and finally the Maurer--Cartan equation~\eqref{eq:MC}. Thus, $\Theta_{MN}{}^\ta$ is the $E_{11}$-covariantisation of $\partial_{[M} \chi_{N]}{}^\ta$ that defines a two-form field strength. Projecting it with $ \Pi_\ta{}^{MN}$ turns it into an $E_{11}$-invariant density.

Another term is required in the topological term for gauge-invariance of the total pseudo-Lagrangian. This term involves the constrained fields $\zeta_M{}^\hL$ and will be needed in order to use the master identity~\eqref{eq:MIDnew}. The total topological term we shall consider is
\begin{align}
\label{eq:Ltop}
\mathcal{L}_{\text{top}} = \frac12 \Pi_\ta{}^{MN} \Theta_{MN}{}^\ta - \frac12 \Omega_{IJ} C^{IM}{}_\wa C^{JN}{}_\hL J_M{}^\wa \zeta_N{}^\hL\,.
\end{align}
We chose the form \eqref{eq:Ltop} to make $E_{11}$ invariance manifest, but note that if we write explicitly the dependence in the current $J_M{}^\alpha$ and the constrained field $\chi_M{}^\ta$ of the second term
\begin{align}
\label{eq:top2}
\Omega_{IJ} C^{IM}{}_\wa C^{JN}{}_\hL J_M{}^\wa \zeta_N{}^\hL &= \Omega_{IJ} C^{IM}{}_\alpha C^{JN}{}_\hL J_M{}^\alpha \zeta_N{}^\hL + \Omega_{IJ} C^{IM}{}_\ta C^{JN}{}_\hL \chi_M{}^\ta \zeta_N{}^\hL 
\nn\\
&= \Omega_{IJ} C^{IM}{}_\alpha C^{JN}{}_\hL J_M{}^\alpha \zeta_N{}^\hL\,,
\end{align}
one obtains that the field $\chi_M{}^\ta$ drops out because this contraction of $C$-tensors vanishes on section according to~\eqref{eq:ID7}.

The topological term~\eqref{eq:Ltop} does not depend on the generalised metric $\cM_{MN}$ explicitly and we have fixed the coefficients by anticipating the gauge invariance that will be verified in Section~\ref{sec:GI}.

\subsubsection*{Summary}

To summarise, the proposed pseudo-Lagrangian is given by the sum \eqref{eq:Lag} of the four terms
\begin{align}\label{AllTermsL} 
&\mathcal{L}_{\text{pot}_1} \hspace{-1.5mm} =  -\frac14 \kappa_{\alpha\beta} \cM^{MN} J_M{}^\alpha J_N{}^\beta + \frac12 J_{M \alpha} T^{\beta M}{}_P \cM^{PQ} T^{\alpha N}{}_Q J_{N\beta}\ ,
\nn\w2
&\mathcal{L}_{\text{pot}_2} \hspace{-1.5mm} = - \frac12 \cM_{\tI\tJ} C^{\tI}{}_{P\widehat\alpha} C^{\tJ}{}_{Q\widehat\beta} \cM^{QM} \cM^{PN} J_M{}^{\widehat\alpha} J_N{}^{\widehat\beta}\ ,
\nn\w2
&\mathcal{L}_{\text{kin}} = \frac14 \cM_{IJ} C^{IM}{}_\wa C^{JN}{}_\wb J_M{}^\wa J_N{}^\wb - \frac12 \cM_{IJ} C^{IM}{}_\wa C^{JN}{}_\hL J_M{}^\wa \zeta_N{}^\hL -\frac14 \cM_{IJ} C^{IM}{}_\hL C^{JN}{}_\hXi \zeta_M{}^\hL \zeta_N{}^\hXi \ ,
\nn\w2
&\mathcal{L}_{\text{top}} = \frac12 \Pi_\ta{}^{MN} \Theta_{MN}{}^\ta - \frac12 \Omega_{IJ} C^{IM}{}_\wa C^{JN}{}_\hL J_M{}^\wa \zeta_N{}^\hL\ ,
\end{align}
where
\begin{align}
\Theta_{MN}{}^\ta = 2\partial_{[M} \chi_{N]}{}^\ta  + J_{[M}{}^{\alpha} T_{\alpha}{}^{\ta}{}_\tb \chi_{N]}{}^\tb + J_{M}{}^\alpha K_{[\alpha}{}^{\ta}{}_{\beta]} J_{N}{}^\beta\ .
\label{Th2}
\end{align}
This decomposition of the pseudo-Lagrangian gives the expected potential term $\mathcal{L}_{\text{pot}_1}+\mathcal{L}_{\text{pot}_2}$~\cite{Cederwall:2019bai} but $\mathcal{L}_{\text{kin}}$ is not the expected kinetic term $\widetilde{\cL}_{\text{kin}}$. 
An alternative form of this pseudo-Lagrangian will be given below in \eq{SL} with the expected kinetic term \eq{tkin}, but an alternative potential term $\widetilde{\cL}_{\text{pot}}$ in  \eq{tpot}. We first present \eqref{eq:Lag} because it is more natural from the point of view of exceptional field theory and because we shall use it to prove the invariance under generalised diffeomorphisms in the next section. The alternative pseudo-Lagrangian \eq{SL} involves fewer terms and we will use it when we vary the pseudo-Lagrangian with respect to the $E_{11} / K(E_{11})$ coset fields in Section~\ref{sec:EOM}.

\subsection{Consistency with the duality equation}
\label{sec:EOMchi}

The way we arrived at the pseudo-Lagrangian~\eqref{eq:Lag} was by considering $E_{11}$ building pieces such that the constrained fields' Euler--Lagrange equations are compatible with the duality equations that we described in~\eqref{eq:DE}, so that both can be imposed consistently. This crucially requires to have both a topological term and a `kinetic term'. For obtaining generalised diffeomorphism invariance one also has to consider terms that are independent of the constrained fields, such as $\mathcal{L}_{\text{pot}_1}$, which is a universal ExFT term.
We postpone the proof of gauge invariance to Section~\ref{sec:GI} and first investigate the consistency with the duality equation. 
This will also provide a first confirmation of the choice of coefficients in the terms of the pseudo-Lagrangian. We shall discuss the Euler--Lagrange equations obtained by varying with respect to the $E_{11}$ generalised metric $\cM$ in Section~\ref{sec:EOM}.

\subsubsection*{Varying with respect to $\chi$}

There is $\chi$-dependence in three of the four terms of the pseudo-Lagrangian, namely the second potential term~\eqref{eq:Lpot2}, the kinetic term~\eqref{eq:Lkin} and the topological term~\eqref{eq:Ltop}, in the former two cases via the component $J_M{}^\ta=\chi_M{}^\ta$. 

Varying $\chi_M{}^\ta$ in the kinetic term leads to
\begin{align}
\label{eq:delchikin}
\delta \mathcal{L}_{\text{kin}} &= \frac12 \cM_{IJ} C^{IM}{}_\ta \Big( C^{JN}{}_\wb J_N{}^\wb  - C^{JN}{}_\hL \zeta_N{}^\hL  \Big) \delta\chi_M{}^\ta
\end{align}
The $\chi$-variation of the second potential term~\eqref{eq:Lpot2} is
\begin{align}
\label{eq:delchipot2}
\delta \mathcal{L}_{\text{pot}_2} &= - \cM_{\tI\tJ} C^\tI{}_{P\ta} C^\tJ{}_{Q\wb} \cM^{QM} \cM^{PN} J_N{}^\wb \delta\chi_M{}^\ta
\nn\\
&= - \cM_{IJ} C^{IM}{}_\ta C^{JN}{}_{\wb}  J_N{}^\wb \delta\chi_M{}^\ta \; . 
\end{align}
Using equation~\eqref{eq:ID6}, that can be written as
\begin{align}
\label{eq:ID6p}
\cM_{\tI\tJ} C^{\tI}{}_{P\ta} C^{\tJ}{}_{Q\wb} \cM^{QM} \cM^{PN} = \cM_{IJ} C^{IM}{}_\ta C^{JN}{}_\wb\, , 
\end{align}
\eqref{eq:delchipot2} combines with the first term in~\eqref{eq:delchikin} to produce a variation proportional to the field strength $F^I$.

The $\chi$-variation of the topological term~\eqref{eq:Ltop} is, up to a total derivative,
\begin{align}
\label{eq:delchitop}
\delta \mathcal{L}_{\text{top}} &= - \frac12 \Pi_\ta{}^{MN}  T_{\alpha}{}^\ta{}_\tb J_N{}^\alpha  \delta \chi_M{}^\tb 
\nn\\
&=  \frac12 \Omega_{IJ} C^{I M}{}_{\tb} C^{JN}{}_\alpha J_N{}^\alpha  \delta \chi_M{}^\tb  
\nn\\
&= \frac12 \Omega_{IJ} C^{I M}{}_{\ta} \Big( C^{JN}{}_\wb J_N{}^\wb  + C^{JN}{}_\hL \zeta_N{}^\hL  \Big)\delta \chi_M{}^\ta \nn\\
&= \frac12 \Omega_{IJ} C^{I M}{}_{\ta} F^J \delta \chi_M{}^\ta \,,
\end{align}
where we have first used \eqref{eq:ID2} since both $J_N{}^\alpha$ and $\delta \chi_M{}^\ta$ are constrained objects. In the next-to-last step we have added many terms that vanish by virtue of~\eqref{eq:ID3} and~\eqref{eq:ID7} in order to identify the field strength.

Putting the variations~\eqref{eq:delchikin}, \eqref{eq:delchipot2} and~\eqref{eq:delchitop} together leads to the $\chi$-variation of the full pseudo-Lagrangian, up to a total derivative,
\begin{align}
\delta \mathcal{L} = -\frac12 C^{IM}{}_\ta \big( \cM_{IJ} -\Omega_{IJ}\big) F^J \delta \chi_M{}^\ta\,.
\end{align}
So the Euler--Lagrange equations for the constrained field $\chi_M{}^\ta$ are a subset of the duality equation~\eqref{eq:DE}.

\subsubsection*{Varying with respect to $\zeta$}

The constrained fields $\zeta_M{}^\hL$ appear in two places, namely in the kinetic term~\eqref{eq:Lkin} and the topological term~\eqref{eq:Ltop}. We immediately write the full variation
\begin{align}
\delta \mathcal{L} &= - \frac12 \Big(\cM_{IJ} C^{IM}{}_\wa J_M{}^\wa +  \cM_{IJ} C^{IM}{}_\hXi  \zeta_M{}^\hXi   +  \Omega_{IJ} C^{IM}{}_\wa  J_M{}^\wa \Big) C^{JN}{}_\hL \delta \zeta_N{}^\hL  \nn\\
 &= - \frac12 \Big(\cM_{IJ} F^J  -  \Omega_{IJ} C^{JM}{}_\wa  J_M{}^\wa \Big)  C^{IN}{}_\hL \delta \zeta_N{}^\hL
 \nn\\
  &= - \frac12 C^{IN}{}_\hL   \big(\cM_{IJ} -\Omega_{IJ}\big) F^J    \delta \zeta_N{}^\hL\ .
\end{align}
In the last step  we have added a term $ -\Omega_{IJ} C^{JM}{}_\hXi  \zeta_M{}^\hXi $ that vanishes due to~\eqref{eq:ID8}, in order to complete the second field strength $F^J$. Therefore, all equations obtained by varying $\mathcal{L}$ with respect to the constrained fields are a subset of the duality equation~\eqref{eq:DE}.

\subsection{Alternative form of the pseudo-Lagrangian}
\label{sec:altL}

In varying the pseudo-Lagrangian \eq{eq:Lag} with respect to the $E_{11} / K(E_{11})$ fields,  it will be convenient to use an alternative decomposition that will exhibit the expected kinetic term. For this purpose we first combine the kinetic term \eqref{eq:Lkin} and the second potential term \eqref{eq:Lpot2} into 
\begin{align}
\label{pot2}
& \hspace{5mm} {\mathcal L}_{\text{kin}}+{\mathcal L}_{\text{pot}_2} \\
  &= - \frac{1}{4} \cM_{IJ} F^I F^J +\frac12 \cM_{IJ} C^{IM}{}_\wa C^{JN}{}_\wb J_M{}^\wa J_N{}^\wb-\frac12 \cM_{\tI \tJ} C^{\tI}{}_{P\wa} C^{\tJ}{}_{Q\wb} \cM^{QM} \cM^{PN} J_M{}^\wa J_N{}^\wb \nn 
\\
&=- \frac{1}{4} \cM_{IJ} F^I F^J+\frac12 \cM_{IJ} C^{IM}{}_\alpha C^{JN}{}_\beta J_M{}^\alpha J_N{}^\beta -\frac12 \cM_{\tI \tJ} C^{\tI}{}_{P\alpha} C^{\tJ}{}_{Q\beta} \cM^{QM} \cM^{PN} J_M{}^\alpha J_N{}^\beta
 \ .\nn
\end{align}
In obtaining \eq{pot2}, we have split the index $\wa=(\alpha,\ta)$ in the last step and used identity \eq{eq:ID6p} to cancel the terms containing summation over the index $\ta$. The first term is the expected kinetic term $\widetilde{\cL}_{\text{kin}}$ in a democratic formulation of the theory with the standard sign. Since the two remaining terms in \eqref{pot2} do not depend on the constrained fields, it is natural to combine them with the first potential term to obtain an alternative potential term $\widetilde{\cL}_{\text{pot}}$. The resulting $\widetilde{\cL}_{\text{pot}}$ simplifies remarkably upon using~\eqref{e12Equation}, which can also be expressed as
\begin{align}
\label{eq:altid}
&\hspace{6mm} \cM_{\tI\tJ} C^{\tI}{}_{P\alpha} C^{\tJ}{}_{Q\beta} \cM^{QM} \cM^{PN}- \cM_{IJ} C^{IM}{}_\alpha C^{JN}{}_\beta  \CR
&=  \cM^{MP}   T_{\beta}{}^{R}{}_P \cM_{RS}T_{\alpha}{}^S{}_Q \cM^{QN}+T_{\alpha}{}^M{}_P  \cM^{PQ} T_{\beta}{}^{N}{}_Q -\cM^{MN}\cM_{\alpha\beta}\,.
 \end{align}
Using this identity,~\eqref{eq:Jherm} and the fact that $\kappa_{\alpha\beta}  J_N{}^\beta = \cM_{\alpha\beta}  J_N{}^\beta$, we find that the pseudo-Lagrangian \eq{eq:Lag} takes the alternative form
\be
\label{SL}
{\mathcal L} =  \widetilde{\mathcal L}_{\text{kin}} + \widetilde{\mathcal L}_{\text{pot}} +  {\mathcal L}_{\text{top}}\ ,
\ee
where
\begin{subequations}
\begin{align}
 \widetilde{\mathcal L}_{\text{kin}}&= - \frac{1}{4} \cM_{IJ} F^I F^J\ ,
 \label{tkin}\w2
 \widetilde{\mathcal L}_{\text{pot}}&= \frac14 \kappa_{\alpha\beta} \cM^{MN} J_{M}{}^\alpha J_N{}^\beta - \frac12 J_{M}{}^\alpha  T_\alpha{}^M{}_P \cM^{PQ} T_\beta{}^N{}_Q J_{N}{}^{\beta} \ ,
 \label{tpot} 
 \end{align}
\end{subequations}
with $F^I$ is defined in \eq{eq:FS} and ${\mathcal L}_{\text{top}}$ as in \eqref{AllTermsL}. Note that the potential term $\widetilde{\mathcal{L}}_{\text{pot}}$ differs from $-\mathcal{L}_{\text{pot}_1}$ in the contraction of the derivative indices.


\section{Gauge invariance of the pseudo-Lagrangian}
\label{sec:GI}

We now show that the $E_{11}$ exceptional field theory pseudo-Lagrangian given in~\eqref{eq:Lag} is gauge-invariant. 
For this we calculate the variation of each term in the pseudo-Lagrangian~\eqref{eq:Lag} under generalised diffeomorphisms and then demonstrate that the combination of  these variations vanishes.  As always in these checks in exceptional field theory it is sufficient to show that the non-covariant gauge variation $\Delta_\xi$ defined in~\eqref{eq:defDxi} vanishes up to total derivatives. The expressions for the non-covariant gauge transformations of the fields were given in~\eqref{eq:NCJ} and~\eqref{eq:DxiCS}.

Our proof proceeds in two steps.  In order to underline the necessity of including the fields $\zeta_M{}^\hL$, we first consider the pseudo-Lagrangian for $\zeta_M{}^\hL=0$ and computes its non-covariant gauge variation. As we shall see there are already many cancellations but some terms are left over. Then we shall show that these terms are exactly cancelled by the $\zeta_M{}^\hL$-dependent terms in~\eqref{eq:Lag}.

\subsection{\texorpdfstring{Gauge variation at $\zeta=0$}{Gauge variation at zeta=0}}

As explained above, we compute first the non-covariant gauge variation of all the pieces of $\mathcal{L}$ at $\zeta_M{}^\hL=0$.

\subsubsection*{First potential term}

The first potential term~\eqref{eq:Lpot1}, does not depend on $\zeta_M{}^\hL$ and we can immediately calculate the full non-covariant gauge variation. 
A standard exceptional field theory calculation involving the definition of the current $J_M{}^\alpha$ and the section constraint gives the first step~\cite{Cederwall:2017fjm}
\begin{align}
\label{eq:Dxipot1}
\Delta_\xi \Big[ \mathcal{L}_{\text{pot}_1} \Big] &= \bigg[-  T_\beta{}^{R}{}_Q \cM^{MN} +T_\beta{}^M{}_P\delta_Q^N \cM^{PR} +  f_{\beta\alpha\gamma} T^{\gamma M}{}_P T^{\alpha R}{}_Q \cM^{NP}
\bigg] \partial_M\partial_R\xi^Q J_N{}^\beta\nn\\
&= \cM_{\tI\tJ}  C^{\tI}{}_{P\wa } 
C^{\tJ}{}_{Q\beta}  T^{\wa R}{}_S  \cM^{QM}
\cM^{PN} \partial_M\partial_R\xi^S J_N{}^\beta - 2 \partial_{[M} \Big( \partial_{N]} \partial_P \xi^M \cM^{NP}\Big) \,.
\end{align}
In the second step we have used the identity~\eqref{eq:ID5} and simplified the terms with a single representation matrix $T_{\beta}$ and a single inverse $\cM^{MN}$ into a total derivative.

It is worthwhile to remark that the $E_{11}$-representation with index $\tI$ has as lowest component $R(\Lambda_3)$ according to~\eqref{eq:tildeI}. When decomposing $E_{11}$ with respect to $GL(11-n)\times E_n$ the first time this representation enters the scalar sector is for $E_8$ which is in agreement with the fact that this is the first time the potential term~\eqref{eq:Lpot1} is not gauge-invariant and also the first time ancillary transformations are needed. We shall show next how the failure of gauge-invariance of the first potential term involving the index $\tI$ is accounted for by the second potential term.

\subsubsection*{Second potential term}

The second potential term~\eqref{eq:Lpot2} does not depend on $\zeta_M{}^\hL$ either. Calculating the full non-covariant gauge transformation yields
\begin{align}
\label{eq:Dxipot2}
\Delta_\xi \Big[ \mathcal{L}_{\text{pot}_2} \Big] &= 
-\cM_{\tI\tJ} C^{\tI}{}_{P\wa} C^{\tJ}{}_{Q\wb} \cM^{QM} \cM^{PN} T^{\wa R}{}_S \Bigl( \partial_M\partial_R\xi^S + \cM_{RU} \cM^{ST} \partial_M \partial_T \xi^U\Bigr) J_N{}^\wb\nn\\
&\quad \quad - \cM_{\tI\tJ} C^{\tI}{}_{P\ta} C^{\tJ}{}_{Q\wb} \cM^{QM} \cM^{PN} \Pi^\ta{}_{RS} \cM^{TR} \partial_M\partial_T \xi^S J_N{}^\wb\nn\\
&= -\cM_{\tI\tJ} C^{\tI}{}_{P\wa} C^{\tJ}{}_{Q\beta} T^{\wa R}{}_S \cM^{QM} \cM^{PN} \partial_M\partial_R\xi^S J_N{}^\beta 
\nn\\
&\quad\quad -\cM_{IJ} C^{IM}{}_\ta C^{JN}{}_\wb \Pi^\ta{}_{QP} \cM^{QR} \partial_M\partial_R\xi^P J_N{}^\wb\,,
\end{align}
where we have first written out the non-covariant variation $\Delta_\xi J_M{}^\wa$ using~\eqref{eq:NCJ} and~\eqref{eq:Dxichi}. In the next step we have distributed the parenthesis on the first line and used the identity \eqref{eq:ID9} to cancel the second contribution 
\bea
 && \left(  \cM_{\tI\tJ} C^\tI{}_{P\wa} T^{\wa R}{}_S\, \cM^{PN}  \cM^{ST}  \cM_{RU}\right)    C^\tJ{}_{Q\wb} \cM^{QM}  \partial_M\partial_T\xi^U  J_N{}^\wb
 \nn\\
 &=&  \left( \eta_{\tI\tJ} C^\tI{}_{Q\wa} T^{\wa P}{}_R\,  \eta^{QM}  \eta^{RN} \eta_{PS}\right)  C^\tJ{}_{T\wb} \cM^{TU} \partial_U  \partial_N \xi^S J_M {}^\wb =0  \; , 
\eea
where we  split the $\wb$ index on the first contribution and used the identities \eqref{eq:ID6p} and~\eqref{eq:THA1} to remove the $\chi_N{}^\tb$ component. For the second term we have simply used~\eqref{eq:ID6p} to convert the $C^{\tilde{I}}$ tensor sum into a $C^I$ tensor sum. 

The first term we obtain in~\eqref{eq:Dxipot1} cancels precisely the contribution from the first potential term. This cancelation is the same one that ensures the invariance of the potential for any finite-dimensional simply laced groups~\cite{Cederwall:2019bai}. Consistently, the identity~\eqref{eq:ID5} that was used in this cancelation is proved in Appendix \ref{Sweds}  using a construction that generalises the one of~\cite{Cederwall:2019bai} to the Kac--Moody algebra ${\mf e}_{11}$. Here, we obtain the combined non-covariant gauge variation
\be \label{Dpotonetwo}
\Delta_\xi \Big[ \mathcal{L}_{\text{pot}_1}\hspace{-1mm}+ \mathcal{L}_{\text{pot}_2} \Big] = -\cM_{IJ} C^{IM}{}_\ta C^{JN}{}_\wb \Pi^\ta{}_{QP} \cM^{QR} \partial_M\partial_R\xi^P J_N{}^\wb - 2 \partial_{[M} \Big( \partial_{N]} \partial_P \xi^M \cM^{NP}\Big)  \; . 
\ee
Thus, compared to~\cite{Cederwall:2019bai} where no $\Pi^\ta{}_{MN}$ appears, the combination for $E_{11}$ is not gauge-invariant and we shall invoke an additional ingredient to arrive at a gauge-invariant pseudo-Lagrangian.

\subsubsection*{Kinetic term at $\zeta=0$}

In order to determine the non-covariant gauge variation of the kinetic term~\eqref{eq:Lkin} we break it up into the parts that contain the constrained fields $\zeta_M{}^\hL$ (before variation) and those that do not, beginning with the latter:
\begin{align}
\label{eq:Dxikin}
&\quad\quad \Delta_\xi \Big[ 
\mathcal{L}_{\text{kin}} \big|_{\zeta=0}
\Big] \nn\\
&= \frac12 \cM_{IJ}  \Big( C^{JM}{}_\wa 
T^{\widehat\alpha S}{}_Q \cM_{SP} \cM^{QR}  + C^{JM}{}_\ta \Pi^{\ta}{}_{QP} \cM^{QR}\Big)  C^{IN}{}_\wb J_N{}^\wb  \partial_M\partial_R\xi^P
\end{align}
where we have used the identity
\eqref{eq:THA1} to cancel the term in $T^{\wa N}{}_P\partial_M \partial_N\xi^P$ from the non-covariant gauge variations~\eqref{eq:NCJ} and~\eqref{eq:Dxichi}.

\subsubsection*{Topological term at $\zeta=0$}

 We first compute the non-covariant gauge transformation of~\eqref{eq:Ltop} at $\zeta_M{}^\hL=0$. An important first observation is that the total derivative $\Pi_\ta{}^{MN} \partial_M\chi_N{}^\ta$ is not invariant under its non-covariant gauge transformation up to a total derivative. 
 To compute $\Delta_\xi=\delta_\xi-\mathcal{L}_\xi$ of $\Pi_\ta{}^{MN} \partial_M\chi_N{}^\ta$ we need to determine the Lie derivative of the combined object $\partial_M\chi_N{}^\ta$ which is given by
\begin{multline} \mathcal{L}_\xi \big( \partial_{M} \chi_{N}{}^\ta\big)  = \xi^P \partial_P \big( \partial_{M} \chi_{N}{}^\ta\big)  + \partial_M \xi^P \partial_P \chi_N{}^\ta +  \partial_N \xi^P \partial_M \chi_P{}^\ta \\ - T_{\alpha}{}^P{}_Q \partial_P \xi^Q \bigl( T^{\alpha\ta}{}_{\tb} \partial_M \chi_N{}^\tb + K^{\alpha\ta}{}_\beta \partial_M J_N{}^\beta \bigr) \,.
\end{multline}
This not a total derivative. 
Therefore the non-covariant gauge variation is 
\begin{align}
\label{eq:NCtop1}
\Delta_\xi \Big[ \Pi_\ta{}^{MN} \partial_{M} \chi_N{}^\ta\Big] &= \Pi_\ta{}^{MN} \Big[  \partial_M \bigl(  \delta_\xi \chi_N^{\tilde{\alpha}} \bigr) - \cL_\xi \bigl(  \partial_{M} \chi_{N}{}^\ta\bigr) \Bigr]  \CR
&=\Pi_\ta{}^{MN}\bigg[- T_{\alpha}{}^R{}_P T^{\alpha \ta}{}_{\tb} \partial_M \partial_R \xi^P  \chi_N{}^\tb \nn\\
&\quad\quad + \Big( - T^{\alpha R}{}_P K_\alpha{}^{\ta}{}_\beta - T_{\beta}{}^U{}_Q T^{\ta Q}{}_S  \cM_{UP} \cM^{SR} + T^{\ta U}{}_{Q}  T_{\beta}{}^Q{}_S \cM_{UP} \cM^{SR}\nn\\
&\hspace{20mm} + \Pi^\ta{}_{QP} T_\beta{}^{Q}{}_S \cM^{SR}
\Big) \partial_M \partial_R \xi^P J_N{}^\beta\bigg]
\end{align}
where we used the section constraint~\eqref{eq:SC} on $\cL_\xi \chi_M{}^\ta$ defined from \eqref{eq:chiGT}. The three last terms come from $  \partial_M(  \Delta_\xi \chi_N^{\tilde{\alpha}} )$ and therefore do combine into a total derivative, but it will be convenient to distribute the derivative as above. 

The remaining terms in $\Theta_{MN}{}^\ta$ defined in~\eqref{eq:Theta} just pick up their non-covariant variations defined in~\eqref{eq:NCJ} and~\eqref{eq:Dxichi}. We organise the calculation by looking first at all terms varying into $\chi$ and then at terms varying into the current $J$. The sum of terms varying into $\chi$  give
\begin{align}
\label{eq:Dxitop1}
\Delta_\xi \Big[\mathcal{L}_{\text{top}}\big|_{\zeta=0}\Big]\Big|_{\chi \partial^2\xi} &= \frac12 \Pi_{\ta}{}^{MN} \bigg[
-2 T_{\alpha}{}^R{}_P T^{\alpha \ta}{}_{\tb} \partial_M \partial_R \xi^P  \chi_N{}^\tb \nn\\
&\quad\quad+ T^{\alpha \ta}{}_\tb T_{\alpha}{}^R{}_P \partial_M \partial_R\xi^P  \chi_N^{\tb} + T^{\alpha\ta}{}_\tb T_\alpha{}^S{}_Q \cM_{SP} \cM^{QR} \partial_M \partial_R\xi^P  \chi_N{}^\tb
\bigg]\nn\\
&= -  \Pi_\ta{}^{U[M} T^{\alpha N]}{}_U T_\alpha{}^S{}_Q \cM_{SP} \cM^{QR} \partial_M \partial_R\xi^P  \chi_N{}^\ta\,,
\end{align}
where we used the identity~\eqref{eq:ID1}  on all terms and the fact that the first two vanish using the section constraint.

The terms whose non-covariant gauge variation contains a current $J$ are
\begin{align}
\label{eq:Dxitop2}
&\quad\quad \Delta_\xi \Big[\mathcal{L}_{\text{top}}\big|_{\zeta=0}\Big]\Big|_{J \partial^2\xi} \nn\\
&=  \frac12 \Pi_{\ta}{}^{MN} \bigg[-2 T^{\alpha R}{}_P K_\alpha{}^{\ta}{}_\beta
- 2T_{\beta}{}^U{}_Q T^{\ta Q}{}_S  \cM_{UP} \cM^{SR} + 2T^{\ta U}{}_{Q}  T_{\beta}{}^Q{}_S \cM_{UP} \cM^{SR}
\nn\\
&\hspace{20mm} +2\Pi^\ta{}_{QP} T_\beta{}^{Q}{}_S \cM^{SR}
- T_\beta{}^{ \ta}{}_\tb T^{\tb R}{}_P - T_\beta{}^{ \ta}{}_{\tb} T^{\tb S}{}_Q \cM_{SP} \cM^{QR} - T_\beta{}^{\ta}{}_\tb  \Pi^\tb{}_{QP} \cM^{QR}\nn\\
&\hspace{20mm}+2 K_{[\alpha}{}^{\ta}{}_{\beta]} \Big( T^{\alpha R}{}_P + T^{\alpha S}{}_Q\cM_{SP} \cM^{QR}\Big) \bigg]\partial_M \partial_R\xi^P J_N{}^\beta\nn\\
&=\frac12 \Pi_{\ta}{}^{MN} \bigg[-2 K_{(\alpha}{}^{\ta}{} _{\beta)} T^{\alpha R}{}_P  -T_{\beta}{}^{\ta}{}_{\tb} T^{\tb R}{}_P+  \bigl( 2 K_{(\alpha}{}^{\ta}{} _{\beta)} T^{\alpha S}{}_Q  +T_{\beta}{}^{\ta}{}_{\tb} T^{\tb S}{}_Q \bigr)  \cM_{SP} \cM^{QR} \nn\\
&\hspace{80mm}     - 2 T_\beta{}^{S}{}_{(Q} \Pi^\ta{}_{P)S} \cM^{QR}\bigg]\partial_M \partial_R\xi^P J_N{}^\beta\,,\\
&= -\frac12 \Omega_{IJ} C^{IM}{}_{\wa} C^{JN}{}_{\beta}  T^{\wa S}{}_Q\cM_{SP} \cM^{QR} \partial_M \partial_R\xi^P J_N{}^\beta    - \Pi_\ta{}^{MN} T_\beta{}^{S}{}_{(Q} \Pi^\ta{}_{P)S} \cM^{QR}\partial_M \partial_R\xi^P J_N{}^\beta\,,\nn
\end{align}
where in the first step we have used the commutation relation~\eqref{eq:THA01} and \eqref{eq:ID1}, in the last step we have used the identities \eqref{eq:ID4} and \eqref{eq:ID2} to write the first line in terms of the $C$-tensors and combined the $\alpha$ and $\ta$ components into an $\wa$ index. We finally used the identity~\eqref{eq:THA1} to cancel the $\cM$ independent term.

\subsubsection*{Combined non-covariant gauge variation at $\zeta=0$}

Collecting all the terms from above we therefore find
\begin{align}
\label{eq:Dxi1}
&\quad\quad \Delta_\xi \Big[ \mathcal{L}\big|_{\zeta=0} \Big]  + 2 \partial_{[M} \Big( \partial_{N]} \partial_P \xi^M \cM^{NP}\Big) \\
&= \frac12 \cM_{IJ}  C^{JN}{}_\wb  \Big( C^{IM}{}_\wa  
T^{\widehat\alpha S}{}_Q \cM_{SP} \cM^{QR}  - C^{IM}{}_\ta   \Pi^{\ta}{}_{QP} \cM^{QR}\Big) \partial_M\partial_R\xi^P   J_N{}^\wb \nn\\
&\quad\quad  -  \Pi_\ta{}^{U[M} T^{\alpha N]}{}_U T_\alpha{}^S{}_Q \cM_{SP} \cM^{QR} \partial_M \partial_R\xi^P  \chi_N{}^\ta\nn\\
&\quad\quad  -\frac12 \Omega_{IJ} C^{IM}{}_{\wa} C^{JN}{}_{\beta}  T^{\wa S}{}_Q \cM_{SP} \cM^{QR} \partial_M \partial_R\xi^P J_N{}^\beta    - \Pi_\ta{}^{MN} T_\beta{}^{S}{}_{(Q} \Pi^\ta{}_{P)S} \cM^{QR}\partial_M \partial_R\xi^P J_N{}^\beta \nn
\end{align}
where the first line combines \eqref{eq:Dxikin} and  \eqref{Dpotonetwo} while the remaining lines come from the variation of the topological term given in~\eqref{eq:Dxitop1} and~\eqref{eq:Dxitop2}.

{\allowdisplaybreaks
So far we have avoided using any identity that mixes  $\adjhat$ and $L(\Lambda_{10})\oplus L(\Lambda_4)$. The only equation that does this is the master identity~\eqref{eq:MIDnew} and we shall apply it now to the first line above. Continuing from~\eqref{eq:Dxi1} we then obtain
\begin{align}
\label{eq:Dxi2}
&\quad\quad \Delta_\xi \Big[ \mathcal{L}\big|_{\zeta=0} \Big]  + 2 \partial_{[M} \Big( \partial_{N]} \partial_P \xi^M \cM^{PN}\Big) \nn\\
&= \frac12 \Omega_{IJ} C^{JN}{}_\wb  \Big( C^{IM}{}_\wa   
T^{\widehat\alpha S}{}_Q \cM_{SP} \cM^{QR}  -C^{IM}{}_\ta     \Pi^{\ta}{}_{QP} \cM^{QR}\Big) \partial_M\partial_R\xi^P   J_N{}^\wb \nn\\*
&\quad\quad + \frac12 \big(\cM_{IJ}+\Omega_{IJ}\big)  C^{IN}{}_\wb  C^{JM}{}_\hL \Pi^\hL{}_{QP} \cM^{QR} \partial_M \partial_R \xi^P J_N{}^\wb\nn\\*
&\quad\quad  -  \Pi_\ta{}^{U[M} T^{\alpha N]}{}_U T_\alpha{}^S{}_Q \cM_{SP} \cM^{QR} \partial_M \partial_R\xi^P  \chi_N{}^\ta\nn\\*
&\quad\quad  -\frac12 \Omega_{IJ} C^{IM}{}_{\wa} C^{JN}{}_{\beta}  T^{\wa S}{}_Q \cM_{SP} \cM^{QR} \partial_M \partial_R\xi^P J_N{}^\beta     - \Pi_\ta{}^{MN} T_\beta{}^{S}{}_{(Q} \Pi^\ta{}_{P)S} \cM^{QR}\partial_M \partial_R\xi^P J_N{}^\beta \nn\\
&=  \frac12 \big(\cM_{IJ}+\Omega_{IJ}\big) C^{IN}{}_\wb C^{JM}{}_\hL  \Pi^\hL{}_{QP} \cM^{QR} \partial_M \partial_R \xi^P J_N{}^\wb\nn\\*
&\quad\quad - \frac12 \Pi_\ta{}^{MN} T_\alpha{}^\ta{}_\tb
T^{\alpha S}{}_Q \cM_{SP} \cM^{QR}   \partial_M\partial_R\xi^P   \chi_N{}^\tb 
+\frac12 \Pi_\tb{}^{MN} T_\beta{}^\tb{}_\ta  \Pi^{\ta}{}_{QP} \cM^{QR} \partial_M\partial_R\xi^P   J_N{}^\beta \nn\\*
&\quad\quad  -  \Pi_\ta{}^{U[M} T^{\alpha N]}{}_U T_\alpha{}^S{}_Q \cM_{SP} \cM^{QR} \partial_M \partial_R\xi^P  \chi_N{}^\ta    - \Pi_\ta{}^{MN} T_\beta{}^{S}{}_{(Q} \Pi^\ta{}_{P)S} \cM^{QR}\partial_M \partial_R\xi^P J_N{}^\beta \nn\\
&=  \frac12 \big(\cM_{IJ}+\Omega_{IJ}\big) C^{JM}{}_\hL \big(\Delta_\xi \zeta_M{}^\hL \big)C^{IN}{}_\wb  J_N{}^\wb - \partial_N \big( \Pi_\ta{}^{MN} \Pi^\ta{}_{RP} \cM^{RQ} \partial_M \partial_Q\xi^P\big)
\end{align}
where we have used the identities \eqref{eq:ID3} and \eqref{eq:ID2} to remove most $\Omega_{IJ}$ terms when going to the second equality. In the final step we have used the identity \eqref{eq:ID1} to cancel two terms and have brought out a total derivative. Moreover we can use the identity \eqref{A66} to obtain that the derivative terms cancel.  The remaining term can be written as the non-covariant variation ~\eqref{eq:Dxizeta} of $\zeta_M{}^\hL$ as shown. This result strongly suggests that one might be able to obtain a pseudo-Lagrangian invariant under generalised diffeomorphisms by adding the relevant $\zeta_M{}^\hL$ dependent terms. This is indeed what we will show next. 
}

\subsection{Gauge invariance}

In order to demonstrate gauge-invariance of $\mathcal{L}$, we now consider the $\zeta_M{}^\hL$ dependent terms. These appear in the kinetic term~\eqref{eq:Lkin} and in the topological term~\eqref{eq:Ltop}. Their non-covariant gauge variation is given by
\begin{align}
\label{eq:Nczetaterms}
\Delta_\xi \Big[ \mathcal{L}- \mathcal{L}|_{\zeta=0} \Big] 
&=  -\frac12 (\cM_{IJ}+\Omega_{IJ}) C^{IM}{}_\wa C^{JN}{}_\hL J_M{}^\wa \Delta_\xi \zeta_N{}^\hL 
\nn\\
& \hspace{-18mm}- \frac12 (   \cM_{IJ}+\Omega_{IJ}  ) \Big( C^{IM}{}_\wa T^{\wa S}{}_Q \cM_{SP}  + C^{IM}{}_\ta \Pi^\ta{}_{QP} + C^{IM}{}_{\hL} \Pi^{\hL}{}_{QP}  \Big) \cM^{QR} \partial_M\partial_R\xi^P  C^{JN}{}_\hL  \zeta_N{}^\hL
\nn\\
&=  -\frac12 (\cM_{IJ}+\Omega_{IJ})  C^{JM}{}_\hL  \bigl( \Delta_\xi \zeta_M{}^\hL\bigr) C^{IN}{}_\wb J_N{}^\wb
\end{align}
where in the first step we have written out the non-covariant variations of $J_M{}^\wa$, cancelled one term using the identity~\eqref{eq:THA1}, and used the identities \eqref{eq:ID8} to add one vanishing term and group terms together into the non-covariant variation of $F^I$. In the second step we have then applied  the master identity~\eqref{eq:MIDnew} twice to cancel the middle line.

Now we can collect all terms contributing to the variation of the pseudo-Lagrangian~\eqref{eq:Lag} and obtain from~\eqref{eq:Dxi2} and~\eqref{eq:Nczetaterms} 
\begin{align}
\delta_\xi \mathcal{L} = \partial_M \Bigl(  \xi^M \cL \Big)
\end{align}
where we used moreover that the total derivative terms in~\eqref{eq:Dxi2} cancel using Equation \eqref{A66}.  We have therefore proved that the pseudo-Lagrangian is gauge-invariant up to a total derivative as claimed. Note moreover that it transforms under generalised diffeomorphisms as a density, whereas the non-covariant variation usually only vanishes up to a total derivative.

\section{Equations of motion for constrained fields}
\label{sec:EOM}

In Section~\ref{sec:EOMchi}, we have already demonstrated that our pseudo-Lagrangian~\eqref{eq:Lag} is consistent with the duality equation 
\begin{align}
\label{eq:SD}
\mathcal{E}_I \coloneqq (\cM_{IJ} -\Omega_{IJ} ) F^J =0
\end{align}
that we impose on top of the Euler--Lagrange equations derived from $\mathcal{L}$. These are the equations of motion that the $E_{11}$ fields have to satisfy and they are obtained by varying with respect to the constrained scalar fields.
The main virtue of the pseudo-Lagrangian is that it also provides equations of motion for the constrained scalar fields and these are obtained by varying with respect to the $E_{11}$ fields in the pseudo-Lagrangian. This is what we shall present in detail in this section, explaining first the procedure and then analysing the result.

\subsection{Covariance of the field equations for indecomposable representations}
\label{sec:eomind}

The first question we address is what implications the indecomposable structure of the representation~\eqref{adjhatDef} involving the $E_{11}$ coset fields has. To this end it suffices to consider the general variation of the pseudo-Lagrangian with respect to $\cM$ and $\chi_M{}^\ta$. We first define the equations of motion following from varying the pseudo-Lagrangian by
\begin{align}
\delta \mathcal{L} = \mathcal{E}_\alpha \delta\phi^\alpha + \mathcal{E}^M_\ta \delta \chi_M{}^\ta\,,
\end{align}
where $\delta\phi^\alpha$ denotes the  left-invariant infinitesimal variation of the $E_{11}$ coset fields:
\begin{align}
\label{eq:dJ}
\cM^{-1} \delta \cM = \delta\phi_\alpha t^\alpha
\quad\quad \Rightarrow \quad\quad
\delta J_M{}^\alpha = \partial_M \delta\phi^\alpha + f_{\beta\gamma}{}^\alpha J_M{}^\beta \delta\phi^\gamma\, . 
\end{align} 
From the dressing by $\cM$, $\delta \phi_\alpha$ is not just the infinitesimal variation of the coset fields but includes an infinite sequence of Baker--Campbell--Hausdorff-like terms. Note that because $\chi_M{}^\ta$ is a constrained field, $ \mathcal{E}^M_\ta$ is only defined modulo a term that vanishes when contracted with a constrained vector. However, since we have obtained in Section~\ref{sec:EOMchi} that $ \mathcal{E}^M_\ta$  is proportional to the duality equation, we know this equation has to be satisfied and we can neglect this subtlety.

The indecomposable representation is such that under an ${\mf e}_{11}$ transformation with parameter $\Lambda_\beta$, one has 
\begin{subequations}
\begin{align}
\delta_\Lambda \delta \chi_M{}^\ta &= \Lambda_\beta \big[ T^{\beta\ta}{}_\tb \delta \chi_M{}^\tb + K^{\beta \ta}{}_{\alpha} \big(\partial_M\delta \phi^\alpha+ f_{\gamma \delta}{}^\alpha J_M{}^\gamma \delta\phi^\delta\big)- T^{\beta N}{}_M \delta\chi_N{}^\ta \big]\,,\\
\delta_\Lambda \delta \phi^\alpha &= \Lambda_\beta f^{\alpha\beta}{}_\gamma \delta \phi^\gamma\,.
\end{align}
\end{subequations}
Here, the definition of $\delta \phi^\alpha$ from~\eqref{eq:dJ} implies that it transforms as above consistently with the condition $\delta \phi_\alpha \cM^{-1}  t^{\alpha \dagger} \cM = \delta \phi_\alpha t^{\alpha}$.

As the pseudo-Lagrangian~\eqref{eq:Lag} is invariant under such rigid ${\mf e}_{11}$ transformations, $\delta_\Lambda \mathcal{L}=0$, we get
\begin{align}
\delta_\Lambda \delta \mathcal{L} 
 &= \delta_\Lambda \Big[ \mathcal{E}_{\tilde\alpha}^M   \delta \chi_M{}^{\tilde\alpha}+ \mathcal{E}_\alpha  \delta\phi^\alpha \Big]\nn\\
&= \Lambda_\beta\Big[ \Big( T^{\beta\ta}{}_{\tb} \delta \chi_M{}^\tb + K^{\beta\ta}{}_\alpha  (\partial_M\delta \phi^\alpha + f_{\gamma\delta}{}^\alpha J_M{}^\gamma \delta\phi^\delta) - T^{\beta N}{}_M \delta \chi_N{}^{\ta} \Big)\mathcal{E}_{\tilde\alpha}^M\nn\\
&\hspace{20mm} + f^{\alpha\beta}{}_\gamma \delta\phi^\gamma \mathcal{E}_\alpha \Big] + \delta\chi_M{}^{\tilde\alpha} \delta_\Lambda\mathcal{E}_{\tilde\alpha}^M + \delta\phi^\alpha \delta_\Lambda \mathcal{E}_\alpha \nn\\
&=0 \; . 
\end{align}
This determines the $E_{11}$-transformations of the equations of motion $\delta_\Lambda\mathcal{E}_{\tilde\alpha}^M$ and $\delta_\Lambda \mathcal{E}_\alpha$ as
\begin{subequations}
\begin{align}
\delta_\Lambda \mathcal{E}_{\tilde\alpha}^M&= -\Lambda_\beta \Big[ T^{\beta\tilde\alpha}{}_{\tilde\beta} \mathcal{E}_{\tilde\alpha}^M - T^{\beta M}{}_N \mathcal{E}_{\tilde\alpha}^N \Big]\,,\\
\label{eq:Ealpha}
\delta_\Lambda \mathcal{E}_\alpha&= -\Lambda_\beta \Big[ -f^{\beta\gamma}{}_{\alpha}\mathcal{E}_\gamma - K^{\beta\tilde\alpha}{}_\alpha \partial_M \mathcal{E}_{\tilde\alpha}^M + K^{\beta\tilde\alpha}{}_{\gamma} f_{\delta\alpha}{}^\gamma J_M^\delta \mathcal{E}_{\tilde\alpha}^M\Big]\,,
\end{align} 
\end{subequations}
from which we see that the equation $\mathcal{E}_\alpha$ obtained by varying the $E_{11}$ coset fields transforms with a cocycle under $E_{11}$. This is to be expected as the Euler--Lagrange equations are in the dual representation to that of the fields and thus the indecomposability is in the other direction.

As we saw in Section~\ref{sec:EOMchi} we have
\begin{align}
\label{eq:covchieom}
\mathcal{E}_\ta^M = -\frac12 C^{IM}{}_\ta \mathcal{E}_I
\end{align}
in terms of the duality equation~\eqref{eq:SD} and this is consistent with being an $E_{11}$-covariant object as the component $C^{IM}{}_\ta$ is a tensor under $E_{11}$. By contrast, $C^{IM}{}_\alpha$ is not a tensor and to obtain a covariant equation for the $\delta\phi^\alpha$ variation we have to combine the non-covariant equation $\mathcal{E}_\alpha$ from~\eqref{eq:Ealpha} with an appropriate projection of the duality equation. The correct choice is
\begin{align}
\label{eq:covSeom}
\widehat{\mathcal{E}}_\alpha = \mathcal{E}_\alpha -\frac12  C^{IM}{}_\alpha \partial_M \mathcal{E}_I +\frac12  f_{\gamma\alpha}{}^\beta J_M{}^\gamma C^{IM}{}_\beta \mathcal{E}_I
\end{align}
and it transforms covariantly under ${\mf e}_{11}$:  $\delta_\Lambda \widehat{\mathcal{E}}_\alpha = \Lambda_\beta f^{\beta\gamma}{}_\alpha \widehat{\mathcal{E}}_\beta$. 

In order to prove this, we use the notation of non-covariant $E_{11}$ transformations introduced in~\eqref{eq:DNC}. First,~\eqref{eq:Ealpha} can be rewritten as 
\begin{align}
\Delta^\beta \mathcal{E}_\alpha =  K^{\beta\ta}{}_\alpha \partial_M \mathcal{E}_\ta^M - K^{\beta\ta}{}_\gamma f_{\delta\alpha}{}^\gamma J_M{}^\delta \mathcal{E}_\ta^M\,,
\end{align}
while the other parts of~\eqref{eq:covSeom} transform as
\begin{align}
&\quad\quad \Delta^\beta \Big[ -\frac12C^{IM}{}_\alpha \partial_M\mathcal{E}_I + \frac12 f_{\gamma\alpha}{}^\delta J_M{}^\gamma C^{IM}{}_\delta \mathcal{E}_I\Big]\nn\\
&= \frac12 K^{\beta\ta}{}_\alpha C^{IM}{}_\ta \partial_M \mathcal{E}_I -\frac12 f_{\gamma\alpha}{}^\delta K^{\beta\ta}{}_\delta C^{IM}{}_\ta J_M{}^\gamma \mathcal{E}_I\nn\\
&= -K^{\beta\ta}{}_\alpha \partial_M \mathcal{E}_\ta^M + K^{\beta\ta}{}_\gamma f_{\delta\alpha}{}^\gamma J_M{}^\delta \mathcal{E}_\ta^M
\end{align}
where we used~\eqref{eq:DeltaC} and~\eqref{eq:covchieom}. Combining this equation with the previous one shows that~\eqref{eq:covSeom} is $E_{11}$-covariant and $\widehat{\mathcal{E}}_\alpha$ is the equation we shall now determine from the pseudo-Lagrangian~\eqref{eq:Lag}.

\subsection{\texorpdfstring{Varying with respect to the $E_{11} / K(E_{11})$ fields}{Varying with respect to the E11/K(E11) fields}}

We consider the variation defined in~\eqref{eq:dJ} that implies in the various $E_{11}$ representations
\begin{align}
\delta \cM^{MN} &= -T^{\alpha M}{}_P \cM^{PN} \delta \phi_\alpha\,,&
\hspace{10mm}
\delta J_{M \alpha} &= \partial_M \delta \phi_\alpha + f^{\beta\gamma}{}_\alpha J_{M \beta} \delta \phi_\gamma\,,
\nn\\
\delta \cM_{IJ} &=  T^{\alpha K}{}_I \cM_{JK} \delta \phi_\alpha\,,&
\hspace{10mm}
\delta \cM_{\tI\tJ} &=  T^{\alpha \tK}{}_\tI \cM_{\tJ\tK} \delta \phi_\alpha\,.
\end{align}
We start with the pseudo-Lagrangian \eq{SL}, and we first look at the combined variation of $\widetilde{\mathcal{L}}_{\text{kin}}$ and $\mathcal{L}_{\text{top}}$ as these are the only parts of the pseudo-Lagrangian where the indecomposable structure enters and for which the considerations from Section~\ref{sec:eomind} have to be taken into account.
The variation of $\widetilde{\mathcal L}_{\text{kin}}$ from~\eqref{tkin} with respect to the $E_{11}$ fields, up to a total derivative terms which we discard,  gives
\bea
\delta \Big[\widetilde{\mathcal{L}}_{\text{kin}}\Big] &=&  \delta\phi^\alpha \bigg[ \,\frac12 C^{IM}{}_\alpha \partial_M \Big( \cM_{IJ} F^J\Big) + \frac12 \cM_{IJ} F^J C^{IM}{}_\gamma f_{\alpha\beta}{}^\gamma J_M{}^\beta  -\frac14 T_\alpha{}^K{}_I \cM_{JK} F^I F^J \bigg]
\nn\\
&=& \delta\phi^\alpha \bigg[ \, \frac12 C^{IM}{}_\alpha \partial_M \mathcal{E}_I + \frac12 f_{\alpha\beta}{}^\gamma  C^{IM}{}_\gamma J_M{}^\beta   \mathcal{E}_I  -\frac14 T_\alpha{}^K{}_I \cM_{JK} F^I F^J \
\nn\\
&& \quad + \frac14\Omega_{IJ} \bigg( 2F^J C^{MI}{}_\gamma f_{\alpha\beta}{}^\gamma J_M{}^\beta 
-C^{IM}{}_\alpha C^{JN}{}_\beta f_{\gamma\delta}{}^\beta J_M{}^\gamma J_N{}^\delta 
+2 C^{IM}{}_\alpha C^{JN}{}_\tb \partial_{[M}\chi_{N]}{}^\tb 
\nn\\
&&  \hspace{20mm}  +2 C^{IM}{}_\alpha C^{JN}{}_\hL \partial_{M}\zeta_{N}{}^\hL \bigg)
\bigg]\ ,
\label{v1}
\eea
where we have used 
 the Bianchi identity from~\cite[Eq.~(3.42)]{Bossard:2019ksx}:
\begin{align}
\label{eq:BI}
\Omega_{IJ} C^{IM}{}_\alpha \partial_M F^J & =- \frac12 \Omega_{IJ} C^{IM}{}_\alpha C^{JN}{}_\beta f_{\gamma\delta}{}^{\beta} J_M{}^\gamma J_N{}^\delta  + \Omega_{IJ} C^{IM}{}_\beta C^{JN}{}_{\tilde\beta} \partial_{[M} \chi_{N]}{}^{\tilde\beta}\nn\\
&\quad +\Omega_{IJ} C^{IM}{}_\alpha C^{JN}{}_\hL \partial_{M} \zeta_{N}{}^\hL\,.
\end{align}
The first two terms in \eq{v1} are the appropriate covariantisation terms in accordance with \eq{eq:covSeom}.

The variation of ${\mathcal L}_{\text{top}}$ from~\eqref{eq:Ltop}, up to total derivative terms which we discard, is given by
\begin{align}
\delta \mathcal{L}_{\text{top}} &=  \delta\phi^\alpha\bigg[ \Pi_\ta{}^{MN} \bigg( -\frac12 T_\alpha{}^ \ta{}_\tb \partial_M \chi_N{}^\tb + \frac12 f_{\gamma\alpha}{}^\beta  T_\beta{}^\ta{}_\tb J_M{}^\gamma \chi_N{}^\tb +\frac12 K_{[\alpha}{}^\ta{}_{\beta]}  f^{\beta}{}_{\gamma\delta} J_M{}^\gamma J_N{}^\delta 
\nn\\
&\qquad   - K_{[\beta}{}^\ta{}_{\gamma]} f_{\delta\alpha}{}^\beta J_M{}^\gamma J_N{}^\delta\bigg)+ \frac12 \Omega_{IJ} C^{IM}{}_\beta C^{JN}{}_\hL\left(    \delta_\alpha^\beta\, \partial_M \zeta_N{}^\hL - f_{\gamma\alpha}{}^\beta J_M{}^\gamma \zeta_N{}^\hL \right) \bigg]\ ,
\end{align}

Combining the variations  $ \delta \widetilde{\mathcal{L}}_{\text{kin}}$ and $ \delta \mathcal{L}_{\text{top}}$ leads to 
\bea
\delta \Big[\widetilde{\mathcal{L}}_{\text{kin}}+ \mathcal{L}_{\text{top}}\Big] &=& \delta\phi^\alpha \bigg[ \, \frac12 C^{IM}{}_\alpha \partial_M \mathcal{E}_I + \frac12 f_{\alpha\beta}{}^\gamma  C^{IM}{}_\gamma J_M^\beta   \mathcal{E}_I  -\frac14 T_\alpha{}^K{}_I \cM_{JK} F^I F^J 
\label{pt}\\
&&
-\frac12 \Pi_\ta{}^{MN} T_\alpha{}^\ta{}_\tb \Theta_{MN}{}^\tb + \Omega_{IJ} C^{IM}{}_\beta C^{JN}{}_\hL \left(    \delta_\alpha^\beta\, \partial_{M} \zeta_{N}{}^\hL - \frac12 f_{\gamma\alpha}{}^\beta J_M{}^\gamma \zeta_N{}^\hL \right)
\nn\\
&& 
 +\frac12 \Pi_\ta{}^{MN} T_\gamma{}^\ta{}_\tb  J_M{}^\gamma \left( T_\alpha{}^\tb{}_\tg J_M{}^\gamma  \chi_N{}^\tg + K_\alpha{}^\tb{}_\delta J_N{}^\delta\right)   + \frac12 \Omega_{IJ} F^J C^{IM}{}_\beta f^\beta{}_{\alpha\gamma} J_M{}^\gamma
\bigg]\ ,
\nn
\eea
where \eq{eq:Ttrm}, \eq{eq:Ktrm}, \eq{eq:ID4} and \eq{eq:ID2} have been used. 

There remains the variation of $\widetilde{\mathcal L}_{\text{pot}}$ defined in \eq{tpot}. Adding its variation to \eq{pt}  gives the field equation $\widehat{\mathcal E}^{\alpha}=0$ for the constrained fields, where
\bea
\widehat{\mathcal{E}}^{\alpha} \hspace{-0.5mm} &=& \hspace{-0.5mm}  \frac12\left(\cM^{\alpha\beta}  \hspace{-0.5mm} +  \hspace{-0.5mm}  \kappa^{\alpha\beta}\right) \bigg[
-\frac12 \Pi_\ta{}^{MN} T_\beta{}^\ta{}_\tb \Theta_{MN}{}^\tb + \Omega_{IJ} C^{IM}{}_\gamma C^{JN}{}_\hL  \Bigl(    \delta_\beta^\gamma\, \partial_{M} \zeta_{N}{}^\hL - \frac12 f_{\delta\beta}{}^\gamma J_M{}^\delta \zeta_N{}^\hL \Bigr) 
\nn\\
&&
\hspace{-4mm} -\frac14 T_\beta{}^K{}_I \cM_{JK} F^I F^J + \frac12 \Omega_{IJ} F^J C^{IM}{}_\gamma f^\gamma{}_{\beta\delta} J_M{}^\delta +\frac12 \Pi_\ta{}^{MN} T_\gamma{}^\ta{}_\tb  J_M{}^\gamma \! \left( T_\beta{}^\tb{}_\tg  \chi_N{}^\tg + K_\beta{}^\tb{}_\delta J_N{}^\delta \right) 
\nn \\
&& 
\hspace{-1mm}-\frac12 \partial_M \left( \kappa^{M \hspace{-0.3mm} N}_{PQ\beta\gamma}\, \cM^{PQ} J_N{}^\gamma \right)
-\frac14 \Big( \kappa^{M \hspace{-0.3mm} N}_{PQ\gamma\delta}\, T_\beta{}^P{}_S \cM^{SQ} -   2\kappa^{M \hspace{-0.3mm} N}_{PQ\epsilon\delta}\,\cM^{PQ} f^\epsilon{}_{\gamma\beta}\Big)  J_M{}^\gamma J_N{}^\delta 
\bigg]  \; ,  \label{ve2}
\eea
and for brevity in notation we have defined the $E_{11}$ invariant tensor 
\be
\kappa_{PQ\alpha\beta}^{M \hspace{-0.3mm} N} \equiv \kappa_{\alpha\beta}\, \delta^M_P \delta^N_Q -2 \,T_\alpha{}^M{}_P T_\beta{}^N{}_Q\ , 
\ee
and the projector is required because $\delta\phi^\alpha$ satisfies 
\be 
\cM_{\alpha\beta}   \delta\phi^\beta  =   \kappa_{\alpha\beta}   \delta\phi^\beta  \; . 
\ee
The equation of motion~\eqref{ve2} is indeed an $E_{11}$ tensor. To see this we note that $\Omega_{IJ} C^{IM}{}_\beta C^{JN}{}_\hL$ is an invariant tensor and one checks that 
\be \Delta^\gamma \Bigl[  \frac12 \Omega_{IJ} F^J C^{IM}{}_\beta f^\beta{}_{\alpha\delta} J_M{}^\delta +\frac12 \Pi_\ta{}^{MN} T_\gamma{}^\ta{}_\tb  J_M{}^\gamma \left( T_\alpha{}^\tb{}_\tg  \chi_N{}^\tg + K_\alpha{}^\tb{}_\delta J_N{}^\delta\right)   \Bigr] = 0 \; ,  \ee
using the identities \eqref{eq:ID3} and \eqref{eq:ID2}.

To understand the implications of Equation~\eqref{ve2}, it is useful to look at its linearised approximation. At linearised order one gets with $J_M{}^\alpha = \partial_M \phi^\alpha$ that \eqref{ve2} simplifies to
\begin{multline}
\left(\eta^{\alpha\beta} + \kappa^{\alpha\beta}\right) \Omega_{IJ} C^{IM}{}_\beta \left( C^{JN}{}_\ta \partial_{[M}\chi_{N]}{}^\tb 
-C^{JN}{}_\hL  \partial_M \zeta_N{}^\hL \right) \\
=   \left(\eta^{\alpha\gamma} + \kappa^{\alpha\gamma}\right) \left( \frac12 \eta^{MN} \kappa_{\gamma\beta} - 
T_{\beta}{}^M{}_P  T_\gamma{}^N{}_Q\,  \eta^{PQ} \right) \partial_M\partial_N \phi^\beta \ ,
\label{LEN}
\end{multline}
where the index $\alpha$ is explicitly projected to $\mf{e}_{11}\ominus K(\mf{e}_{11})$. This equation, together with the linearised duality equation
\begin{align} \label{LinearisedDuality}
(\eta_{IJ}-\Omega_{IJ}) \big(C^{JN}{}_\beta \partial_N \phi^\beta + C^{JN}{}_\ta \chi_N{}^\ta  +C^{JN}{}_\hL \zeta_N{}^\hL \big) = 0 \ ,
\end{align}
define the linearised equations of $E_{11}$ exceptional field theory. One can anticipate that the right-hand side of \eqref{LEN} looks like a propagating equation for the $E_{11}$ coset fields, whereas its left-hand side takes the form of an integrability condition for the constrained fields. One may have hoped that the linearised duality equation \eqref{LinearisedDuality} allowed setting both sides of \eqref{LEN} to zero, but the structure of these equations is more complicated than this. To extract the propagating degrees of freedom of the theory one needs to consider a particular solution to the section constraint and analyse these equations in an appropriate level decomposition. We will carry out this analysis for the $GL(11)$ level decomposition in Section \ref{HigherLevelSection}.

\section{Analysing the pseudo-Lagrangian in level decomposition}
\label{sec:SF}

The definition of a Kac--Moody Lie group from a Kac--Moody algebra is subtle because the formal exponential of generic Lie algebra elements diverges. In particular, the current $J_M = \cM^{-1} \partial_M \cM$ defining the theory is not well-defined for $\cM = \cV^\dagger \eta \cV$ and $\cV$ a product of exponentials of generators in the positive Borel subgroup. Likewise, the na\"ive duality equation~\eqref{eq:DE} is ill-defined since the matrix $\cM_{IJ}$ is the product of an infinite lower- and an infinite upper-triangular matrix, leading to infinite sums that do not converge. We will see that the theory is nonetheless well-defined in the unendlichbein formulation, modulo mathematical subtleties related to the definition of the Kac--Moody group that we discuss below. Since we want to recover (exceptional) field theory in $D$ dimensions it will be more convenient to use a hybrid formulation that was introduced in~\cite{Bossard:2019ksx} and that we shall refer to as the `semi-flat formulation' here. In this section, we review this formulation and write the $E_{11}$ pseudo-Lagrangian \eqref{eq:Lag} in a way that is appropriate for level decomposition.  

\subsection{The semi-flat formulation}
\label{sec:semiflat}

For any Cartan generator of $\mf{e}_{11}$ with integer eigenvalues we can decompose the algebra $\mf{e}_{11}$ into eigenspaces of fixed (adjoint action) eigenvalue $k\in\ints$ that we call \textit{level}
\begin{align}
\label{LevelAlgebra}  
\mf{e}_{11} = \bigoplus_{k=-\infty}^{-1} \overline{\mathfrak{u}}^\ord{-k} \oplus \mf{l}^\ord{0} \oplus \bigoplus_{k=1}^\infty \mathfrak{u}^\ord{k} \; , 
\end{align}
where $\mf{l}^\ord{0}$ is a reductive Levi subalgebra and $\mf{u} = \bigoplus_{k=1}^\infty \mathfrak{u}^\ord{k}$ a `nilpotent' subalgebra including all strictly positive levels $k$.\footnote{The algebra $\bigoplus_{k>1} \mf{u}^\ord{k}$ is not strictly nilpotent as the level goes to infinity. However, this terminology for the algebra is convenient and suggestive, as is `unipotent' for its exponential image.} The typical example we have in mind is when the Cartan generator is $H_{\Lambda_D}$ for the fundamental weight associated with some node $D$ of the Dynkin diagram and then~\eqref{LevelAlgebra} is the level decomposition of the type studied in~\cite{West:2002jj,Nicolai:2003fw}.

We work in an Iwasawa patch\footnote{Since the subalgebra $K(\mf{e}_{11})$ is fixed by the temporal involution, there is no global Iwasawa decomposition. In particular, time-like U-dualities, mediated by Weyl reflections associated to compact coset elements, do not preserve the Iwasawa patch connected to the identity. The resulting group element under such an action has to be decomposed in a different patch which can also correspond to a different signature of space-time~\cite{Keurentjes:2004bv}. An analogy with pure gravity is to write the metric in ADM form everywhere which is possible only for globally hyperbolic space-times.} where any element of the Kac--Moody group can be decomposed as
\be  g = h\hspace{0.1mm} l u \; , \qquad \text{for } h \in K(E_{11})\; , \quad l \in L\; , \quad u \in U\; , \ee
with $L$ the  Levi subgroup with Lie algebra $\mf{l}=\mf{l}^\ord{0}$ (that we assume finite-dimensional in this section, for example $GL(11)$ or $GL(3)\times E_8$) and $U$ the unipotent subgroup associated to $\mf{u}$. We write $\cV$ in this gauge as
\begin{align}
\label{eq:pargauge}
\cV = v\, \cU
\end{align}
with $v\in L$  the coset representative of the finite-dimensional symmetric space $K(L) / L $ and $\cU \in U$. From the Levi element $v$ we define
\begin{align}
\label{eq:LeviJ}
m = v^\dagger \eta v\,,\quad\quad  m^{-1} \partial_M m = v^{-1} \big(\partial_M v v^{-1} +\eta^{-1} (\partial_M v v^{-1})^\dagger \eta \big)v\,.
\end{align}
Recalling that $\cM = \cV^\dagger \eta \cV = \cU^\dagger  m \cU$, the $E_{11}$ current $J_M$ introduced in~\eqref{eq:curdef} then becomes
\begin{align}
J_M = \cM^{-1} \partial_M \cM = \cU^{-1} \big(m^{-1} \partial_M m+ \partial_M \cU \cU^{-1} + m^{-1}  (\partial_M\cU \cU^{-1})^\dagger  m\big) \cU
\end{align}
and is still ill-defined. Conjugating with $\cU$, however, produces the well-defined object
\begin{align}
\label{eq:SFcur}
\tilde{\cJ}_M = \cU J_M \cU^{-1} =m^{-1} \partial_M m +  \cN_M + m^{-1}\cN_M^\dagger m
\end{align}
with $\cN_M = \partial_M \cU \cU^{-1}$. This is well-defined since $\cN_M$ can be expanded to any order in the level associated to $L$ using only polynomials and $m$ acts on each level component as a finite-dimensional $L$ matrix. All the $E_{11}$ modules we consider in this paper are integrable, and the components of $\cN_M$ on an integrable module vanish at some finite order by definition. In components we write 
\begin{align}
J_{M}{}^\alpha = \cU^{-1\alpha}{}_\beta \tilde{\cJ}_M{}^\beta\ .
\end{align}
The corresponding field redefinition of the constrained fields is
\begin{align}
\label{eq:SFCF}
\chi_M{}^\ta = \cU^{-1\ta}{}_\tb \tilde{\chi}_M{}^\tb + \omega_\beta^\ta(\cU^{-1}) \tilde{\cJ}_M{}^\beta\,,\quad \zeta_M{}^\Lambda = \cU^{-1\Lambda}{}_\Xi\tilde{\zeta}_M{}^\Xi\ ,\quad   \zeta_M{}^\tL = \cU^{-1\tL}{}_\tXi \tilde\zeta_M{}^\tXi \,,
\end{align}
with the indecomposable structure under the action of (the unipotent subgroup) of $E_{11}$ entering for $\chi$ through the group cocycle $\omega$ whose infinitesimal version is $K^{\alpha\ta}{}_\beta$ introduced in~\eqref{eq:T0CR} (see~\cite[Eq.~(2.4)]{Bossard:2019ksx}). Note, in particular, that one can write $J_M{}^\wa = \cU^{-1\wa}{}_\wb  \tilde\cJ_M{}^\wb$ with $\tilde{\cJ}_M{}^\ta=\tilde{\chi}_M{}^\ta$. The semi-flat field strength,  defined by $\tilde{\cF}^I=\cU^{-1I}{}_J F^J$,  is
\begin{align}
\label{eq:FSSF}
\tilde{\cF}^I = \cU^{-1N}{}_M \left( C^{IM}{}_\wa  \tilde{\cJ}_N{}^\wa + C^{IM}{}_\hL  \tilde\zeta_N{}^\hL\right)\ .
\end{align}
The point of all these redefinitions is that the duality equation~\eqref{eq:DE} becomes
\begin{align}
\label{eq:DElev}
m_{IJ} \tilde{\cF}^J = \Omega_{IJ} \tilde{\cF}^J
\end{align}
and the bilinear form $m_{IJ}$ now acts as a well-defined block diagonal matrix level-by-level on the level decomposition of the field strength representation $\cT_{-1}$. Recall that $m=\cU^{-1\dagger} \cM \cU^{-1}=v^\dagger \eta v$ can be simply computed level by level using the bilinear form $\eta$ and we have used invariance of the symplectic form $\cU \Omega \cU^{\dagger}=\Omega$ here. 

\subsection{The pseudo-Lagrangian}

The redefinition applied to the  pseudo-Lagrangian~\eqref{eq:Lag} also gives a well-defined pseudo-Lagran\-gian for $m$, $\cU$, $\tilde{\chi}_{M}{}^{\tilde{\alpha}}$ and $\tilde{\zeta}_M{}^{\hL}$.  For most terms of the  pseudo-Lagrangian, $E_{11}$ invariance allows us to simply substitute the tilde fields for the original ones, but some more care is needed to obtain the topological term because its $E_{11}$ invariance is not manifest (see~\eqref{NonManifestE11}). The main part~\eqref{eq:Theta} of the topological term involves the field strength 
\begin{align}
\label{eq:topSF}
\Theta_{MN}{}^\ta &\underset{[MN]}{=} 2 \partial_{M} \big( \cU^{-1\ta}{}_\tb \tilde{\chi}_N{}^\tb + \omega_\beta^\ta(\cU^{-1}) \tilde{\cJ}_N{}^\beta\big) +  \cU^{-1\alpha}{}_\beta \tilde{\cJ}_M{}^\beta T_\alpha{}^{ \ta}{}_\tg  \big( \cU^{-1\tg}{}_\tb \tilde{\chi}_N{}^\tb + \omega_\gamma^\tg(\cU^{-1}) \tilde{\cJ}_N{}^\gamma\big)\nn\\
&\hspace{10mm} + \tilde{\cJ}_M{}^\gamma \cU^{-1 \alpha}{}_\gamma K_{[\alpha}{}^\ta{}_{\beta]}\cU^{-1 \beta}{}_\delta \tilde{\cJ}_N{}^\delta\,,
\end{align}
where the right-hand side is to be anti-symmetrised in $[MN]$.
To simplify this we note that the indecomposable action of $E_{11}$ implies from the non-tensorial nature of the cocycle that
\begin{align}
\label{eq:DKE11}
\cU^{-1 \ta}{}_\tb \cU^{-1\alpha}{}_\gamma \cU^{\delta}{}_\beta K^{\gamma \tb}{}_\delta &= K^{\alpha \ta}{}_\beta +  \omega^\tb_\delta(\cU^{-1}) U^\delta{}_\beta T^{\alpha\ta}{}_\tb + \omega^\ta_\gamma (\cU^{-1})\cU^\gamma{}_\delta f^{\alpha \delta}{}_\beta\nn\\
\Longleftrightarrow \quad\quad \cU^{-1 \ta}{}_\tb K^{\alpha \tb}{}_\tb &= U^{\alpha}{}_\gamma K^{\gamma \ta}{}_\delta \cU^{-1 \delta}{}_\beta + \cU^\alpha{}_\gamma T^{\gamma \ta}{}_\tb \omega^\tb_\beta(\cU^{-1}) + \omega^\ta_\gamma(\cU^{-1}) f^{\alpha\gamma}{}_\beta\,.
\end{align}
This follows from exponentiation of~\eqref{eq:Dcoc}.\footnote{   We recall that the action of the group $E_{11}$ on $\mf{e}_{11}\oleft R(\Lambda_2)$ is defined by $g^{-1} t^\alpha g = g^\alpha{}_\beta t^\beta\,,\quad g^{-1} \tilde{t}^{\tilde\alpha} g = g^{\tilde\alpha}{}_{\tilde\beta} \tilde{t}^{\tilde\beta} + \omega^{\tilde\alpha}_\beta(g) t^\beta\,,$ such that $g_1^\alpha{}_\beta g_2^\beta{}_\gamma = (g_1g_2)^\alpha{}_\gamma$ and where $\omega^{\tilde\alpha}_\beta(g)$ is a group 1-cocycle satisfying
$\omega^{\tilde\alpha}_\beta(g_1g_2) = \omega^{\tilde\alpha}_\gamma(g_1) g_2^\gamma{}_\beta + g_1^{\tilde\alpha}{}_{\tilde\gamma} \omega^{\tilde\gamma}_\beta(g_2)$.}

One can also check that this is the identity that is needed for $\Theta_{MN}{}^\ta$ to transform covariantly when acting on the $\ta$ index, using
\begin{align}
\chi_M{}^\ta \to g^{-1 \ta}{}_\tb \chi_M{}^\tb + \omega^\ta_\beta(g^{-1}) J_M{}^\beta\,,\quad\quad J_M{}^\alpha \to g^{-1 \alpha}{}_\beta J_M{}^\beta
\end{align}
and the Maurer--Cartan equation, similar to the infinitesimal calculation in \eqref{NonManifestE11}. 

Using~\eqref{eq:DKE11} in the topological term~\eqref{eq:topSF} leads to
\begin{align}
\label{eq:topSF1}
\Theta_{MN}{}^\ta 
&\underset{[MN]}{=} 2 \partial_{M} \big( \cU^{-1\ta}{}_\tb \tilde{\chi}_N{}^\tb \big) + 2 \partial_M \big(\omega_\beta^\ta(\cU^{-1})  \cU^\beta{}_\gamma \big) \cU^{-1 \gamma}{}_\delta\tilde{\cJ}_N{}^\delta\nn\\
&\hspace{10mm} +\cU^{-1\ta}{}_\tb \tilde{\cJ}_M{}^\alpha \big( T_\alpha{}^\tb{}_\tg \tilde{\chi}_N{}^\tg  + K_\alpha{}^\ta{}_\beta \tilde{\cJ}_N{}^\beta\big)\,.
\end{align}
This can be further simplified by computing the derivative of the group cocycle as follows
\begin{align}
\partial_M \big(\omega_\beta^\ta(\cU^{-1})\big) 
&=
\frac{d}{dt} \omega_\beta^\ta\big( \cU^{-1} \exp(-t \cN_M)\big)\big|_{t=0}
\nn\\
&=\frac{d}{dt} \Big[ \omega^\ta_\gamma(\cU^{-1}) \exp(-t\cN_M)^\gamma{}_\beta + \cU^{-1\ta}{}_\tb \omega_\beta^\tb\big(\exp(-t\cN_M)\big)\Big]_{t=0}\ .
\end{align}
Using $\cN_M = \partial_M \cU \cU^{-1}$ in the first term, and that $\omega_\beta^\ta \left(\exp(-t\cN)\right) = -t \Lambda_\gamma K^{\gamma\ta}{}_\beta + O(\Lambda^2)$ in the second term, it follows that
\be
\partial_M \big(\omega_\beta^\ta(\cU^{-1})  \cU^\beta{}_\gamma \big)\, U^{-1\gamma}{}_\delta = - \cU^{-1 \ta}{}_\tb K_\gamma{}^\tb{}_\delta \cN_M{}^\gamma\ .
\ee
Substituting this into~\eqref{eq:topSF1} leads to
\begin{align}
\label{eq:topSF2}
\Theta_{MN}{}^\ta &\underset{[MN]}{=} 2 \partial_{M} \big( \cU^{-1\ta}{}_\tb \tilde{\chi}_N{}^\tb \big) +\cU^{-1\ta}{}_\tb \Big[ \tilde{\cJ}_M{}^\alpha T_\alpha{}^\tb{}_\tg \tilde{\chi}_N{}^\tg  + ( \tilde{\cJ}_M{}^\alpha - 2\cN_M{}^\alpha) K_\alpha{}^\tb{}_\beta \tilde{\cJ}_N{}^\beta\Big]\,.
\end{align}
 
The similar analysis for the other constrained fields $\zeta_M{}^\hL$ is simpler since they are not in an indecomposable representation. Including their semi-flat version~\eqref{eq:SFCF} leads to the following semi-flat topological term
\begin{align}
\mathcal{L}_{\text{top}} &= \frac12 \Pi_\ta{}^{PQ} \cU^{-1 M}{}_P \cU^{-1 N}{}_Q
\Big[ \tilde{\cJ}_M{}^\alpha T_\alpha{}^\ta{}_\tb \tilde{\chi}_N{}^\tb  + ( \tilde{\cJ}_M{}^\alpha - 2\cN_M{}^\alpha) K_\alpha{}^\ta{}_\beta \tilde{\cJ}_N{}^\beta\Big]\nn\\
&\quad 
-\frac12 \Omega_{IJ} C^{IP}{}_\wa C^{JQ}{}_\hL \cU^{-1 M}{}_P \cU^{-1 N}{}_Q \tilde{\cJ}_M{}^\wa \tilde{\zeta}_N{}^\hL+\Pi_\ta{}^{MN} \partial_{M} \big( \cU^{-1\ta}{}_\tb \tilde{\chi}_N{}^\tb \big)\,,
\label{rtop}
\end{align}
where we have used the $E_{11}$ invariance of the tensors appearing in the pseudo-Lagrangian.
 Note that we could drop the last total derivative and the resulting pseudo-Lagrangian would still be gauge invariant up to a total derivative.\footnote{If one instead drops $2\partial_{[M}\chi_{N]}{}^\ta$, gauge invariance is only true up to a more complicated total derivative mixing linear and non-linear terms.} We shall nonetheless keep this term such that the pseudo-Lagrangian transforms as a density under generalised diffeomorphisms. 
Because
\begin{align}\label{JminusN} 
\tilde{\cJ}_M - 2\cN_M = m^{-1} \partial_M m - \cN_M + m^{-1} \cN_M^\dagger m
\end{align}
compared to~\eqref{eq:SFcur}, the sign of the positive level components changes and in this way the cocycle $K_{\alpha}{}^{\ta}{}_\beta$ appears symmetrised in the adjoint indices for high enough levels, rather than antisymmetrised as in~\eqref{eq:Theta}.

The rewriting of all other terms in the pseudo-Lagrangian is straightforward and similar to what we just saw for the fields $\zeta_M{}^\hL$.

We note that there is still an action of the unipotent matrix $\cU$ on the constrained space-time indices of the current and the constrained fields. $\cU^{-1 N}{}_M \partial_N $ gives infinite sums for an arbitrary unconstrained derivative, but we shall always consider the class of solutions to the section constraint associated to a finite-dimensional Levi subgroup $L$, such that only finitely many components of $\partial_M$ of bounded level are non-zero and $\cU^{-1 N}{}_M \partial_N $ is well-defined.\footnote{One can define the subgroup $U_k \subset U$ associated to the Lie algebra $\mf{u}_{\ge k} = \bigoplus_{\ell=k}^\infty \mf{u}_\ell$, and for some finite $k = k_{\rm s}$ we will have that $\cU^{-1 N}{}_M \partial_N $  only depends on the finite-dimensional equivalence class of $\cU$ in $U / U_{k_{\rm s}{+}1}$.} This matrix action is important when considering the embedding of $GL(D) \times E_{11-D}$ exceptional field theory in $E_{11}$ exceptional field theory as we shall see in more detail in Section~\ref{sec:E8}. For $GL(11)$ and the $D=11$ solution of the section constraint for which the fields only depend on the eleven coordinates $x^m$, it has no effect since $\cU$ does not contain any $GL(11)$ component and so leaves $\partial_m$ invariant and we have in this case
\be 
\cU^{-1 N}{}_M \tilde{\cJ}_N{}^\alpha = \delta_M^m  \tilde{\cJ}_m{}^\alpha \; , \quad \cU^{-1 N}{}_M \tilde{\chi}_N{}^{\tilde{\alpha}} = \delta_M^m   \tilde{\chi}_m{}^{\tilde{\alpha}}  \; , \quad \cU^{-1 N}{}_M \tilde{\zeta}_N{}^{\widehat{\Lambda}} = \delta_M^m  \tilde{\zeta}_m{}^{\widehat{\Lambda}}\; . 
\ee

\subsection{Gauge transformations and compensators}
\label{sec:GT}

The action of the generalised diffeomorphisms on the fields $m$ and $\cU$ requires a compensating $K({\mf e}_{11})$ rotation as in \cite{Bossard:2019ksx}, such that it preserves the parabolic gauge condition \eqref{eq:pargauge}. The general formula can be given in level decomposition~\eqref{LevelAlgebra} of ${\mf e}_{11}$. We write the generators appearing at level $k$ as $t^{\alpha_\dgr{k}}$ and similarly for the associated structure constants. The generalised diffeomorphism acting on $\cV$ with a $K({\mf e}_{11})$ compensator to restore the parabolic gauge~\eqref{eq:pargauge} is then 
\begin{align} \delta_\xi \cV &= \xi^M \partial_M \cV +  T_{\alpha}{}^M{}_N \partial_M \xi^N \cV t^\alpha -\sum_{k=1}^\infty T_{\alpha_{\dgr{-k}}\!\!}{}^M{}_N \cV^{-1P}{}_M \cV^N{}_Q \partial_P \xi^Q  ( t^{\alpha_\dgr{-k}} -  \eta^{-1}  t^{\alpha_\dgr{-k} \dagger} \eta ) \cV\\
&= \xi^M \partial_M \cV +  \cU^{-1P}{}_M \cU^N{}_Q \partial_P \xi^Q    \biggl( \sum_{k=0}^\infty T_{\alpha_\dgr{k} }{}^M{}_N  v  t^{\alpha_\dgr{k}}  +\sum_{k=1}^\infty T_{\alpha_{\dgr{-k}}\!\!}{}^M{}_N  \eta^{-1} v^{-1\dagger} t^{\alpha_\dgr{-k} \dagger} m \biggr) \cU \; . \nn  
\end{align}
For $k\ge 1$, both $t^{\alpha_\dgr{k}} \in \mf{u}_k$ and $ \eta^{-1}  t^{\alpha_\dgr{-k} \dagger} \eta \in  \mf{u}_k$.  For a choice of section adapted to the Levi subgroup $L$, $T^{\alpha_\dgr{k} N}{}_M \partial_N = 0 $ for $k>k_{\rm s}$ for some finite $k_{\rm s}$, and the compensating $K({\mf e}_{11})$ transformation in the first line is a finite sum with $k\le k_{\rm s}$. For $GL(11)$, $k_{\rm s}=0$ and for $GL(D) \times E_{11-D}$, $k_{\rm s}=1$.  

This gives the gauge transformations 
\begin{subequations}
\begin{align}
\delta_\xi m &= \xi^M \partial_M m  + T_{\alpha_{\dgr{0}}\!\!}{}^M{}_N \cU^{-1P}{}_M \cU^N{}_Q \partial_P \xi^Q \bigl( m t^{\alpha_{\dgr{0}}} +t^{\alpha_{\dgr{0}} \dagger } m\bigr)  \; , \\
\delta_\xi \cU  &= \xi^M \partial_M \cU + \cU^{-1P}{}_M \cU^N{}_Q \partial_P \xi^Q  \sum_{k=1}^\infty   \Bigl(T_{\alpha_{\dgr{k}}\!\!}{}^M{}_N t^{\alpha_{\dgr{k}}} + T_{\alpha_{\dgr{-k}}\!\!}{}^M{}_N m^{-1}  t^{\alpha_{\dgr{-k}}\dagger} m \Bigr) \cU  \; . 
\end{align}
\end{subequations}
One obtains as well for the current and the constrained fields 
\begin{subequations}
\label{gaugeflatten}
\begin{align}
\label{eq:currentxiflatten}
\delta_\xi \tilde{\cJ}_M{}^\alpha &=  \xi^N \partial_N \tilde{\cJ}_M{}^{\alpha}  + \partial_M \xi^N \tilde{\cJ}_N{}^{\alpha} + \cU^{-1 Q}{}_N \cU^P{}_R \partial_Q \xi^R  T_{\gamma_{\dgr{0}}\!\!}{}^N{}_P    f^{\gamma_{\dgr{0}}\alpha}{}_{\beta} \tilde{\cJ}_M{}^\beta   
\\
& \quad  + \sum_{k=1}^{k_{\rm s}}  \cU^{-1 Q}{}_N \cU^P{}_R \partial_Q \xi^R  \Bigl( T_{\gamma_{\dgr{-k}}\!\!}{}^N{}_P     f^{\gamma_{\dgr{-k}}\alpha}{}_{\beta} -m^{NS} m_{PU} T_{\gamma_{\dgr{k}}\!\!}{}^U{}_S      f^{\gamma_{\dgr{k}}\alpha}{}_{\beta} \Bigr)  \tilde{\cJ}_M{}^{\beta} 
\nn\\
&\qquad  + T^{\alpha R}{}_S  \cU^{-1 N}{}_R \cU^S{}_P \left( \partial_M\partial_N \xi^P + m_{NQ} m^{PU} \partial_M\partial_U\xi^Q\right) \,, \nn  \\
\label{eq:chiGTflatten}
\delta_\xi \tilde{\chi}_M{}^\ta &=  \xi^N \partial_N \tilde{\chi}_M{}^{\tilde\alpha}  + \partial_M \xi^N \tilde{\chi}_N{}^{\tilde\alpha} - \cU^{-1 Q}{}_N \cU^P{}_R \partial_Q \xi^R  T_{\alpha_{\dgr{0}}\!\!}{}^N{}_P    T^{\alpha_{\dgr{0}}\tilde\alpha}{}_{\tilde\beta} \tilde{\chi}_M{}^\tb   
\\
& \quad  -\sum_{k=1}^{k_{\rm s}} \cU^{-1 Q}{}_N \cU^P{}_R \partial_Q \xi^R  \Bigl[ T_{\alpha_{\dgr{-k}}\!\!}{}^N{}_P   \Bigl(    T^{\alpha_{\dgr{-k}}\tilde\alpha}{}_{\tilde\beta} \tilde{\chi}_M{}^\tb + K^{\alpha_{\dgr{-k}}\tilde\alpha}{}_{\beta} \tilde{\cJ}_M{}^\beta  \Bigr)  \nn\\
& \hspace{60mm} -m^{NS} m_{PU} T_{\alpha_{\dgr{k}}\!\!}{}^U{}_S   \Bigl(    T^{\alpha_{\dgr{k}}\tilde\alpha}{}_{\tilde\beta} \tilde{\chi}_M{}^\tb + K^{\alpha_{\dgr{k}}\tilde\alpha}{}_{\beta} \tilde{\cJ}_M{}^\beta  \Bigr) \Bigr]
\nn\\
& + T^{\tilde\alpha R}{}_S  \cU^{-1 N}{}_R \cU^S{}_P \left( \partial_M\partial_N \xi^P + m_{NQ} m^{PU} \partial_M\partial_U\xi^Q\right)+ {\Pi^{\tilde\alpha}{}_{ Q S }} m^{RQ} \cU^{-1 N}{}_R \cU^S{}_P \partial_M \partial_N \xi^P \,,\nn \\
\delta_\xi \tilde{\zeta}_M{}^\hL &=  \xi^N \partial_N \tilde{\zeta}_M{}^{\hL}  + \partial_M \xi^N \tilde{\zeta}_N{}^{\hL} - \cU^{-1 Q}{}_N \cU^P{}_R \partial_Q \xi^R  T_{\alpha_{\dgr{0}}\!\!}{}^N{}_P    T^{\alpha_{\dgr{0}}\hL}{}_{\widehat{\Sigma}} \tilde{\chi}_M{}^{\widehat{\Sigma}}   
\\
& \quad  - \sum_{k=1}^{k_{\rm s}}  \cU^{-1 Q}{}_N \cU^P{}_R \partial_Q \xi^R  \Bigl( T_{\alpha_{\dgr{-k}}\!\!}{}^N{}_P     T^{\alpha_{\dgr{-k}}\hL}{}_{\widehat{\Sigma}} -m^{NS} m_{PU} T_{\alpha_{\dgr{k}}\!\!}{}^U{}_S      T^{\alpha_{\dgr{k}}\hL}{}_{\widehat{\Sigma}} \Bigr)  \tilde{\zeta}_M{}^{\widehat{\Sigma}} 
\nn\\
&\quad\quad  +  {\Pi^{\hL}{}_{ Q S }} m^{RQ} \cU^{-1 N}{}_R \cU^S{}_P \partial_M \partial_N \xi^P \,. \nn
\end{align}
\end{subequations}
Here, we assumed that the sum over $k$ can be bounded by some $k_{\rm s}$ associated to a choice of section, but the same formula holds generally if one takes $k_{\rm s}\rightarrow \infty$. We also used that one can always choose $K^{\alpha_{\dgr{0}}\tilde\alpha}{}_{\beta} =0$ for a finite-dimensional $L$.\footnote{This would not be the case for $L=SL(2) \times E_9$ or $GL(1)\times E_{10}$, where the indecomposability already occurs at level zero~\cite{Bossard:2018utw}.} One obtains in this way that the transformations of the fields $\cU^{-1 N}{}_M \tilde{\cJ}^{\wa}$ and  $\cU^{-1 N}{}_M \tilde{\zeta}^{\hL}$ that appear in the pseudo-Lagrangian are finite and well-defined. The proof of the gauge invariance of the pseudo-Lagrangian in Section~\ref{sec:GI} is not affected by the similarity transformation with respect to $\cU$, so the formal manipulations carried out there apply to the well-defined pseudo-Lagrangian in the semi-flat formulation. 

\subsection{Mathematical subtleties}
\label{sec:subtle}

The Lie algebra $\mf{e}_{11}$ is an infinite-dimensional (Lorentzian) Kac--Moody algebra that is defined from its Chevalley--Serre presentation in terms of simple root generators and Cartan generators satisfying a set of relations~\cite{Kac}. It is not known how to obtain a closed list of all the infinitely generators with their multiplicities and structure constants from this implicit definition. There are calculational methods to probe the structure of $\mf{e}_{11}$ at low levels in a level decomposition and the most comprehensive results were obtained in~\cite{West:2002jj,Nicolai:2003fw}. In the same way that the algebra is defined, the existence and uniqueness of irreducible lowest and highest modules can be proved from the implicit definition of $\mf{e}_{11}$. For our purposes we typically only require existence and uniqueness of the modules and certain associated invariant tensors.

In order to properly define the theory we need to be more precise about the Kac--Moody group $E_{11}$ and its modules. There are two natural Kac--Moody groups associated to the Kac--Moody algebra ${\mf e}_{11}$: the minimal group $E_{11}^{\rm m}$ and its maximal positive completion $E_{11}^{{\rm c}+}$ \cite[\S7, \S8]{Marquis:2018}.\footnote{In fact, there are different definitions of each of these groups but they are equivalent in our split real form~\cite{Marquis:2018}.}
Any element $g^{\rm }\in E_{11}^{\rm m}$ of the minimal group can be written as a \textit{finite} product of one-parameter subgroup elements associated to real roots $g = \prod_{\alpha\text{ real}} \exp( x_\alpha t^\alpha)$ (with $x_\alpha\in \reals$) acting on an integrable module. By contrast, elements $g^{{\rm c}+} \in E_{11}^{{\rm c}+}$  of the (positive Borel) maximal group $E_{11}^{{\rm c}+}$ are defined as products of one-parameter group elements associated to roots $g^{{\rm c}+} = \prod_{\alpha } \exp( x_\alpha t^\alpha )$ in an integrable module, such that the product involves finitely many negative real roots $\alpha \in \Delta_-\cap \Delta_{\rm Re} $, but possibly \textit{infinitely} many positive (real or imaginary) roots $\alpha \in \Delta_+ $.\footnote{\label{fn:affan}An analogy to keep in mind for the definition of these groups is the affine case. The minimal loop group $\widehat{G}^{\rm m}_\ell$ is defined as the group of Laurent polynomials in the spectral parameter $w$ valued in the group $G$, whereas its maximal positive extension $ \widehat{G}^{\rm {\rm c}+}_\ell$ is defined as the group of formal Laurent series  $\sum_{n=-m}^\infty c_n w^n$ in the spectral parameter $w$ valued in the group $G$.} 
Similar to the different notions of the Kac--Moody group we can also consider different types of modules. Any integrable module of the Kac--Moody Lie algebra $\mf{e}_{11}$ has a weight space decomposition into finite-dimensional weight spaces. The two different types of `modules' now differ by allowing only finite or also specific infinite linear combinations of vectors from these weight spaces. To be more precise, we define  a minimal integrable module, such as ${\mf e}_{11}^{\rm m}$ or $R(\lambda)^{\rm m}$, as the set of elements in these vector spaces that have only finitely many non-zero components in a weight space decomposition. This means that, grading the weights by height, any element of a minimal module has a bounded height. Equivalently, grading the weights by the level associated to a finite-dimensional Levi subgroup as in \eqref{LevelAlgebra}, any element of a minimal module has a bounded level.
When one talks about the Kac--Moody algebra $\mf{e}_{11}$ one often means $\mf{e}_{11}^{\rm m}$ and this vector space carries the Lie bracket defined in~\cite{Kac}.
In particular, all elements of the Lie algebra $\mf{e}_{11}^{\rm m}$ act finitely on minimal modules.\footnote{`Acting finitely on a module' here means that any weight vector in the representation is mapped to a finite linear combination of weight vectors in the representation. We assume a basis for the module adapted to the weights with finite-dimensional weight spaces.} 
One can define the exponential map for the real root generators and the group generated by these one-parameter group elements acting on all minimal modules, such as ${\mf e}_{11}^{\rm m}$ or $R(\lambda)^{\rm m}$ for any dominant weight $\lambda$, defines the minimal group $E_{11}^{\rm m}$. The minimal group $E_{11}^{\rm m}$ therefore acts on minimal modules. 
The minimal lowest-weight module $\overline{R(\lambda)}^{\rm m}$ is defined in the same way as a module of the minimal group $E_{11}^{\rm m}$.

Let us consider on which minimal modules one can act with the maximal group $E_{11}^{{\rm c}+}$. Since the imaginary root generators are not locally nilpotent, an exponential of a positive imaginary root generator can and will take a finite linear combination in $\mf{e}_{11}^{\rm m}$ to an infinite linear combination, going infinitely far in the direction of positive height. Therefore, $E_{11}^{{\rm c}+}$ does not act (finitely) on the (minimal) Lie algebra $\mf{e}_{11}^{\rm m}$. In the same way, $E_{11}^{{\rm c}+}$ does not act (finitely) on minimal lowest weight modules $\overline{R(\lambda)}^{\rm m}$ since it can produce infinite combinations of weight vectors (for positive weight) from exponentials of imaginary roots. However, it does act on highest weight modules $R(\lambda)^{\rm m}$ since the heights of the weights in $R(\lambda)^{\rm m}$  are bounded above and therefore all exponential of positive (imaginary or real) root generators evaluate to finite sums. 

In order to define an action of the completed group $E_{11}^{{\rm c}+}$ on more modules one can consider positively completed modules where infinite linear combinations of weight vectors of positive weights are allowed. We shall write these modules as $\mf{e}_{11}^{{\rm c}+}$ and $\overline{R(\lambda)}^{{\rm c}+}$ and now the group $E_{11}^{{\rm c}+}$ can act on them formally, since in order to compute the result of the action of an exponential of a positive imaginary root generator on an element of $\overline{R(\lambda)}^{{\rm c}+}$ is a finite calculation for each fixed weight space. The group $E_{11}^{{\rm c}+}$ \textit{cannot} act on completed highest weight modules $R(\lambda)^{{\rm c}-}$  (including infinite linear combinations of weight vectors of negative weights) since there would be infinitely many contributions to a fixed weight space.\footnote{In the analogy of Footnote~\ref{fn:affan}, the action of $E_{11}^{{\rm c}+}$ on $\overline{R(\lambda)}^{{\rm c}+}$ is like multiplying two formal Laurent series around zero which is well-defined, whereas the action of $E_{11}^{{\rm c}+}$ on $R(\lambda)^{{\rm c}-}$ would be like multiplying a Laurent series around zero with one around infinity which is not well-defined.} The module $\overline{R(\lambda)}^{{\rm c}+}$ is the (algebraic) dual of $R(\lambda)^{\rm m}$ and the pairing is invariant under $E_{11}^{{\rm c}+}$. 

\medskip 

With this more refined notion of group and modules, we can now be more precise as to where the fields live and what the symmetry group of our model is.

If one defines the fields and the coordinates of the theory in minimal modules, and $\cV \in E_{11}^{\rm m}$, all the algebraic identities used in this paper are well-defined. However, including all supergravity fields in a given Iwasawa gauge for the coset $E_{11} / K(E_{11})$ requires to consider also the exponential of imaginary (or null) roots generators. This is the case for example for the dual graviton field with all $GL(11)$ indices different in eleven dimensions~\cite{West:2002jj,Nicolai:2003fw,Houart:2011sk}.
The physical fields must therefore be defined in the (positive Borel) completed group $E_{11}^{{\rm c}+}$.\footnote{When using the affine symmetry to study solutions, one also requires elements of a completed group to describe black hole solutions~\cite{Maison:1988zx}, so physically the minimal group appears too small. But there are more completions available in the affine case, and what appears relevant for black hole solutions is the group of meromorphic functions of the spectral parameter on a Riemann surface valued in the  group $G$ and not the full group of formal Laurent series.}

We thus define the fields of the theory such that
\begin{align}
\label{eq:gps}
 \cV \in E_{11}^{{\rm c}+}\; , \quad \chi  \in R(\Lambda_1)^{\rm m} \otimes \overline{L(\Lambda_2)}^{{\rm c}+}\; , \quad  \zeta  \in R(\Lambda_1)^{\rm m} \otimes \overline{L(\Lambda_{10})}^{{\rm c}+}\ , \quad  \tilde{\zeta}  \in R(\Lambda_1)^{\rm m} \otimes \overline{L(\Lambda_{4})}^{{\rm c}+} 
\end{align}
and the derivative $d = P^m \partial_m \in R(\Lambda_1)^{\rm m}$ in general. With this definition, the field components can be non-zero to arbitrarily large height (or level in a level decomposition), whereas the coordinate dependence, prior to a choice of section, must be such that there is always a maximal height for which all the derivative components vanish, and as well for the constrained indices $M$ of the constrained fields. This definition ensures that $\cV$ acts on the fields of the theory and their derivatives. It also ensures that the action of generalised diffeomorphisms is well-defined in the semi-flat formulation, because for any field configuration there exists a finite maximal level $k_{\rm s}$ in \eqref{gaugeflatten}.

However, the Cartan involution is not defined on  ${\mf e}_{11}^{{\rm c}+}$, because it maps an element of ${\mf e}_{11}^{{\rm c}+}$ to an element of ${\mf e}_{11}^{{\rm c}-}$, the Lie algebra extended along the negative Borel. For $\cV\in E_{11}^{{\rm c}+}$, we have by construction that $d \cV \cV^{-1} \in R(\Lambda_1)^{\rm m} \otimes {\mf e}_{11}^{{\rm c}+}$, but according to \eqref{eq:SFcur}, $\tilde{\cJ}_M $ takes values in the doubly extended vector space ${\mf e}_{11}^{{\rm c}+-}$, in which elements can have infinitely many non-zero components for both positive and negative roots. This vector space ${\mf e}_{11}^{{\rm c}+-}$ is not a Lie algebra, because commutators are not well-defined, and it is not a module of the completed group $E_{11}^{{\rm c}+}$ either. Considering the minimal Kac--Moody group $E_{11}^{{\rm m}}$, it is nevertheless well-defined as the co-adjoint $E_{11}^{{\rm m}}$-module ${\mf e}_{11}^{{\rm c}+-}\cong {\mf e}_{11}^{\rm m *}$, and there is a well-defined  $E_{11}^{{\rm m}}$-homomorphism from $R(\Lambda_1)^{\rm m} \otimes {\mf e}_{11}^{{\rm c}+-}\rightarrow R(\Lambda_1)^{{\rm c} -}$. The field strength $\tilde{F}^I$ belongs to the doubly extended module $\cT_{-1}^{{\rm c}+-}$ of $E_{11}^{\rm m}$. 

The  coset representative $\cV$ transforms under \eqref{Vrightleft} under the constant $g \in E_{11}^{{\rm c}+}$ on the right, and the compensating $k(z) \in K(E_{11}) \subset E_{11}^{{\rm m}}\subset E_{11}^{{\rm c}+}$  on the left, so by construction the coset projection of $d \cV \cV^{-1}$ transforms under $K(E_{11}) \subset E_{11}^{{\rm m}}$. It follows that $E_{11}^{{\rm c}+}$ is a symmetry of  $E_{11}$ exceptional field theory. 

\medskip

With the definitions~\eqref{eq:gps}, the pseudo-Lagrangian \eqref{eq:Lag} remains a formal object. If we consider for example the first term in \eqref{eq:Lpot1}, it involves the Killing--Cartan contraction of two currents $\tilde{\cJ}_M{}^\alpha$, which  generally produces infinitely many terms for an element  $\tilde{\cJ}_M{}^\alpha\in {\mf e}_{11}^{{\rm c}+-}$. In the same way, the second term  in \eqref{eq:Lpot1} involves the scalar product of an element in $R(\Lambda_1)^{{\rm c} -}$ with an element in $\overline{R(\Lambda_1)}^{{\rm c} +}$, which is again a formal infinite sum. Nevertheless, this is not a problem for the definition of the theory. The pseudo-Lagrangian  \eqref{eq:Lag} is an infinite sum of terms that depend on infinitely many fields and is well-defined as a formal pseudo-Lagrangian that generates well-defined Euler--Lagrange equations.\footnote{To give an illustrative example, this is the same infinity of terms that would occur in the Lagrangian of infinitely many free fields, which have well-defined equations of motion. For the $E_{11}$ ExFT pseudo-Lagrangian as well, the Euler--Lagrange equations are well-defined but it does not  make sense a priori to evaluate the pseudo-Lagrangian  \eqref{eq:Lag} for a particular solution of the equations of motion.}  In the next section, we shall show explicitly how to use the pseudo-Lagrangian and the duality equations to obtain well-defined equations in a level decomposition when a solution to the section constraint is chosen. The level decompositions in Sections~\ref{sec:GL11} and~\ref{sec:E8} use finite-dimensional Levi subgroups. The analysis for infinite-dimensional Levi subgroups, such as $E_{10}\subset E_{11}$, is yet more subtle and we consider this in Appendix~\ref{app:E10}.

To avoid unnecessary complications in the notation we chose to never specify in which precise module or group the various objects used in the paper lie in. The reader can refer to this section to obtain this information, but we shall ignore these precisions elsewhere in this paper. 

\section{Recovering eleven-dimensional supergravity}
\label{sec:GL11}

In this section, we study the abstract pseudo-Lagrangian~\eqref{eq:Lag} of $E_{11}$ exceptional field theory in a level decomposition appropriate to the bosonic part of $D=11$ supergravity. This means that we branch all representations of $E_{11}$ with respect to the subgroup $GL(11)$~\cite{West:2002jj,Nicolai:2003fw,West:2003fc,Kleinschmidt:2003jf,Kleinschmidt:2003mf} and also consider the solution of the section constraint~\eqref{eq:SC} that corresponds to $D=11$ space-time, so that we only retain derivatives $\partial_m$ with $m=0,\ldots,10$ out of the infinitely may $\partial_M$. We have collected results on $E_{11}$ in $GL(11)$ decomposition in Appendix~\ref{app:GL11}. The main result of this section is that the metric and the three-form gauge potential, that appear at level 0 and 1,  satisfy the eleven-dimensional supergravity equations of motion \cite{Cremmer:1978km}.

\subsection{Taming the infinity of fields on section}
\label{sec:extall}

Before working out the individual terms in the  pseudo-Lagrangian explicitly, we first discuss the general structure of all terms in the level decomposition. As we shall see there are (infinitely) many simplifications which make the explicit form amenable to study. 

The pseudo-Lagrangian contains an infinite sum of terms, involving both the $E_{11}$ coset fields through the generalised metric $\cM$ as well as the constrained fields $\chi_M{}^\ta$ and $\zeta_M{}^\hL$.
In the $GL(11)$ level decomposition this infinite sum is ordered by the $GL(11)$ level and
we shall use a notation similar to the one in Section~\ref{sec:GT}, where we append a subscript to (most) indices to keep track of their $GL(11)$ level, sometimes with a shift to ease notation. The level assignments we use are displayed in Table~\ref{tab:11levs}.

\begin{table}[t!]
\begin{tabular}{c|c|l}
\text{upper index} & \text{$GL(11)$ level} & \hspace{10mm} \text{fields and representations}\\\hline\hline
 $\alpha_\dgr{k}$ & $k$ & \text{adjoint of $E_{11}$, labelling the current}\\
$\ta_\dgr{k}$ & $k$ & \text{$L(\Lambda_2)$ module for constrained field $\tilde{\chi}$}\\
$\hL_\dgr{k}$ & $k$ &\text{constrained field $\tilde{\zeta}$ in $L(\Lambda_{10})\oplus L(\Lambda_4)$}\\
$m$ & $\frac32$ & \text{only non-trivial part $\partial_m$ of derivatives $\partial_M$  on $D=11$ section}\\
$I_\dgr{k}$ & $-\frac32 + k$ & \text{field strength components in $\cT_{-1}$}\\
$\tI_\dgr{k}$ & $\frac32 + k$ & \text{$L(\Lambda_3)$ representation}
\end{tabular}
\caption{\label{tab:11levs}\sl Index assignments for the indices of the various representations in $GL(11)$ level decomposition. See also Appendix~\ref{app:GL11} for details on the level decomposition.}
\end{table}

 We recall that the $GL(11)$ level is the eigenvalue of the action of the Cartan subalgebra element $H_{\Lambda_{11}}$ where $\Lambda_{11}$ is the fundamental weight of node $11$ in Figure~\ref{fig:e11dynk}. The level for an index upstairs or downstairs are opposite and we have indicated the counting for upper indices only.
All $E_{11}$-invariant tensors preserve the $GL(11)$ level, so that for instance $C^{I_\dgr{k}m}{}_{\wa_\dgr{\ell}}$ is only non-zero for $k=\ell$. The cocycle $K_\alpha{}^\ta{}_\beta$ is not $E_{11}$-invariant but $GL(11)$-invariant and therefore also preserves the $GL(11)$ level.

With this notation, the kinetic term~\eqref{eq:Lkin} expands as
\begin{align}
\label{eq:Lkinall}
\mathcal{L}_{\text{kin}} &= \frac14 \cM_{IJ} F^I F^J - \cM_{IJ} C^{IM}{}_\wa C^{JN}{}_\hL J_M{}^\wa \zeta_N{}^\hL -\frac12 \cM_{IJ} C^{IM}{}_\hL C^{JN}{}_\hXi \zeta_M{}^\hL \zeta_N{}^\hXi
\nn\\
&=  \frac14 \sum_{k\in \mathds{Z}} m_{I_\dgr{k} J_\dgr{k}} \tilde{\cF}^{I_\dgr{k}} \tilde{\cF}^{J_\dgr{k}} 
\\
 & \quad  -\sum_{k= 4}^\infty  m_{I_\dgr{k}J_\dgr{k}} C^{I_\dgr{k} m}{}_{\wa_\dgr{k}}  C^{J_\dgr{k} n}{}_{\hL_\dgr{k}}  \tilde{\mathcal{J}}_m{}^{\wa_\dgr{k}} \tilde{\zeta}_n{}^{\hL_\dgr{k}} -\frac12 \sum_{k=4}^\infty m_{I_\dgr{k}J_\dgr{k}} C^{I_\dgr{k}m}{}_{\hL_\dgr{k}} C^{J_\dgr{k}n}{}_{\hXi_\dgr{k}} \tilde{\zeta}_m{}^{\hL_\dgr{k}} \tilde{\zeta}_{n}{}^{\hXi_\dgr{k}} \,,
 \nn
\end{align} 
where in the last line we have used that the $\hL$ index starts at $GL(11)$ level four, see~\eqref{eq:L1011} below.

To write the topological term \eq{rtop}, we first observe that the last  term in \eqref{eq:topSF2} expands according to 
\begin{align}
  &\quad \Pi_{\ta_\dgr{3}\!}{}^{mn}   \sum_{k\in \mathds{Z} } (  \tilde{\mathcal{J}}_m{}^{\alpha_\dgr{-k}} -2 {\mathcal{N}}_m{}^{\alpha_\dgr{-k}} ) K_{\alpha_\dgr{-k}}{}^{\tilde{\alpha}_\dgr{3}}{}_{\beta_\dgr{3+k}}  \tilde{\mathcal{J}}_n{}^{\beta_\dgr{3+k}}\CR
&=
\Pi_{\ta_\dgr{3}\!\!}{}^{mn}   \Bigl( \sum_{k=0 }^\infty    \tilde{\mathcal{J}}_m{}^{\alpha_\dgr{-k}}  K_{\alpha_\dgr{-k}}{}^{\tilde{\alpha}_\dgr{3}}{}_{\beta_\dgr{3+k}}  \tilde{\mathcal{J}}_n{}^{\beta_\dgr{3+k}} - \sum_{k=-1 }^{-\infty}  \tilde{\mathcal{J}}_m{}^{\alpha_\dgr{-k}} K_{\alpha_\dgr{-k}}{}^{\tilde{\alpha}_\dgr{3}}{}_{\beta_\dgr{3+k}}  \tilde{\mathcal{J}}_n{}^{\beta_\dgr{3+k}} \Bigr) \CR
&=  -\Pi_{\ta_\dgr{3}\!\!}{}^{mn}  \Bigl(  \tilde{\mathcal{J}}_m{}^{\alpha_\dgr{1}} K_{\alpha_\dgr{1}}{}^{\tilde{\alpha}_\dgr{3}}{}_{\beta_\dgr{2}}  \tilde{\mathcal{J}}_n{}^{\beta_\dgr{2}} +  \tilde{\mathcal{J}}_m{}^{\alpha_\dgr{2}} K_{\alpha_\dgr{2}}{}^{\tilde{\alpha}_\dgr{3}}{}_{\beta_\dgr{1}}  \tilde{\mathcal{J}}_n{}^{\beta_\dgr{1}} \Bigr)  \CR
&\quad   + \Pi_{\ta_\dgr{3}\!}{}^{mn}  \sum_{k=0 }^\infty   \Bigl(   \tilde{\mathcal{J}}_m{}^{\alpha_\dgr{-k}}  K_{\alpha_\dgr{-k}}{}^{\tilde{\alpha}_\dgr{3}}{}_{\beta_\dgr{3+k}}  \tilde{\mathcal{J}}_n{}^{\beta_\dgr{3+k}} - \tilde{\mathcal{J}}_m{}^{\alpha_\dgr{k+3}} K_{\alpha_\dgr{k+3}}{}^{\tilde{\alpha}_\dgr{3}}{}_{\beta_\dgr{-k}}  \tilde{\mathcal{J}}_n{}^{\beta_\dgr{-k}} \Bigr) 
\CR
&=  -2 \Pi_{\ta_\dgr{3}\!}{}^{mn}  \tilde{\mathcal{J}}_m{}^{\alpha_\dgr{1}} K_{[\alpha_\dgr{1}}{}^{\tilde{\alpha}_\dgr{3}}{}_{\beta_\dgr{2}]}  \tilde{\mathcal{J}}_n{}^{\beta_\dgr{2}}  +2 \Pi_{\ta_\dgr{3}\!}{}^{mn}  \sum_{k=0 }^\infty    \tilde{\mathcal{J}}_m{}^{\alpha_\dgr{-k}}  K_{(\alpha_\dgr{-k}}{}^{\tilde{\alpha}_\dgr{3}}{}_{\beta_\dgr{3+k})}  \tilde{\mathcal{J}}_n{}^{\beta_\dgr{3+k}} \; , 
\end{align}
using that ${\mathcal{N}}_m{}^{\alpha_\dgr{k}} = \tilde{\mathcal{J}}_m{}^{\alpha_\dgr{k}} $ for $k\ge 1$ and zero otherwise. Moreover, we have used $E_{11}$-invariance of $\Pi_\ta{}^{MN}$ and the $D=11$ solution to the section constraint to simplify the unipotent matrices $\cU$, see the discussion below~\eqref{rtop}. Using this one obtains for the complete topological term \eqref{rtop}
\bea
\label{eq:Ltopall}
\mathcal{L}_{\text{top}} &=& - \Pi_{\ta_\dgr{3}\!}{}^{mn}  \tilde{\mathcal{J}}_m{}^{\alpha_\dgr{1}} K_{[\alpha_\dgr{1}\!\!}{}^{\tilde{\alpha}_{\dgr{3}}}{}_{\!\beta_\dgr{2}]}  \tilde{\mathcal{J}}_n{}^{\beta_\dgr{2}}  -\frac12 \sum_{k= 4}^\infty  
 \Omega_{I_\dgr{3-k}J_\dgr{k}} C^{I_\dgr{3-k} m}{}_{\wa_\dgr{3-k}}  \tilde{\mathcal{J}}_m{}^{\wa_\dgr{3-k}}  C^{J_\dgr{k} n}{}_{\hL_\dgr{k}} \tilde{\zeta}_n{}^{\hL_\dgr{k}} 
\nn\\
&&  + \frac{1}{2} \Pi_{\ta_\dgr{3}\!}{}^{mn}   \sum_{k=0}^\infty  \tilde{\mathcal{J}}_m{}^{\alpha_\dgr{-k}} \bigl( T_{\alpha_\dgr{-k}\!\!}{}^{\tilde{\alpha}_\dgr{3}}{}_{\!\tilde{\beta}_\dgr{3+k}} \tilde{\chi}_n{}^{\tilde{\beta}_\dgr{3+k}} + 2 K_{(\alpha_\dgr{-k}\!\!}{}^{\tilde{\alpha}_\dgr{3}}{}_{\! \beta_\dgr{3+k})}  \tilde{\mathcal{J}}_n{}^{\beta_\dgr{3+k}} \bigr) +\Pi_{\ta_\dgr{3}\!\!}{}^{mn}   \partial_{m}  \tilde{\chi}_n{}^{\ta_\dgr{3}}
\nn\\
&=&-\Pi_{\ta_\dgr{3}\!\!}{}^{mn}  \tilde{\mathcal{J}}_m{}^{\alpha_\dgr{1}} K_{[\alpha_\dgr{1}\!\!}{}^{\tilde{\alpha}_{\dgr{3}}}{}_{\!\beta_\dgr{2}]}  \tilde{\mathcal{J}}_n{}^{\beta_\dgr{2}}  -\frac12 \sum_{k= 0}^\infty  \Omega_{I_\dgr{-k}J_\dgr{3+k}} \tilde{\cF}^{I_\dgr{-k}}\tilde{\cF}^{J_\dgr{3+k}} +\Pi_{\ta_\dgr{3}\!\!}{}^{mn}   \partial_{m}  \tilde{\chi}_n{}^{\ta_\dgr{3}}\,,
\eea
where in the second step we used \eqref{eq:ID4} and \eqref{eq:ID2}, and the property that the constrained fields only contribute to $\tilde{\cF}^{J_\dgr{k}}$ for $k\ge 3$.

It turns out that computing the pseudo-Lagrangian in level decomposition produces an infinite number of terms that are  quadratic in the duality equations at level $k$
\be   
{\mathcal E}_{I_\dgr{k}} \equiv m_{I_\dgr{k}J_\dgr{k}} \tilde{\cF}^{J_\dgr{k}} -  \Omega_{I_\dgr{k}  J_\dgr{3-k}}   \tilde{\cF}^{J_\dgr{3-k}} =0\ .
\label{DualityLevelDecompose} 
\ee
For $k\geq 2$, we define $\mathcal{O}_k$ as the square of the duality equation in the form
\begin{align}
\label{eq:Lk}
\mathcal{O}_k  &= -\frac14 m^{I_\dgr{k} J_\dgr{k}}  {\mathcal E}_{I_\dgr{k}}{\mathcal E}_{J_\dgr{k}}
\nn\\
&= - \frac14  m_{I_\dgr{k} J_\dgr{k}} \tilde{\cF}^{I_\dgr{k}} \tilde{\cF}^{J_\dgr{k}} + \frac14  m_{I_\dgr{3-k} J_\dgr{3-k}} \tilde{\cF}^{I_\dgr{3-k}} \tilde{\cF}^{J_\dgr{3-k}}   -\frac12 \Omega_{I_\dgr{3-k}J_\dgr{k}} \tilde{\cF}^{I_\dgr{3-k}}\tilde{\cF}^{J_\dgr{k}}\; .
\end{align}
For explicit expressions for $\mathcal{O}_2$ and $\mathcal{O}_3$ see \eq{cO2} and \eq{cO3} below. Varying $\mathcal{O}_k$ in a pseudo-Lagrangian will not change the field equations since their variation is proportional to the duality equations that are imposed in addition to the Euler--Lagrange equations.
The contribution of these terms 
for all $k\ge 3$ will be seen to contain the full dependence on the auxiliary fields  $\tilde{\chi}_m{}^{\tilde{\alpha}},\, \tilde{\zeta}_m{}^{\widehat{\Lambda}}$.

{\allowdisplaybreaks
The Euler--Lagrange equations for the fields $\tilde{\chi}_m{}^{\tilde{\alpha}},\, \tilde{\zeta}_m{}^{\widehat{\Lambda}}$  give (at least a subset of) the duality equations \eqref{DualityLevelDecompose} for  all $k\ge 3$ and $k\le 0$, see Section~\ref{sec:EOMchi}. Because the section condition is completely solved in eleven dimensions, the components of the constrained fields $\tilde{\chi}_m{}^{\tilde{\alpha}},\, \tilde{\zeta}_m{}^{\widehat{\Lambda}}$ are independent and appear algebraically, hence we can integrate them out by applying their equations of motion, which should be equivalent to subtracting the pseudo-Lagrangian $\sum_{k\ge 3} \mathcal{O}_k$ from the pseudo-Lagrangian $\mathcal{L}$.  Combining the first term in $\mathcal{L}_{\text{kin}}$ from~\eqref{eq:Lkinall} with the last term in $\mathcal{L}_{\text{top}}$ from~\eqref{eq:Ltopall} 
gives
\begin{align} 
 &\quad\, \frac14 \sum_{k\in \mathds{Z}} m_{I_\dgr{k} J_\dgr{k}} \tilde{\cF}^{I_\dgr{k}} \tilde{\cF}^{J_\dgr{k}} -\frac12 \sum_{k= 0}^\infty  \Omega_{I_\dgr{-k}J_\dgr{3+k}} \tilde{\cF}^{I_\dgr{-k}}\tilde{\cF}^{J_\dgr{3+k}}  \CR
&= \frac14 \sum_{k=1}^\infty  m_{I_\dgr{k} J_\dgr{k}} \tilde{\cF}^{I_\dgr{k}} \tilde{\cF}^{J_\dgr{k}}+\frac14 \sum_{k=0}^\infty   m_{I_\dgr{3+k}J_\dgr{3+k}} \tilde{\cF}^{I_\dgr{3+k}}\tilde{\cF}^{J_\dgr{3+k}} +  \sum_{k=0}^\infty \mathcal{O}_{3+k}\CR
&= \frac14 \sum_{k=1}^2  m_{I_\dgr{k} J_\dgr{k}} \tilde{\cF}^{I_\dgr{k}} \tilde{\cF}^{J_\dgr{k}}+\frac12  \sum_{k=3}^\infty  m_{I_\dgr{k} J_\dgr{k}} \tilde{\cF}^{I_\dgr{k}} \tilde{\cF}^{J_\dgr{k}}+  \sum_{k=0}^\infty \mathcal{O}_{3+k}\; . 
 \end{align}
The second term cancels precisely the dependence in $\tilde{\zeta}_m{}^{\widehat{\Lambda}}$ in $\mathcal{L}_{\text{kin}}$, and one gets in total 
\begin{align}\label{kt}
\mathcal{L}_{\text{kin}} + \mathcal{L}_{\text{top}}  &=  \frac14 \sum_{k=1}^2  m_{I_\dgr{k} J_\dgr{k}} \tilde{\cF}^{I_\dgr{k}} \tilde{\cF}^{J_\dgr{k}}    -\Pi_{\ta_\dgr{3}\!\!}{}^{mn}  \tilde{\mathcal{J}}_m{}^{\alpha_\dgr{1}} K_{[\alpha_\dgr{1}}{}^{\tilde{\alpha}_{\dgr{3}}}{}_{\beta_\dgr{2}]}  \tilde{\mathcal{J}}_n{}^{\beta_\dgr{2}}   \\
& \qquad +\frac12 \sum_{k= 3}^\infty  m_{I_\dgr{k}J_\dgr{k}} C^{I_\dgr{k} m}{}_{\wa_\dgr{k}}  C^{J_\dgr{k} n}{}_{\wb_\dgr{k}}  \tilde{\mathcal{J}}_m{}^{\wa_\dgr{k}} \tilde{\mathcal{J}}_n{}^{\wb_\dgr{k}} + \sum_{k=3}^\infty \mathcal{O}_{k} +\Pi_{\ta_\dgr{3}\!\!}{}^{mn}   \partial_{m}  \tilde{\chi}_n{}^{\ta_\dgr{3}}\; . \nonumber
\end{align}
Note that the first line contains only the field strengths and currents of levels $1$ and $2$. 
The first term in the second line can be combined with $\mathcal{L}_{\text{pot}_2}$ and identity \eqref{e12Equation} to give 
\begin{align} 
\label{eq:Lpot2all}
&\quad   \mathcal{L}_{\text{pot}_2}+\frac12 \sum_{k= 3}^\infty  m_{I_\dgr{k}J_\dgr{k}} C^{I_\dgr{k} m}{}_{\wa_\dgr{k}}  C^{J_\dgr{k} n}{}_{\wb_\dgr{k}}  \tilde{\mathcal{J}}_m{}^{\wa_\dgr{k}} \tilde{\mathcal{J}}_n{}^{\wb_\dgr{k}}  
\nn\\
 &=  \frac{1}{2}  \sum_{k= 3}^\infty  \Bigl(  m_{I_\dgr{k}J_\dgr{k}}  C^{I_\dgr{k} m}{}_{\wa_\dgr{k}}  C^{J_\dgr{k} n}{}_{\wb_\dgr{k}}   - m_{\tilde{I}_\dgr{k}\tilde{J}_\dgr{k}} C^{\tilde{I}_\dgr{k}}{}_{p \wa_\dgr{k}} C^{\tilde{J}_\dgr{k}}{}_{q \wb_\dgr{k}} m^{qm} m^{pn} \Bigr)  \tilde{\mathcal{J}}_m{}^{\wa_\dgr{k}} \tilde{\mathcal{J}}_n{}^{\wb_\dgr{k}}  
 \CR
  &=  \frac{1}{2}  \sum_{k= 3}^\infty  \Bigl(  m_{\alpha_\dgr{k}\beta_\dgr{k}} m^{mn}  -m_{\alpha_\dgr{k}\gamma_\dgr{k}}  T^{\gamma_\dgr{k} n}{}_{Q} T_{\beta_\dgr{k}}{}^Q{}_p m^{pm}  -m_{\alpha_\dgr{k}\gamma_\dgr{k}}  T^{\gamma_\dgr{k} Q}{}_{p} T_{\beta_\dgr{k}}{}^n{}_Q m^{pm}  \Bigr)  \tilde{\mathcal{J}}_m{}^{\alpha_\dgr{k}} \tilde{\mathcal{J}}_n{}^{\beta_\dgr{k}} 
  \nn\\*
  &=   \frac{1}{2}  \sum_{k= 3}^\infty  \Bigl(  m_{\alpha_\dgr{k}\beta_\dgr{k}} m^{mn}  -m_{\alpha_\dgr{k}\gamma_\dgr{k}}  T^{\gamma_\dgr{k} n}{}_{Q} T_{\beta_\dgr{k}}{}^Q{}_p m^{pm}  \Bigr)  \tilde{\mathcal{J}}_m{}^{\alpha_\dgr{k}} \tilde{\mathcal{J}}_n{}^{\beta_\dgr{k}}  \; ,  
\end{align}
where we used the highest weight property  $ T^{\gamma_\dgr{k} Q}{}_{p}=0=T_{\beta_\dgr{k}}{}^m{}_Q $ for all $k\ge 1$ in the last step. 
}

Similarly, one  obtains
\begin{align}
 \mathcal{L}_{\text{pot}_1} &= - \frac{1}{4} \sum_{k \in \mathds{Z}} \kappa_{\alpha_\dgr{-k}\beta_{\dgr{k}}} m^{mn} \tilde{\mathcal{J}}_m{}^{\alpha_\dgr{-k}} \tilde{\mathcal{J}}_n{}^{\beta_\dgr{k}}  + \frac12 \sum_{k \in \mathds{Z}}   \tilde{\mathcal{J}}_{m}{}^{\alpha_\dgr{k}} T_{\beta_\dgr{k}}{}^{m}{}_P m^{PQ} T_{\alpha_\dgr{k}}{}^{n}{}_Q \tilde{\mathcal{J}}_{n}{}^{\beta\dgr{k}}    
\CR
&=  - \frac{1}{4} m_{\alpha_\dgr{0}\beta_{\dgr{0}}} m^{mn} \tilde{\mathcal{J}}_m{}^{\alpha_\dgr{0}} \tilde{\mathcal{J}}_n{}^{\beta_\dgr{0}}  - \frac12   \sum_{k=1}^\infty m_{\alpha_\dgr{k}\beta_{\dgr{k}}} m^{mn} \tilde{\mathcal{J}}_m{}^{\alpha_\dgr{k}} \tilde{\mathcal{J}}_n{}^{\beta_\dgr{k}} \CR
& \qquad + \frac12 \sum_{k =0}^\infty m_{\alpha_\dgr{k}\gamma_\dgr{k}}  T^{\gamma_\dgr{k} n}{}_{Q} T_{\beta_\dgr{k}}{}^Q{}_p m^{pm}   \tilde{\mathcal{J}}_m{}^{\alpha_\dgr{k}} \tilde{\mathcal{J}}_n{}^{\beta_\dgr{k}}    \,,
\label{p1}
\end{align}
where we have used $m^{-1} \tilde{\mathcal{J}}_m^\dagger m = \tilde{\mathcal{J}}_m$ to get the factor of $2$ for the second term, and the $E_{11}$-invariance of the structure constants to convert $m^{PQ}$  in $m^{pm}$. We see that there will be cancellations when combining these terms with those from~\eqref{eq:Lpot2all}.

Combining all the terms together we obtain eventually that the $E_{11}$ exceptional field theory pseudo-Lagrangian reduces to
\begin{align}
\label{LagrangianReduced11D} 
\mathcal{L} &=  - \frac{1}{4} m_{\alpha_\dgr{0}\beta_{\dgr{0}}} m^{mn} \tilde{\mathcal{J}}_m{}^{\alpha_\dgr{0}} \tilde{\mathcal{J}}_n{}^{\beta_\dgr{0}}  - \frac12   \sum_{k=1}^2 m_{\alpha_\dgr{k}\beta_{\dgr{k}}} m^{mn} \tilde{\mathcal{J}}_m{}^{\alpha_\dgr{k}} \tilde{\mathcal{J}}_n{}^{\beta_\dgr{k}} 
\CR
& \quad + \frac12 \sum_{k =0}^2 m_{\alpha_\dgr{k}\gamma_\dgr{k}}  T^{\gamma_\dgr{k} n}{}_{Q} T_{\beta_\dgr{k}}{}^Q{}_p m^{pm}   \tilde{\mathcal{J}}_m{}^{\alpha_\dgr{k}} \tilde{\mathcal{J}}_n{}^{\beta_\dgr{k}}  +\frac14  \sum_{k=1}^2m_{I_\dgr{k} J_\dgr{k}} \tilde{\cF}^{I_\dgr{k}} \tilde{\cF}^{J_\dgr{k}} 
\CR
& \qquad-\Pi_{\ta_\dgr{3}\!\!}{}^{mn}  \tilde{\mathcal{J}}_m{}^{\alpha_\dgr{1}} K_{[\alpha_\dgr{1}\!\!}{}^{\tilde{\alpha}_{\dgr{3}}}{}_{\!\beta_\dgr{2}]}  \tilde{\mathcal{J}}_n{}^{\beta_\dgr{2}} + \Pi_{\ta_\dgr{3}\!}{}^{mn}   \partial_{m}  \tilde{\chi}_n{}^{\ta_\dgr{3}} + \sum_{k=3}^\infty \mathcal{O}_{k}\; .
\end{align}
We stress that the terms $\mathcal{O}_k$ are not $E_{11}$-invariant since we have chosen the $GL(11)$-invariant solution to the section constraint. Note also that the $\mathcal{O}_k$, being proportional to squares of the duality equations, do not modify the dynamics implied by the pseudo-Lagrangian, since by construction the Euler--Lagrange equations for the pseudo-Lagrangians $\mathcal{O}_k$ are integrability conditions for the duality equations  \eqref{DualityLevelDecompose}. It follows that there is no loss of information in integrating out the fields $\tilde{\chi}_m{}^{\tilde{\alpha}},\, \tilde{\zeta}_m{}^{\widehat{\Lambda}}$, as long as we keep the duality equations. In particular, the Euler--Lagrange equations of the $E_{11}$ coset fields $ \phi^{\alpha_\dgr{k}}$ for $k\ge 3$ are integrability conditions for the duality equations, and are consequences of the constrained fields' Euler--Lagrange equations. In what follows, we shall next show that  $ {\cal L}- \sum_{k=2}^\infty \mathcal{O}_{k} $ on section is the (bosonic part of the) eleven-dimensional supergravity Lagrangian \cite{Cremmer:1978km}. We shall also analyse the duality equations at the linearised level, their integrability conditions, and the r\^ole of the 
constrained fields. 
To this end, in Section~\ref{sec:GL11d} below, we shall begin with some preliminaries on the $GL(11)$ decomposition.

We conclude this section by expressing the result \eqref{LagrangianReduced11D} in an alternative form which will be useful in the following. To do so, we define the restrictions of the  pseudo-Lagrangian components of \eqref{eq:Lag} to level $k$ as follows
\bea 
\cL_{{\rm kin }}|_k &\underset{k\le 3}{=}& \frac{1}{4} m_{I_\dgr{k} J_\dgr{k}} \tilde{\cF}^{I_\dgr{k}} \tilde{\cF}^{J_\dgr{k}} \; , 
 \CR 
\cL_{{\rm kin }}|_k &\underset{k\ge 4}{=}& \frac{1}{4} m_{I_\dgr{k} J_\dgr{k}} \tilde{\cF}^{I_\dgr{k}} \tilde{\cF}^{J_\dgr{k}} 
 \CR 
 &&  -  m_{I_\dgr{k}J_\dgr{k}} C^{I_\dgr{k} m}{}_{\wa_\dgr{k}}  C^{J_\dgr{k} n}{}_{\hL_\dgr{k}}  \tilde{\mathcal{J}}_m{}^{\wa_\dgr{k}} \tilde{\zeta}_n{}^{\hL_\dgr{k}} -\frac12  m_{I_\dgr{k}J_\dgr{k}} C^{I_\dgr{k}m}{}_{\hL_\dgr{k}} C^{J_\dgr{k}n}{}_{\hXi_\dgr{k}} \tilde{\zeta}_m{}^{\hL_\dgr{k}} \tilde{\zeta}_{n}{}^{\hXi_\dgr{k}} 
 \nn \\
\cL_{{\rm top }}|_2 &=& - \Pi_{\ta_\dgr{3}\!}{}^{mn}  \tilde{\mathcal{J}}_m{}^{\alpha_\dgr{1}} K_{[\alpha_\dgr{1}}{}^{\tilde{\alpha}_{\dgr{3}}}{}_{\beta_\dgr{2}]}  \tilde{\mathcal{J}}_n{}^{\beta_\dgr{2}} 
\; , 
\CR
\cL_{{\rm top }}|_k &\underset{k\ge3}{=}& -\frac12  \Omega_{I_\dgr{3-k}J_\dgr{k}} \tilde{\cF}^{I_\dgr{3-k}}\tilde{\cF}^{J_\dgr{k}} \; , \CR 
\cL_{{\rm pot}_1}|_{0} &=&   - \frac{1}{4} m_{\alpha_\dgr{0}\beta_{\dgr{0}}} m^{mn} \tilde{\mathcal{J}}_m{}^{\alpha_\dgr{0}} \tilde{\mathcal{J}}_n{}^{\beta_\dgr{0}} + \frac12  m_{\alpha_\dgr{0}\gamma_\dgr{0}}  T^{\gamma_\dgr{0} n}{}_{Q} T_{\beta_\dgr{0}}{}^Q{}_p m^{pm}   \tilde{\mathcal{J}}_m{}^{\alpha_\dgr{0}} \tilde{\mathcal{J}}_n{}^{\beta_\dgr{0}}  \; , \CR
\cL_{{\rm pot}_1}|_{k} &\underset{k\ge 1}{=}&- \frac12   m_{\alpha_\dgr{k}\beta_{\dgr{k}}} m^{mn} \tilde{\mathcal{J}}_m{}^{\alpha_\dgr{k}} \tilde{\mathcal{J}}_n{}^{\beta_\dgr{k}}  + \frac12  m_{\alpha_\dgr{k}\gamma_\dgr{k}}  T^{\gamma_\dgr{k} n}{}_{Q} T_{\beta_\dgr{k}}{}^Q{}_p m^{pm}   \tilde{\mathcal{J}}_m{}^{\alpha_\dgr{k}} \tilde{\mathcal{J}}_n{}^{\beta_\dgr{k}}    \,, \CR
\cL_{{\rm pot}_2}|_{k} &\underset{k\ge 3}{=}& - \frac{1}{2} m_{\tilde{I}_\dgr{k}\tilde{J}_\dgr{k}} C^{\tilde{I}_\dgr{k}}{}_{p \wa_\dgr{k}} C^{\tilde{J}_\dgr{k}}{}_{q \wb_\dgr{k}} m^{qm} m^{pn} \tilde{\mathcal{J}}_m{}^{\wa_\dgr{k}} \tilde{\mathcal{J}}_n{}^{\wb_\dgr{k}} \; . 
\eea
It follows that
\be
\cL_{\rm kin} = \sum_{k \in \mathds{Z}} \cL_{{\rm kin }}|_k \;, \quad
\cL_{\rm top} = \sum_{k =2}^\infty \cL_{{\rm top }}|_k \;, \quad
\cL_{{\rm pot}_1} = \sum_{k =0}^\infty \cL_{{{\rm pot}_1 }}|_k \;, \quad
\cL_{{\rm pot}_2} = \sum_{k =3}^\infty \cL_{{{\rm pot}_2 }}|_k \; . 
\ee
Using this notation, one can now resume the computation carried out in this section to the property that for $k\ge 3$ we have 
\bea \label{OfromL}  &&   \cL_{\rm kin} \big|_{3-k} + \cL_{\rm kin} \big|_{k} + \mathcal{L}_{\text{pot}_1} \big|_{k}  + \mathcal{L}_{\text{pot}_2}\big|_{k} +\mathcal{L}_{\text{top}}\big|_{k} \CR
&=&   \cL_{\rm kin} \big|_{k} - \cL_{\rm kin} \big|_{3-k} +\mathcal{L}_{\text{top}}\big|_{k}  = \mathcal{O}_k \; , 
\eea
such that \eqref{LagrangianReduced11D} can be written as 
\be
\mathcal{L}  =  \sum_{k=1,2}  \mathcal{L}_{\rm kin}\big|_{k} + \sum_{k=0,1,2} \mathcal{L}_{{\rm pot}_1}\big|_{k} + \mathcal{L}_{\rm top}\big|_{2} + \Pi_{\ta_\dgr{3}\!}{}^{mn}   \partial_{m}  \tilde{\chi}_n{}^{\ta_\dgr{3}} +\sum_{k=3}^\infty \mathcal{O}_{k} \ .
\label{sugra}
\ee

It is remarkable that the infinite number of terms in the total pseudo-Lagrangian needed for the $E_{11}$ symmetry boils down to the infinite sum over the `squares' of the duality equations, see \eq{eq:Lk}, and only a finite number of terms remains outside this sum. The emergence of the duality-equations squared terms is due to the relation \eq{OfromL}. Since this relation is central to the result \eq{sugra}, we illustrate how it works for $k=3$  in detail in Appendix \ref{CancelO3}, as an example.

\subsection{\texorpdfstring{$GL(11)$ decomposition}{GL(11) decomposition}}
\label{sec:GL11d}

For the $GL(11)$ level decomposition we parametrise the unipotent element as
\begin{align}
\label{eq:U11}
\cU =   \exp\left( \tfrac1{3!} A_{n_1n_2n_3} E^{n_1n_2n_3} \right)  \exp\left( \tfrac1{6!} A_{n_1\dots n_6} E^{n_1\dots n_6} \right) \exp\left( \tfrac1{8!} h_{n_1\dots n_8,m} E^{n_1\dots n_8,m} \right) \cdots \,.
\end{align}
This is the usual $GL(11)$ level decomposition of $E_{11}$ up to level three involving the three-form, its dual six-form and the dual graviton~\cite{West:2002jj,Nicolai:2003fw,West:2003fc,Kleinschmidt:2003jf,Kleinschmidt:2003mf}.\footnote{The $GL(11)$ level equals the difference between the number of upper and lower $GL(11)$ indices divided by three, for the adjoint representation, and this number plus $\frac{11}{6}$ in the $R(\Lambda_1)$ representation of $E_{11}$. For more details, see \cite{West:2002jj,Nicolai:2003fw,West:2003fc,Bossard:2017wxl}. This rule is only applicable if one keeps sets of $11$ antisymmetric indices.} 

The metric $m$ on the Levi factor $GL(11)$ is nothing but the eleven-dimensional space-time metric $g_{mn}$. From the way the semi-flat formulation was set up, the duality equation~\eqref{eq:DElev} contains the $m_{IJ}$ which is block diagonal in $GL(11)$ level decomposition. Therefore it acts just like the $GL(11)$ metric on a given $GL(11)$ field strength component by raising/lowering its indices, with an additional factor of $\sqrt{-g}$ because the $GL(11)$ components of $\tilde{\cF}^I$ are $GL(11)$ tensor densities of weight one-half. The duality equation~\eqref{eq:DElev}, however, becomes a set of $GL(11)$ tensorial equations precisely because of this factor.

In the parametrisation~\eqref{eq:U11} we obtain
\begin{align}
\label{eq:NM11}
\cN_M &=  \partial_M \cU  \cU^{-1} = \frac{1}{3!} \partial_M A_{n_1n_2n_3} E^{n_1n_2n_3} + \frac{1}{6!} \Big( \partial_M A_{n_1\ldots n_6}  + 10 A_{[n_1n_2n_3} \partial_M A_{n_4n_5n_6]}\Big) E^{n_1\ldots n_6}\nn\\
& \hspace{30mm} + \frac1{8!} \Big( \partial_M h_{n_1\ldots n_8,m} +56  A_{\langle n_1n_2n_3} \partial_M A_{n_4\ldots n_8,m\rangle} \\
&\hspace{50mm}+140 A_{\lsharp n_1n_2n_3} A_{n_4n_5n_6} \partial_M A_{n_7n_8,m\rsharp}  \Big) E^{n_1\ldots n_8,m} + \ldots\,,\nn
\end{align}
where $\lsharp \cdots,\cdot\rsharp$ denotes the projection on the $(8,1)$ Young symmetry:
\be
A_{\lsharp n_1n_2n_3} A_{n_4n_5n_6} \partial_M A_{n_7n_8,m\rsharp} = A_{[n_1n_2n_3} A_{n_4n_5n_6} \partial_M A_{n_7n_8]m} - A_{[ n_1n_2n_3} A_{n_4n_5n_6} \partial_M A_{n_7n_8m]} . 
\ee
The $R(\Lambda_1)$ index $M$ is not included in the anti-symmetrisation in this equation.

Besides the $E_{11}$ fields $v$ and $\cU$ we also require the constrained fields~\eqref{eq:CF}. The $L(\Lambda_2)$ index $\ta$ on the field $\chi_M{}^\ta$ decomposes as~\cite{Bossard:2019ksx}
\begin{align}
\label{eq:L211}
\chi_M{}^\ta \to \Big( \chi_{M;n_1\ldots n_9}, \chi_{M;n_1\ldots n_{10},p_1p_2},\chi_{M;n_1\ldots n_{11},p},\ldots\Big)\,,
\end{align}
where we use the same notation as in~\cite{Bossard:2017wxl,Bossard:2019ksx} that a comma separates different columns of an irreducible $GL(11)$ tableau, while a semi-colon separates different columns of a reducible tensor product. 

The $L(\Lambda_{10})$ index of the constrained field $\zeta_M{}^\Lambda$ decomposes as 
\begin{align}
\label{eq:L1011}
\zeta_M{}^\Lambda \to \Big( \zeta_{M;n_1\ldots n_{11},p}, \zeta_{M;n_1\ldots n_{11},p_1\ldots p_4},\ldots\Big)\,.
\end{align}
The $L(\Lambda_4)$ index of the constrained field $\zeta_M{}^\tL$ decomposes as
\begin{align}
\label{eq:L411}
\zeta_M{}^\tL \to \Big( \zeta_{M;n_1\ldots n_{11},p_1\ldots p_7}, \zeta_{M;n_1\ldots n_{11},p_1\ldots p_8;q_1q_2},\ldots\Big)\,.
\end{align}

\medskip

We also choose the $D=11$ solution of the section constraint~\eqref{eq:SC} and therefore only keep the ordinary space-time derivatives $\partial_m$. Since the fields in~\eqref{eq:L211} and~\eqref{eq:L1011} are also constrained, we also keep only their first component $\tilde{\chi}_m{}^{\tilde{\alpha}}$, $\tilde{\zeta}_m{}^{\hL}$, e.g.
\begin{align}
\label{eq:chi11}
\chi_{m;n_1\ldots n_9}\,,
\end{align}
where we recall that the semi-colon indicates that this is a reducible representation of $GL(11)$ obtained as a tensor product of a one-form with the components of a nine-form and therefore has the irreducible pieces with Young symmetries $(10)$ and $(9,1)$. 

\subsubsection*{Currents, field strengths and tensors}

As a preparation to the evaluation of the pseudo-Lagrangian, we first describe how to obtain its various ingredients in $GL(11)$ level decomposition. We shall from now on leave out all tildes on the components of the redefined fields, currents and field strengths for readability. However, we shall denote the components in level decomposition with calligraphic latin letters.

Working out the components of the Levi current~\eqref{eq:LeviJ} and the semi-flat current~\eqref{eq:SFcur} using~\eqref{eq:NM11}, we find (at non-negative $GL(11)$ levels $\alpha_{(k)}$)
\begin{align}
\label{eq:J11}
k=0:&&
\cJ_{p;n}{}^m&= g^{ms} \partial_p g_{ns}\,,\nn\\
k=1:&&\cJ_{p;n_1n_2n_3} &= \partial_p A_{n_1n_2n_3}\,,\nn\\
k=2:&&\cJ_{p;n_1\ldots n_6} &= \partial_p A_{n_1\ldots n_6}+10 A_{[n_1n_2n_3} \partial_{|p|} A_{n_4n_5n_6]} \,,\\
k=3:&&\cJ_{p;n_1\ldots n_8,m} &= \partial_p h_{n_1\ldots n_8,m}+ \frac{56}3 \big( A_{[n_1n_2n_3} \partial_{|p|} A_{n_4\ldots n_8]m} - A_{m[n_1n_2} \partial_{|p|} A_{n_3\ldots n_8]}\big)\nn\\
&&&\quad \quad- \frac{280}3 A_{m[n_1n_2} A_{n_3n_4n_5} \partial_{|p|} A_{n_6n_7n_8]} \,.\nn
\end{align}
All the indices occurring here are curved with respect to the Levi factor $GL(11)$ which is why we call the formulation semi-flat. We shall use the $GL(11)$ metric $g_{mn}$ to raise and lower indices on these and similar $GL(11)$ tensor (densities). 

The components of the semi-flat field strength can be obtained from the current components~\eqref{eq:J11} and the structure constants $C^{IM}{}_{\widehat{\alpha}}$ given in \eqref{F} as
\begin{align}
\label{eq:JF11}
 \cF_{n_1n_2}{}^m&= 2\cJ_{[n_1;n_2]}{}^m  \,,\nn\\
\cF_{n_1\ldots n_4}  &= 4 \cJ_{[n_1;n_2n_3n_4]}\,,\nn\\
 \cF_{n_1\ldots n_7}  &=7 \cJ_{[n_1;n_2\ldots n_7]}\,,\\
\cF_{n_1\ldots n_9;m}  &=  9 \cJ_{[n_1;n_2\ldots n_9],m} + \chi_{m;n_1\ldots n_9}\,,\nn
\end{align}
where in the last equation we have defined the $GL(11)$ reducible field strength
\begin{align}
\label{eq:F91red}
\cF_{n_1\ldots n_9;m} = \cF_{n_1\ldots n_9,m} - \cF_{n_1\ldots n_9m}  
\,.
\end{align}
Separating its relation~\eqref{eq:JF11} to the current components into irreducible constituents yields
\begin{align}
\cF_{n_1\ldots n_9,m} &= 9\cJ_{[n_1;n_2\ldots n_9],m} + \chi_{m;n_1\ldots n_9} - \chi_{[m;n_1\ldots n_9]} \,,\nn\\
\cF_{n_1\ldots n_{10}} &= \chi_{[n_1;n_2\ldots n_{10}]}\,.
\end{align}
These field strengths components take the explicit form
\begin{align}
\label{eq:FS11}
 \cF_{n_1n_2}{}^m &=  2g^{mp} \partial_{[n_1} g_{n_2]p}    \,,\nn\\
 \cF_{n_1\dots n_4} &= 4 \partial_{[n_1} A_{n_2n_3n_4]} \,,\\
 \cF_{n_1\dots n_7} &= 7 \partial_{[n_1} A_{n_2\dots n_7]}+ 70 A_{[n_1n_2n_3} \partial_{n_4} A_{n_5n_6n_7]} \,,\nn\\
 \cF_{n_1\dots n_9;m}  &= 9 \partial_{[n_1} h_{n_2n_3n_9],m}  + 840 A_{m[n_1n_2} A_{n_3n_4n_5} \partial_{n_6} A_{n_7n_8n_9]}  \nn\\
  &\qquad + 168\bigl( A_{[n_1n_2n_3} \partial_{n_4} A_{n_5n_6n_7n_8n_9]m} + A_{m[n_1n_2}  \partial_{n_3} A_{n_4n_5n_6n_7n_8n_9]}\bigr)  + \chi_{m;n_1\dots n_9} \,.\nn
\end{align}
As we have seen in the previous section, the pseudo-Lagrangian \eqref{LagrangianReduced11D} only involves the three first field strengths above. 

\medskip

The $E_{11}$ representation matrices are most easily determined by reading them off from the rigid $\mf{e}_{11}$ transformations of the fields and derivatives. These transformations are displayed in Appendix~\ref{GL11E11tr} in the conventions of~\cite{Bossard:2017wxl}. In particular the components of the matrices $T^{\alpha M}{}_N$ can be obtained from \eq{eq:L1mod}.

The components of $\Pi_\ta{}^{MN}$ can be  determined from the explicit gauge transformation of $\chi_M{}^\ta$  as given in~\cite[Eq.~(4.17e)]{Bossard:2019ksx}. The relevant lowest component  of $\Pi_\ta{}^{MN}$ is fixed by comparing~\cite{Bossard:2019ksx}\footnote{Our convention is that $\varepsilon^{0\hspace{0.1mm}1\,\ldots\,1\hspace{-0.2mm}0}=+1$ is the Levi--Civita symbol (with constant components) and its indices are lowered with the metric, so that $\varepsilon^{n_1\ldots n_{1\hspace{-0.2mm}1}} \varepsilon_{m_1\ldots m_{1\hspace{-0.2mm}1}}= 11! \det(g) \delta^{[n_1}_{[m_1} \cdots \delta^{n_{1\hspace{-0.2mm}1}]}_{m_{1\hspace{-0.2mm}1}]}$.}  
\begin{align}
\label{eq:Pi11}
\delta_\xi \chi_{m; n_1\ldots n_9} &= 24 \partial_m \partial_{[n_1} \lambda_{n_2\dots n_8]}  - \tfrac{1}{\sqrt{-g}}g_{n_1 p_1} \dots g_{n_9p_9} g_{rq}  \varepsilon^{p_1\dots p_9 p r}  \partial_m\partial_p \xi^q \; , 
\end{align}
to~\eqref{eq:chiGT}. This implies that the relevant component $\Pi_\ta{}^{MN} $ is $\Pi^{n_1\ldots n_9}{}^{p_1p_2} =- \varepsilon^{n_1\ldots n_9p_1p_2}$ and we take the summation over this $\ta$ component to include the canonical combinatorial factor $\frac1{9!}$.\footnote{In general, the combinatorial factors are those in the $K(E_{11})$ invariant bilinear form on the $R(\Lambda_2)$ representation.}

For the components of the cocycle $K^{\alpha\ta}{}_\beta$ we need to extend the results on the rigid $\mf{e}_{11}$ transformation given in~\cite[Eq.~(4.33h)]{Bossard:2017wxl} to a few more components. We write this here in terms of the commutator~\eqref{eq:T0CR} in the form
\begin{align}\label{topoKT} 
&\quad -\left[ h_m{}^n K^m{}_n + \frac1{3!} e_{n_1n_2n_3} E^{n_1n_2n_3} + \frac1{6!}e_{n_1\ldots n_6} E^{n_1\ldots n_6} + \frac1{8!}e_{n_1\ldots n_8,m} E^{n_1\ldots n_8,m}   , \tilde{F}_{n_1\ldots n_9} \right]\nn\\
&= 9 h_{[n_1}{}^p \tilde{F}_{n_2\ldots n_9]p}   -28  e_{[n_1n_2n_3} F_{n_4\ldots n_9]} - 56 e_{[n_1\ldots n_6} F_{n_7n_8n_9]} + 9 e_{[n_1\ldots n_8,|p|} h^p{}_{n_9]} \,,
\end{align}
where we recall that the lowest component $t^\ta$ in $R(\Lambda_2)\subset \cT_0$ is a nine-form $\tilde{F}_9$, see~\eqref{eq:L211}.

The tensors $C^{\tI}{}_{P\wa}$ can be determined conveniently from the auxiliary field strength
\begin{align}
\label{eq:FM3}
G^{\tI} = C^{\tI}{}_{M \widehat\alpha} \cM^{MN} \cJ_N{}^{\widehat\alpha} = C^{\tI}{}_{M \alpha} \cM^{MN} \cJ_N{}^{\alpha} + C^{\tI}{}_{M \ta} \cM^{MN} \chi_N{}^{\ta}
\end{align}
 that was defined in~\cite{Bossard:2017wxl}. Expanding it out to lowest orders in the $GL(11)$ level decomposition leads to~\cite[Eq.~(5.26)]{Bossard:2017wxl}
\begin{align}
G_{n_1\ldots n_8} &=  g^{pq} ( \cJ_{p;n_1\dots n_8,q}  +\chi_{p; n_1\dots n_8q} ) 
\end{align}
where we have given only the components that are non-zero when choosing the $D=11$ solution to the section constraint.
From this we find the following components of $C^\tI{}_{P\wa}$
\begin{align}
\label{eq:CtIGL11}
C_{n_1\ldots n_8;p}{}^{q_1\ldots  q_8,r} = 8!\,\left( \delta_{n_1\ldots n_8}^{q_1\ldots q_8 }\delta_p^{r} -\delta_{n_1\ldots n_8}^{[q_1\ldots q_8 }\delta_p^{r]} \right)\,,\quad\quad
C_{n_1\ldots n_8;p}{}^{q_1\ldots  q_9} = 9!\,\delta_{n_1\ldots n_8p}^{q_1\ldots q_8 q_9}\,.
\end{align}
Note that we will not need the components of $\Pi_\ta{}^{MN}$, $K^{\alpha\ta}{}_\beta$ and $C^{\tI}{}_{P\wa}$ defined above in Section~\ref{sec:D11dec} below, because all the corresponding contributions are already canceled in \eqref{sugra}. Nevertheless, they are important when considering the dual graviton field and they will be used in Section \ref{DualGravitonIn11D} and in Appendices \ref{LowLevelFather} and \ref{CancelO3}.

\subsubsection*{Expansion of the duality equation}

Before considering the pseudo-Lagrangian~\eqref{eq:Lag} in level decomposition, let us briefly comment on the expansion of the duality equations as this underscores the r\^ole of the constrained fields.

The duality equations~\eqref{eq:DElev} become for the first two instances~\cite[Eq.~(4.14)]{Bossard:2019ksx}
\begin{subequations}
  \label{eq:DE11}
 \begin{align}
   \cF_{n_1\dots n_7}  &= \frac{1}{24\sqrt{-g}} g_{n_1p_1} \dots g_{n_7p_7} \varepsilon^{p_1\dots p_7 q_1\dots q_4 } \cF_{q_1q_2q_3q_4} \,,
   \label{fe}\\
 \cF_{n_1n_2}{}^m  &=     \frac{1}{9!\sqrt{-g}} g_{n_1q_1} g_{n_2q_2}  \varepsilon^{q_1q_2p_1\dots p_9 } g^{mq}  \cF_{p_1\dots p_9;q} \,. \label{eq:chi11EOM}
\end{align}
\end{subequations}
The first equation does not contain the constrained fields and is recognised as the usual matter duality equation of $D=11$ supergravity. An integrability condition of this equation is $\partial_m\big(\sqrt{-g} \cF^{mn_1n_2n_3}\big) = \frac1{1152} \varepsilon^{n_1n_2n_3p_1\ldots p_8} \cF_{p_1\ldots p_4} \cF_{p_5\ldots p_8}$, where the familiar non-linear term  \cite{Cremmer:1978km} is due to the Bianchi identity $\partial_{[n_1} \cF_{n_2\ldots n_8]} = \frac{35}{8} \cF_{[n_1\ldots n_4} \cF_{n_5\ldots n_8]}$ of the seven-form field strength that is implied by~\eqref{eq:FS11}. 
The second equation is akin to the first-order dual gravity relation~\cite{Nieto:1999pn,Hull:2001iu,West:2002jj,Boulanger:2008nd,Henneaux:2019zod,Boulanger:2020yib}. It contains the additional field $\chi_{m;n_1\dots n_9}$ that plays the same r\^ole as the St\"{u}ckelberg gauge field in the vielbein formulation \cite{West:2002jj,Boulanger:2008nd}. Equation~\eqref{eq:DE11} can be seen as simply determining the field $\chi_{m;n_1\dots n_9}$ in terms of the other fields and therefore imposes no condition on the gravitational dynamics. This is remedied by the pseudo-Lagrangian~\eqref{eq:Lag} that we have put forward in this paper: It implies additional equations for the constrained fields that we derived in Section~\ref{sec:EOM}. As we shall see in this section, we can alternatively use the pseudo-Lagrangian together with a choice of section condition to integrate out the constrained fields, thereby solving their equations of motion. The remaining dynamics for the $E_{11}$ coset  fields will then be seen to be equivalent to the bosonic sector of $D=11$ supergravity.

\subsection{\texorpdfstring{$D=11$ supergravity from the $GL(11)$ expanded pseudo-Lagrangian}{D=11 supergravity from the GL(11) expanded pseudo-Lagrangian}}
\label{sec:D11dec}

We shall now explicitly write out the various terms in~\eqref{LagrangianReduced11D}, using the notation introduced in \eq{sugra}.  
The resulting pseudo-Lagrangian $ {\cal L}- \sum_{k=2}^\infty \mathcal{O}_{k}$ is the (bosonic part of the) eleven-dimensional supergravity Lagrangian  \cite{Cremmer:1978km}, where we recall from~\eqref{eq:Lk} that $\mathcal{O}_k {=} -\frac14 m^{I_\dgr{k}J_\dgr{k}} \mathcal{E}_{I_\dgr{k}}\mathcal{E}_{J_\dgr{k}}$.

Evaluating the first term in \eq{sugra} requires knowledge of the representation-theoretic contraction that can be extracted from the linearised analysis in~\cite[Eq.~(5.39)]{Bossard:2017wxl}. From this we can immediately write down the resulting expression in $GL(11)$ level decomposition for the components of the field strengths deduced above using \eqref{eq:LLT0}
\bea
\sum_{k=1,2} \mathcal{L}_{\text{kin}}\big|_{k} &=& \frac14  \sum_{k=1}^2m_{I_\dgr{k} J_\dgr{k}} \tilde{\cF}^{I_\dgr{k}} \tilde{\cF}^{J_\dgr{k}}
\label{eq:KT111}\\
&=&  \frac14 \sqrt{-g} \Big( \frac{1}{7!} \cF_{n_1\cdots n_7}  
\cF^{n_1\cdots n_7} 
+ \frac{1}{4!} \cF_{n_1\cdots n_4}  \cF^{n_1\cdots n_4}  
\Big) \ .
\nn
\eea
The $\sqrt{-g}$ comes from the weight of the $R(\Lambda_1)$ coordinate module~\eqref{eq:L1wt}. This contains all the contributions to the kinetic term that  remain after integrating out all the constrained fields. Indices are raised and lowered with the $GL(11)$ metric $g_{mn}$. In particular the matrix components $m_{I_\dgr{1}J_\dgr{1}}$ and $m_{I_\dgr{2}J_\dgr{2}}$ are given respectively by  
\bea
m^{n_1n_2n_3n_4;p_1p_2p_3p_4} &=& 4 ! \sqrt{-g} g^{[n_1|p_1} g^{|n_2|p_2} g^{|n_3|p_3} g^{|n_4]p_4} \; , \CR
 m^{n_1\dots n_7;p_1\dots p_7} &=& 7 ! \sqrt{-g} g^{[n_1|p_1} g^{|n_2|p_2} \dots  g^{|n_7]p_7}\; .
\eea
Next, we compute the terms from $\mathcal{L}_{{\rm pot}_1}$.  For the first line of  \eqref{p1} we get 
\begin{align}
\label{eq:GL11pot11}
&\quad - \frac{1}{4} m_{\alpha_\dgr{0}\beta_{\dgr{0}}} m^{mn} \tilde{\mathcal{J}}_m{}^{\alpha_\dgr{0}} \tilde{\mathcal{J}}_n{}^{\beta_\dgr{0}}  - \frac12   \sum_{k=1}^2 m_{\alpha_\dgr{k}\beta_{\dgr{k}}} m^{mn} \tilde{\mathcal{J}}_m{}^{\alpha_\dgr{k}} \tilde{\mathcal{J}}_n{}^{\beta_\dgr{k}} \\
&=-\frac14 \sqrt{-g} g^{mn} \Big( \cJ_{m;p}{}^q \cJ_{n;q}{}^p - \frac12 \cJ_{m;p}{}^p \cJ_{n;q}{}^q + \frac{2}{3!} \cJ_{m;p_1p_2p_3} \cJ_{n;}{}^{p_1p_2p_3} + \frac{2}{6!} \cJ_{m;p_1\ldots p_6} \cJ_{n;}{}^{p_1\ldots p_6} \Bigr)\,,\nn
\end{align}
where we used $m^{mn} = \sqrt{-g} g^{mn}$ and the bilinear form $m_{\alpha_\dgr{k}\beta_\dgr{k}}$ in the same normalisation as in~\cite[App.~A.1]{Bossard:2017wxl}, see also~\eqref{KCG11}.
For the second line of  \eqref{p1} we use~\eqref{eq:L1mod} and~\eqref{eq:L1wt} with the rule 
\be  \Lambda_{\beta} T^{\beta Q}{}_p  \partial_Q =  \delta_\Lambda  \partial_m \; , \qquad \Lambda_\gamma T^{\gamma m}{}_{Q}    \partial_m =  \delta_\Lambda  \partial_Q \; , \ee
so that for any parameters $\xi_m, \zeta_m, \psi^{\alpha}, \phi^{\beta}$
\be m_{\alpha_\dgr{k}\gamma_\dgr{k}}  T^{\gamma_\dgr{k} n}{}_{Q} T_{\beta_\dgr{k}}{}^Q{}_p m^{pm}  \xi_m \psi^{\alpha_\dgr{k}}  \zeta_n \phi^{\beta_\dgr{k}}   = m^{pm} \xi_m\delta_{m^{-1} \psi^\dagger m}  \delta_{\phi} \zeta_p \; , \ee
where $\delta_\phi$ and $\delta_{m^{-1} \psi^\dagger m} $ act only on $\zeta_p$ as $\delta_\Lambda$ in~\eqref{eq:L1mod} and~\eqref{eq:L1wt}. 
Performing these steps for all terms,  we find the following expression 
\bea
\label{eq:GL11pot12}
 &&   \sum_{k =0}^2 m_{\alpha_\dgr{k}\gamma_\dgr{k}}  T^{\gamma_\dgr{k} n}{}_{Q} T_{\beta_\dgr{k}}{}^Q{}_p m^{pm}   \tilde{\mathcal{J}}_m{}^{\alpha_\dgr{k}} \tilde{\mathcal{J}}_n{}^{\beta_\dgr{k}} \\
&= & \sqrt{-g}  g^{mn}  \Big( \cJ_{q;n}{}^p \cJ_{m;p}{}^q - \cJ_{p;q}{}^q \cJ_{m;n}{}^p {+}\tfrac14 \cJ_{m;p}{}^p \cJ_{n;q}{}^q {+} \tfrac12 \cJ_{q;mp_1p_2} \cJ_{n;}{}^{qp_1p_2} {+} \tfrac{1}{5!} \cJ_{q;mp_1\ldots p_5} \cJ_{n;}{}^{qp_1\ldots p_5} \Bigr) 
\nn
\eea
where we used $\cJ_{m;q}{}^r  g^{pq} g_{rn} =\cJ_{m;n}{}^p$. To combine these terms we use moreover the expansion of the Einstein--Hilbert Lagrangian 
\begin{align} \label{HEder} 
\sqrt{-g} R &= \partial_p \Big[ \sqrt{-g} g^{mn} g^{pq} (\partial_m g_{nq} -\partial_q g_{mn})\Big]
\nn\\
&\quad +\sqrt{-g} g^{mn} \Big( \frac12 \cJ_{p;m}{}^q \cJ_{n;q}{}^p -\frac14 \cJ_{m;p}{}^q \cJ_{n;q}{}^p + \frac14 \cJ_{m;p}{}^p \cJ_{n;q}{}^q -\frac12 \cJ_{p;q}{}^q \cJ_{m;n}{}^p \Big)
\end{align}
that can be checked by expanding everything out in the metric and its derivative. Combining \eq{eq:GL11pot11} with \eq{eq:GL11pot12}, and using \eq{HEder} we find 
\bea
\sum_{k=0,1,2}  \mathcal{L}_{{\rm pot}_1}\big|_{k}
&=& \sqrt{-g} \Big( R - \frac{1}{2\cdot 4!} \cF_{n_1\ldots n_4} \cF^{n_1\ldots n_4} - \frac1{2\cdot 7!} \cF_{n_1\ldots n_7} \cF^{n_1\ldots n_7}  \Bigr) \nn\\
&& \hspace{50mm} -\partial_p \Big[ \sqrt{-g} g^{mn} g^{pq} (\partial_m g_{nq} -\partial_q g_{mn})\Big]\ .
\label{eq:pot11}
\eea
where we have used the relation~\eqref{eq:JF11} for expressing the current components via field strength components. 

Finally, we  use \eqref{topoKT} to write \eqref{rtop}
\be
\mathcal{L}_{\text{top}} \big|_{2} = \frac{1}{2\cdot 9!} \varepsilon^{n_1\ldots n_{11}} \cF_{n_1\ldots n_4} \cF_{n_5\ldots n_{11}} \ .
\label{eq:top11}
\ee

Summing up the results for~\eqref{eq:KT111}, \eqref{eq:pot11} and~\eqref{eq:top11}, we get
\bea
\mathcal{L} &=& \sqrt{-g} \Big( R - \frac{1}{4\cdot 4!} \cF_{n_1\ldots n_4} \cF^{n_1\ldots n_4} - \frac1{4\cdot 7!} \cF_{n_1\ldots n_7} \cF^{n_1\ldots n_7} \Big)+ \frac{1}{2\cdot 9!} \varepsilon^{n_1\ldots n_{11}} \cF_{n_1\ldots n_4} \cF_{n_5\ldots n_{11}} 
\nn\\
&&\quad - \partial_p \Big[ \sqrt{-g} g^{mn} g^{pq} (\partial_m g_{nq} -\partial_q g_{mn}) + \frac1{9!} \varepsilon^{pn_1\dots n_{10}} \chi_{n_1;n_2\dots n_{10}} \Big]  +\sum_{k=3}^\infty \mathcal{O}_{k}\ .
\label{eq:L116form}
\eea

The first line describes the bosonic sector of eleven-dimensional supergravity in the democratic formulation in which both  three-form and six-form potentials occur.\footnote{Though we are not aware of a worked out democratic formulation of $D=11$ supergravity, it is similar in spirit to the democratic formulation of supergravities in $D=10$~ \cite{Bergshoeff:2001pv}.}  In order to compare with the standard formulation of eleven-dimensional supergravity it is convenient to exhibit that the terms involving the six-form potential also combine in a term proportional to the square of its defining duality equation \eqref{fe}
\be 
\mathcal{O}_2 = - \frac{\sqrt{-g}}{4\cdot 7!} \Big( \cF_{n_1\ldots n_7} -\frac{1}{4!\sqrt{-g}} \varepsilon_{n_1\ldots n_7}{}^{p_1\ldots p_4} \cF_{p_1\ldots p_4}\Big) \Big( \cF^{n_1\ldots n_7} - \frac{1}{4!\sqrt{-g}} \varepsilon^{n_1\ldots n_7q_1\ldots q_4} \cF_{q_1\ldots q_4} \Big) \; . 
\label{cO2}
\ee
Although $\mathcal{O}_2$ cannot be canceled by integrating out an auxiliary field, it does not affect the equations of motion that include the duality equations. We obtain finally
\bea
\mathcal{L} 
&= & {\mathcal L}_{\text{sugra}}+ \partial_p ( \mathcal{U}^p + \mathcal{V}^p) +\sum_{k=2}^\infty \mathcal{O}_{k}\ ,
\label{eq:L11}
\eea
where
\begin{subequations}
\begin{align}
\label{Lsugra} 
{\mathcal L}_{\text{sugra}}&= \sqrt{-g} \Big( R - \frac{1}{2\cdot 4!} \cF_{n_1\ldots n_4} \cF^{n_1\ldots n_4}\Big)- \frac1{144^2} \varepsilon^{n_1\ldots n_{11}}  A_{n_1n_2n_3} \cF_{n_4\ldots n_7}\cF_{n_8\ldots n_{11}} \ ,
\w2
\mathcal{U}^p &\equiv - \sqrt{-g} g^{mn} g^{pq} (\partial_m g_{nq} -\partial_q g_{mn})\ ,
\w2
\mathcal{V}^p &\equiv -  \frac{1}{9!} \varepsilon^{pn_1\ldots n_{10}}  \bigl( 4 A_{n_1n_2n_3} \cF_{n_4\ldots n_{10}}+ \chi_{n_1;\dots n_{10}}\bigr) \ ,
\end{align}
\end{subequations}
and we recall that ${\mathcal O}_k$ represents the square of the duality equations, as defined in \eq{eq:Lk}. This result shows that  $\mathcal{L}-\sum_{k=2}^\infty \mathcal{O}_{k}$  is the bosonic part of the eleven-dimensional supergravity Lagrangian, up to total derivative terms, as claimed at the beginning of this section. We have denoted this  Lagrangian by ${\mathcal L}_{\text{sugra}}$,  even though it represents only the bosonic sector, for brevity in notation. Given that $\sum_{k=2}^\infty \mathcal{O}_{k}$ does not affect the equations of motion, as a sum of squares of the duality equations, we proved that $E_{11}$ exceptional field theory reproduces eleven-dimensional supergravity dynamics on section. 

Moreover, this result implies that the whole dynamics of  $E_{11}$ exceptional field theory on the eleven-dimensional section is described by the duality equation \eqref{eq:DE} and the supergravity Einstein equation deriving from ${\mathcal L}_{\text{sugra}}$. We shall now describe the dynamics of the higher-level fields.

\section{Higher level dualities and the constrained fields}
\label{HigherLevelSection}

In this section, we describe how the Euler--Lagrange equations of the pseudo-Lagrangian  provide integrability conditions for the constrained fields, such that the higher-level components of the duality equation  \eqref{eq:DE} are dynamical and describe the propagation of an infinite sequence of dual fields, providing a rather explicit realisation of the proposal of~\cite{Riccioni:2006az}.  As explained in the last section, this is equivalent  to considering the duality equation \eqref{eq:DE} together with the eleven-dimensional Einstein equation from \eqref{Lsugra}, nevertheless it will sometimes be more convenient to consider other pseudo-Lagrangians that are equivalent to the $E_{11}$ exceptional field theory pseudo-Lagrangian \eqref{eq:Lag}. 

\subsection{Preliminaries}
\label{sec:prel}

The Einstein equation that follows from the full pseudo-Lagrangian $\mathcal{L}$ includes by construction an energy-momentum tensor with an infinite sum of contributions from all positive level fields. Using all the infinitely many duality equations~\eqref{eq:DE} these infinitely many contributions reduce to a finite expression. If one wants to obtain a finite energy-momentum for a specific finite set of fields without using the duality equations, one has to consider
pseudo-Lagrangians of the form ${\mathcal L}_{\text{sugra}} + \sum_{k \in I} c_k \mathcal{O}_k   $, where $I$ is a finite set of integers $k\geq 2$. By construction, any such pseudo-Lagrangian produces an equivalent set of Euler--Lagrange equations together with the duality equation~\eqref{eq:DE}. For some specific choice of $c_k \in \{0,1,2\}$ for $k\in I$, one can moreover obtain an actual Lagrangian for a subset of the fields, as ${\mathcal L}_{\text{sugra}} $ for the metric and the three-form potential. Although these pseudo-Lagrangians are not invariant under generalised diffeomorphisms (up to a total derivative), the set of equations of motion they produce transform into themselves and into the duality equations under generalised diffeomorphisms. The reason for this is that they define the same set of equations as the Euler--Lagrange equations of the invariant pseudo-Lagrangian $\mathcal{L}$ and the duality equations~\eqref{eq:DE}.

The field content of the $E_{11}$ coset representative was analysed to all $GL(11)$ levels in~\cite{Riccioni:2006az} with the result that all generators fall into one of the following four classes of $GL(11)$ Young tableaux\footnote{Since all generators are in representations of $GL(11)$, any tensor necessarily will involve root spaces that only exist for $\mf{e}_{11}$ and not for its $\mf{e}_n$ subalgebra with $n=8,9,10$. In the table, we refer to the lowest weight element of any $GL(11)$ tensor generator, which is then a root of $\mf{e}_n$. Equivalently one can think of the $GL(11)$ tensor with indices restricted to run from $1$ to $n$.}
\begin{align}
\label{eq:gens}
\text{longest column has length $\leq 8$:}&\quad \text{generators that are part of $E_8$}\nn\\
\text{longest column has length $9$:}&\quad \text{generators that are part of $E_9$ but not of $E_8$}\nn\\
\text{longest column has length $10$:}&\quad \text{generators that are part of $E_{10}$ but not of $E_9$}\\
\text{longest column has length $11$:}&\quad \text{generators that are part of $E_{11}$ but not of $E_{10}$}\nn
\end{align}
The fields that include at least one column of either eleven or ten antisymmetrised indices, i.e. belong to $E_{10}$ or $E_{11}$, appear in exponentially growing number in the $GL(11)$ decomposition, but they are non-propagating. The fields with one single column of ten antisymmetrised indices correspond for example to non-geometric deformations of the theory, such as massive type IIA \cite{Bergshoeff:2007qi,Tumanov:2016dxc}. All the propagating fields instead appear with (outer) multiplicity one at each level and stem from $E_9$. These are the metric $g_{\mu\nu}$ at level $0$,  the tensor of Young symmetry $(9^n,3)$ at level $1+3n$, of Young symmetry $(9^n,6)$ at level $2+3n$ and $(9^n,8,1)$ at level $3+3n$, for all $n\in \mathds{N}$. Here we use the notation that $(9^n,p,q)$ is the partition of $ 9n + p + q$ whose Young tableau includes $n$ columns of $9$ antisymmetrised indices. The fields with $n=0$ are the standard three-form, the six-form and the dual graviton field. 

It was shown in \cite{Henneaux:2011mm} that the fields with $n\ge 1$ cannot be  described in the general framework of \cite{Labastida:1987kw,Bekaert:2006ix} for unitary representations of the Lorentz group, in which the field equations are in the same $SO(1,10)$ representation as the gauge field. This difficulty was circumvented  in  \cite{Boulanger:2015mka}, in which it was shown that there are consistent duality equations for the fields of Young symmetry $(9^n,3)$, such that they propagate the same degrees of freedom as the three-form gauge field. The field strength $G_{10^n,4} \sim d^{n+1} A_{9^n,3}$ for the potential of Young symmetry $(9^n,3)$ is defined to be in the irreducible $GL(11)$ representation of Young symmetry $(10^n,4)$~\cite{Bekaert:2002dt}. This representation decomposes under the Lorentz group $SO(1,10)$ into a tensor with Young symmetry  $(4,1^n)$ (i.e. of $B_5$ weight $n \Lambda_1 + \Lambda_4$ in Bourbaki numbering conventions) plus all possible traces. The next-largest representation in the branching has Young symmetry $(5,1^{n-3})$ ($B_5$ highest weight $(n-1)\Lambda_1+2\Lambda_5$) and corresponds to taking four traces.\footnote{That this is the first non-trivial trace can be seen for example for $G_{10,4}$ from the identity 
$$
G_{a_1\dots a_{10},b_1\dots b_4} = -\frac1{3!7!}  \varepsilon_{a_1\dots a_{10}c} \varepsilon_{b_1\dots b_4}{}^{d_1\dots d_7} G_{d_1\dots d_7e_1e_2e_3,}{}^{e_1e_2e_3c} - \frac1{4!6!} \varepsilon_{a_1\dots a_{10}c} \varepsilon_{b_1\dots b_4}{}^{d_1\dots d_6 c} G_{d_1\dots d_6e_1\dots e_4,}{}^{e_1\dots e_4}
$$ 
which implies that one does not project out any component of a $(10,4)$ tensor of $GL(11)$ by contracting up to three pairs of indices.}
The field equation in~\cite{Boulanger:2015mka} for the $(9^n,3)$ potential is that all $SO(1,10)$ components that are not in the largest $SO(1,10)$ representation $(4,1^n)$ have to vanish and this is the appropriate alternative generalisation of the usual `Riemann equal Weyl' equation for mixed symmetry fields. However, it is important that this requires taking several traces and not only a single trace, so that the Weyl tensor is not defined by the property that any possible trace vanishes. Therefore the field equation in~\cite{Boulanger:2015mka} differs from that in~\cite{Labastida:1987kw,Bekaert:2006ix,Henneaux:2011mm}. Similarly, for the potentials $(9^n,6)$ and $(9^n,8,1)$ the field equation requires keeping only the largest $SO(1,10)$ component of the corresponding field strengths. The fact that the propagating degrees of freedom can be described alternatively using these higher potentials is also in agreement with the fact that the $E_9$ linear system for $D=2$ supergravity provides an infinite cascade of fields dual to the propagating ones using $E_9$ generators~\cite{Julia,Nicolai:1987kz}.

The linearised duality equation \eqref{LinearisedDuality} was recognised in \cite{Bossard:2017wxl} to relate fields of level shifted by $3$, 
\be 
m_{I_\dgr{k} J_\dgr{k}} \tilde{\cF}^{J_\dgr{k}}   = \Omega_{I_\dgr{k} J_\dgr{3-k}} \tilde{\cF}^{J_\dgr{3-k}} 
 \ee
such that a field of Young symmetry $(9^{n+1},3)$ is indeed dual to the one of Young symmetry $(9^n,3)$ 
\bea 
&&  m_{I_\dgr{4+3n} J_\dgr{4+3n}} \bigl( C^{J_\dgr{4+3n} m}{}_{\alpha_\dgr{4+3n}} J_m{}^{\alpha_\dgr{4+3n}} +C^{J_\dgr{4+3n} m}{}_{\tilde{\alpha}_\dgr{4+3n}} \chi_m{}^{\tilde{\alpha}_\dgr{4+3n}}   +C^{J_\dgr{4+3n} m}{}_{\widehat{\Lambda}_\dgr{4+3n}} \zeta_m{}^{\widehat{\Lambda}_\dgr{4+3n}}  \bigr)    \CR
&=& \Omega_{I_\dgr{4+3n} J_\dgr{-1-3n}} \bigl( C^{J_\dgr{-1-3n} m}{}_{\beta_\dgr{-1-3n}}   m^{\beta_\dgr{-1-3n} \gamma_\dgr{-1-3n}} \kappa_{\alpha_\dgr{1+3n}\gamma_\dgr{-1-3n}} J_m{}^{\alpha_\dgr{1+3n}}  \CR
&& \hspace{40mm}  +C^{J_\dgr{-1-3n} m}{}_{\tilde{\alpha}_\dgr{-1-3n}} \chi_m{}^{\tilde{\alpha}_\dgr{-1-3n}}   +C^{J_\dgr{-1-3n} m}{}_{\widehat{\Lambda}_\dgr{-1-3n}} \zeta_m{}^{\widehat{\Lambda}_\dgr{-1-3n}}  \bigr) \; ,
\eea
and similarly for $(9^n,6)$ and $(9^n,8,1)$. Moreover, assuming the constrained fields are total derivatives, the duality equations reduce to
\bea && \eta_{I_\dgr{4+3n} J_\dgr{4+3n}} \bigl( C^{J_\dgr{4+3n} m}{}_{\alpha_\dgr{4+3n}} \partial_m \phi^{\alpha_\dgr{4+3n}} \!+\!C^{J_\dgr{4+3n} m}{}_{\tilde{\alpha}_\dgr{4+3n}} \partial_m X^{\tilde{\alpha}_\dgr{4+3n}}  \! +\!C^{J_\dgr{4+3n} m}{}_{\widehat{\Lambda}_\dgr{4+3n}} \partial_m Y^{\widehat{\Lambda}_\dgr{4+3n}}  \bigr)    \CR
&=& \Omega_{I_\dgr{4+3n} J_\dgr{-1-3n}} \bigl( C^{J_\dgr{-1-3n} m}{}_{\beta_\dgr{-1-3n}}   \eta^{\beta_\dgr{-1-3n} \gamma_\dgr{-1-3n}} \kappa_{\alpha_\dgr{1+3n}\gamma_\dgr{-1-3n}} \partial_m \phi^{\alpha_\dgr{1+3n}}  \CR
&& \hspace{40mm}  +C^{J_\dgr{-1-3n} m}{}_{\tilde{\alpha}_\dgr{-1-3n}} \partial_m X^{\tilde{\alpha}_\dgr{-1-3n}}   +C^{J_\dgr{-1-3n} m}{}_{\widehat{\Lambda}_\dgr{-1-3n}} \partial_m Y^{\widehat{\Lambda}_\dgr{-1-3n}}  \bigr) \; ,  \eea
and the fields $X^{\tilde{\alpha}}$ and $Y^{\widehat{\Lambda}}$ provide the correct set of fields to write gauge-invariant  linearised duality equations \cite{Bossard:2019ksx} consistently with \cite{Boulanger:2015mka}. In this section we shall describe how the use of the pseudo-Lagrangian  \eqref{eq:Lag} allows us to define integrability conditions for the constrained fields $\chi_m{}^{\tilde{\alpha}}$ and $\zeta_m{}^{\widehat{\Lambda}}$ to obtain the  non-linear  duality equations.

We shall find in particular that ${\mathcal L}_{\text{sugra}}+ {\mathcal O}_3$ defines a Lagrangian for the metric, the three-form potential, the dual graviton and the associated constrained field $\chi_{1;9}$, which describes dual gravity at the non-linear level. The Euler--Lagrange equation for the metric can then be interpreted as an integrability condition for the constrained field $\chi_{1;9}$, while the constrained field Euler--Lagrange equation is the duality equation \eqref{eq:chi11EOM} for the metric that is algebraic in $\chi_{1;9}$. Injecting the algebraic solution of the latter equation into the former gives back the Einstein equation, while solving first the integrability condition for the constrained field produces a dynamical duality equation for the dual graviton. This duality equation reproduces the expected dual graviton propagation in the linearised approximation. As an aside, we shall show the relation between this dual graviton action and that of  \cite{West:2001as,West:2002jj,Boulanger:2008nd} which gives an equivalent description of the dual graviton in the vielbein formulation. 

After this, we shall examine ${\mathcal L}_{\text{sugra}}+ 2 {\mathcal O}_4$, which provides a Lagrangian for the metric, the three-form potential, the level 4 potential $A_{9,3}$ and the associated constrained field $\chi_{1;10;2}$. As for the dual graviton, the Euler--Lagrange equation of the three-form potential gives an integrability condition for the constrained field  $\chi_{1;10;2}$, while the Euler--Lagrange equation of  $\chi_{1;10;2}$ gives the level 4 duality equation. Solving first the duality equation for $\chi_{1;10;2}$ one recovers the eleven-dimensional supergravity equation for the three-form, while solving first the integrability condition for $\chi_{1;10;2}$ gives a dynamical duality equation for the gradient dual $A_{9,3}$. 

We expect that this procedure can be generalised to all levels to give the infinite chains of gradient dual fields as described originally in \cite{Boulanger:2015mka}, along the lines of Section 4.3 in \cite{Bossard:2019ksx}. The missing step for showing this equivalence to the chain of linearised duality equations in~\cite{Boulanger:2015mka} is to prove that the constrained fields $\chi_m{}^{\tilde{\alpha}}$ that appear in the propagating duality equations must be total derivatives $\partial_m X^{\tilde{\alpha}}$  in the linearised approximation, up to terms that do not contribute to the duality equations.  The generalisation to the $GL(11)$ level 5 field $A_{9,6}$ is straightforward, but things are more subtle starting from $GL(11)$ level 6.  We shall only describe schematically the generalisation to higher levels, and explain in particular that one must take additional curls of the gauge-invariant Euler--Lagrange equations as their integrability conditions to deduce $\chi_m{}^\ta=\partial_m X^\ta$. In~\cite{Tumanov:2016abm,Tumanov:2016dxc,Glennon:2020uov} additional derivatives were required to get gauge-invariant higher-order duality equations, whereas here the higher derivatives are rather a technical tool for proving from the duality equations and the integrability conditions that the constrained fields are effectively total derivatives in the way they appear in the field strengths to all levels. For the analysis of our system the equivalence to the chain of linearised duality equations~\cite{Boulanger:2015mka} is not a necessary property since we already proved that our model propagates the correct degrees of freedom non-linearly.

\subsection{The dual graviton}
\label{DualGravitonIn11D}

We will first analyse the dual graviton. We shall write down the full nonlinear Lagrangian, the resulting equations of motion and relation to the vielbein formulation. In this section we show that the $E_{11}$-invariant pseudo-Lagrangian entails a Lagrangian  for the (non-linear) dual graviton in the metric formulation, which we will show, upon truncation to the gravitational sector, to agree  with the Lagrangian derived in \cite{West:2001as,West:2002jj,Boulanger:2008nd} in the vielbein formulation.

\subsubsection*{The Lagrangian}

\vskip 2mm


As explained above, we shall now compute the Lagrangian
\be 
\cL_{\text{dual-gr}} \equiv  {\mathcal L}- \partial_p  \mathcal{V}^p -  \mathcal{O}_2 -\sum_{k=4}^\infty \mathcal{O}_{k} 
 =  {\mathcal L}_{\text{sugra}}+ \partial_p \, \mathcal{U}^p + {\mathcal O}_3 \ , 
\ee
that is appropriate for the description of the dual graviton. The explicit expression of ${\mathcal O}_3$ is
\be  
\mathcal{O}_3 = - \frac{\sqrt{-g}}{4\cdot 8!}  \Bigl(  \cF_{n_1\dots n_{8}p;q} + \frac{1}{2 \sqrt{-g}} \varepsilon_{n_1\dots n_8p}{}^{s_1s_2} g_{qm} \cF_{s_1s_2}{}^m \Bigr) \Bigl( \cF^{n_1\cdot n_8q;p} + \frac{1}{2 \sqrt{-g}} \varepsilon^{n_1\dots n_8qr_1r_2} \cF_{r_1r_2}{}^p   \Bigr) \; . 
\label{cO3}
\ee
Using the identity
\bea 
\label{metricTopoTerm} 
 && \sqrt{-g} \Big(  R +\frac{1}{8} \cF_{n_1n_2}{}^p \cF^{n_1n_2}{}_p -\frac14 \cF_{np}{}^p \cF^{nq}{}_q \Bigr) - \partial_p \Big( \sqrt{-g} g^{mn} g^{pq} (\partial_m g_{nq} -\partial_q g_{mn})\Big)\CR 
 &=& - \frac12 \sqrt{-g} g^{p[m} g^{n]q} g^{rs} \partial_{m} g_{pr} \partial_{n} g_{qs} \; , 
 \eea
we can now write the Lagrangian
\bea
\label{DualGravityL} 
\cL_{\text{dual-gr}} &=& \sqrt{-g} \Big(   - \frac12 g^{p[m} g^{n]q} g^{rs} \partial_{m} g_{pr} \partial_{n} g_{qs} - \frac{1}{2\cdot 4!} \cF_{n_1\ldots n_4} \cF^{n_1\ldots n_4} - \frac{1}{4\cdot 8!} \cF_{n_1\dots n_{8}p;q} \cF^{n_1\dots n_8q;p}  \Big)
\CR
& &\quad - \frac1{144^2} \varepsilon^{n_1\ldots n_{11}}  A_{n_1n_2n_3} \cF_{n_4\ldots n_7} \cF_{n_8\ldots n_{11}} - \frac{1}{2\cdot 8!} \varepsilon^{n_1\dots n_{11}} g^{pq} \partial_{n_1} g_{n_2p} \cF_{qn_3\dots n_{10};n_{11}}\; . 
\eea
This Lagrangian depends only on the metric $g_{mn}$, the three-form potential $A_{n_1n_2n_3}$ and the field strength $\cF_{n_1\dots n_9;m}$. Although the explicit expression of the field strength \eqref{eq:FS11} depends on  both the three-form and the six-form potential through the current $\mathcal{J}_{m;n_1\dots n_8,p}$, all these terms can be absorbed in the constrained field $\chi_{m;n_1\dots n_9}$ by a field redefinition. So we can indeed consider that \eqref{DualGravityL} is a Lagrangian for $g_{mn}$,  $A_{n_1n_2n_3}$ and $\chi_{m;n_1\dots n_9}$.

\subsubsection*{Equations of motion}

The Euler--Lagrange equation for the constrained field  $\chi_{m;n_1\ldots n_9}$ is algebraic and gives 
\begin{align} \label{DualGraEL} 
-\frac1{2\cdot 8!} \cF^{m[n_1\ldots n_8; n_9]}  =\frac1{4\cdot 8!\sqrt{-g}} \varepsilon^{q_1q_2m[n_1\ldots n_8} \cF_{q_1q_2}{}^{n_9]}\,,
\end{align}
which is equivalent to the duality equation \eqref{eq:chi11EOM}, as has to be the case on the basis of our general analysis in Section~\ref{sec:EOMchi}. Integrating out the constrained field is by construction equivalent  to removing the term in $\mathcal{O}_3$, and so gives back the  bosonic component of the eleven-dimensional supergravity Lagrangian \eqref{Lsugra}.  

The Euler--Lagrange equation for the metric field gives 
\bea \label{IntegrChi} 
&& \frac{1}{2\cdot 8! \sqrt{-g}} \varepsilon^{p_1\dots p_{11}} \Bigl( g_{(m|p_1} \partial_{p_2} \cF_{|n)p_3\dots p_{10};p_{11}} \!- g_{(m|p_1} \partial_{p_2} g_{|n)q} \cF^q{}_{p_3\dots p_{10};p_{11}}\! -\partial_{p_1} g_{p_2(m} \cF_{n)p_3\dots p_{10};p_{11}} \Bigr)   \CR
&=&  \frac{1}{12} \cF_{p_1p_2p_3(m} \cF_{n)}{}^{p_1p_2p_3} -\frac{1}{96} g_{mn} \cF_{p_1p_2p_3p_4} \cF^{p_1p_2p_3p_4}  \CR
&& + \frac{1}{4\cdot 7!} \cF_{(m|p_1\dots p_7q;r}\cF_{|n)}{}^{p_1\dots p_7r;q} + \frac{1}{2\cdot 8!} \cF_{p_1\dots p_8q;(m}\cF_{n)}{}^{p_1\dots p_8;q}  - \frac{1}{8\cdot 8!} g_{mn} \cF_{p_1\dots p_{8}q;r} \cF^{p_1\cdot p_8r;q} \CR
&& \hspace{10mm} - \frac{1}{4} \bigl( 2g^{pr} g^{qs} - g^{pq} g^{rs} \bigr) \bigl( \partial_{(m|} g_{rs} \partial_p g_{|n)q} - \partial_{p} g_{rs} \partial_{(m} g_{n)q}\bigr) \CR
&& \hspace{30mm} - \frac12 g^{r[p} g^{q]s} \partial_{p} g_{r(m|} \partial_{q} g_{|n)s} - \frac14 g_{mn} g^{r[p} g^{q]s} g^{tu} \partial_p g_{rt} \partial_q g_{su} \; . 
\eea
Note that because  \eqref{metricTopoTerm} is a total derivative in the linearised approximation, this equation does not depend on the metric in the linearised approximation and should be interpreted instead as an integrability condition for the dual graviton  field strength $\cF_{9;1}$. 
 
If one uses the Euler--Lagrange equation~\eqref{DualGraEL} for the constrained field  to solve the field strength $\cF_{n_1\dots n_9;m} = - \frac{1}{2\sqrt{-g}} \varepsilon_{n_1\dots n_9}{}^{p_1p_2} g_{mq} \cF_{p_1p_2}{}^q$ in the   Euler--Lagrange equation for the metric field, one can combine all the terms involving the derivative of the metric on the left-hand side to obtain the Einstein equation 
\be 
R_{mn} - \frac12 g_{mn} R =  \frac{1}{12} \cF_{p_1p_2p_3(m} \cF_{n)}{}^{p_1p_2p_3} -\frac{1}{96} g_{mn} \cF_{p_1p_2p_3p_4} \cF^{p_1p_2p_3p_4} \; . 
\ee
This can be seen  more easily in the linearised approximation starting from the left-hand side of \eqref{IntegrChi} and substituting \eqref{DualGraEL}
\bea  \frac{1}{2\cdot 8!} \varepsilon^{p_1\dots p_{11}}  \eta_{(m|p_1} \partial_{p_2} \cF^\lin_{|n)p_3\dots p_{10};p_{11}}  &=&-\frac1{4\cdot 8!} \varepsilon_{r(m}{}^{p_2\dots p_{10}}\varepsilon^{q_1q_2}{}_{n)p_3\ldots p_{10}} \partial_{p_2}   \cF^\lin_{q_1q_2}{}^{r}\CR
&=&  -\frac12 \eta^{qr} \partial_r \cF^\lin_{q(m}{}^p \eta_{n)p}^{\phantom{[}} -\frac12\partial_{(m}^{\phantom{[}} \cF^\lin_{n)p}{}^p + \frac12 \eta_{mn} \eta^{rq} \partial_r \cF^\lin_{qp}{}^p \CR
&=& R_{mn}^\lin -\frac12 \eta_{mn} R^\lin\; .  \label{LinearisedEinstein} \eea

To analyse the dual graviton we must first solve the integrability condition \eqref{IntegrChi} for the constrained field. This is a non-linear equation that one cannot solve in general, but one may solve it perturbatively starting from a background metric solution to Einstein equation. Here, we shall only consider the linearised approximation around Minkowski space-time. In the linearised approximation, \eqref{IntegrChi} reduces to the curl-free equation 
\be 
\partial_{[n_1} \cF^\lin_{|m|n_2\dots n_9;n_{10}]} +5 \delta_{m[n_1} \bigl( \partial_{n_2} \cF^\lin_{n_3\dots n_{10}]p;}{}^p - \partial^p \cF^\lin_{p|n_2\dots n_9;n_{10]}} \bigr) =  0\; .   
\ee
Using moreover the trace and the divergence of \eqref{DualGraEL} one obtains that 
\bea
g^{pq} \cF_{n_1\dots n_8p;q} &=& - \frac{1}{\sqrt{-g}} \varepsilon_{n_1\dots n_8}{}^{p_1p_2p_3} \partial_{p_1} g_{p_2p_3} = 0 \; , \CR
 \partial_p \cF^{\lin p[n_1\dots n_8;n_9]}  &=& - \frac12  \varepsilon^{p_1p_2p_3[n_1\dots n_8} \partial_{p_1} \cF^\lin_{p_2p_3}{}^{n_9]} = 0 \; , 
\eea
by symmetry of the metric and the Bianchi identity for $\cF_{n_1n_2}{}^m$. We have therefore the integrability condition 
\be \label{IntegrChi9}
\partial_{[n_1} \cF^\lin_{|m|n_2\dots n_9;n_{10}]} = \partial_{[n_1} \chi^\lin_{|m|n_2\dots n_9;n_{10}]}=  0\; .   
\ee

According to the generalised Poincar\'e lemma  \cite{Bekaert:2002dt},\footnote{For this we split  $\chi_{1;9} = \chi_{10} + \chi_{1,9}$ into irreducible components. One can take an additional curl  on \eqref{IntegrChi9} to obtain $(\partial \partial \chi)_{10,2} = 0$ in which $\chi_{10}$ drops out and the generalised Poincar\'e lemma implies that $\chi_{1,9}$ takes the form \eqref{PoincareChi9} projected to the irreducible component. The component $(\partial \chi)_{11}$ implies that $\chi_{10}$ is also a total derivative, so that we get \eqref{PoincareChi9}.} $\chi^\lin_{m;n_1\dots n_9}$ must be of the form 
\be \label{PoincareChi9} \chi^\lin_{m;n_1\dots n_9} = \partial_m X_{n_1\dots n_9} + 9 \partial_{[n_1|} \Sigma_{m;|n_2\dots n_9]} \; , \ee
and checking the first-order constraint  \eqref{IntegrChi9} one finds that
\be \label{Sigma8trace} \partial_{[n_1}  \Sigma_{n_2;n_3\dots n_{10}]} = 0  \qquad  \Rightarrow \qquad \Sigma_{m;n_1\dots n_8}  = \Sigma_{m,n_1\dots n_8} + \partial_m \lambda_{n_1\dots n_8} \; . \ee
But $\lambda_{n_1\dots n_8}$ can be absorbed in a gauge transformation of $X_{n_1\dots n_9}$, and $ \Sigma_{m,n_1\dots n_8}$ can be cancelled by a shift of $h_{n_1\dots n_8,m}\rightarrow h_{n_1\dots n_8,m}-\Sigma_{m,n_1\dots n_8}$ so we get the linearised field strength 
\be \label{chiX3} \cF^\lin_{n_1\dots n_9;m} = 9 \partial_{[n_1} h_{n_2\dots n_9],m} + \partial_m X_{n_1\dots n_9} \;  , \ee
as in \cite{Bossard:2017wxl}. The trivial solution $\Sigma_{1,8}$ can be traced back to the ancillary gauge transformation \eqref{eq:extGL} of parameter $\Sigma_M{}^{\tilde{I}}$, and condition \eqref{Sigma8trace} is the first component of \eqref{eq:sigmatr}. The field $X_9$ is then interpreted as the exact derivative that appears in integrating the standard second-order duality equation for the linearised Riemann tensor~\cite{Nieto:1999pn,Hull:2001iu}
\bea \label{LinearisedRTD}
R^{\lin n_1n_2}{}_{p_1p_2} = - \partial^{[n_1} \cF^\lin_{p_1p_2}{}^{n_2]}   &=&- \frac1{9!} \varepsilon_{p_1p_2}{}^{m_1\ldots m_9}  \partial^{[n_1} \cF^\lin_{m_1\ldots m_9;}{}^{n_2]}\CR
&=&- \frac1{8!} \varepsilon_{p_1p_2}{}^{m_1\ldots m_9}  \partial^{[n_1} \partial_{m_1} h_{m_2 \ldots m_9;}{}^{n_2]} \, ,
\eea
which follows from \eqref{DualGraEL}, and does not depend on the field $X_{n_1\dots n_9}$. If one uses the Bianchi identity $R^q{}_{[p_1p_2p_3]}=0$, one gets 
\bea 
0 &=& \varepsilon_{n_1\dots n_8}{}^{p_1p_2p_3} \eta_{q[m} \partial_{p_3] } \cF^\lin_{p_1p_2}{}^q = \partial^q \cF^\lin_{n_1\dots n_8 q;m} -  \partial_{m} \cF^\lin_{n_1\dots n_8 q;p} \eta^{qp}\CR
&=& \partial^p \partial_p h_{n_1\dots n_8 ,m} + 8 \partial^p \partial_{[n_1} h_{n_2\dots n_8]p,m} - 8 \partial_m \partial_{[n_1} h_{n_2\dots n_8]p,q} \eta^{pq} - \partial_m \partial^p h_{n_1\dots n_8 ,p} \; , 
\label{BI}
\eea
which is the propagating equation for the dual graviton. One can in principle solve the non-linear equations \eqref{IntegrChi} and \eqref{DualGraEL} iteratively oder by order in the number of fields. Because \eqref{IntegrChi} depends on both the metric and the dual graviton, the dual graviton propagating equation is highly non-local, and there is therefore no contradiction with the no-go theorem of \cite{Bekaert:2002uh} that assumes locality. Note that although a natural guess for the non-linear duality equation would have been to replace the left-hand side of \eqref{LinearisedRTD} by the non-linear Riemann tensor, this is not what shows up in this non-linear equation in which the dual graviton field strength is not a tensor. Its gauge transformation 
includes the non-covariant variation \eqref{xt3}
\be 
\Delta_\xi F_{n_1\dots n_9;m} = - \frac{1}{\sqrt{-g}} \varepsilon_{n_1\dots n_9pq} g^{pr} \partial_m \partial_r \xi^q \; , \ee
that compensates the non-tensoriality of the gravitational flux $F_{n_1n_2}{}^m = 2 g^{mp} \partial_{[n_1} g_{n_2]p} $.

\subsubsection*{Relation to the vielbein formalism}


The non-linear metric formulation of dual gravity, following from the Lagrangian \eq{DualGravityL}, is very similar to the vielbein formulation of dual gravity \cite{West:2001as,West:2002jj,Boulanger:2003vs,Boulanger:2008nd}. The precise relation is obtained via the change of variable 
\begin{multline} 
\label{ChangeFY}  
\cF_{n_1\dots n_9;m} = e_m{}^b Y_{n_1\dots n_9;b} - \frac{10}{9} e_{[m}{}^b Y_{n_1\dots n_9];b}  - 9 e^{bp}  g_{m[n_1}  Y_{n_2\dots n_9] p;b}
\\
+ \frac{1}{\sqrt{-g}} g_{n_1p_1} \dots g_{n_9p_9} \varepsilon^{p_1\dots p_9}{}^{pq} ( e_m{}^a \partial_p e_{q a} - e_q{}^a \partial_p e_{m a} ) \; , 
\end{multline}
where $Y_{n_1\dots n_9;b}$ is the field strength defined in~\cite{West:2001as,West:2002jj,Boulanger:2003vs,Boulanger:2008nd}, 
that satisfies the duality equation 
\be \label{Boulanger} \Omega_{ab;c} -2 \Omega_{c[a;b]}  + 4 \eta_{c[a} \Omega_{b]d;}{}^d = \frac{1}{9!} \varepsilon_{ab}{}^{d_1\dots d_9} Y_{d_1\dots d_9;c} \ee
with the anholonomy coefficients 
\be  \Omega_{ab;}{}^{c} = e_a{}^m e_b{}^n ( \partial_m e_n{}^c - \partial_n e_m{}^c) \; . \ee
The second line in the redefinition \eqref{ChangeFY} is linear in the composite Maurer--Cartan $\mathfrak{so}(1,10)$ connection 
\be Q_m{}^{ab} = - e^{n [a} \partial_m e_n{}^{b]} \; , \ee
of the $GL(11) / SO(1,10)$ symmetric space, and compensates for the property that $Y_{n_1\dots n_9;b}$ does not transform homogeneously under $SO(1,10)$ gauge transformations. We have
\be 
e_a{}^{n_1} e_b{}^{n_2} e_m{}^c \cF_{n_1n_2}{}^m = 2 \Omega_{ab}{}^c + 4 Q_{[a;b]}{}^c \; . 
\ee
Note that the change of variables \eqref{ChangeFY} is a redefinition of the field $\chi_{m;n_1\dots n_9}$. We find therefore that $\chi_{m;n_1\dots n_9}$ can be identified with the St\"{u}ckelberg gauge field constituting $Y_{n_1\dots n_9;b}$ in \cite{Boulanger:2008nd}. 

Substituting \eqref{ChangeFY} into the gravitational part of \eqref{DualGravityL} one obtains after a tedious but straightforward computation 
\begin{align}
\label{DualGraViel} 
&\quad \sqrt{-g} \Big(   - \frac12 g^{p[m} g^{n]q} g^{rs} \partial_{m} g_{pr} \partial_{n} g_{qs}  - \frac{1}{4\cdot 8!} \cF_{n_1\dots n_{8}p;q} \cF^{n_1\cdot n_8q;p}  \Big) \CR
& \hspace{60mm} - \frac{1}{2\cdot 8!} \varepsilon^{n_1\dots n_{11}} g^{pq} \partial_{n_1} g_{n_2p} \cF_{qn_3\dots n_{10};n_{11}} \CR
&= - \frac{e}{4\cdot 9!} \Bigl( \frac{8}{9} Y_{a_!\dots a_9;b} Y^{a_1\dots a_9;b} - 9 Y_{a_1\dots a_8b;}{}^b Y^{a_1\dots a_8c;}{}_c + Y_{a_1\dots a_8b;c} Y^{a_1\dots a_8c;b} \Bigr) \\
&  \hspace{60mm} - \frac{1}{2\cdot 9!} \varepsilon^{n_1\dots n_{11}} \Omega_{n_1n_2}{}^b Y_{n_3\dots n_{11};b} - 2 \partial_m ( e e_a{}^m e_b{}^n Q_{n}{}^{ab} )  \CR
&\quad - \frac{e}{2\cdot 8!} \bigl( Y_{a_1\dots a_8b;}{}^b + \tfrac12 \varepsilon_{a_1\dots a_8}{}^{b_1b_2b_3}\Omega_{b_1b_2;b_3} \bigr) \bigl( Y^{a_1\dots a_8c;}{}_c + \tfrac12 \varepsilon^{a_1\dots a_8c_1c_2c_3}\Omega_{c_1c_2;c_3} \bigr) \; , \nn
\end{align}
which we recognise as the Lagrangian of \cite{Boulanger:2008nd} plus a total derivative and the last line that is quadratic in the 3-form component of the duality equation \eqref{Boulanger}. This last term does not modify the equations of motion.\footnote{One may think that the last term in the Lagrangian can be reabsorbed by a redefinition of the St\"{u}ckelberg gauge field constituting $Y_{n_1\dots n_9;b}$, however, this is only possible for a complex coefficient. Still the Lagrangian~\eqref{DualGraViel} is fully equivalent to the one in~\cite{Boulanger:2008nd}.} By construction integrating out the St\"{u}ckelberg gauge field gives back Einstein--Hilbert Lagrangian, just as in \cite{West:2001as,West:2002jj,Boulanger:2008nd}.  The equation of motion of the vielbein $e_m{}^a$ gives the same integrability condition as in \cite{Boulanger:2008nd}, up to a term proportional to the 3-form component of the duality equation \eqref{Boulanger}.

We conclude that $E_{11}$ exceptional field theory includes a non-linear dual graviton with the expected dynamics. The Lagrangian \eqref{DualGravityL} is not completely obvious to derive from the Einstein--Hilbert Lagrangian by dualisation, because of its non-trivial dependence on the metric. It is not possible to obtain the vielbein formulation~\cite{West:2001as} of the dual graviton from the coset component of the Maurer--Cartan form, since the vielbein form requires the anholonomy coefficients that involve the full Maurer--Cartan form with its inhomogeneous $K(E_{11})$ transformation    \cite{Tumanov:2016abm}. Using the coset component of the Maurer--Cartan form makes it clear how to write $E_{11}$ invariant equations  and we have shown in this section how to obtain the dual gravity Lagrangian~\eqref{DualGravityL} from $E_{11}$ exceptional field theory.

\subsection{The gradient dual of the three-form}
\label{SectionGradientDual} 

Similarly as for the dual graviton, one can derive a Lagrangian for the gradient dual $A_{n_1\dots n_9;p_1p_2p_3}$ of the 3-form gauge potential that has the same equations of motion as the $E_{11}$ exceptional field theory pseudo-Lagrangian using the duality equation.

In this case one considers the Lagrangian 
\bea
&& {\mathcal L}_{\text{sugra}}+ 2{\mathcal O}_4\CR
&=& \sqrt{-g} \Big( R + \frac{1}{2} g^{q[m} g^{n]p} g^{r_1s_1} g^{r_2s_2}  \partial_m A_{pr_1r_2} \partial_{n} A_{qs_1s_2}   + \frac{1}{ 9!} \cF_{n_1\dots n_7p_1p_2p_3;q_1q_2q_3}  \cF^{n_1\dots n_7q_1q_2q_3;p_1p_2p_3}  
\CR
&&\hspace{60mm}  - \frac{1}{2 \cdot 11!} \cF_{p_1\dots p_{11},m,n}  \cF^{p_1\dots p_{11},m,n}\Big) 
\nn\\
&& \; - \frac1{144^2} \varepsilon^{n_1\ldots n_{11}}  A_{n_1n_2n_3} \cF_{n_4\ldots n_7}\cF_{n_8\ldots n_{11}}- \frac{2}{9!} \varepsilon^{n_1\dots n_{11}} \partial_{n_1} A_{p_1p_2p_3} \cF^{p_1p_2p_3}{}_{n_2\dots n_8;n_9n_{10}n_{11}} \ ,
\label{eq:LA9}
\eea
where the field strengths with $13$ indices above and their duals $\cF^{m,n}$ and $\cF_m{}^{n_1n_2n_3}$ are defined at the linearised level in~\eqref{F}. The Lagrangian \eqref{eq:LA9} only depends on the metric $g_{mn}$, the three-form potential $A_{n_1n_2n_3}$ and the field strengths $ \cF_{n_1\dots n_{10};p_1p_2p_3}  $ and  $\cF_{p_1\dots p_{11},m,n} $. Although the definition of these field strengths involves the six-form potential and the dual graviton field through the level 4 component of the current $\tilde{\cJ}_M{}^{\alpha}$, these terms can be eliminated by a field redefinition of the constrained fields  $\chi_{m;n_1\dots n_{10},p_1p_2}$ and $\chi_{m;n_1\dots n_{11},p}$, so  \eqref{eq:LA9} defines a Lagrangian for the metric, the three-form potential and these constrained fields.

The Euler--Lagrange equations for the constrained fields $\chi_{m;n_1\dots n_{10},p_1p_2}$ and $\chi_{m;n_1\dots n_{11},p}$ are the duality equations 
\begin{subequations}
\bea \label{Dualitytenthree}
 \cF_{n_1\dots n_{10};p_1p_2p_3} &=& - \frac{1}{\sqrt{-g}} \varepsilon_{n_1\dots n_{10}}{}^m g_{p_1q_1} g_{p_2q_2} g_{p_3q_3}  \cF_{m}{}^{q_1q_2q_3} =  \frac{1}{\sqrt{-g}}  \varepsilon_{n_1\dots n_{10}}{}^m \partial_m A_{p_1p_2p_3} \; , \qquad \\
  \label{Dualityeleventwo}
  \cF_{n_1\dots n_{11};m,p} &=& \frac1{2\sqrt{-g}} \varepsilon_{n_1\dots n_{11}} \cF_{m,p} = 0 \; ,
 \eea
 \end{subequations}
where we used the property that the field strength $\cF^{m,n} = 0 $ on section.\footnote{One can consider $\cF^{m,n} \ne 0$ in a non-geometric background that gives rise to massive type IIA similarly as in \cite{Hohm:2011cp,Ciceri:2016dmd}. This requires however to use the semi-flat formulation associated to $L= GL(1)\times GL(10) \subset GL(11)$, and checking the consistency of this mild violation of the section constraint is beyond the scope of this paper.}  
As the fields appear algebraically in the Lagrangian they can be integrated out to give the eleven-dimensional supergravity Lagrangian. In particular, the field strength $ \cF_{p_1\dots p_{11},m,n}$ decouples and can be eliminated by integrating out the constrained field $\chi_{(m|;p_1\dots p_{11},|n)}$.  The three-form gauge field Euler--Lagrange equation
\begin{multline} 11 \partial_{[n_1} \cF^{p_1p_2p_3}{}_{n_2\dots n_8;n_9n_{10}n_{11}]} =  - \frac{1}{20 g} \varepsilon_{n_1\dots n_{11}} \partial_{m} \bigl( \sqrt{-g} g^{p_1[m} g^{n]q_1} g^{p_2q_2} g^{p_3q_3} \bigr) \partial_{n} A_{q_1q_2q_3}  \\ +  \frac{1155}{4} \delta^{p_1p_2p_3}_{[n_1n_2n_3} \cF_{n_4n_5n_6n_7} \cF_{n_8n_9n_{10}n_{11}]}  \; , \label{Intchi102}  \end{multline}
is not propagating and should rather be interpreted as an integrability condition for the constrained fields. 

As for the dual graviton, substituting the solution to \eqref{Dualitytenthree} into the integrability condition \eqref{Intchi102} gives 
\bea 
&&  11 \partial_{[n_1} \cF^{p_1p_2p_3}{}_{n_2\dots n_8;n_9n_{10}n_{11}]} \\
&=&  \frac{1}{5! g} \varepsilon_{n_1\dots n_{11}} \partial_{m} \Bigl( \sqrt{-g} g^{mn} g^{p_1q_1} g^{p_2q_2} g^{p_3q_3} \partial_n A_{q_1q_2q_3} - 3\sqrt{-g}  g^{nq_1} g^{p_1q_2} g^{p_2q_3} g^{p_3 m }\partial_n A_{q_1q_2q_3} \Bigr)\, ,\nn
\eea
for the left-hand side and one obtains the three-form potential equation of motion of eleven-dimensional supergravity.  

To solve the integrability condition for the constrained fields first, we must analyse the linearised approximation. The field strengths then reduce to
\bea \cF^\lin_{n_1\dots n_{10};p_1p_2p_3} \hspace{-1mm}  &=&\hspace{-1mm}  10 \partial_{[n_1} A_{n_2\dots n_{10}],p_1p_2p_3} \!+ \! 3 \chi^\lin_{[p_1|;n_1\dots n_{10},|p_2p_3]} \! + \! \tfrac32\bigl(  \chi^\lin_{[p_1|;n_1\dots n_{10}|p_2,p_3]}\! +\! \zeta^\lin_{[p_1|;n_1\dots n_{10}|p_2,p_3]} \bigr) \CR
 \cF^\lin_{n_1\dots n_{11};m,p} \hspace{-1mm}  &=& \hspace{-1mm}   11 \partial_{[n_1} B_{n_2\dots n_{11}],m,p} + 2 \partial_{(m|} C_{n_1\dots n_{11},|p)} + \tfrac52   \chi^\lin_{(m|;n_1\dots n_{11},|p)} + \tfrac12   \zeta^\lin_{(m|;n_1\dots n_{11},|p)}\; . \label{linearisedF103}
\eea
They satisfy the duality equations \eqref{Dualitytenthree} and \eqref{Dualityeleventwo}. The second equation \eqref{Dualityeleventwo} is a flat curvature equation and implies that all the fields appearing in $ \cF_{n_1\dots n_{11};m,p}$ are pure gauge. As expected, the $E_{11} / K(E_{11}) $ fields $B_{n_1\dots n_{10},m,p} $ and $C_{n_1\dots n_{11},m}$ do not propagate degrees of freedom. We shall therefore concentrate on the first duality equation  \eqref{Dualitytenthree}. 

To simplify the discussion we recombine the $\chi$ and $\zeta$ fields in $\cF_{10;3}$ in the reducible field 
\be
 \chi_{m;n_1\dots n_{10};p_1p_2} = \chi_{m;n_1\dots n_{10},p_1p_2} +\tfrac32  \chi_{m;n_1\dots n_{10}[p_1,p_2]}+ \tfrac32  \zeta_{m;n_1\dots n_{10}[p_1,p_2]} \; .
  \ee
and we note that the field strength 
\be  \cF^{\scalebox{0.6}{lin}}_{n_1\dots n_{10};p_1p_2p_3} =   10 \partial_{[n_1} A_{n_2\dots n_{10}],p_1p_2p_3} + 3 \chi^\lin_{[p_1|;n_1\dots n_{10};|p_2p_3]} \ee
does not depend on the component of $\chi_{1;10;2}$ with $(10,2,1)$ Young symmetry. Therefore this component $\chi_{10;2,1}$ in 
\be \chi_{m;n_1\dots n_{10};p_1p_2} = \tfrac13  \chi_{n_1\dots n_{10};mp_1p_2} + \chi_{n_1\dots n_{10};p_1p_2,m}\; ,  \ee
 is pure gauge and we can write the integrability condition for $\chi_{n_1\dots n_{10};mp_1p_2}$. The integrability condition gives in the linearised approximation 
\be 
 11 \partial_{[n_1} \cF^\lin_{|p_1p_2p_3|n_2\dots n_8;n_9n_{10}n_{11}]} =   11 \partial_{[n_1} \chi^\lin_{|p_1p_2p_3|n_2\dots n_8;n_9n_{10}n_{11}]}  = 0 \label{IntegrTentwo}  \; . 
\ee
Taking the curl of this equation in the $(11,4)$ Young tableau representation one obtains from the generalised Poincar\'e lemma  \cite{Bekaert:2002dt} that 
 \be
  \chi^\lin_{n_1\dots n_{10};p_1p_2p_3}   = 10 \partial_{[n_1} \Sigma_{n_2\dots n_{10}];p_1p_2p_3} + 3 \partial_{[p_1} X_{n_1\dots n_{10};|p_2p_3]}\; , 
  \ee
 which reinserted in \eqref{IntegrTentwo} implies that $\Sigma_{9;3}$ is the sum of an arbitrary $\Sigma_{9,3}$ and the curl of a $(9;2)$ form. But the second can be absorbed in $X_{10;2}$ so we get 
 \be
  \chi^\lin_{n_1\dots n_{10};p_1p_2p_3}   = 10 \partial_{[n_1} \Sigma_{n_2\dots n_{10}],p_1p_2p_3]} + 3 \partial_{[p_1} X_{n_1\dots n_{10};|p_2p_3]}\; , 
  \ee
 and the linearised field strength reduces to 
 \be 
  \cF^\lin_{n_1\dots n_{10};p_1p_2p_3} =  10 \partial_{[n_1} ( A_{n_2\dots n_{10}],p_1p_2p_3}+ \Sigma_{n_2\dots n_{10}],p_1p_2p_3} ) + 3  \partial_{[p_1} X_{n_1\dots n_{10};|p_2p_3]}\; .
 \ee
 The $\Sigma_{9,3}$ is reabsorbed in $A_{9,3}$ and we get the expected linearised field strength in which 
\be \label{chiX4} 
 \chi^\lin_{m;n_1\dots n_{10};p_1p_2}  = \partial_m X_{n_1\dots n_{10};p_1p_2} \; ,
 \ee
 is a total derivative. 
 
Using \eqref{xt4}, \eqref{xt5} and \eqref{xt6} in \eqref{linearisedF103}, one obtains the gauge transformation 
 \be 
 \Delta_\xi  \cF^\lin_{n_1\dots n_{10};p_1p_2p_3}  = 3 \varepsilon_{n_1\dots n_{10}q} \partial^q \partial_{[p_1} \lambda_{p_2p_3]} \; , \ee
 that can be ascribed to the gauge transformation of $X_{10;2}$ and which ensures gauge invariance of the first order duality equation \eqref{Dualitytenthree}. The dependence in the field $X_{10;2}$ drops out in the second-order duality equation 
\be \label{SecondOrderA93}
4 \partial_{[p_1|}  \cF^\lin_{n_1\dots n_{10};|p_2p_3p_4]} =4 \varepsilon_{n_1\dots n_{10}}{}^m \partial_m\,   \partial_{[p_1} A_{p_2p_3p_4]}  =\varepsilon_{n_1\dots n_{10}}{}^m \partial_m\,   \cF_{p_1p_2p_3p_4}   \; , 
 \ee
which reproduces the duality equation introduced in \cite{Boulanger:2015mka}, and $X_{10;2}$ can be interpreted as a constant of integration in integrating this second-order equation to \eqref{Dualitytenthree}.

 The Bianchi identity for the gradient of the gauge field gives the wave equation for the $(9,3)$ form
\bea 
0 &=& \partial^q  \varepsilon_{n_1\dots n_{9}q}{}^m \partial_m A_{p_1p_2p_3} = \partial^q  \cF^\lin_{n_1\dots n_{9}q;p_1p_2p_3}  \CR
&=& - \partial^q \partial_q A_{n_1\dots n_9,p_1p_2p_3} + 9 \partial^q \partial_{[n_1} A_{n_2\dots n_9]q,p_1p_2p_3} +3 \partial^q \partial_{[p_1|} X_{n_1\dots n_9q;|p_2p_3]} \; , 
\eea
 which depends on $X_{10;2}$ because the second-order Bianchi identity is not  manifestly gauge invariant. One can write a propagating equation for the field $A_{9,3}$ that does not depend on $X_{10;2}$, but it is then third order in derivatives 
 \be
  4 \partial_{[ p_1|} \bigl(  \partial^q \partial_q A_{n_1\dots n_9,|p_2p_3p_4]} - 9 \partial^q \partial_{[n_1} A_{n_2\dots n_9]q,|p_2p_3p_4]}\bigr) = 0 \; . 
\ee
We find therefore that the $E_{11}$ exceptional field theory reproduces the expected equations for $A_{9,3}$.

\subsection{Higher levels}
\label{sec:HighLev}

According to the analysis of Section 4.3 in \cite{Bossard:2019ksx}, we expect that the higher level dual fields can be described in the linearised approximation by the field strength
\be
F^{\lin I} =  C^{I M}{}_\alpha\, \partial_M \phi^\alpha + C^{IM}{}_{\ta}\, \partial_M X^\ta +  C^{IM}{}_\Lambda\, \partial_M Y^\Lambda + C^{IM}{}_\tL\,  \partial_M Y^\tL\ ,
\label{eq:FSlin}
\ee
and the gauge transformations 
\begin{subequations}
\begin{align}
\delta_\xi \phi^\alpha &= T^{\alpha N}{}_P \left( \partial_N \xi^P + \eta_{NQ} \eta^{PR} \partial_R\xi^Q\right) \; , \\
\delta_\xi X^\ta &= T^{\tilde\alpha N}{}_P \left( \partial_N \xi^P + \eta_{NQ} \eta^{PR} \partial_R\xi^Q\right)+ {\Pi^{\tilde\alpha}{}_{ Q P }} \eta^{NQ} \partial_N \xi^P \,,\\
\delta_\xi Y_M{}^\hL &= \Pi^\hL{}_{ Q P } \eta^{NQ} \partial_N \xi^P \, ,
\end{align}
\end{subequations}
that leave the linearised duality equation 
\be 
\Omega_{IJ} F^{\lin J} = \eta_{IJ} F^{\lin J} \;  
\ee
invariant, according to the analysis of Section \ref{sec:DE}. To get these linearised equations from the non-linear duality equations, we need to show that the pseudo-Lagrangian Euler--Lagrange equations imply through the generalised Poincar\'e lemma that
\be
 \label{DerivativeLin} 
  \chi_M{}^{\ta} \approx  \partial_M X^\ta \; , \qquad \zeta_M{}^\Lambda \approx \partial_M Y^\Lambda \; , \qquad \zeta_M{}^{\tL} \approx \partial_M Y^{\tL}\; ,
\ee
up to terms that do not contribute to the components of the field strengths $F^{\lin I}$ corresponding to propagating degrees of freedom. We know that the linearised $E_{11}$ exceptional field theory equations do not impose directly that the constrained fields are total derivatives, but we expect that this should be true up to $\Sigma_M{}^{\tilde{I}}$ gauge transformations and possibly higher level shift symmetries that leave $F^{\lin I}$ invariant.  We have indeed found in \eqref{chiX3} and \eqref{chiX4} that $F^{\lin I}$ takes this form \eqref{eq:FSlin} at levels $k=3$ and $4$.

To derive this result for all $k$, we could in principle analyse the pseudo-Lagrangians associated to the higher level fields in the same way, with for any $n\ge 0$
\bea 
\label{Lagrangians} 
\text{for } A_{9^n,3} &:& \qquad \cL_{1,n} = \cL_{\text{sugra}} + 2 \sum_{j=1}^n \mathcal{O}_{1+3j} \; , \CR
\text{for } A_{9^n,6} &:& \qquad \cL_{2,n} =  \cL_{\text{sugra}} + 2 \sum_{j=0}^n \mathcal{O}_{2+3j} \; , \\
\text{for } h_{9^n,8,1} &:& \qquad \cL_{3,n} = \cL_{\text{sugra}} +  \sum_{j=0}^n \mathcal{O}_{3+3j} \; . \nn
\eea
These pseudo-Lagrangians are not invariant under generalised diffeomorphisms, but since they only differ from the invariant Lagrangian $\cL$ by a total derivative and terms quadratic in the duality equation, their Euler--Lagrange equations together with the duality equation transform into each other under generalised diffeomorphisms. 
They are defined such that for the fields $A_{9^n,3}$, $A_{9^n,6}$ or $h_{9^n,8,1}$ of level $k= i + 3n$, the pseudo-Lagrangians $\cL_{i,n}$, with $i=1,2,3$, respectively, only depend on the fields of level $\ell$ between $0$ and $k$. Moreover, in the linearised approximation,  $\cL_{1,n} $ only depends on the metric and the fields of level $k = 1+3 j$ (as the propagating fields $A_{9^j,3}$), $\cL_{2,n}$ on the metric and the fields of level $k=2+3j$ (as the propagating fields $A_{9^j,6}$) and $ \cL_{3,n} $ on the three-form potential and the fields of level $k = 3+3j$ (as the propagating fields $h_{9^j,8,1}$).\footnote{The property that one needs more and more fields to obtain parent actions for the infinite sequence of gradient duals was already observed in \cite{Boulanger:2012df}.}

Using \eqref{eq:Lk} and \eqref{eq:altid} one obtains that the bilinear terms in the $E_{11}$ fields of level $\ell$ almost cancel in $\mathcal{O}_\ell + \mathcal{O}_{\ell+3}$, and one is left with a term in $ - \frac14 m_{\tilde{I}_\dgr{\ell}\tilde{J}_\dgr{\ell}} C^{\tilde{I}_\dgr{\ell}}{}_{p \wa_\dgr{\ell}} C^{\tilde{J}_\dgr{\ell}}{}_{q \wb_\dgr{\ell}} m^{qm} m^{pn}  \tilde{\cJ}_m{}^{{\wa}_\dgr{\ell}}  \tilde{\cJ}_n{}^{{\wb}_\dgr{\ell}}$ for $\ell\geq 3$. The computation follows the one below in \eqref{SymmetryLinearised}. Schematically, this remaining term is quadratic in the divergence of the level $\ell$ fields rather than their curl, so that the corresponding equations of motion are not propagating. This is because for a fields  $\phi^{\alpha_\dgr{\ell}}$ of level $\ell$  that carry  $3\ell$ $GL(11)$ indices, the components of $ C^{I_\dgr{\ell} m}{}_{\alpha_\dgr{\ell}}  \tilde{\cJ}_m{}^{\alpha_\dgr{\ell}}$ have $3\ell + 1$ free $GL(11)$ indices, as expected for a field strength, whereas the components of  $ C^{\tilde{I}_\dgr{\ell}}{}_{n \alpha_\dgr{\ell}}m^{mn} \tilde{\cJ}_m{}^{\alpha_\dgr{\ell}}$ have only $3\ell-1$ free $GL(11)$ indices, corresponding to a divergence with the explicit inverse metric $m^{mn}= \sqrt{-g} g^{mn}$. The sums in \eqref{Lagrangians} eliminate in this way the dependence on the propagating components\footnote{Here, `propagating' generalises the notion of transverse traceless degrees of freedom. See also the beginning of Section~\ref{HigherLevelSection} and~\cite{Boulanger:2015mka} for a related discussion.} of all fields of intermediate level in the linearised approximation, and $\cL_{1,n}$ only depends on the propagating degrees of freedom of the metric and $A_{9^n,3}$,  $\cL_{2,n}$ on the propagating degrees of freedom of the metric and $A_{9^n,6}$, and  $\cL_{3,n}$ on the propagating degrees of freedom of  $A_3$ and $h_{9^n,8,1}$. 

 At the non-linear level these pseudo-Lagrangians still depend on all lower level fields through the non-abelian terms in the current $\tilde{\cJ}_m{}^\alpha$. Nonetheless, the pseudo-Lagrangians $\cL_{i,n}$ are defined such that their equations of motion are by construction linear combinations of eleven-dimensional supergravity equations of motions and the duality equations $\cE_{I_\dgr{i+3j}}$. It follows that the Euler--Lagrange equations for the lower level fields give a projection of the  Euler--Lagrange equations for the constrained fields, and therefore contain no new information. In this way $\cL_{1,n}$ can be considered as Lagrangians for the metric, and all the fields at level $\ell = 1 + 3 j$ for $0\le j\le n$, and $\cL_{3,n}$ as Lagrangians for the metric, the three-form gauge field, and all the fields at level $\ell =3 + 3 j$ for $0\le j\le n$.

The pseudo-Lagrangians $\cL_{2,n}$ associated to $ A_{9^n,6}$ instead  are not Lagrangians for a subset of the fields. In particular for the six-form potential already, $\cL_{\text{sugra}} + 2 \mathcal{O}_{2}$ does not determine the dynamics of the three-form potential and one must keep the duality equation \eqref{fe} as given. But this is not a major difficulty since  we know how to describe the dual six-form and we expect that the same analysis as in the preceding subsection would allow us to prove  \eqref{DerivativeLin} at level $5$ and to get the correct non-linear equations for the gradient dual field  $A_{9,6}$. 

However, a more striking difficulty occurs starting from level $6$ and above, for dual potentials with a Young tableau with more than three columns. For a dual field at level $k$, with a  Young tableau of  $K{=}1{+}\lfloor k/3 \rfloor$ columns, the standard field strength is obtained by taking $K$ derivatives~\cite{Bekaert:2002dt}.
The expected duality equation for this field strength will then have  $K$ derivatives. The associated integrability conditions can either be a projection, therefore involving $K$ derivatives in total, or a divergence, therefore involving $K+1$ derivatives. On the contrary,  the Euler--Lagrange equations of the fields at level $k{-}3$ will only be first order in derivatives.  So we expect that one needs to take $K{-}1$ or $K$ derivatives of the Euler--Lagrange equations projected to the  appropriate representations to obtain the relevant integrability conditions. We shall now describe succinctly how we anticipate this to work.

In this discussion we shall work directly with the duality equation in the linearised approximation, including the linearised Einstein equation. The equations we derive are also Euler--Lagrange equations for the pseudo-Lagrangians \eqref{Lagrangians}. We drop the ${}^\lin$ superscript to simplify the notation, but all the equations below are understood to be in the linearised approximation. The starting point is to use 
\be 
\Omega_{IJ} C^{Im}{}_\alpha \partial_m \tilde{\cF}^J = \eta_{IJ}  C^{Im}{}_\alpha \partial_m \tilde{\cF}^J \label{DualityDer} 
 \ee
as an integrability condition. From level $6$ and higher, the field $\zeta_M{}^{\tL}$ can contribute a term on the left-hand side that is not a curl, starting with level 6, as follows
\be 
\Omega_{IJ} C^{Im}{}_\alpha \partial_m \tilde{\cF}^J = \Omega_{IJ} C^{Im}{}_\alpha \bigl( C^{Jn}{}_{\tilde{\beta}} \partial_{[m}  \chi_{n]}{}^{\tilde{\beta}} +C^{Jn}{}_{\Lambda} \partial_{[m}  \zeta_{n]}{}^{\Lambda} + C{}^{Jn}{}_{\tL} \partial_{m}  \zeta_{n}{}^{\tL} \bigr) \; .
 \ee
 Note that the symmetric derivative vanishes for the field $\zeta_{n}{}^{\Lambda} $ because of equation \eqref{asforTHA}. 
Moreover, the right-hand side of \eqref{DualityDer} is not an equation of motion for level higher than $6$. To see this,  let us analyse the right-hand side  of \eqref{DualityDer} for different $k$. Because the field strength $ \tilde{\cF}^{I_\dgr{k}}  $ does not involve constrained fields for level $k<3$, we have that for $k>0$
\bea \eta_{I_\dgr{k}J_\dgr{k}} C^{I_\dgr{k} m}{}_{\alpha_\dgr{k}} \partial_m   \tilde{\cF}^{J_\dgr{k}}  &=&  \Omega_{I_\dgr{k}J_\dgr{3-k}} C^{I_\dgr{k} m}{}_{\alpha_\dgr{k}}   \partial_m \tilde{\cF}^{J_\dgr{3-k}}  \CR
& =&  \Omega_{I_\dgr{k}J_\dgr{3-k}} C^{I_\dgr{k} m}{}_{\alpha_\dgr{k}}  C^{J_\dgr{3-k} n}{}_{\beta_\dgr{3-k}}  \partial_{m} \partial_{n}  \phi^{\beta_{\dgr{3-k}}}  =  0 \; . \label{LinearisedNegEq}
\eea
The last equality is due to $d^2=0$ as implied by the tensor hierarchy algebra, and the property that $\tilde{\cF}^{J_\dgr{3-k}} $ is the total derivative of the $E_{11} / K(E_{11})$ coset fields for $3-k<3$ in eleven dimensions. For $k=0$ we need the Euler--Lagrange equation to get the linearised Einstein equation \eqref{LinearisedEinstein}, which gives \eqref{LinearisedNegEq} for $k=0$, i.e.
\be \eta_{I_\dgr{0}J_\dgr{0}} C^{I_\dgr{0} m}{}_{\alpha_\dgr{0}} \partial_m   \tilde{\cF}^{J_\dgr{0}}  =  0 \; .
\ee
For $k<0$, $\tilde{\cF}^{J_\dgr{3-k}} $  does depend on the constrained field $\chi_m{}^{\tilde{\alpha}}$ and \eqref{LinearisedNegEq} does not vanish.

One can nonetheless use \eqref{eq:almID5} to show that 
\bea \label{SymmetryLinearised}
 &&  \Bigl( \eta_{I_\dgr{k}J_\dgr{k}}  C^{I_\dgr{k} m}{}_{\alpha_\dgr{k}}  C^{J_\dgr{k} n}{}_{\wb_\dgr{k}}   - \eta_{\tilde{I}_\dgr{k}\tilde{J}_\dgr{k}} C^{\tilde{I}_\dgr{k}}{}_{p \alpha_\dgr{k}} C^{\tilde{J}_\dgr{k}}{}_{q \wb_\dgr{k}} \eta^{qm} \eta^{pn} \Bigr)  \tilde{\cJ}_n{}^{{\wb}_\dgr{k}}  \\
 &=&  \Bigl(  \eta_{\alpha_\dgr{k}\beta_\dgr{k}} \eta^{mn}  -\eta_{\alpha_\dgr{k}\gamma_\dgr{k}}  T^{\gamma_\dgr{k} m}{}_{Q} T_{\beta_\dgr{k}}{}^Q{}_p \eta^{pn} \Bigr)   \tilde{\cJ}_n{}^{{\beta}_\dgr{k}}    \CR
  &=& \eta_{\alpha_\dgr{k}}{}^{\delta_\dgr{-k}}  \Bigl(  \eta_{\delta_\dgr{-k}\beta_\dgr{-k}} \eta^{mn}  -\eta_{\beta_\dgr{-k}\gamma_\dgr{-k}}T_{\delta_\dgr{-k}}{}^m{}_Q   T^{\gamma_\dgr{-k} Q}{}_{p}  \eta^{pn}\Bigr)   \tilde{\cJ}_n{}^{{\beta}_\dgr{-k}}  \CR
  &=&  \eta_{\alpha_\dgr{k}}{}^{\gamma_\dgr{-k}}  \Bigl( \eta_{I_\dgr{-k}J_\dgr{-k}}  C^{I_\dgr{-k} n}{}_{\gamma_\dgr{-k}}  C^{J_\dgr{-k} m}{}_{\wb_\dgr{-k}}   - \eta_{\tilde{I}_\dgr{-k}\tilde{J}_\dgr{-k}} C^{\tilde{I}_\dgr{-k}}{}_{p \gamma_\dgr{-k}} C^{\tilde{J}_\dgr{-k}}{}_{q \wb_\dgr{-k}} \eta^{qn} \eta^{pm} \Bigr)  \tilde{\cJ}_n{}^{{\wb}_\dgr{-k}}   \; , \nonumber
\eea
where we introduced 
\be  
\eta_{\alpha}{}^\gamma = \eta_{\alpha \delta} \kappa^{\delta\gamma} \; , 
\ee
and used $\eta_{\alpha\beta}  \tilde{\cJ}_n{}^{{\beta}} = \kappa_{\alpha\beta} \tilde{\cJ}_n{}^{{\beta}}$ in the third step. For $k\ge 0$ one has $ C^{\tilde{I}_\dgr{-k}}{}_{n \alpha_\dgr{-k}}  =0 $ in eleven dimensions (the non-zero components are $ C^{\tilde{I}_\dgr{k}}{}_{n \alpha_\dgr{k}}  $ for $k\ge 3$), so one can use the curl of \eqref{SymmetryLinearised} to get 
\bea &&  \eta_{I_\dgr{-k}J_\dgr{-k}}  C^{I_\dgr{-k} m}{}_{\alpha_\dgr{-k}}    \partial_m \tilde{\cF}^{J_\dgr{-k}} =  \eta_{I_\dgr{-k}J_\dgr{-k}}  C^{I_\dgr{-k} m}{}_{\alpha_\dgr{-k}}  C^{J_\dgr{-k} n}{}_{\beta_\dgr{-k}}  \partial_m \partial_n \phi^{\beta_\dgr{-k}} \CR
 &=& \Bigl( \eta_{I_\dgr{-k}J_\dgr{-k}}  C^{I_\dgr{-k} n}{}_{\alpha_\dgr{-k}}  C^{J_\dgr{-k} m}{}_{\wb_\dgr{-k}}   - \eta_{\tilde{I}_\dgr{-k}\tilde{J}_\dgr{-k}} C^{\tilde{I}_\dgr{-k}}{}_{p \alpha_\dgr{-k}} C^{\tilde{J}_\dgr{-k}}{}_{q \wb_\dgr{-k}} \eta^{qn} \eta^{pm} \Bigr)  \partial_m \tilde{\cJ}_n{}^{{\wb}_\dgr{-k}}\CR
 &=&  \eta_{\alpha_\dgr{-k}}{}^{\gamma_\dgr{k}} \Bigl( \eta_{I_\dgr{k}J_\dgr{k}}  C^{I_\dgr{k} m}{}_{\gamma_\dgr{k}}  C^{J_\dgr{k} n}{}_{\wb_\dgr{k}}   - \eta_{\tilde{I}_\dgr{k}\tilde{J}_\dgr{k}} C^{\tilde{I}_\dgr{k}}{}_{p \gamma_\dgr{k}} C^{\tilde{J}_\dgr{k}}{}_{q \wb_\dgr{k}} \eta^{qm} \eta^{pn} \Bigr) \partial_m \tilde{\cJ}_n{}^{{\wb}_\dgr{k}} \CR
&=& - \eta_{\alpha_\dgr{-k}}{}^{\gamma_\dgr{k}} \partial_m \Bigl( \eta_{I_\dgr{k}J_\dgr{k}} C^{I_\dgr{k}m}{}_{\gamma_\dgr{k}} C^{J_\dgr{k} n}{}_{\hL} \zeta_n{}^{\hL}+  \eta_{\tilde{I}_\dgr{k}\tilde{J}_\dgr{k}} C^{\tilde{I}_\dgr{k}}{}_{p \gamma_\dgr{k}} C^{\tilde{J}_\dgr{k}}{}_{q \wb_\dgr{k}} \eta^{qm} \eta^{pn}  \tilde{\cJ}_n{}^{{\wb}_\dgr{k}} \Bigr) \; ,  \eea
where we used \eqref{LinearisedNegEq} in the last step. Therefore, the integrability condition for the field strength of level $k\ge 0$ gives
\bea \label{GeneralIntegrability}
&&   \Omega_{I_\dgr{-k}J_\dgr{3+k}} C^{I_\dgr{-k}m}{}_{\alpha_\dgr{-k}}   \bigl( C^{J_\dgr{3+k}n}{}_{\tilde{\beta}_\dgr{3+k}} \partial_{[m}  \chi_{n]}{}^{\tilde{\beta}_\dgr{3+k}} +   C^{J_\dgr{3+k}n}{}_{\Lambda_\dgr{3+k}} \partial_{[m}  \chi_{n]}{}^{\Lambda_\dgr{3+k}}   \bigr) \\
 \hspace{-4mm}&=& \hspace{-1mm} - \partial_m \Bigl( \eta_{\alpha_\dgr{-k}}{}^{\hspace{-2mm}\gamma_\dgr{k}}  \eta_{\tilde{I}_\dgr{k}\tilde{J}_\dgr{k}} C^{\tilde{I}_\dgr{k}}{}_{p \gamma_\dgr{k}} C^{\tilde{J}_\dgr{k}}{}_{q \wb_\dgr{k}} \eta^{qm} \eta^{pn} \tilde{\cJ}_n{}^{{\wb}_\dgr{k}} +\eta_{I_\dgr{k}J_\dgr{k}} C^{I_\dgr{k}m}{}_{\gamma_\dgr{k}} C^{J_\dgr{k} n}{}_{\hL} \zeta_n{}^{\hL} \CR
 && \hspace{70mm} +  \Omega_{I_\dgr{-k}J_\dgr{3+k}} C^{I_\dgr{-k}m}{}_{\alpha_\dgr{-k}}  C{}^{J_\dgr{3+k}n}{}_{\tL_\dgr{3+k}}   \zeta_{n}{}^{\tL_\dgr{3+k}}  \Bigr) \; , \nn
 \eea
 where the last line vanishes for $k=0,1,2$ but does not for $k\ge 3$. Note that for $k=0$ we used the linearised Einstein equation. At this stage it is not yet clear that the solution to equation \eqref{GeneralIntegrability} will be  \eqref{DerivativeLin} as we would like. We shall now argue that we must take additional curls to get an integrability condition for the constrained field  $ \zeta_{n}{}^{\tL_\dgr{3+k}}$. 
   
Let us look at the first non-trivial example. At level $6$ we get for $k=3$ that the right-hand side of  \eqref{GeneralIntegrability} gives 
 \be 
 \partial_m \Bigl( \varepsilon_{n_1\dots n_{11}} \partial^q (h_{p_1\dots p_8,q} + X_{p_1\dots p_8q}) +8 \zeta^\lin_{[p_1|;n_1\dots n_{11},|p_2\dots p_8]} \Bigr)\Bigr|_{(8,1)} \; , 
 \ee
which is in the irreducible $(8,1)$ representation in the indices $\lsharp p_1 \ldots p_8,m\rsharp$. Using $\eta_{m[n_1} F^\lin_{n_2n_3]}{}^m=0$ one gets that 
\be \eta^{qr} F^\lin_{p_1\dots p_{8}q;r}=0 \; ,  \ee 
so that this term can be rewritten as 
 \be \label{RightDualDualGravity} 
 8\, \partial_m \Bigl(- \varepsilon_{n_1\dots n_{11}} \eta^{qr}  \partial_{[p_1} h_{p_2\dots p_8]q,r} + \zeta^\lin_{[p_1|;n_1\dots n_{11},|p_2\dots p_8]} \Bigr) \; .
 \ee
This suggests that one must take an additional curl in $\partial_{p_9}$ to get an integrability condition for the field $\zeta_{1;11,7}$ above. We conclude that rather than taking the component of Young symmetry $(8,1)$ from \eqref{GeneralIntegrability} as a first-order Bianchi identity at level 6, one must take a second-order Bianchi identity of Young symmetry $(9,1)$ to derive that the constrained fields are total derivatives as in \eqref{GeneralIntegrability}. 
This is something that one may expect because the gauge-invariant duality equation for the $E_{11} / K(E_{11})$ fields are third order at level $k=6,7,8$. We have for example for $k=6$
\be
 \cF^\lin_{n_1\dots n_{10};p_1\dots p_8,m} = 10 \partial_{[n_1} h_{n_2\dots n_{10}],p_1\dots p_8,m} + \chi^\lin_{n_1\dots n_{10};p_1\dots p_8,m} = \varepsilon_{n_1\dots n_{10}}{}^q \partial_q\,  h _{p_1\dots p_8],m}\; ,  \ee
where $\chi_{10;8,1}$ combines all the constrained fields that contribute to this field strength. Using \eqref{BI} one obtains the integrability condition for $\cF_{10;8,1}$ 
\begin{multline}
 \partial_{[n_1} \cF^\lin_{n_2\dots n_{11}];p_1\dots p_8,m} + 8 \partial_{[p_1|} \cF^\lin_{[n_1\dots n_{10};n_{11}]|p_2\dots p_8],m} \\ + 8 \eta_{m[n_1} \partial_{[p_1} \cF^\lin_{|n_2\dots n_{11}];|p_2\dots p_8]q,r} \eta^{qr} -  \partial_m \cF^\lin_{[n_1\dots n_{10}|;p_1\dots p_8,|n_{11}]} = 0 \; .
\end{multline}
However, this integrability condition still depends on $h_{9,8,1}$ so in order to eliminate it, one needs to take a second-order integrability condition by taking an additional curl
\be 
\label{BianchiSix} \partial_{[p_1|}  \partial_{[n_1} \cF^\lin_{n_2\dots n_{11}];|p_2\dots p_9],m} - \partial_m \partial_{[p_1|} \cF^\lin_{[n_1\dots n_{10}|;p_2\dots p_9],|n_{11}]} = 0 \; .
\ee
This gives indeed  an integrability condition of Young symmetry $(11,9,1)$ as for the curl of \eqref{RightDualDualGravity}.

We expect similarly that the fields at level $k$ will generally involve order $\lfloor k/3 \rfloor$ Bianchi identities for their field strengths. Further analysis is required to understand how these higher order integrability conditions can systematically be constructed at all levels and how one can prove \eqref{eq:FSlin} from the pseudo-Lagrangian Euler--Lagrange equations in the linearised approximation. 
Note that since the Bianchi identity \eqref{BianchiSix} is not the lowest weight component of a lowest weight $E_{11}$ representation, these higher-order integrability conditions cannot be organised in lowest weight representations of $E_{11}$. As we discussed in Section~\ref{sec:prel}, the need for these higher order equations appears to be unrelated to the approach in~\cite{Tumanov:2016abm,Tumanov:2017whf,Glennon:2020qpt}.

\section{\texorpdfstring{Relation to $E_8$ exceptional field theory}{Relation to E8 exceptional field theory} }
\label{sec:E8}

In this section, we shall show how $E_8$ exceptional field theory~\cite{Hohm:2014fxa} can be recovered from the $E_{11}$ pseudo-Lagrangian~\eqref{eq:Lag} and the self-duality equation~\eqref{eq:DE}. To this end we will consider the level decomposition of all $E_{11}$ objects under $GL(3)\times E_8$ in Section~\ref{sec:E8levdec} and in Section~\ref{sec:e8tame} demonstrate a simplification pattern similar to the one exhibited in~\ref{sec:extall}.

\subsection{\texorpdfstring{$GL(3)\times E_8$ level decomposition}{GL(3)xE8 level decomposition}}
\label{sec:E8levdec}

The $GL(3)\times E_{8}$ level decompositions of the adjoint representation of $E_{11}$ and its $R(\Lambda_1)$ representation were deduced originally in~\cite{Riccioni:2007au,Bergshoeff:2007qi,Cook:2008bi}. The tensor hierarchy algebra $\cT(\mf{e}_{11})$ and the duality equations~\eqref{eq:DE} in $GL(3)\times E_8$ basis were studied in~\cite{Bossard:2019ksx} and we briefly recall the salient points to fix the notation. A short summary of objects is in Table~\ref{tab:e8dec}, for more details see Appendix~\ref{app:E8} and~\cite{Bossard:2019ksx}.

\begin{table}[t!]
\centering
\renewcommand{\arraystretch}{1.2}
\begin{tabular}{c|c|c|c|c|c|c|c|c}
$E_{11}$  & \multicolumn{8}{c}{object}\\\cline{2-9}
rep. & $\ell{=}{-}\tfrac32$ & $\ell{=}{-}1$ &  $\ell{=}{-}\tfrac12$ & $\ell{=}0$ &  $\ell{=}\tfrac12$ & $\ell{=}1$ & $\ell{=}\tfrac32$ & $\ell{=}2$ \\\hline\hline
$\mf{e}_{11}$ & & $\cdots$ && $g_{\mu\nu}$, $M_{AB}$ & &  $A_\mu^A$ &&  $\acute{B}_{\mu\nu}^{AB}$, $B_{\mu\nu}$, $h_{\mu,\nu}^A$ \\
$R(\Lambda_1)$ &  $\partial_A$ & & $\partial_\mu$ && --- && ---  \\
$\overline{L(\Lambda_2)}$ && --- && --- &&  $X_{\nu}$ && $X_{\mu\nu}^A$\\
$\cT_{-1}$ & $\cdots$ && $\cF_{\mu\nu}{}^\rho$, $\cF_{\mu A}$ && $\cF_{\mu;\nu}$, $\cF_{\mu\nu}^A$ && $\cdots$ &
\end{tabular}
\caption{\label{tab:e8dec}\sl The lowest level components of the various objects of $E_{11}$ exceptional field theory in $GL(3)\times E_8$ decomposition. Dashes indicate that there is no $GL(3)\times E_8$ representation at that level (since these are parts of highest/lowest  weight $\mf{e}_{11}$ representations), while ellipses indicate that there are representations but they play no r\^ole in our discussion.}
\renewcommand{\arraystretch}{1}
\end{table}

In this section, we use the indices $\mu,\nu,\ldots$ to denote $(2+1)$-dimensional external space-time indices and indices $A,B,\ldots$ to denote internal coordinate indices that are valued in the adjoint of $E_8$ and thus take $248$ different values. The corresponding derivatives $\partial_A$ satisfy the usual $E_8$ section constraints~\cite{Hohm:2014fxa}
\begin{align}
\label{eq:SCE8}
\kappa^{AB} \partial_A \otimes \partial_B = 0 \,,\quad f^{BC}{}_A \partial_B \otimes \partial_C = 0 \,,\quad 
P^{CD}{}_{AB} \partial_{C} \otimes \partial_{D} = 0 \,,
\end{align}
where $\kappa^{AB}$ is $E_8$ Killing--Cartan form and $f^{AB}{}_C$ the $E_8$ structure constants. The $E_8$-invariant tensor (see~\cite{Koepsell:1999uj} for conventions)
\begin{align}
P^{AB}{}_{CD} = \frac{1}{7} \delta^{(A}_C \delta^{B)}_D - \frac1{56} \kappa^{AB} \kappa_{CD} + \frac{1}{14} f^{AE}{}_{(C} f_{D)E}{}^B 
\end{align}
is the projector onto the ${\bf 3875}$ representation in the symmetric tensor product of two adjoint ${\bf 248}$. One finds indeed that if the fields are not constant in the  $(2+1)$-dimensional external space-time, the $E_{11}$ section condition \eqref{eq:SC} implies that their derivatives of levels $\ell \le - \frac52$ vanish identically while their level $\ell = - \frac32$ derivative $\partial_A$ satisfy the section constraint \eqref{eq:SCE8}. Throughout this section, we shall adopt this $GL(3)\times E_8$ partial solution to the $E_{11}$ section constraint~\eqref{eq:SC}, so we shall only consider the derivatives $\partial_\mu$ and $\partial_A$, where $\partial_A$ still needs to satisfy~\eqref{eq:SCE8}.   

The $E_8$ scalar matrix is written as $M_{AB}$ and the external space-time metric $g_{\mu\nu}$ parametrises the $GL(3)$ subgroup. The $\ell$-form fields at level $1\le \ell\le 3$ in the decomposition of the adjoint of $E_{11}$ appear also in the usual tensor hierarchy algebra of $D=3$ maximal supergravity~\cite{deWit:2008ta,Palmkvist:2013vya}. The one-form field $A_\mu^A$ are dual to the scalar field $E_8$ currents. The two-form field $\acute{B}_{\mu\nu}^{AB}$ at level $\ell=2$ is in the ${\bf 3875}$ representation and sometimes we will combine it with the singlet $B_{\mu\nu}$ into the reducible field $B_{\mu\nu}^{AB} = \acute{B}_{\mu\nu}^{AB} + \kappa^{AB} B_{\mu\nu}$.
We note that in $D=3$ there is no dual graviton since the metric does not propagate degrees of freedom. The field $h_{\mu,\nu}^A$ sitting at level $\ell=2$ in the adjoint is symmetric in $\mu$ and $\nu$ and is a gradient dual to the vector field $A_\mu^A$ similar to the eleven-dimensional field discussed in Section \ref{SectionGradientDual}.

We shall use the semi-flat formulation introduced in Section~\ref{sec:semiflat} with $m(M_{AB} ,g_{\mu\nu})\in GL(3) \times E_8$ and $\cU(A_\mu^A,B_{\mu\nu}^{AB},h_{\mu,\nu}^A,\ldots)$ in the positive level components. As alluded to in that section, the formalism involves derivatives in the form $\cU^{-1 N}{}_M \partial_N$, where $\cU$ is the unipotent matrix associated with the positive level fields in the level decomposition under consideration, see~\eqref{eq:pargauge}. While for the $GL(11)$ level decomposition, where the section constraint was fully solved, the effect of $\cU$ could be ignored since $\cU^{-1 N}{}_m \partial_N = \partial_m$, we here have to take it into account which means that the external space-time derivatives will typically appear in the combination 
\begin{align}
\cU^{-1 N}{}_\mu \partial_N = \partial_\mu - A^A_\mu \partial_A\; , \qquad \cU^{-1 N}{}_A \partial_N = \partial_A \ . \end{align}
We shall use the same convention as in Section~\ref{sec:GL11} that the components of the semi-flat currents and field strengths are written with calligraphic letters and we omit the tilde on the semi-flat components of the constrained fields  
\be  \cU^{-1 N}{}_{M} \tilde{\cJ}_N{}^\alpha = ( \dots , \cJ_{\mu;\nu}{}^\sigma, \cJ_{A;\nu}{}^\sigma, \cJ_{\mu;A} , \cJ_{A;B} , \cJ_{\mu;\nu}^{\hspace{2mm}A},   \cJ_{A;\nu}^{\hspace{2mm}B} , \dots )\ee
where we also use the convention that these components include the factor $\cU^{-1 N}{}_M \partial_N$ on the derivatives. In particular 
\be \cJ_{\mu;\nu}{}^\sigma =  g^{\sigma\rho} (\partial_{\mu}  - A_{\mu}^A \partial_A ) g_{\nu\rho} \; , \qquad  \cJ_{\mu; A} f^{AB}{}_C = -M^{BD} ( \partial_\mu - A_\mu^E \partial_E) M_{CD} \; .\label{E8LeviCurrent} 
\ee
The current $  \cJ_{\mu; A}$ should not be confused with the external $E_8$ current $j_{\mu A}$ of~\cite{Hohm:2014fxa} that we shall define below.

The constrained fields $\chi_M{}^\ta$ are described in the table using the elements $X^\ta \in L(\Lambda_2)$, which is also used in Appendices~\ref{sec:GL11adj} and~\ref{app:E8}.  According to the partial solution to the section constraint, the constrained index $M$ of $\chi_M{}^{\tilde{\alpha}}$ can be either $\mu$ at level $-\frac12$ or $A$ at level $-\frac32$, which leads to the constrained field components 
\begin{align} 
\label{eq:chiE8}
\cU^{-1 N}{}_M \tilde{\chi}_N{}^{\tilde{\alpha}} = (\chi_{\mu;\nu}; \chi_{A;\mu}, \chi_{\mu;\nu\rho}^{\,\,\,\,A}; \chi_{A;\mu\nu}^{\,\,\,\,\,\, B},\ldots )\,.
\end{align}
We shall not display the level decomposition of the  constrained fields $\zeta_M{}^{\widehat{\Lambda}}$, because their components only appear at higher level. 

\medskip

The field strengths given in Table~\ref{tab:e8dec} at $\ell=-\tfrac12$ were worked out in~\cite{Bossard:2019ksx} and take the form
\begin{align}
\cF_{\mu\nu}{}^\sigma &= 2 \cJ_{[\mu ; \nu]}{}^\sigma + 2  \delta^\sigma_{[\mu} \cJ_{A;\nu]}^{\hspace{2mm} A} = 2 g^{\sigma\rho} (\partial_{[\mu}  - A_{[\mu}^A \partial_A ) g_{\nu]\rho} +2 \delta^\sigma_{[\mu} \partial_A A_{\nu]}^A  \,, \nn\\
\cF_{\mu A} &= \cJ_{\mu; A}   +  f_{AC}{}^D  \partial_D A_\mu^C +  \chi_{A;\mu}  \, . 
\end{align}
Similarly, the field strengths at $\ell=+\tfrac12$ are
\begin{align}
\cF_{\mu\nu}^A&=2 \cJ_{[\mu;\nu]}^{\hspace{2mm} A} - \cJ_{B;\mu\nu}^{\hspace{2mm} AB} -f^{AB}{}_{C} \chi_{B;}{}_{\mu\nu}^C \,\nn\\
&=  2 \partial_{[\mu} A_{\nu]}^A- 2 A_{[\mu}^B \partial_B A_{\nu]}^A 
 - \partial_B B_{\mu\nu}^{AB} +\bigl(  14 P^{AB}{}_{CD} + \tfrac{1}{4} \kappa^{AB} \kappa_{CD} \bigr) A_{[\mu}^C \partial_B A_{\nu]}^D-f^{AB}{}_{C} \chi_{B;}{}_{\mu\nu}^C \,,\nn\\
\cF_{\mu;\nu} &=- \cJ_{A;\mu\nu}^{\hspace{2mm} A} +\chi_{\mu;\nu}  + \chi_{A;}{}^A_{\mu\nu}= - \partial_A h^A_{\mu,\nu}  - \tfrac12 f_{BC}{}^A A_{(\mu}^B  \partial_A A_{\nu)}^C  +\chi_{\mu;\nu}  + \chi_{A;}{}^A_{\mu\nu}\,.
\end{align} 
The duality equation~\eqref{eq:DE} implies now that~\cite{Bossard:2019ksx}
\begin{subequations}
\label{eq:E8dual}
\begin{align}
\label{eq:DualYMscalar} 
\cF_{\mu\nu}^A &= \frac{1}{\sqrt{-g}} g_{\mu\sigma} g_{\nu\rho} \varepsilon^{\sigma\hspace{-0.3mm}\rho\lambda} M^{AB}\cF_{\lambda B}\ , \\
\label{eq:E8DG}
\cF_{\mu;\nu} &= - \frac{1}{2\sqrt{-g}} g_{\mu\kappa}   g_{\nu\lambda}   \varepsilon^{\lambda\sigma\rho} \cF_{\sigma\hspace{-0.3mm}\rho}{}^\kappa  \ . 
\end{align}
\end{subequations}

Let us first discuss how the duality relation~\eqref{eq:DualYMscalar} between scalars and vectors relates to the one given in~\cite[Eq.~(3.26)]{Hohm:2014fxa}. In order to do this we need to first identify our constrained field with the ones that appear in~\cite{Hohm:2014fxa} according to\footnote{Note that $B_{\mu A}$ has the opposite sign as in~\cite{Hohm:2014fxa}.}
\begin{subequations}
\begin{align}
\chi_{A;\mu} &= B_{\mu A}\,,\\
\label{eq:HohmChi} 
\chi_{B;}{}_{\mu\nu}^A &= C_{\mu\nu B}{}^A+\tfrac12 f^A{}_{CD} A_{[\mu}^C \partial_B A_{\nu]}^D +\frac{1}{\sqrt{-g}} g_{\mu\sigma} g_{\nu\rho} \varepsilon^{\sigma\hspace{-0.3mm}\rho\lambda} \partial_B A^A_\lambda \ . 
\end{align}
\end{subequations}
Identifying moreover the external  $E_8$ current  covariant under internal $E_8$ diffeormorphisms defined in~\cite{Hohm:2014fxa}  as
\begin{align}
\label{eq:E8curr}
 j_{\mu}{}^A &= M^{AB} \cF_{\mu B} + \kappa^{AB} \bigl( f_{BC}{}^D  \partial_D A_\mu^C +  \chi_{B;\mu}  \bigr) \nn\\
&= \kappa^{AB} \cJ_{\mu; B}   + (\kappa^{AB} + M^{AB} ) (f_{BC}{}^D  \partial_D A_\mu^C + B_{\mu B})  \ ,
 \end{align}
and the $E_8$ two-form field strength as\footnote{Note that we use $F_{\mu\nu}^A$ for the $E_8$ field strength including 2-form components, which was written calligraphic $\cF$ in \cite{Hohm:2014fxa}, whereas here $\cF_{\mu\nu}^A$ refers  to the component of the $E_{11}$ field strength.}
\bea F_{\mu\nu}^A &=& \cF_{\mu\nu}^A-f^{AB}{}_{C}\frac{1}{\sqrt{-g}} g_{\mu\sigma} g_{\nu\rho} \varepsilon^{\sigma\hspace{-0.3mm}\rho\lambda} \partial_B A^C_\lambda \; \\
&=&   2 \partial_{[\mu} A_{\nu]}^A-  A_{[\mu}^B \partial_B A_{\nu]}^A  + f_{CE}{}^A f^{BE}{}_D A_{[\mu}^C \partial_B A_{\nu]}^D + A_{[\mu}^A \partial_B A_{\nu]}^C - \partial_B B_{\mu\nu}^{AB} -f^{AB}{}_{C} C_{B;}{}_{\mu\nu}^C \; , \nn \eea
we see that~\eqref{eq:DualYMscalar} can be written as
\begin{align}
\label{eq:E8vs}
F_{\mu\nu}^A =  \frac{1}{\sqrt{-g}} g_{\mu\sigma} g_{\nu\rho} \varepsilon^{\sigma\hspace{-0.3mm}\rho\lambda} ( j_\lambda{}^A- \kappa^{AB} B_{\lambda B}) \; . 
\end{align}
This equation is consistent with the Euler--Lagrange equation of the constrained field $B_{\mu A}$ in $E_8$ exceptional field theory~\cite[Eq.~(3.26)]{Hohm:2014fxa}, and is satisfied without further projection in  $E_{11}$ exceptional field theory. The results of Section \ref{sec:GL11} imply that it is also consistent with eleven-dimensional supergravity.\footnote{If one further breaks $E_8$ covariance to a Levi subgroup one can show that this additional term in $B_{\mu A}$ can be reabsorbed by an additional redefinition of $C_{\mu\nu B}{}^A$ such that one gets $F^A = \star j^A$.}

The other duality equation~\eqref{eq:E8DG} can be identified with the dual graviton equation, even though there is no dual graviton field.
Similar to the $GL(11)$ decomposition, this equation is not dynamical by itself and only determines the field $\chi_{\mu;\nu}$ algebraically. As we saw in Section~\ref{DualGravitonIn11D} there are two ways of looking at this. Either the integrability condition for the  Einstein equation of~\cite{Hohm:2014fxa} to be satisfied determines the first-order equation for $\chi_{\mu;\nu}$. Or one can derive the first-order equation for $\chi_{\mu;\nu}$ from the pseudo-Lagrangian of $E_{11}$ exceptional field theory derived in this paper. 

The only other duality equation that cannot be solved algebraically for the constrained fields is 
\begin{align}
\label{eq:E8emb}
\cF^{AB}_{\mu\nu\sigma} =  \frac{1}{\sqrt{-g}} g_{\mu\rho} g_{\nu\lambda} g_{\nu\kappa} \varepsilon^{\rho\lambda\kappa} M^{AC} M^{BD} \cF_{CD}\; ,  
\end{align}
where 
\be \cF_{AB} = \bigl(  14 P_{AB}{}^{CD} + \tfrac{1}{4} \kappa_{AB} \kappa^{CD} \bigr) \cJ_{C;D} \; , \ee
and $\cF^{AB}_{\mu\nu\sigma} $ is the field strength of the two-form field $B_{\mu\nu}^{AB}$ that involves the three-form potential in the $ {\bf 248}\oplus {\bf 3875}\oplus {\bf 147250}$ and the constrained fields in the ${\bf 1}\oplus  {\bf 248}\oplus {\bf 3875}\oplus {\bf 30380}$. 

\subsection{From infinitely many to finitely many fields}
\label{sec:e8tame}

Before analysing the pseudo-Lagrangian in $GL(3)\times E_8$ level decomposition in detail, we first repeat the general consideration of Section~\ref{sec:extall} to show that one only has to consider a finite number of terms on the chosen solution of the section condition. In particular, this illustrates how to treat the infinitely many constrained fields in the $GL(3)\times E_8$ decomposition of $E_{11}$. In this section we choose to use the alternative form of the Lagrangian \eqref{SL}. This turns out to slightly simplify the computation.

We must now treat the two derivatives $\partial_\mu$ and $\partial_A$ of $GL(3)\times E_8$ level $-\frac12$ and $-\frac32$, separately (see Table~\ref{tab:e8dec}). We first introduce a convenient index $k$ on $E_{11}$ indices referring to the $GL(3)\times E_8$ level. We shall use $\wa_{(k)}$ to refer to an index of $\adjhat$ at level $k$, as does $\hL_{(k)}$ for $L(\Lambda_{10})\oplus L(\Lambda_4)$. The lowest level for a constrained field $\chi_M{}^\ta$ is $k=1$ on the $\ta$ index while for $\zeta_M{}^\hL$ it is $k=3$ on the $\hL$ index. The (upper) index $\tI$ labelling the $L(\Lambda_3)$ components (see~\eqref{eq:tildeI}) will be written as $\tI_{(k)}$ when referring to a component at level $\tfrac{1}{2}+k$ with  $k\geq 1$. 
An (upper) field strength index $I_{(k)}$ denotes a field strength component at level $-\tfrac12+k$. These two one-half shifts are defined such that the non-vanishing invariant tensors components are $C^{I_\dgr{k} \mu}{}_{\wa_\dgr{k}}$ and $C^{\tilde{I}_\dgr{k}}{}_{\mu \wa_\dgr{k}}$ for an external derivative index. The non-vanishing components with an internal derivative index are instead $C^{I_\dgr{k} A}{}_{\wa_\dgr{k+1}}$ and $C^{\tilde{I}_\dgr{k}}{}_{A \wa_\dgr{k-1}}$. Note that although $K_\alpha{}^\ta{}_\beta$ is not an $E_{11}$ invariant tensor alone (only $f_\alpha{}^\wa{}_\wb$ is) its components under level decomposition are 
 $GL(3) \times E_8$ invariant tensors and in particular it preserves the level. 

According to the convention of Section~\ref{sec:SF} we shall also absorb the $\cU$ matrix in the definition of $\cJ_\mu{}^{\alpha_\dgr{k}}$ and $\chi_\mu{}^{\alpha_\dgr{k}}$ according to 
\be 
\cU^{-1 N}{}_\mu \tilde{\cJ}_{N}{}^\alpha  t_\alpha  = \sum_{k\in \mathds{Z}}  \tilde{\cJ}_{\mu}{}^{\alpha_\dgr{k}}  t_{\alpha_\dgr{k}} \; , \qquad  \cU^{-1 N}{}_\mu \tilde{\chi}_{N}{}^\ta  \bar t_\ta  = \sum_{k=1}^\infty  \tilde{\chi}_{\mu}{}^{\ta_\dgr{k}}  \bar t_{\ta_\dgr{k}} \; ,
\ee 
and identically for $\zeta_\mu{}^{\alpha_\dgr{k}}$.

Because the Lagrangian is second order in derivatives, there are three classes of terms according to the derivative (or constrained indices) $MN$ taking the values $\mu\nu$, $\mu A$ and $AB$.  
The topological term~\eqref{rtop} involving the projector $\Pi_{\ta}{}^{MN}$ expands accordingly into three blocks with $\ta_\dgr{k}$ with $k=1,2,3$. We write first the term involving $K_\alpha{}^\ta{}_\beta$ in the topological term~\eqref{eq:topSF2} and expand it using \eqref{JminusN} 
\begin{align}
&\quad\quad \left(\tilde{\cJ}_M{}^\alpha -2\cN_M{}^\alpha\right) \cU^{-1 M}{}_P \cU^{-1 N}{}_Q \Pi_\ta{}^{PQ} K_\alpha{}^\ta{}_\beta \tilde{\cJ}_N{}^\beta\nn\\
& =  \sum_{k\in\ints} \Pi_{\ta_{(1)}}{}^{\mu\nu} \left(\tilde{\cJ}_\mu{}^{\alpha_\dgr{-k}} - 2\cN_\mu{}^{\alpha_\dgr{-k}}\right) K_{\alpha_\dgr{-k}}{}^{\ta_\dgr{1}}{}_{\beta_\dgr{1+k}} \tilde{\cJ}_\nu{}^{\beta_\dgr{1+k}}\nn\\
&\quad + \sum_{k\in\ints} \Pi_{\ta_{(2)}}{}^{\mu A}  \left(\tilde{\cJ}_\mu{}^{\alpha_\dgr{-k}} - 2\cN_\mu{}^{\alpha_\dgr{-k}}\right) K_{\alpha_\dgr{-k}}{}^{\ta_\dgr{2}}{}_{\beta_\dgr{2+k}} \tilde{\cJ}_A{}^{\beta_\dgr{2+k}} \nn\\
&\quad +  \sum_{k\in\ints} \Pi_{\ta_{(2)}}{}^{A \nu} \left(\tilde{\cJ}_A{}^{\alpha_\dgr{-k}} - 2\cN_A{}^{\alpha_\dgr{-k}}\right)   K_{\alpha_\dgr{-k}}{}^{\ta_\dgr{2}}{}_{\beta_\dgr{2+k}} \tilde{\cJ}_\nu{}^{\beta_\dgr{2+k}} \nn\\
&\quad +  \sum_{k\in\ints} \Pi_{\ta_{(3)}}{}^{A B}  \left(\tilde{\cJ}_A{}^{\alpha_\dgr{-k}} - 2\cN_A{}^{\alpha_\dgr{-k}}\right)  
K_{\alpha_\dgr{-k}}{}^{\ta_\dgr{3}}{}_{\beta_\dgr{3+k}} \tilde{\cJ}_B{}^{\beta_\dgr{3+k}} \nn\\
&=   -  2 \Pi_{\ta_\dgr{2}}{}^{\mu A} \tilde{\cJ}_\mu{}^{\alpha_\dgr{1}} K_{[\alpha_\dgr{1}}{}^{\ta_\dgr{2}}{}_{\beta_\dgr{1}]} \tilde{\cJ}{}_A{}^{\beta_\dgr{1}} -    2\Pi_{\ta_\dgr{3}}{}^{A B} \tilde{\cJ}_A{}^{\alpha_\dgr{1}} K_{[\alpha_\dgr{1}}{}^{\ta_\dgr{3}}{}_{\beta_\dgr{2}]} \tilde{\cJ}{}_B{}^{\beta_\dgr{2}} \\
&\quad + 2 \sum_{k=0}^\infty  \Pi_{\ta_{(1)}}{}^{\mu\nu} \tilde{\cJ}_\mu{}^{\alpha_\dgr{-k}} K_{(\alpha_\dgr{-k}}{}^{\ta_\dgr{1}}{}_{\beta_\dgr{1+k})} \tilde{\cJ}_\nu{}^{\beta_\dgr{1+k}} 
+2 \sum_{k=0}^\infty  \Pi_{\ta_\dgr{2}}{}^{\mu A} \tilde{\cJ}_\mu{}^{\alpha_\dgr{-k}} K_{(\alpha_\dgr{-k}}{}^{\ta_\dgr{2}}{}_{\beta_\dgr{2+k})} \tilde{\cJ}_A{}^{\beta_\dgr{2+k}} \nn\\
&\quad  +
2 \sum_{k=0}^\infty  \Pi_{\ta_\dgr{2}}{}^{A \nu} \tilde{\cJ}_A{}^{\alpha_\dgr{-k}} K_{(\alpha_\dgr{-k}}{}^{\ta_\dgr{2}}{}_{\beta_\dgr{2+k})} \tilde{\cJ}_\nu{}^{\beta_\dgr{2+k}} +2 \sum_{k=0}^\infty  \Pi_{\ta_\dgr{3}}{}^{A B} \tilde{\cJ}_A{}^{\alpha_\dgr{-k}} K_{(\alpha_\dgr{-k}}{}^{\ta_\dgr{3}}{}_{\beta_\dgr{3+k})} \tilde{\cJ}_B{}^{\beta_\dgr{3+k}}\,.\nn
\end{align}
We find therefore that for all but finitely many levels, the tensor $K_{\alpha}{}^{\ta}{}_{\beta}$ appears symmetrised on its adjoint indices $\alpha$ and $\beta$, such that we will be able to use~\eqref{eq:ID4}. Using this equation and~\eqref{eq:ID2} and the fact that $\tilde{\cF}^{I_\dgr{-k}}$ does not depend on the constrained field $\zeta_M{}^{\hL}$ for $k\geq 0$, one obtains eventually that the total topological term can then be written as {\allowdisplaybreaks
\begin{align}
\label{eq:Ltope8}
\mathcal{L}_{\text{top}} 
&=-   \Pi_{\ta_\dgr{2}}{}^{\mu A} \tilde{\cJ}_\mu{}^{\alpha_\dgr{1}} K_{[\alpha_\dgr{1}}{}^{\ta_\dgr{2}}{}_{\beta_\dgr{1}]} \tilde{\cJ}{}_A{}^{\beta_\dgr{1}} -    \Pi_{\ta_\dgr{3}}{}^{A B} \tilde{\cJ}_A{}^{\alpha_\dgr{1}} K_{\alpha_\dgr{1}}{}^{\ta_\dgr{3}}{}_{\beta_\dgr{2}} \tilde{\cJ}{}_B{}^{\beta_\dgr{2}} \nn\\*
&\quad + \Omega_{I_\dgr{0}J_\dgr{1}} C^{I_\dgr{0} A}{}_{\ta_\dgr{1}} \tilde{\chi}_A{}^{\ta_\dgr{1}} \left(C^{J_\dgr{1}\nu}{}_{\wb_\dgr{1}} J_\nu{}^{\wb_\dgr{1}} +C^{J_\dgr{1}B}{}_{\wb_\dgr{2}} J_B{}^{\wb_\dgr{2}}\right) 
+\Pi_\ta{}^{MN} \partial_{M} \big( \cU^{-1\ta}{}_\tb \tilde{\chi}_N{}^\tb \big) \nn\\*
&\quad + \frac12 \Omega_{I_\dgr{0}J_\dgr{1}} C^{I_\dgr{0} A}{}_{\alpha_\dgr{1}} \tilde{\cJ}_A{}^{\alpha_\dgr{1}} C^{J_\dgr{1} \nu}{}_{\beta_\dgr{1}}  \tilde{\cJ}_\nu{}^{\beta_\dgr{1}}- \frac12 \sum_{k=0}^\infty \Omega_{I_\dgr{-k}J_\dgr{k+1}}\tilde{\cF}^{I_\dgr{-k}} \tilde{\cF}^{J_\dgr{k+1}}\nn\\
&=  \Pi_{\ta_\dgr{2}}{}^{\mu A} \tilde{\cJ}_\mu{}^{\alpha_\dgr{1}} K_{\beta_\dgr{1}}{}^{\ta_\dgr{2}}{}_{\alpha_\dgr{1}} \tilde{\cJ}{}_A{}^{\beta_\dgr{1}} -    \Pi_{\ta_\dgr{3}}{}^{A B} \tilde{\cJ}_A{}^{\alpha_\dgr{1}} K_{\alpha_\dgr{1}}{}^{\ta_\dgr{3}}{}_{\beta_\dgr{2}} \tilde{\cJ}{}_B{}^{\beta_\dgr{2}} \\*
&\quad +  \Omega_{I_\dgr{0}J_\dgr{1}} C^{I_\dgr{0} A}{}_{\ta_\dgr{1}} \tilde{\chi}_A{}^{\ta_\dgr{1}} \tilde{\cF}^{I_\dgr{1}} -  \frac12 \sum_{k=0}^\infty \Omega_{I_\dgr{-k}J_\dgr{k+1}}\tilde{\cF}^{I_\dgr{-k}} \tilde{\cF}^{J_\dgr{k+1}} +  \Pi_\ta{}^{MN} \partial_{M} \big( \cU^{-1\ta}{}_\tb \tilde{\chi}_N{}^\tb \big)\,,\nn
\end{align}
where we used~\eqref{eq:ID4} and~\eqref{eq:ID2}. The total derivative evaluates explicitly to
\begin{align}
 \Pi_\ta{}^{MN} \partial_{M} \big( \cU^{-1\ta}{}_\tb \tilde{\chi}_N{}^\tb \big) &=  \partial_M \Bigl( \cU^{-1 M}{}_\mu  \bigl( \Pi_{\ta_\dgr{1}}{}^{\mu\nu} \tilde{\chi}_\nu{}^{\tb_\dgr{1}} +\Pi_{\ta_\dgr{2}}{}^{\mu B} \tilde{\chi}_B{}^{\tb_\dgr{2}}  \bigr)  \Big)\nn\\
 &\hspace{15mm}+ \partial_A \Bigl(  \Pi_{\ta_\dgr{2}}{}^{A \nu} \tilde{\chi}_\nu{}^{\tb_\dgr{2}} +\Pi_{\ta_\dgr{3}}{}^{A B} \tilde{\chi}_B{}^{\tb_\dgr{3}}  \Big)\,,
 \end{align}
 showing that it evaluates to a finite number of terms on section that can be discarded safely.
 } 
 
We define now the square of the duality equation at $GL(3)\times E_8$ level $k\geq 1$  as (cf.~\eqref{eq:Lk})
\begin{align}
\label{eq:LKe8}
\mathcal{O}_k &= -\frac14 m_{I_\dgr{k}J_\dgr{k}} \left( \tilde{\cF}^{I_\dgr{k}} - m^{I_\dgr{k} K'_\dgr{k}} \Omega_{K'_\dgr{k} K_\dgr{1-k}} \tilde{\cF}^{K_\dgr{1-k}}\right)\left( \tilde{\cF}^{J_\dgr{k}} - m^{J_\dgr{k} L'_\dgr{k}} \Omega_{L'_\dgr{k} L_\dgr{1-k}} \tilde{\cF}^{L_\dgr{1-k}}\right)\nn\\
&= - \frac14  m_{I_\dgr{k} J_\dgr{k}} \tilde{\cF}^{I_\dgr{k}} \tilde{\cF}^{J_\dgr{k}} + \frac14  m_{I_\dgr{1-k} J_\dgr{1-k}} \tilde{\cF}^{I_\dgr{1-k}} \tilde{\cF}^{J_\dgr{1-k}}   -\frac12 \Omega_{I_\dgr{1-k}J_\dgr{k}} \tilde{\cF}^{I_\dgr{1-k}}\tilde{\cF}^{J_\dgr{k}}\; .
\end{align}
With this notation we can combine the last term of the topological term~\eqref{eq:Ltope8} with the alternative kinetic term $\widetilde{\mathcal{L}}_{\text{kin}}$ of~\eqref{tkin} as
\begin{align}
\underbrace{-\frac14 \sum_{k\in\ints} m_{I_\dgr{k}J_\dgr{k}} \tilde{\cF}^{I_\dgr{k}} \tilde{\cF}^{J_\dgr{k}}}_{\widetilde{\mathcal{L}}_{\text{kin}}} - \frac12\sum_{k=0}^\infty \Omega_{I_\dgr{-k}J_\dgr{1+k}} \tilde{\cF}^{I_\dgr{-k}} \tilde{\cF}^{J_\dgr{1+k}} 
= -\frac12  \sum_{k=0}^\infty  m_{I_\dgr{-k} J_\dgr{-k}} \tilde{\cF}^{I_\dgr{-k}} \tilde{\cF}^{J_\dgr{-k}}+  \sum_{k=1}^\infty \mathcal{O}_{k}\; . 
\end{align}
The field strengths $\tilde{\cF}^{I_\dgr{-k}}$ simplify for $k\geq 0$ as follows. From the general formula restricted to the $E_8$ solution of the section condition we have 
\begin{align}
\tilde{\cF}^{I_\dgr{-k}} &= C^{I_\dgr{-k} \mu}{}_{\wa_\dgr{-k}} \tilde{\cJ}_\mu{}^{\wa_\dgr{-k}}+C^{I_\dgr{-k} A}{}_{\wa_\dgr{-k+1}} \tilde{\cJ}_A{}^{\wa_\dgr{-k+1}} + C^{I_\dgr{-k} \mu}{}_{\hL_\dgr{-k}} \tilde{\zeta}_\mu{}^{\hL_\dgr{-k}} +  C^{I_\dgr{-k} A}{}_{\hL_\dgr{-k+1}} \tilde{\zeta}_A{}^{\hL_\dgr{-k+1}}\nn\\
&= \left\{ \begin{array}{cl}
C^{I_\dgr{-k} \mu}{}_{\alpha_\dgr{-k}} \tilde{\cJ}_\mu{}^{\alpha_\dgr{-k}}+C^{I_\dgr{-k} A}{}_{\alpha_\dgr{-k+1}} \tilde{\cJ}_A{}^{\alpha_\dgr{-k+1}} & \quad\quad\text{, \; for $k>0$,}\\[4mm]
C^{I_\dgr{0} \mu}{}_{\alpha_\dgr{0}} \tilde{\cJ}_\mu{}^{\alpha_\dgr{0}}+C^{I_\dgr{0} A}{}_{\alpha_\dgr{1}} \tilde{\cJ}_A{}^{\alpha_\dgr{1}} + C^{I_\dgr{0} A}{}_{\ta_\dgr{1}} \tilde{\chi}_A{}^{\ta_\dgr{1}}& \quad\quad\text{, \; for $k=0$.}\\
\end{array}\right.
\end{align}
Thus, there is only the single component $\tilde\chi_A{}^{\ta_\dgr{1}}$ of all the constrained fields remaining in the infinite sum over field strengths squared. We can rewrite this sum using~\eqref{eq:altid}, where $\tI$ only contributes for $k\le 1$ in one of the following three combinations
\begin{align}  
& m_{I_\dgr{-k}J_\dgr{-k}} C^{I_\dgr{-k} \mu }{}_{\alpha_\dgr{-k}} C^{J_\dgr{-k} \nu }{}_{\beta_\dgr{-k}}  - m_{\tilde{I}_\dgr{-k}\tilde{J}_\dgr{-k}} C^{\tilde{I}_\dgr{-k} }{}_{\sigma \alpha_\dgr{-k}} C^{\tilde{J}_\dgr{-k} }{}_{\rho \beta_\dgr{-k}}  m^{\sigma\nu} m^{\rho\mu}\; ,  \\
 & m_{I_\dgr{-k}J_\dgr{-k}} C^{I_\dgr{-k} \mu }{}_{\alpha_\dgr{-k}} C^{J_\dgr{-k} B }{}_{\beta_\dgr{-k+1}}  - m_{\tilde{I}_\dgr{-k+1}\tilde{J}_\dgr{-k+1}} C^{\tilde{I}_\dgr{-k+1} }{}_{C \alpha_\dgr{-k}} C^{\tilde{J}_\dgr{-k+1} }{}_{\rho \beta_\dgr{-k+1}}  m^{CB} m^{\rho\mu}\; ,  \CR
  & m_{I_\dgr{-k}J_\dgr{-k}} C^{I_\dgr{-k} A }{}_{\alpha_\dgr{-k+1}} C^{J_\dgr{-k} B }{}_{\beta_\dgr{-k+1}}  - m_{\tilde{I}_\dgr{-k+2}\tilde{J}_\dgr{-k+2}} C^{\tilde{I}_\dgr{-k+2} }{}_{C \alpha_\dgr{-k+1}} C^{\tilde{J}_\dgr{-k+2} }{}_{D \beta_\dgr{-k+1}}  m^{CB} m^{DA} \;  \nn
\end{align}
as
\begin{align}
&\quad \,  -\frac12  \sum_{k=0}^\infty  m_{I_\dgr{-k} J_\dgr{-k}} \tilde{\cF}^{I_\dgr{-k}} \tilde{\cF}^{J_\dgr{-k}} \nn\\
&=
-\frac12    m_{I_\dgr{0} J_\dgr{0}} \tilde{\cF}^{I_\dgr{0}} \tilde{\cF}^{J_\dgr{0}}-\frac12  \sum_{k=1}^\infty  m_{I_\dgr{-k} J_\dgr{-k}} \bigg[ C^{I_\dgr{-k} \mu}{}_{\alpha_\dgr{-k}} C^{J_\dgr{-k} \nu}{}_{\beta_\dgr{-k}} \tilde{\cJ}_\mu{}^{\alpha_\dgr{-k}}\tilde{\cJ}_\nu{}^{\beta_\dgr{-k}}\nn\\
&\hspace{10mm}
+2C^{I_\dgr{-k} \mu}{}_{\alpha_\dgr{-k}} C^{J_\dgr{-k} A}{}_{\beta_\dgr{-k+1}} \tilde{\cJ}_\mu{}^{\alpha_\dgr{-k}}\tilde{\cJ}_A{}^{\beta_\dgr{-k+1}}
+C^{I_\dgr{-k} A}{}_{\alpha_\dgr{-k+1}} C^{J_\dgr{-k} B}{}_{\beta_\dgr{-k+1}} \tilde{\cJ}_A{}^{\alpha_\dgr{-k+1}}\tilde{\cJ}_B{}^{\beta_\dgr{-k+1}}
\bigg]\nn\\
&=
-\frac12    m_{I_\dgr{0} J_\dgr{0}} \tilde{\cF}^{I_\dgr{0}} \tilde{\cF}^{J_\dgr{0}}- \frac12  m_{\tilde{I}_\dgr{1}\tilde{J}_\dgr{1}} C^{\tilde{I}_\dgr{1} }{}_{C \alpha_\dgr{0}} C^{\tilde{J}_\dgr{1} }{}_{D \beta_\dgr{0}}  m^{CB} m^{DA}\tilde{\cJ}_A{}^{\alpha_\dgr{0}}  \tilde{\cJ}_B{}^{\beta_\dgr{0}}  \nn\\
&\quad 
+ T_{\beta_\dgr{0}}{}^\kappa{}_\sigma m^{AB} m_{\kappa\lambda}m^{\mu\sigma} T_{\alpha_\dgr{-1}}{}^\lambda{}_B \tilde{\cJ}{}_\mu{}^{\alpha_\dgr{-1}} \tilde{\cJ}_A{}^{\beta_\dgr{0}}\nn\\
&\quad
+\frac12 T_{\beta_\dgr{0}}{}^E{}_D m^{BC} m_{EF} m^{AD} T_{\alpha_\dgr{0}}{}^F{}_C  \tilde{\cJ}_A{}^{\alpha_\dgr{0}}\tilde{\cJ}_B{}^{\beta_\dgr{0}}
+\frac12 T_{\beta_\dgr{-1}}{}^\mu{}_D m^{BC} m_{\mu\nu} m^{AD} T_{\alpha_\dgr{0}}{}^\nu{}_C  \tilde{\cJ}_A{}^{\alpha_\dgr{-1}}\tilde{\cJ}_B{}^{\beta_\dgr{-1}}
\nn\\
&\quad 
-\frac12 \sum_{k=1}^\infty \bigg[ m_{\alpha_\dgr{-k} \beta_\dgr{-k}} m^{\mu\nu}   \tilde{\cJ}_\mu{}^{\alpha_\dgr{-k}}\tilde{\cJ}_\nu{}^{\beta_\dgr{-k}}  + m_{\alpha_\dgr{-k} \beta_\dgr{-k}} m^{AB} \tilde{\cJ}_A{}^{\alpha_\dgr{-k}}\tilde{\cJ}_B{}^{\beta_\dgr{-k}}\bigg] -\frac12 m_{\alpha_\dgr{0}\beta_\dgr{0}} \tilde{\cJ}_A{}^{\alpha_\dgr{0}}\tilde{\cJ}_A{}^{\beta_\dgr{0}}\nn\\
&\quad 
+\frac12 \sum_{k=1}^\infty \bigg[  T_{\beta_\dgr{-k}}{}^\nu{}_Q m^{PQ} T_{\alpha_\dgr{-k}}{}^\mu{}_P \tilde{\cJ}_\mu{}^{\alpha_\dgr{-k}}\tilde{\cJ}_\nu{}^{\beta_\dgr{-k}} +2 T_{\beta_\dgr{1-k}}{}^B{}_Q m^{PQ} T_{\alpha_\dgr{-k}}{}^\mu{}_P \tilde{\cJ}_\mu{}^{\alpha_\dgr{-k}}\tilde{\cJ}_A{}^{\beta_\dgr{1-k}} \nn\\
&\hspace{30mm}  + T_{\beta_\dgr{1-k}}{}^B{}_Q m^{PQ} T_{\alpha_\dgr{1-k}}{}^A{}_P \tilde{\cJ}_A{}^{\alpha_\dgr{1-k}}\tilde{\cJ}_B{}^{\beta_\dgr{1-k}} \bigg]\,,
\end{align}
where we used that $T_{\alpha_\dgr{-k}}{}^Q{}_\mu=0$ for $k\geq 1$ and $T_{\alpha_\dgr{-k}}{}^Q{}_A=0$ for $k\geq 2$, as well as~\eqref{eq:altid}.

{\allowdisplaybreaks
The final term to be added to this is the alternative potential term $\widetilde{\mathcal{L}}_{\text{pot}}$ of~\eqref{tpot} that evaluates in $E_8$ decomposition to
\begin{align}
\widetilde{\mathcal{L}}_{\text{pot}} 
&= \frac14 \sum_{k\in\ints} \bigg[ m_{\alpha_\dgr{k} \beta_\dgr{k}} m^{\mu\nu} \tilde{\cJ}_\mu{}^{\alpha_\dgr{k}} \tilde{\cJ}_\nu{}^{\beta_\dgr{k}} + m_{\alpha_\dgr{k} \beta_\dgr{k}} m^{AB} \tilde{\cJ}_A{}^{\alpha_\dgr{k}} \tilde{\cJ}_B{}^{\beta_\dgr{k}} \bigg]\nn\\*
&\quad
-\frac12 \sum_{k\in\ints} \bigg[T_{\alpha_\dgr{k}}{}^\mu{}_R m^{RQ} T_{\beta_\dgr{k}}{}^\nu{}_Q \tilde{\cJ}_\mu{}^{\alpha_\dgr{k}} \tilde{\cJ}_\nu{}^{\beta_\dgr{k}} + 2T_{\alpha_\dgr{k}}{}^\mu{}_R m^{RQ} T_{\beta_\dgr{k+1}}{}^A{}_Q \tilde{\cJ}_\mu{}^{\alpha_\dgr{k}} \tilde{\cJ}_A{}^{\beta_\dgr{k+1}}
\nn\\*
&\hspace{20mm} + T_{\alpha_\dgr{k+1}}{}^A{}_R m^{RQ} T_{\beta_\dgr{k+1}}{}^B{}_Q \tilde{\cJ}_A{}^{\alpha_\dgr{k+1}} \tilde{\cJ}_B{}^{\beta_\dgr{k+1}} \bigg]\nn\\
&= \frac14 m_{\alpha_\dgr{0}\beta_\dgr{0}} m^{\mu\nu} \tilde{\cJ}_\mu{}^{\alpha_\dgr{0}} \tilde{\cJ}_\nu{}^{\beta_\dgr{0}}
+\frac14 m_{\alpha_\dgr{0}\beta_\dgr{0}} m^{AB} \tilde{\cJ}_A{}^{\alpha_\dgr{0}} \tilde{\cJ}_B{}^{\beta_\dgr{0}}
\nn\\*
&\quad  +\frac12 \sum_{k=1}^\infty \bigg[ m_{\alpha_\dgr{k} \beta_\dgr{k}} m^{\mu\nu} \tilde{\cJ}_\mu{}^{\alpha_\dgr{k}} \tilde{\cJ}_\nu{}^{\beta_\dgr{k}} + m_{\alpha_\dgr{k} \beta_\dgr{k}} m^{AB} \tilde{\cJ}_A{}^{\alpha_\dgr{k}} \tilde{\cJ}_B{}^{\beta_\dgr{k}} \bigg]\nn\\*
&\quad 
-\frac12 \sum_{k=0}^\infty  \bigg[T_{\alpha_\dgr{-k}}{}^\mu{}_P m^{PQ} T_{\beta_\dgr{-k}}{}^\nu{}_Q \tilde{\cJ}_\mu{}^{\alpha_\dgr{-k}} \tilde{\cJ}_\nu{}^{\beta_\dgr{-k}} + 2T_{\alpha_\dgr{-k}}{}^\mu{}_P m^{PQ} T_{\beta_\dgr{1-k}}{}^A{}_Q \tilde{\cJ}_\mu{}^{\alpha_\dgr{-k}} \tilde{\cJ}_A{}^{\beta_\dgr{1-k}}\nn\\*
& \hspace{20mm}
+ T_{\alpha_\dgr{1-k}}{}^A{}_P m^{PQ} T_{\beta_\dgr{1-k}}{}^B{}_Q \tilde{\cJ}_A{}^{\alpha_\dgr{1-k}} \tilde{\cJ}_B{}^{\beta_\dgr{1-k}} \bigg]\,,
\end{align}
where we again used the highest weight property of the module $R(\Lambda_1)$.
}

Combining all the terms therefore leads to
\begin{align}
\mathcal{L} &= \mathcal{L}_{\text{top}} + \widetilde{\mathcal{L}}_{\text{kin}} + \widetilde{\mathcal{L}}_{\text{pot}}\nn\\
&= \Pi_{\ta_\dgr{2}}{}^{\mu A} \tilde{\cJ}_\mu{}^{\alpha_\dgr{1}} K_{\beta_\dgr{1}}{}^{\ta_\dgr{2}}{}_{\alpha_\dgr{1}} \tilde{\cJ}{}_A{}^{\beta_\dgr{1}} -    \Pi_{\ta_\dgr{3}}{}^{A B} \tilde{\cJ}_A{}^{\alpha_\dgr{1}} K_{\alpha_\dgr{1}}{}^{\ta_\dgr{3}}{}_{\beta_\dgr{2}} \tilde{\cJ}{}_B{}^{\beta_\dgr{2}} +  \Omega_{I_\dgr{0}J_\dgr{1}} C^{I_\dgr{0} A}{}_{\ta_\dgr{1}} \chi_A{}^{\ta_\dgr{1}} \tilde{\cF}^{J_\dgr{1}}  \nn\\
&\quad-\frac12    m_{I_\dgr{0} J_\dgr{0}} \tilde{\cF}^{I_\dgr{0}} \tilde{\cF}^{J_\dgr{0}} 
+ \frac14 m_{\alpha_\dgr{0}\beta_\dgr{0}} m^{\mu\nu} \tilde{\cJ}_\mu{}^{\alpha_\dgr{0}} \tilde{\cJ}_\nu{}^{\beta_\dgr{0}} \nn\\
&\qquad  - \frac12  m_{\tilde{I}_\dgr{1}\tilde{J}_\dgr{1}} C^{\tilde{I}_\dgr{1} }{}_{C \alpha_\dgr{0}} C^{\tilde{J}_\dgr{1} }{}_{D \beta_\dgr{0}}  m^{CB} m^{DA}\tilde{\cJ}_A{}^{\alpha_\dgr{0}}  \tilde{\cJ}_B{}^{\beta_\dgr{0}}   -\frac14 m_{\alpha_\dgr{0}\beta_\dgr{0}} m^{AB} \tilde{\cJ}_A{}^{\alpha_\dgr{0}} \tilde{\cJ}_B{}^{\beta_\dgr{0}} \nn\\
&\quad
+\frac12 \bigg[ T_{\beta_\dgr{0}}{}^E{}_D  m^{AD} T_{\alpha_\dgr{0}}{}^B{}_E  \tilde{\cJ}_A{}^{\alpha_\dgr{0}}\tilde{\cJ}_B{}^{\beta_\dgr{0}}
+2T_{\beta_\dgr{0}}{}^\rho{}_\sigma  T_{\alpha_\dgr{1}}{}^{A}{}_\rho m^{\mu\sigma}  \tilde{\cJ}{}_\mu{}^{\alpha_\dgr{1}} \tilde{\cJ}_A{}^{\beta_\dgr{0}}\nn\\
&\hspace{90mm}
+T_{\beta_\dgr{1}}{}^A{}_\mu m^{\mu\nu} T_{\alpha_\dgr{1}}{}^B{}_\nu  \tilde{\cJ}_A{}^{\alpha_\dgr{1}}\tilde{\cJ}_B{}^{\beta_\dgr{1}}\bigg]
\nn\\
&\quad 
-\frac12 \bigg[T_{\alpha_\dgr{0}}{}^\mu{}_\rho m^{\rho\sigma} T_{\beta_\dgr{0}}{}^\nu{}_\sigma \tilde{\cJ}_\mu{}^{\alpha_\dgr{0}} \tilde{\cJ}_\nu{}^{\beta_\dgr{0}} + 2T_{\alpha_\dgr{0}}{}^\mu{}_\rho m^{\rho\sigma} T_{\beta_\dgr{1}}{}^A{}_\sigma \tilde{\cJ}_\mu{}^{\alpha_\dgr{0}} \tilde{\cJ}_A{}^{\beta_\dgr{1}}\nn\\
&\hspace{90mm}
+ T_{\alpha_\dgr{1}}{}^A{}_\mu m^{\mu\nu} T_{\beta_\dgr{1}}{}^B{}_\nu \tilde{\cJ}_A{}^{\alpha_\dgr{1}} \tilde{\cJ}_B{}^{\beta_\dgr{1}} \bigg]\nn\\
&\quad 
+ \sum_{k=1}^\infty \mathcal{O}_{k}+\Pi_\ta{}^{MN} \partial_{M} \big( \cU^{-1\ta}{}_\tb \tilde{\chi}_N{}^\tb \big) 
\end{align}
where we used $E_{11}$ invariance of the structure constants and $m^{-1} \tilde{\cJ}^\dagger m = \tilde{\cJ}$ which amounts to $m_{\alpha\beta} \tilde{\cJ}^\beta = \kappa_{\alpha\beta} \tilde{\cJ}^\beta$.

Anticipating the comparison with $E_8$ exceptional field theory, we now group the terms according to
\be
\cL =  \mathcal{L}_{\scalebox{0.7}{pot}}^{\scalebox{0.5}{3D}} + \mathcal{L}_{\scalebox{0.7}{top}}^{\scalebox{0.5}{3D}} + \mathcal{L}_{\scalebox{0.7}{kin}}^{\scalebox{0.5}{3D}}  + \sum_{k=1}^\infty \mathcal{O}_{k}+\Pi_\ta{}^{MN} \partial_{M} \big( \cU^{-1\ta}{}_\tb \tilde{\chi}_N{}^\tb \big) \; , 
\ee
with
\begin{subequations}
\label{eq:LE8dec}
\begin{align}
\label{eq:L8pot}
 \mathcal{L}_{\scalebox{0.7}{pot}}^{\scalebox{0.5}{3D}} &= - \frac14 m_{\alpha_\dgr{0}\beta_\dgr{0}} m^{AB} \tilde{\cJ}_A{}^{\alpha_\dgr{0}} \tilde{\cJ}_B{}^{\beta_\dgr{0}} +\frac12 T_{\beta_\dgr{0}}{}^E{}_D  m^{AD} T_{\alpha_\dgr{0}}{}^B{}_E  \tilde{\cJ}_A{}^{\alpha_\dgr{0}}\tilde{\cJ}_B{}^{\beta_\dgr{0}} \nn\\
& \qquad - \frac12  m_{\tilde{I}_\dgr{1}\tilde{J}_\dgr{1}} C^{\tilde{I}_\dgr{1} }{}_{C \alpha_\dgr{0}} C^{\tilde{J}_\dgr{1} }{}_{D \beta_\dgr{0}}  m^{CB} m^{DA}\tilde{\cJ}_A{}^{\alpha_\dgr{0}}  \tilde{\cJ}_B{}^{\beta_\dgr{0}}\; ,  \\
\label{eq:L8top}
 \mathcal{L}_{\scalebox{0.7}{top}}^{\scalebox{0.5}{3D}} &=  \Pi_{\ta_\dgr{2}}{}^{\mu A} \tilde{\cJ}_\mu{}^{\alpha_\dgr{1}} K_{\beta_\dgr{1}}{}^{\ta_\dgr{2}}{}_{\alpha_\dgr{1}} \tilde{\cJ}{}_A{}^{\beta_\dgr{1}} -    \Pi_{\ta_\dgr{3}}{}^{A B} \tilde{\cJ}_A{}^{\alpha_\dgr{1}} K_{\alpha_\dgr{1}}{}^{\ta_\dgr{3}}{}_{\beta_\dgr{2}} \tilde{\cJ}{}_B{}^{\beta_\dgr{2}} \nn\\
 & \qquad+ \Omega_{I_\dgr{0}J_\dgr{1}} C^{I_\dgr{0} A}{}_{\ta_\dgr{1}} \tilde{\chi}_A{}^{\ta_\dgr{1}} \tilde{\cF}^{J_\dgr{1}}\; ,  \\
 \label{eq:L8kin}
 \mathcal{L}_{\scalebox{0.7}{kin}}^{\scalebox{0.5}{3D}} &=- \frac14 m_{\alpha_\dgr{0}\beta_\dgr{0}} m^{\mu\nu} \tilde{\cJ}_\mu{}^{\alpha_\dgr{0}} \tilde{\cJ}_\nu{}^{\beta_\dgr{0}} +\frac12 T_{\beta_\dgr{0}}{}^\mu{}_\rho m^{\rho\sigma} T_{\alpha_\dgr{0}}{}^\nu{}_\sigma \tilde{\cJ}_\mu{}^{\alpha_\dgr{0}} \tilde{\cJ}_\nu{}^{\beta_\dgr{0}} \nn\\
&\quad
-    m_{I_\dgr{0} J_\dgr{0}} C^{I_\dgr{0} \mu}{}_{\alpha_\dgr{0}} C^{J_\dgr{0} B}{}_{\wb_\dgr{1}} \tilde{\cJ}_\mu{}^{\alpha_\dgr{0}} \tilde{\cJ}_B{}^{\wb_\dgr{1}}-    \tfrac12 m_{I_\dgr{0} J_\dgr{0}} C^{I_\dgr{0} A}{}_{\wb_\dgr{1}} C^{J_\dgr{0} B}{}_{\wb_\dgr{1}} \tilde{\cJ}_A{}^{\wa_\dgr{1}} \tilde{\cJ}_B{}^{\wb_\dgr{1}}  \nn\\
&\quad 
+T_{\beta_\dgr{0}}{}^\rho{}_\sigma  T_{\alpha_\dgr{1}}{}^{A}{}_\rho m^{\mu\sigma}  \tilde{\cJ}{}_\mu{}^{\alpha_\dgr{1}} \tilde{\cJ}_A{}^{\beta_\dgr{0}} - T_{\alpha_\dgr{0}}{}^\mu{}_\rho m^{\rho\sigma} T_{\beta_\dgr{1}}{}^A{}_\sigma \tilde{\cJ}_\mu{}^{\alpha_\dgr{0}} \tilde{\cJ}_A{}^{\beta_\dgr{1}}
 \nn\\
&\quad 
+\frac12 T_{\beta_\dgr{1}}{}^A{}_\mu m^{\mu\nu} T_{\alpha_\dgr{1}}{}^B{}_\nu  (  \tilde{\cJ}_A{}^{\alpha_\dgr{1}}\tilde{\cJ}_B{}^{\beta_\dgr{1}} -\tilde{\cJ}_B{}^{\alpha_\dgr{1}}\tilde{\cJ}_A{}^{\beta_\dgr{1}} ) \; , 
 \end{align}
\end{subequations}
where  we expanded $-\frac12    m_{I_\dgr{0} J_\dgr{0}} \tilde{\cF}^{I_\dgr{0}} \tilde{\cF}^{J_\dgr{0}} $ and used~\eqref{eq:altid} on the terms $\cJ_\mu{}^{\alpha_\dgr{0}} \cJ_\nu{}^{\beta_\dgr{0}} $. 
The first term $ \mathcal{L}_{\scalebox{0.7}{pot}}^{\scalebox{0.5}{3D}} $ contains the terms in $\tilde{\cJ}_A{}^{\alpha_\dgr{0}}$, i.e. internal derivatives of either the $E_8$ scalar fields or the external metric. The second term $ \mathcal{L}_{\scalebox{0.7}{top}}^{\scalebox{0.5}{3D}}$ contains all terms involving the Levi-Civita symbol (signalled by either $\Pi$ or $\Omega$) and do not depend on the $E_8$ scalar fields and the external metric. The last term includes all terms involving the external derivatives of the $E_8$ scalar fields or the external metric,  as well as the internal derivative of the vector field and the constrained field $\tilde{\chi}_A{}^{\ta_\dgr{1}}$ through $\tilde{\cJ}_A{}^{\wa_\dgr{1}}$. Varying with respect to $\tilde{\chi}_A{}^{\ta_\dgr{1}}$ leads to the equation of motion
\begin{align}
 C^{I_\dgr{0} A}{}_{\ta_\dgr{1}} \left[\Omega_{I_\dgr{0}J_\dgr{1}} \tilde{\cF}^{J_\dgr{1}} - m_{I_\dgr{0}J_\dgr{0}} \tilde{\cF}^{J_\dgr{0}}\right] =0\,,
\end{align}
which is a projection of the duality equation~\eqref{eq:E8dual}, consistently with the general consideration in Section~\ref{sec:EOMchi}.

\subsection{\texorpdfstring{Recovering $E_8$ exceptional field theory}{Recovering E8 exceptional field theory}}

We now give the various parts of~\eqref{eq:LE8dec} in an explicit parametrisation. Many steps parallel those in Section~\ref{sec:D11dec}, so we shall be rather brief. Expressions for the various tensors can be found in Appendix~\ref{app:E8}. Starting with the kinetic term $ \mathcal{L}_{\scalebox{0.7}{kin}}^{\scalebox{0.5}{3D}} $, the first line of~\eqref{eq:L8kin} becomes
\begin{align}
 &\quad - \frac14 m_{\alpha_\dgr{0}\beta_\dgr{0}} m^{\mu\nu} \tilde{\cJ}_\mu{}^{\alpha_\dgr{0}} \tilde{\cJ}_\nu{}^{\beta_\dgr{0}} +\frac12 T_{\beta_\dgr{0}}{}^\mu{}_\rho m^{\rho\sigma} T_{\alpha_\dgr{0}}{}^\nu{}_\sigma \tilde{\cJ}_\mu{}^{\alpha_\dgr{0}} \tilde{\cJ}_\nu{}^{\beta_\dgr{0}} \nn\\
&= \sqrt{-g} \Bigl(  - \frac14 g^{\mu\nu} \cJ_{\mu;\sigma}{}^\rho \cJ_{\nu;\rho}{}^\sigma + \frac12 g^{\sigma\mu} {\cJ_{\mu;\rho}{}^\nu \cJ_{\nu;\sigma}{}^\rho} + \frac14 g^{\mu\nu} \cJ_{\mu;\sigma}{}^\sigma \cJ_{\nu;\rho}{}^\rho - {\frac12} g^{\nu\rho} \cJ_{\mu;\sigma}{}^\sigma \cJ_{\nu;\rho}{}^\mu  \nn\\
& \hspace{70mm} - \frac14 g^{\mu\nu} \kappa^{AB} \cJ_{\mu;A} \cJ_{\nu;B}\Bigr)  \; ,
\end{align}
using \eqref{E8LeviCurrent}, see Table~\ref{tab:e8dec} for the fields on different levels.

The second line is more conveniently combined with the last term above to give
\begin{align}
&\quad
-    m_{I_\dgr{0} J_\dgr{0}} C^{I_\dgr{0} \mu}{}_{\alpha_\dgr{0}} C^{J_\dgr{0} B}{}_{\wb_\dgr{1}} \tilde{\cJ}_\mu{}^{\alpha_\dgr{0}} \tilde{\cJ}_B{}^{\wb_\dgr{1}}-    \frac12 m_{I_\dgr{0} J_\dgr{0}} C^{I_\dgr{0} A}{}_{\wa_\dgr{1}} C^{J_\dgr{0} B}{}_{\wb_\dgr{1}} \tilde{\cJ}_A{}^{\wa_\dgr{1}} \tilde{\cJ}_B{}^{\wb_\dgr{1}} \nn\\
& \hspace{70mm} -  \frac14 \sqrt{-g} g^{\mu\nu} \kappa^{AB} \cJ_{\mu;A} \cJ_{\nu;B} \\
&= \sqrt{-g} \Bigl( - \frac12 M^{AB}g^{\mu\nu} \cF_{\mu A} \cF_{\nu B} \! + \frac14 g^{\mu\nu} \kappa^{AB} \cJ_{\mu;A} \cJ_{\nu;B}\! + \! \bigl( g^{\mu\sigma} \cJ_{\mu;\sigma}{}^\nu\!\! -\! g^{\mu\nu} \cJ_{\mu;\sigma}{}^\sigma \bigr) \cJ_{A;\nu}^{\hspace{2.5mm} A}  \! + g^{\mu\nu}J_{A;\mu}^{\hspace{2.5mm} A} \cJ_{B;\nu}^{\hspace{2.5mm} B} \Bigr) \nn\\
&= \sqrt{-g} \Bigl(  - \frac14 \kappa_{AB} g^{\mu\nu} j_\mu{}^A j_\nu{}^B + \frac12  {\kappa^{AB}} g^{\mu\nu}( f_{AC}{}^D \partial_D A_\mu^C + \chi_{A;\mu}) ( f_{BE}{}^F \partial_F A_\nu^E + \chi_{B;\nu}) \nn\\
& \hspace{20mm} +  \bigl( g^{\mu\sigma} \cJ_{\mu;\sigma}{}^\nu - g^{\mu\nu} \cJ_{\mu;\sigma}{}^\sigma \bigr) \cJ_{A;\nu}^{\hspace{2.5mm} A}  + g^{\mu\nu}\cJ_{A;\mu}^{\hspace{2.5mm} A} \cJ_{B;\nu}^{\hspace{2.5mm} B} \Bigr)   \nn\\
&= \sqrt{-g} \Bigl(  - \frac14 \kappa_{AB} g^{\mu\nu} j_\mu{}^A j_\nu{}^B \!+\! \bigl( g^{\mu\sigma} \cJ_{\mu;\sigma}{}^\nu\!\! -\! g^{\mu\nu} \cJ_{\mu;\sigma}{}^\sigma \bigr)  \cJ_{A;\nu}^{\hspace{2.5mm} A} \!+ \frac12   g^{\mu\nu} (3   \partial_A A_\mu^A \partial_B A_\nu^B +\partial_B A_\mu^A \partial_A A_\nu^B)  \Bigr)  \,, \nn
\end{align}
where we introduced the $E_8$ current~\eqref{eq:E8curr} and the $E_8$ section constraint~\eqref{eq:SCE8}. Note also that $\cJ_{A;\mu}^{\hspace{2.5mm} B} = \partial_A A_\mu^B$ for the level one current component along the adjoint $\alpha_\dgr{1}$.

The last two lines of~\eqref{eq:L8kin} give
\begin{align}
&\quad \; T_{\beta_\dgr{0}}{}^\rho{}_\sigma  T_{\alpha_\dgr{1}}{}^{A}{}_\rho m^{\mu\sigma}  \tilde{\cJ}{}_\mu{}^{\alpha_\dgr{1}} \tilde{\cJ}_A{}^{\beta_\dgr{0}} - T_{\alpha_\dgr{0}}{}^\mu{}_\rho m^{\rho\sigma} T_{\beta_\dgr{1}}{}^A{}_\sigma \tilde{\cJ}_\mu{}^{\alpha_\dgr{0}} \tilde{\cJ}_A{}^{\beta_\dgr{1}}
 \nn\\
&\hspace{10mm}
+\frac12 T_{\beta_\dgr{1}}{}^A{}_\mu m^{\mu\nu} T_{\alpha_\dgr{1}}{}^B{}_\nu  (  \tilde{\cJ}_A{}^{\alpha_\dgr{1}}\tilde{\cJ}_B{}^{\beta_\dgr{1}} -\tilde{\cJ}_B{}^{\alpha_\dgr{1}}\tilde{\cJ}_A{}^{\beta_\dgr{1}} ) \; \nn\\
&=\sqrt{-g}  \Bigl(  \bigl( g^{\mu\sigma} \cJ_{A;\sigma}{}^\nu - \tfrac12 g^{\mu\nu} \cJ_{A;\sigma}{}^\sigma \bigr) \cJ_{\mu;\nu}^{\hspace{2.5mm} A} -    \bigl( g^{\mu\sigma} \cJ_{\mu;\sigma}{}^\nu - \tfrac12 g^{\mu\nu} \cJ_{\mu;\sigma}{}^\sigma \bigr) \cJ_{A;\nu}^{\hspace{2.5mm} A} \nn\\
&\quad  + \frac12 g^{\mu\nu} \bigl( \cJ_{A;\mu}^{\hspace{2.7mm} B}   \cJ_{B;\nu}^{\hspace{2.5mm} A}-  \cJ_{A;\mu}^{\hspace{2.7mm} A}   \cJ_{B;\nu}^{\hspace{2.5mm} B}      \bigr)  \Bigr) 
\end{align}
such that in total 
\begin{align}
\label{eq:Lkin3}
  \mathcal{L}_{\scalebox{0.7}{kin}}^{\scalebox{0.5}{3D}} &=  \sqrt{-g} \Bigl(  - \frac14 \kappa_{AB} g^{\mu\nu} j_\mu{}^A j_\nu{}^B  \nn\\
& \hspace{20mm}  - \frac14 g^{\mu\nu} \cJ_{\mu;\sigma}{}^\rho \cJ_{\nu;\rho}{}^\sigma + \frac12 g^{\sigma\mu} {\cJ_{\mu;\rho}{}^\nu \cJ_{\nu;\sigma}{}^\rho} + \frac14 g^{\mu\nu} \cJ_{\mu;\sigma}{}^\sigma \cJ_{\nu;\rho}{}^\rho - {\frac12} g^{\nu\rho} \cJ_{\mu;\sigma}{}^\sigma \cJ_{\nu;\rho}{}^\mu   \nn\\
&\quad  +  \bigl( g^{\mu\sigma} \cJ_{A;\sigma}{}^\nu - \tfrac12 g^{\mu\nu} \cJ_{A;\sigma}{}^\sigma \bigr) \cJ_{\mu;\nu}^{\hspace{2.5mm} A}-    \tfrac12 g^{\mu\nu} \cJ_{\mu;\sigma}{}^\sigma \cJ_{A;\nu}^{\hspace{2.5mm} A}  +g^{\mu\nu} \bigl( \cJ_{A;\mu}^{\hspace{2.7mm} B}   \cJ_{B;\nu}^{\hspace{2.5mm} A}+  \cJ_{A;\mu}^{\hspace{2.7mm} A}   \cJ_{B;\nu}^{\hspace{2.5mm} B}      \bigr)  \Bigr) \nn\\
&=   \sqrt{-g} \Bigl(  - \frac14 \kappa_{AB} g^{\mu\nu} j_\mu{}^A j_\nu{}^B  + \widehat{R}   \Bigr)  \CR
&\hspace{10mm}-D_\mu \Bigl( \sqrt{-g}g^{\mu\nu} \bigl( g^{\sigma\rho}   \bigl( D_\sigma g_{\rho\nu} - D_\nu g_{\sigma\rho} \bigr)-  \cJ_{A;}{}_\nu^A \bigr) \Bigr) - \partial_A \bigl(  \sqrt{-g} g^{\mu\nu} \cJ_{\mu;\nu}^{\hspace{1.5mm} A}\bigr) 
\end{align}
where $\widehat{R}$ defines the covariant Einstein--Hilbert Lagrangian as
\begin{multline}  \label{EinsteinHilbert}
\sqrt{-g} \widehat{R}  =  \sqrt{-g} \Bigl( - \tfrac14 g^{\mu\nu} g^{\sigma\rho} g^{\kappa\lambda} D_\mu g_{\sigma\kappa} D_\nu g_{\rho\lambda}  + \tfrac{1}{2} g^{\mu\sigma} g^{\nu\rho} g^{\kappa\lambda} D_\mu g_{\rho\kappa} D_\nu g_{\sigma\lambda} \\
+\tfrac14 g^{\mu\nu} g^{\sigma\rho} g^{\kappa\lambda} D_\mu g_{\sigma\rho} D_\nu g_{\kappa\lambda} - \tfrac12 g^{\mu\sigma} g^{\nu\rho} g^{\kappa\lambda} D_\mu g_{\kappa\lambda} D_\nu g_{\sigma\rho} \Bigr) \\
+ D_\mu \Bigl( \sqrt{-g}g^{\sigma\rho} g^{\mu\nu}  \bigl( D_\sigma g_{\rho\nu} - D_\nu g_{\sigma\rho} \bigr) \Bigr) \; , 
\end{multline} 
and $D_\sigma g_{\mu\nu}=  \partial_\sigma g_{\mu\nu} - A_\sigma^A \partial_A g_{\mu\nu} - 2 g_{\mu\nu} \partial_A A_\sigma^A$. The  kinetic term $\mathcal{L}_{\scalebox{0.7}{kin}}^{\scalebox{0.5}{3D}} $ therefore matches the kinetic terms in~\cite[Eqs.~(3.2) and (3.5)]{Hohm:2014fxa} up to the last line of~\eqref{eq:Lkin3} which is a total derivative.\footnote{It follows from the weight $2$ of the metric and the weight $0$ of $\partial_A A_\mu^A$ that $D_\mu ( \sqrt{-g} X^\mu) = \partial_\mu  ( \sqrt{-g} X^\mu)  - \partial_A ( A_\mu^A \sqrt{-g} X^\mu)$ in the last lines of \eqref{eq:Lkin3} and \eqref{EinsteinHilbert}.}

The topological term~\eqref{eq:L8top} gives using \eqref{deltaX}
\begin{align}
 \mathcal{L}_{\scalebox{0.7}{top}}^{\scalebox{0.5}{3D}} &= \Omega_{I_\dgr{0}J_\dgr{1}} C^{I_\dgr{0} A}{}_{\ta_\dgr{1}} \chi_A{}^{\ta_\dgr{1}} \tilde{\cF}^{J_\dgr{1}} +   \Pi_{\ta_\dgr{2}}{}^{\!\mu A} \tilde{\cJ}_\mu{}^{\alpha_\dgr{1}} K_{\beta_\dgr{1}\!}{}^{\ta_\dgr{2}}{}_{\!\alpha_\dgr{1}} \tilde{\cJ}{}_A{}^{\beta_\dgr{1}} -    \Pi_{\ta_\dgr{3}}{}^{\!\!A B} \tilde{\cJ}_{\!A}{}^{\alpha_\dgr{1}} K_{\alpha_\dgr{1}\!}{}^{\ta_\dgr{3}}{}_{\!\beta_\dgr{2}} \tilde{\cJ}{}_B{}^{\beta_\dgr{2}}  \nn\\
&=\! - \tfrac12 \varepsilon^{\mu\nu\sigma}\!  \Bigl( F_{\mu\nu}^A \chi_{A;\sigma} +f_{AB}{}^C  \cJ_{\mu;\nu}^{\hspace{2mm} A} \cJ_{C;}{}_\sigma^B + \tfrac13 f_{CD}{}^{[A} \cJ_{A;}{}_{\mu\nu}^{B]C} \cJ_{B;}{}_\sigma^D \Bigr) \nn\\
&=\! - \tfrac12 \varepsilon^{\mu\nu\sigma}\! \Bigl( F_{\mu\nu}^A  B_{\sigma A}  +f_{AB}{}^C  ( \partial_\mu A_\nu^A \!-\! A_\mu^D \partial_D A_\nu^A ) \partial_C A_\sigma^B + \tfrac13 f_{GEF} f^{EA}{}_C   f^{FB}{}_D A_\mu^G \partial_{[A} A_\nu^C \partial_{B]} A_\sigma^D \Bigr) \nn\\
&\quad- \tfrac16   \varepsilon^{\mu\nu\sigma} f_{CD}{}^{[A} \partial_{A} \Bigl( B_{\mu\nu}^{B]C}  \partial_{B} A_\sigma^D - A_\mu^C A_\nu^D \partial_B A_\sigma^{B]} \Bigr)  \,,
\end{align}
which reproduces  the Chern--Simons term of  \cite[Eq.~(3.8)]{Hohm:2014fxa} up to a total derivative using  \cite[Eq. (A.1)]{Hohm:2014fxa}. 
 
 The potential term~\eqref{eq:L8pot} produces immediately~\cite[Eq.~(3.18)]{Hohm:2014fxa}: The component of $\tilde{\cJ}_A{}^{\beta_\dgr{0}}$ along $\mf{e}_8$ is the $E_8$ internal current $\cJ_A{}^B $ and the one along $\mf{gl}(3)$ the internal derivative of the external metric $\cJ_{A \mu}{}^\nu = g^{\nu\sigma} \partial_A g_{\mu\sigma}$, while $C^{\tI_\dgr{1}}{}_{A\beta_\dgr{0}}$ reduces to $C^{1}{}_{AB} = \kappa_{AB}$ since $\tI_\dgr{1}=1$ belongs to a singlet under $E_8$  from~\eqref{eq:GM3a}. Consequently, the potential term~\eqref{eq:L8pot} takes the form
 \begin{multline} \mathcal{L}_{\scalebox{0.7}{pot}}^{\scalebox{0.5}{3D}}  = \sqrt{-g}\Bigl(   \frac14 M^{AB}  g^{\mu\sigma}g^{\nu\rho} \bigl( \partial_A g_{\mu\sigma}  \partial_B g_{\nu\rho} - \partial_A g_{\nu\sigma}  \partial_B g_{\mu\rho} \bigr)  + \frac12  M^{AB} g^{\mu\nu} \partial_A g_{\mu\nu}  f^{CD}{}_B \cJ_{C;D} \\
 - \frac14 M^{AB} \kappa^{CD} \cJ_{A;C} \cJ_{B;D} + \frac12 M^{EF} f^{AD}{}_E f^{BC}{}_F  \cJ_{A;C} \cJ_{B;D}- \frac12 \cJ_{A;}{}^B \cJ_{B;}{}^A  \Bigr)  \; . \end{multline}

In summary, our $E_{11}$ exceptional field theory pseudo-Lagrangian and duality equations produce exactly the Lagrangian of $E_8$ exceptional field theory~\cite{Hohm:2014fxa}. In addition to the $E_8$ exceptional field theory Euler--Lagrange equations, $E_{11}$ exceptional field theory includes an infinite series of duality equations, starting with \eqref{eq:E8DG}, \eqref{eq:E8vs} and \eqref{eq:E8emb}, whose r\^ole is to determine redundant or non-propagating higher-form fields.

\section{Conclusions}
\label{sec:concl}

In this paper, we have constructed a pseudo-Lagrangian~\eqref{eq:Lag}, consisting of the terms displayed in~\eqref{AllTermsL}, that is invariant under rigid $E_{11}$ transformation and that complements a set of $E_{11}$-invariant first-order duality equations that were given in a previous publication~\cite{Bossard:2019ksx}. A summary of all the fields and the relevant $E_{11}$ representations is given in Table~\ref{tab:sum1}. Imposing the $E_{11}$-covariant section condition~\eqref{eq:SC}, makes the pseudo-Lagrangian transform as a density under $E_{11}$ generalised diffeomorphisms such that the Euler--Lagrange equations derived from the pseudo-Lagrangian are gauge-invariant, as are the first-order duality equations. This theory therefore deserves to be called $E_{11}$ exceptional field theory. 

We stress that, while our pseudo-Lagrangian, the duality equation and the section constraint have rigid $E_{11}$ symmetry, choosing a particular solution of the section constraint breaks the $E_{11}$ symmetry to a subgroup, so that there is no rigid $E_{11}$ symmetry in $D=11$ supergravity for example. This is a property of all exceptional field theories, namely that $E_n$ is only a rigid symmetry for field configurations that do not depend on the internal coordinates and then agrees with the Cremmer--Julia symmetry of ungauged maximal supergravity in $11-n$ dimensions.

A crucial ingredient in our construction is the appearance of constrained fields that go beyond the usual tensor hierarchy of fields that are predicted by $E_{11}$. These constrained fields are familiar from $E_n$ exceptional field theory for $n\leq 9$~\cite{Hohm:2014fxa,Bossard:2018utw,Bossard:2021jix} and, importantly and in analogy with the $E_9$ case, some of them transform indecomposably with the tensor hierarchy fields under rigid $E_{11}$~\cite{Bossard:2017wxl}, see~\eqref{eq:T0CR}. The construction of the topological term~\eqref{eq:Ltop} in our pseudo-Lagrangian, as the $E_{11}$-invariant derivative of the constrained field $\chi_M{}^\ta$, was in particular inspired by a similar construction in $E_9$ exceptional field theory~\cite{Bossard:2021jix}. 

To prove gauge-invariance of our theory, a number of group-theoretic identities are required to hold, as summarised in Table~\ref{tab:ids} in Appendix~\ref{app:ids}. Many of these identities can be deduced either from $E_{11}$ representation theory or from using an enveloping tensor hierarchy algebra structure~\cite{Palmkvist:2013vya,Cederwall:2019qnw,Cederwall:2019bai}, based on $\mf{e}_{11}$ or even $\mf{e}_{12}$. The tensor hierarchy algebra $\cT(\mf{e}_{11})$ provides naturally the aforementioned indecomposable representation. Among the required identities there is one, namely the master identity~\eqref{NewMaster}, that we have only been able to derive partially for some of the $E_{11}$ representations involved. Its full proof remains an outstanding problem of our derivation. Another conjecture that needs to be proved is the existence of the non-degenerate $K(E_{11})$-invariant bilinear form $\eta_{IJ}$, which is crucial for the definition of the duality equation and the pseudo-Lagrangian, see the discussion after~\eqref{eq:Tminus1}. We have not seen any sign of reducibility of the module $\cT_{-1}$ to the levels we have checked, which include all levels up to the level of $D$-form potentials in the tensor hierarchy in $D\geq 3$. If $\cT_{-1}$ turned out to be completely reducible, it would follow then that $\eta_{IJ}$ exists and that the master identity~\eqref{NewMaster} is satisfied. However, complete reducibility is not a necessary condition for its existence.
Let us note moreover that our checks of the master identity cover the infinitely many components of the sub-module $R(\Lambda_{10}) \oplus R(2\Lambda_{3}) \subset L(\Lambda_{10})$ and in particular all possible checks for any $GL(D)\times E_{11-D}$  level decomposition up to the level of the $D$-form potentials for $D\geq 3$ (i.e. of fields with $D$ external indices in general, including in particular the dual graviton). 
It is worth noting that while we do expect that our pseudo-Lagrangian is the unique one compatible with both $E_{11}$ generalised diffeomorphism invariance and the duality equation, our incomplete knowledge of $E_{11}$ tensor calculus precludes a proof of uniqueness.

\medskip 

We have presented two main checks of $E_{11}$ exceptional field theory. The first one, described in Section~\ref{sec:GL11} is that, upon choosing the $D=11$ solution of the section constraint and performing the associated $GL(11)$ level decomposition, we recover exactly the bosonic part of $D=11$ supergravity at the non-linear level. No level truncations by hand are necessary for this analysis as the higher level fields arrange themselves  automatically as squares of duality equations in the pseudo-Lagrangian so that their contribution to the equations of motion can be ignored consistently. The second check is a similar analysis for the $GL(3)\times E_8$ decomposition that is performed in Section~\ref{sec:E8} where we show that our theory also contains the well-known $E_8$ exceptional field theory of~\cite{Hohm:2014fxa}.

The pseudo-Lagrangian~\eqref{eq:Lag} can be considered as a master Lagrangian since it contains \textit{all} $E_n$ exceptional field theories ($n\leq 8$) for maximal supersymmetry upon choosing appropriate level decompositions and associated solutions to the section condition. We expect that it also reproduces the (minimal) pseudo-Lagrangian of $E_9$ exceptional field theory \cite{Bossard:2021jix}. 
This behaviour is well-known from $E_{11}$ level decompositions at the kinematic level from  previous investigations~\cite{Schnakenburg:2001he,Schnakenburg:2002xx,Riccioni:2007au,Bergshoeff:2007qi} and,  based on our examples, we expect this to hold dynamically and non-linearly when using the semi-flat formulation of Section~\ref{sec:SF}. Of particular interest might be to see the relation to type IIB supergravity~\cite{Schnakenburg:2001he,Bossard:2017wxl} or massive type IIA supergravity~\cite{Schnakenburg:2002xx,Tumanov:2016dxc}. Massive type IIA supergravity will require a mild violation of the section constraint along the lines of~\cite{Ciceri:2016dmd,Hohm:2011cp
}. The mild violation will only occur in the construction of the semi-flat pseudo-Lagrangian in intermediate steps and the final massive theory will not violate the section constraint.

We also consider the decomposition of $E_{11}$ exceptional field theory to its $GL(1)\times E_{10}$ subgroup in Appendix~\ref{app:E10}. This defines $E_{10}$ ExFT and we discuss how it relates to the $E_{10}$ sigma model introduced in~\cite{Damour:2002cu} that emerged from considerations of the Belinskii--Khalatnikov--Lifshitz limit of $D=11$ supergravity~\cite{Damour:2000hv}. This one-dimensional sigma-model only depends on a worldline parameter (to be thought of as time) and is conjectured to encode the spatial dependence of all supergravity fields via `gradient representations'~\cite{Damour:2002cu} that are the $E_{10}$-analogues of the $E_{11}$ dual fields of type $A_{9^n,3}$, $A_{9^n,6}$ and $A_{9^n,8,1}$. The $E_{10}$ sigma model does not contain any constrained fields. 
One may have na\"ively thought that the restriction of $E_{10}$ ExFT to fields that do not depend on the $E_{10}$ internal coordinates would reproduce the $E_{10}$ sigma model. 
However, we know already from $E_9$ ExFT that it is necessary to keep the constrained field $\chi_{M}{}^\ta$ with the constrained index in the $E_9$ internal coordinate module in order to obtain ungauged maximal supergravity \cite{Bossard:2021jix}.\footnote{There is a component $B_{\mu A}{}^B$ of $ \chi_M{}^\ta$ that cannot be set to zero. In particular $B_{\mu A}{}^A = 0 $ would imply that the two-dimensional dilaton $\rho$ is constant, whereas it is an arbitrary harmonic function.} We argue in Appendix \ref{app:E10} that considering the $D=1$ exceptional field theory in which the $E_{11}$ ExFT fields only depend on the time coordinate, but with non-zero constrained field $\chi_M{}^{N}$ subject to the $E_{10}$ section constraint (with $M$ and $\ta = N$ in the coordinate module $R_\ten(\Lambda_1)$ of $E_{10}$), one indeed reproduces the $E_{10}$ sigma model, with an additional algebraic constraint on the $E_{10}$ current. We suggest that this algebraic constraint may be useful for resolving  some of the puzzles in the $E_{10}$ sigma model conjecture~\cite{Damour:2002cu}. 

As mentioned in the introduction, one can also perform a $GL(1)\times Spin_+(10,10)\subset E_{11}$ level decomposition. This should produce the double field theory (DFT) formulation of type II theories~\cite{Hohm:2011dv} and in particular includes the Ramond--Ramond fields in a spinor representation of $Spin_+(10,10)$~\cite{Kleinschmidt:2004dy,Hohm:2011dv}. One can also envisage performing a construction similar to the one of the present paper for theories without maximal supersymmetry. This would involve DFT itself, but also cases such as pure general relativity in $D=4$ space-time dimensions by replacing $E_{11}$ by other very-extended Kac--Moody algebras such as $A_1^{+++}$~\cite{Kleinschmidt:2003mf,Glennon:2020qpt}. For previous work on non-pure supergravity theories see for example~\cite{Riccioni:2008jz,Kleinschmidt:2008jj}. For subalgebras ${\mf g}$ of ${\mf e}_{11}$, one obtains a consistent truncation of $E_{11}$ exceptional field theory by restricting all fields and tensors to the singlets of a commuting subgroup, as e.g.  $E_7$ for ${\mf g}= A_1^{+++}$ or more generally $E_{8-n}$ for ${\mf g}= A_n^{+++}$ with $n\le 5$.

\medskip

Besides reproducing the known exceptional field theories, our system also provides an explicit form of the duality equations for the infinitely many higher level fields. This is discussed in detail in Section~\ref{HigherLevelSection}. The constrained fields are again central for this mechanism and we have exemplified this in detail for the dual graviton and the first gradient dual of the three-form gauge field in eleven dimensions. We have in particular derived non-linear Lagrangians for these fields that descend directly from the $E_{11}$ pseudo-Lagrangian. Our analysis of the higher duality equations in eleven dimensions remains nonetheless incomplete and we have only described schematically the equations for all higher level fields that are dual to the propagating degrees of freedom. It would be interesting to analyse the complete set of gauge transformations to solve systematically the linearised duality equations. We have argued in particular that the $\Sigma$ gauge parameters mentioned in Section \ref{sec:gendiff} generalise the St\"{u}ckelberg shift gauge parameters introduced in \cite{Boulanger:2008nd} for the dual graviton to all higher dual fields. A complete analysis of the gauge transformations and the propagating degrees of freedom in $E_{11}$ exceptional field theory should permit us to derive a completely explicit realisation of the proposal \cite{Riccioni:2006az}, with all duality equations for the higher gradient duals described as in \cite{Boulanger:2015mka}. The precise relation to the modulo equation of e.g.~\cite{Glennon:2020uov} remains open.  Note nevertheless that all higher dual fields with more than one column can be consistently eliminated as they are algebraic equations for the constrained fields components that define St\"{u}ckelberg potentials similarly as in  \cite{West:2001as,West:2002jj,Boulanger:2008nd}. Only the tensor-hierarchy $p$-forms in exceptional field theory satisfy equations that are not tautological in this sense. The higher $p$-forms are not normally considered in exceptional field theories, but their duality equations are by construction components of the $E_{11}$ duality equation. The simplest instances were given in~\eqref{eq:E8vs} and \eqref{eq:E8emb} for the $GL(3)\times E_8$ decomposition, as well as the equivalent of the dual graviton equation in~\eqref{eq:E8DG}. Although these higher forms do not need to be introduced in exceptional field theory in order to write a (pseudo)-Lagrangian for the dynamical fields, they are naturally defined as dual potentials. The three-form and four-form field strengths were for example introduced in $E_7$ exceptional field theory in \cite{Butter:2018bkl}, and it is clear from representation theory that the corresponding duality equations would be part of the $E_{11}$ exceptional field theory duality equation in the $GL(4) \times E_7$ decomposition.  

\medskip

While the tensor hierarchy algebra $\cT(\mf{e}_{11})$ is an excellent tool for deriving results about the algebraic structure of $E_{11}$ exceptional field theory and in particular for demonstrating group-theoretic identities, the symmetries of the theory are the usual rigid $E_{11}$, local $K(E_{11})$ and $E_{11}$ generalised diffeomorphisms. Thus, there are no symmetries associated with the tensor hierarchy algebra itself. Understanding whether it can be made to play a more direct r\^ole in the theory is an interesting question. As was hinted at in~\cite{Bossard:2019ksx} and commented on in Section~\ref{sec:gendiff}, the extension to a supersymmetric theory seems to require making more direct use of the tensor hierarchy algebra. 

The first steps for the inclusion of fermions into the duality equations were undertaken in~\cite{Bossard:2019ksx}, building on the finite-dimensional spinor representations of $\widetilde{K(E_{11})}$, the double cover of $K(E_{11})$~\cite{Harring:2019}, that were found in~\cite{Kleinschmidt:2006tm,Steele:2010tk}. While we have proposed a supersymmetric version of the duality equation extended by fermion bilinears in~\cite{Bossard:2019ksx}, adding a Rarita--Schwinger-like term to the pseudo-Lagrangian and employing a Noether procedure appears to be an interesting challenge. As there does not appear to be a $\widetilde{K(E_{11})}$-invariant quartic fermion term, the quartic term in the fermions of eleven-dimensional supergravity would have to come from the infinite sum of quadratic terms $\mathcal{O}_n$ in the duality equation in \eqref{eq:L11}, as anticipated in~\cite{Bossard:2019ksx}. However, because the spinor is in a finite-dimensional representation, the sum of the quartic terms in the fermion diverges and it is unclear to us whether a pseudo-Lagrangian with finite order fermionic terms exists or can be defined unambiguously. 
The resulting theory should have local supersymmetry on the extended space-time $R(\Lambda_1)$ modulo the section constraint  and we hope to report on this in the future.

One main property of the $\widetilde{K(E_{11})}$ representation is that the bilinear in the superymmetry parameter and the gravitino field is not in ${\mf e}_{11}\ominus K(\mf{e}_{11})$ but in the quotient of the module $L(\Lambda_2) \oleft \mf{e}_{11} \oleft L(\Lambda_2)$ by $K(\mf{e}_{11})$~\cite{Bossard:2019ksx}.\footnote{Note that the same doubly indecomposable (`socle length 3') representation of $E_{11}$ also features in our derivation of some $E_{11}$ identities in~\eqref{DoubleExt}, therefore proving its existence which was conjectured in~\cite{Bossard:2019ksx}.} To define the supersymmetry transformations one must therefore  extend the $E_{11}$ coset fields to include a new field $\phi^\ta \in L(\Lambda_2)$ for the second $L(\Lambda_2)$, that may be consistently set to zero using field dependent  (and traceful) $\Sigma_M{}^\tI$ ancillary gauge transformations. One needs to analyse the extension of the theory including these fields to derive the correct supersymmetry transformations.

This problem is also related to the definition of the general ancillary gauge transformation of the constrained fields that we have not derived in this paper. By construction, the pseudo-Lagrangian and the duality equation are invariant under the ancillary gauge transformations generated by the commutator of two generalised diffeormorphisms. In particular it is invariant under the ancillary gauge transformations of parameter $\Sigma_M{}^\tI = C^{\tI}{}_{P \wa} T^{\wa N}{}_Q \Sigma_{(MN)}{}^{[PQ]}$ with $MN$ constrained, as in  the second term of \eqref{eq:sigma12}. But this does not cover the whole space of traceless parameters $\Sigma_M{}^\tI$. The general ancillary gauge transformation of the \textit{constrained} fields should be defined such that it reproduces the known one when it is generated by the commutator of two generalised diffeomorphisms and such that the duality equation is invariant. The commutator of two generalised diffeomorphisms acting on the constrained fields does not only generate a gauge transformation of parameter $\Sigma_M{}^\tI$, but also an additional ancillary gauge transformation of parameter $  \Upsilon_{MN}{}^{P\widehat{\Lambda}}$ with two constrained indices $M$ and $N$, which remains to be defined as a symmetry of the duality equation. Note moreover that the algebra of generalised diffeomorphisms is infinitely reducible, and only closes on itself up to trivial parameters. In $D\ge 2$ exceptional field theory, these  trivial parameters give rise to 1-form gauge transformations that act on the gauge fields of the theory \cite{Hohm:2014fxa,Bossard:2021jix}. Since there is no external $p$-form in $E_{11}$  exceptional field theory, the trivial parameters do not define gauge symmetries of the theory. It would only be necessary to introduce them to determine the infinite chain of ghosts for ghosts of  the BRST algebra of $E_{11}$ generalised diffeomorphisms. 

$E_9$ exceptional field theory admits a Virasoro-extended formulation, which involves all negative Virasoro generators $L_{-n}$ that transform indecomposably with the adjoint of $E_9$~\cite{Bossard:2021jix}. This formulation allows reproducing the $E_9$ linear system of two-dimensional supergravity as a consistent truncation of exceptional field theory, whereas the relation to the linear system remains unclear in the non-extended (minimal) formulation. As mentioned above, the construction necessary for the supersymmetric theory includes an additional field  $\phi^\ta \in L(\Lambda_2)$. In the case of $E_9$, this fields is commonly called  $\tilde{\rho}$ and is the one associated to the Virasoro generator $L_{-1}$ in $E_9$ exceptional field theory. The Virasoro-extended formulation of~\cite{Bossard:2021jix} has similar fields for all $L_{-n}$. Whether a `Virasoro-extended form' of $E_{11}$ exceptional field theory, including all fields $\phi^{\ta_1\dots \ta_n} \in L(n\Lambda_2)$, exists, is not clear to us. We have verified in Appendix~\ref{app:ext} using local algebra techniques,  see \eqref{VirE11}, that one can adjoin the representation $\bigoplus_{n>0} L(n\Lambda_2)$ to $E_{11}$ in an indecomposable manner, such that the algebra extension exists.  If such a Virasoro-extended version of $E_{11}$ exceptional field theory existed, it might permit understanding more systematically the infinite chain of dualities with higher level fields and possibly lead to new integrability structures.

Another approach to supersymmetrisation of the $E_{11}$ exceptional field theory may be a suitable extended superspace approach.  This has been studied for the $E_7$ exceptional field theory in \cite{Butter:2018bkl} where only the four-dimensional (external) geometry was elevated to $(4|32)$-dimensional superspace, see also~\cite{Cederwall:2013oaa,Cederwall:2016ukd,Butter:2021dtu}. 
Should a suitable generalisation of this construction be found for the $E_{11}$ exceptional field theory, it could provide a framework for the construction of an action for M-branes propagating in the resulting generalised target superspace. An additional and very powerful method that could be utilised is based on the superembedding  principle for which superspace is essential \cite{Howe:1998tsa}. 
 
\medskip

The previous points addressed mainly the formal development of $E_{11}$ exceptional field theory and we now turn to some potential applications. 

One interesting avenue is to explore the connection to exotic branes and non-geometric backgrounds~\cite{Kumar:1996zx,Dabholkar:2005ve,Hull:2006qs,Kleinschmidt:2011vu,deBoer:2012ma,Sakatani:2014hba,Gunaydin:2016axc,Lombardo:2016swq,Berman:2018okd,Plauschinn:2018wbo,Arvanitakis:2018hfn,Otsuki:2019owg,Fernandez-Melgarejo:2019mgd,Otsuki:2020qhb}.
Depending on the co-dimension of the exotic brane, it couples to different generators from the list~\eqref{eq:gens}, where we restrict to the $GL(11)$ decomposition and an M-theory discussion for concreteness. Ordinary branes correspond to the generators of $E_8$ and are the KK-momentum, the M2, the M5 and the KK-monopole solutions. Their co-dimension is greater than two. Co-dimension two objects are associated with genuine $E_9$ generators, co-dimension one objects with genuine $E_{10}$ generators and space-filling objects with genuine $E_{11}$ generators. As these are related by rigid $E_{11}$ symmetries one can consider the transformation of the solutions. But, as the rigid $E_{11}$ transformation typically changes the solution of the section constraint, the transformed solutions tend to be solutions of a different model. Exceptions to this statement occur when the solution has isometries that are preserved by the rigid $E_{11}$ transformation and this is the framework of U-duality as a solution generating technique~\cite{Obers:1998fb}. The non-geometricity of a transformed solution arises when higher-level fields and coordinates are formally turned on in the solution.  A famous example in the context of DFT is the twisted torus~\cite{Shelton:2005cf} and we refer the reader to~\cite{Hull:2006va,Plauschinn:2018wbo} for more examples and references. These non-geometric solutions can in principle be thought of as  coupling electrically to higher level dual fields~\cite{Englert:2007qb}, and it would be interesting to see whether our model provides a means to constructing the corresponding geometric theory. The considerations of Section~\ref{sec:HighLev} appear relevant for this. 

DFT can be understood as a low-energy limit of the string field theory classical action \cite{Hull:2009mi}. One may therefore hope to get some hints on what should be the M-theory effective action from $E_{11}$ exceptional field theory.  Our construction is limited to the bosonic two-derivative sector. The tree-level string theory effective action is expected to admit the continuous $Spin_+(10,10)$ symmetry to all orders in $\alpha^\prime$ \cite{Sen:1991zi}, so one can possibly construct higher-derivative couplings in DFT~\cite{Hohm:2013jaa,Godazgar:2013bja,Marques:2015vua,Codina:2020yma,Codina:2020kvj}, see however~\cite{Hronek:2020xxi} for a recent puzzle in this context. On the contrary, U-duality is a symmetry of the quantum theory~\cite{Hull:1994ys}, and there is no higher-derivative coupling with the continuous symmetry $E_{11}(\reals)$. It was already understood in the original conjecture of \cite{West:2001as} that only a discrete symmetry could remain at the quantum level. One may only expect to be able to construct higher derivative couplings with the discrete symmetry~\cite{Bao:2007er}, which would necessarily involve $E_{11}(\mathds{Z})$-invariant Kac--Moody automorphic forms~\cite{Fleig:2012xa}.  Moreover, the M-theory low energy effective actions strongly depends on the background. So an effective action with $E_{11}(\mathds{Z})$ symmetry could only be viewed as a formal object, which upon reduction by taking various limits would reproduce the effective action of string theory on different backgrounds. One may expect for example the effective action of string theory on $\mathds{R}^{1,D-1} \times T^{10-D}$ (see e.g \cite{Green:2010wi}) to be captured by such a formal $E_{11}(\mathds{Z})$-invariant effective action.  It seems unlikely that the requirement of $E_{11}(\mathds{Z})$ symmetry, together with supersymmetry, will determine such a formal effective action uniquely to all orders in derivatives, but a more detailed study of these questions could provide interesting insights. An alternative option would be to try to quantise directly the effective theory. However it is not clear that one can bypass the problem of the strong section constraint. Double field theory, for example, is not a consistent truncation of string field theory that can be quantised independently. In exceptional field theory one has the same problem, there is no integrated action because of the strong section constraint, and quantum effects allow two $1/2$-BPS excitations to produce $1/4$-BPS and non-BPS states in string theory. The effective theory quantum loops considered in \cite{Bossard:2015foa} are in this spirit defined such that the internal space is a generalised torus. For $E_{11}$ one must necessarily consider a non-compact space-time, so a discrete Fourier expansion on the module $R(\Lambda_1)$ is not a physically meaningful option.

\subsubsection*{Acknowledgements}

We thank Martin Cederwall, Alex J. Feingold, Ralf K\"ohl and Jakob Palmkvist for correspondence on Kac--Moody algebras and Franz Ciceri, Thibault Damour, Marc Henneaux, Gianluca Inverso, Hermann Nicolai and Henning Samtleben for useful discussions. GB and AK are grateful to Texas A\&M University and the Mitchell Foundation for its warm hospitality. The work of GB was partially supported by the ANR grant Black-dS-String (ANR-16-CE31-0004). The work of AK has received funding from the European Research  Council (ERC) under the  European Union's Horizon 2020 research and  innovation programme (grant agreement No 740209).
The work of ES is supported in part by the NSF grant PHY-1803875. 


\appendix

\section{\texorpdfstring{Identities for $E_{11}$ tensors}{Identities for E11 tensors}}
\label{app:ids}

In this appendix, we summarise and prove the identities of the various $E_{11}$ objects that appear in the pseudo-Lagrangian and the check of gauge-invariance. These objects were summarised in Table~\ref{tab:sum2} and we collect the salient identities in Table~\ref{tab:ids}. The only identity for which we do not have a full proof is~\eqref{NewMaster}, although in Appendix~\ref{app:MID} we show it on an infinite-dimensional $E_{11}$ representation.

\begin{table}[t]
\centering
\begin{tabular}{c|c|c}
& Identity & Proof\\\hline\hline&&\\[-2mm]
\eqref{eq:ID5} & $C^{\tilde{I}}{}_{P \widehat{\beta}} T^{\widehat{\beta} (M}{}_Q \overline C_{\tilde{I}}{}^{N)\alpha} \partial_M  \partial_N{=}  \Bigl( f^{\alpha\beta}{}_\gamma T^{\gamma (M}{}_P T_\beta{}^{N)}{}_Q {-}2 \delta^{(M}_{[P} T^{\alpha N)}{}_{Q]} \Bigr) \partial_M  \partial_N$  & \S\ref{Sweds}, THA\\[2mm]
\eqref{NewMaster} &$\Omega_{IJ} C^{JM}{}_{\widehat{\alpha}} T^{\widehat{\alpha} N}{}_P   = \overline C_{IP}{}^{\tilde{\alpha}} \Pi_{\tilde{\alpha}}{}^{MN} +  \overline C_{IP}{}^\hL \Pi_\hL{}^{MN}$ & [\S\ref{app:MID}, THA]
\\[2mm]
\eqref{eq:ID3} & $  \Omega_{IJ} C^{IM}{}_{\tilde{\alpha}}  C^{IN}{}_{\tilde{\beta}} \partial_M \otimes \partial_N = 0$ & \S\ref{app:Moth}, $E_{11}$
\\[2mm]
\eqref{eq:ID7} & $ \Omega_{IJ} C^{IM}{}_{\tilde{\alpha}} C^{IN}{}_\hL\, \partial_M \otimes \partial_N =  0$ & \S\ref{app:Moth}, $E_{11}$
\\[2mm]
\eqref{eq:ID8} & $ \Omega_{IJ} C^{IM}{}_\hL  C^{IN}{}_\hXi\, \partial_M \otimes \partial_N = 0 $ & \S\ref{app:Moth}, $E_{11}$
\\[2mm]
\eqref{eq:ID2} & $ \Bigl(  \Pi_\tb{}^{MN} T_\beta{}^{\tb}{}_\ta +  \Omega_{IJ} C^{IM}{}_{\tilde{\alpha}} C^{JN}{}_{\beta} \Bigr)\, \partial_M \otimes \partial_N =0$ &  \S\ref{app:Moth}, $E_{11}$
\\[2mm]
\eqref{eq:ID4} & $\Bigl(  \Pi_\ta{}^{MN} K_{(\alpha}{}^\ta{}_{\beta)} + \frac{1}{2} \Omega_{IJ} C^{IM}{}_{(\alpha} C^{JN}{}_{\beta)} \Bigr)\, \partial_M \otimes \partial_N = 0$ &  \S\ref{app:Moth}, $E_{11}$
\\[2mm]
\eqref{eq:ID6} & $\overline{C}_{IP}{}^{\ta} C^{IQ}{}_{\wb} = \Pi_{\tI}{}^{Q\ta} C^{\tI}{}_{P\wb} $ & \S\ref{Sweds}, $E_{11}$
\\[2mm]
\eqref{eq:ID9} & $ C^{\tilde{I}}{}_{M \widehat{\alpha}} T^{\widehat{\alpha} P}{}_{N} \bar \partial^M \otimes \bar \partial^N = 0 $  & \S\ref{Sweds}, $E_{11}$ 
\end{tabular}
\caption{\label{tab:ids}{\sl Summary of identities satisfied by the various $E_{11}$ objects. Identities that are only valid on a solution to the section constraint~\eqref{eq:SC} are shown contracted with partial derivatives. The identity \eq{NewMaster} plays a crucial r\^ole in gauge invariance of the duality equation, and we sometimes refer to it as the master identity. It is the only identity in the table for which we do not have a complete proof.}}
\end{table}

In the table, we have also indicated the method of proof that will be used for demonstrating the identities and there are two principal methods
\begin{itemize}
\item \underline{THA techniques}: This method relies on constructing a consistent (super-)algebra extending $\mf{e}_{11}$. This can be either the tensor hierarchy algebra $\cT(\mf{e}_{11})$ already encountered in Section~\ref{sec:THA} or the tensor hierarchy algebra $\cT(\mf{e}_{12})$ based on $\mf{e}_{12}$. The identity will then be implied by Jacobi identities in the tensor hierarchy algebra. An important additional identity that follows from $\cT(\mf{e}_{12})$  is~\eqref{e12Equation}.
\item \underline{$E_{11}$ representation theory}: This means that one can analyse the identity either using highest weight methods for $E_{11}$ or by constructing appropriate homomorphisms between $E_{11}$ representations. Some of the explicit calculations for these proofs will be done in $GL(11)$ level decomposition which is reviewed in Appendix~\ref{app:GL11}.
\end{itemize}

Some proofs also combine both methods and in the table we have indicated the one that is more prominent in the given proof.

\subsection{\texorpdfstring{$\cT(\mf{e}_{12})$ and the proof of (\ref{eq:ID5})}{T(e12) and the proof of (2.52)}}
\label{Sweds} 

An important ingredient in the second potential term~\eqref{eq:Lpot2} and the ancillary gauge transformations~\eqref{eq:extGL} is the tensor $C^\tI{}_{M\wa}$ and its relation to the tensor $C^{IM}{}_\wa$ introduced in~\eqref{eq:FS} and that appears as structure constants of the tensor hierarchy algebra $\cT(\mf{e}_{11})$. In this section we will find that $C^\tI{}_{M\wa}$ is defined as structure constants of the tensor hierarchy algebra $\cT(\mf{e}_{12})$. Using this observation, we  prove identity~\eqref{eq:ID5} that we reproduce here for convenience
\begin{align}
\label{Father} 
C^{\tilde{I}}{}_{P \widehat{\beta}} T^{\widehat{\beta} (M}{}_Q \eta^{N)R} \eta_{\tI\tJ} \eta^{\alpha\gamma}
C^{\tilde{J}}{}_{R\gamma} \partial_M  \partial_N =  \Bigl( f^{\alpha\beta}{}_\gamma T^{\gamma (M}{}_P T_\beta{}^{N)}{}_Q -2 \delta^{(M}_{[P} T^{\alpha N)}{}_{Q]} \Bigr) \partial_M  \partial_N \; . 
\end{align}
We have written the partial derivatives without $\otimes$ to emphasise that $M$ and $N$ are symmetric. Our proof uses the tensor hierarchy algebra $\cT(\mf{e}_{12})$ extending the Kac--Moody algebra $\mf{e}_{12}$ and follows ideas developed in~\cite{Cederwall:2019bai}. In~\cite{Cederwall:2019bai}, it was shown that for a finite-dimensional algebra $\mf{g}$ and an irreducible coordinate representation $R(\lambda)$, one can construct a Lagrangian for the fields in $G/ K(G)$ that is invariant under the generalised diffeomorphisms for coordinates in $R(\lambda)$. This Lagrangian takes the form of~\eqref{eq:Lpot1} plus~\eqref{eq:Lpot2} and its invariance uses the equivalent of \eqref{Father}, which is shown to be a Jacobi identity for the tensor hierarchy algebra $\cT(\mf{g}_\lambda)$ for the Kac--Moody extension $\mf{g}_\lambda$ of $\mf{g}$ by the weight $\lambda$. In our context, the further Kac--Moody extension of ${\mf e}_{11}$ by the weight $\Lambda_1$ is $\mf{e}_{12}$~\cite{Kleinschmidt:2003jf}, in which an extra node is attached at node $1$ in Figure~\ref{fig:e11dynk}.

Indeed by using the tensor hierarchy algebra $\cT(\mf{e}_{12})$ we can realise the $\mf{e}_{11}$ tensors  appearing in~\eqref{Father} as structure constants in a Lie super-algebra and use its Jacobi identities to deduce new identities for the $\mf{e}_{11}$ tensors. 
We first note that the proof of the existence of the tensor hierarchy algebra $\cT(\mathfrak{e}_n)$ in \cite{Bossard:2017wxl} applies to any $n\ge 3$, and therefore to $\cT(\mathfrak{e}_{12})$. This proof is based on a local super-algebra construction as defined in \cite{Kac2}. We can either use the local super-algebra associated to the $GL(12)$ decomposition introduced in \cite{Bossard:2017wxl}, or the one associated to the $GL(4) \times E_8$ decomposition introduced in \cite{Bossard:2019ksx} and further analysed in Section \ref{app:e8tha} below.\footnote{The generalisation from $\cT(\mf{e}_{11})$ to $\cT(\mf{e}_{n})$ is simply obtained by considering superfields depending on $n$ Grassmann variables instead of $11$. One easily verifies that none of the computations depend on the number of Grassmann variables.}

The algebra $\cT(\mf{e}_{12})$ has $\mf{e}_{12}$ as a subalgebra and a grading analogous to~\eqref{eq:Tdec}. One finds that $\cT_0(\mathfrak{e}_{12})$ decomposes as
\begin{align}
 \label{Te12} 
 \cT_0(\mathfrak{e}_{12}) = \mathfrak{e}_{12}\oleft L_{\twelve}(\Lambda_3)  \oplus L_{\twelve}(\Lambda_{1}+\Lambda_{11})   
\end{align}
with the bounded weight modules 
 \begin{subequations}
\begin{align}
 \label{L12} 
L_{\twelve}(\Lambda_3) &= R_{\twelve}(\Lambda_3)\oplus \dots \\
L_{\twelve}( \Lambda_{1}+\Lambda_{11}) &= R_{\twelve}(\Lambda_{1}+\Lambda_{11}) \oplus R_{\twelve}(\Lambda_{12}) \oplus R_{\twelve}(\Lambda_1 + 2\Lambda_4) \oplus  \dots   
\end{align}
\end{subequations}
This decomposition is analogous to~\eqref{conjectured}, \eqref{adjhatDef} and we have written $R_{\twelve}$ (resp. $L_{\twelve}$) for all highest weight (resp. bounded weights) representations to distinguish $\mf{e}_{12}$-representations from the $\mf{e}_{11}$-representations.  By the same mechanism as  \cite[Eq. (B.50)]{Bossard:2017wxl}  we deduce that  $L_{\twelve}(\Lambda_3)\supset R_{\twelve}(\Lambda_3)$ mixes under $\mathfrak{e}_{12}$ with $\mathfrak{e}_{12}$ in an indecomposable manner, while the other highest weight modules in $L_{\twelve}(\Lambda_1 + \Lambda_{11})$ are proper submodules that appear as a direct sum. This is also discussed in~\eqref{Reducibility} below. Similar to $\cT_0(\mf{e}_{11})$, it is plausible that $L_{\twelve}(\Lambda_3) = R_{\twelve}(\Lambda_3)$. 

Branching these $\mf{e}_{12}$-representations in $\cT_0(\mf{e}_{12})$ under $\mf{e}_{11}$, one has\footnote{We note that the complete reducibility of a representation of $E_{12}$ implies the complete reducibility under $E_{11}$ as can be seen by fixing the $GL(1)$ eigenvalue as we do here.}
\begin{align}
\label{eq:T0branch}
 \mathfrak{e}_{12} &= \dots \oplus R(\Lambda_1)^\ord{-1} \oplus ( \mathfrak{gl}(1) \oplus \mathfrak{e}_{11})^\ord{0} \oplus \overline{R(\Lambda_1)}^\ord{1} \oplus \dots \,,\CR
R_{\twelve}(\Lambda_3) &= \ldots
\oplus \big(R(\Lambda_3)\oplus R(\Lambda_1{+}\Lambda_{10}) \oplus R(\Lambda_{11})\oplus R(\Lambda_1{+}\Lambda_4)\oplus R(\Lambda_5) \oplus R(\Lambda_1{+}\Lambda_2{+}\Lambda_{10}) \nn\\
&\quad\quad\oplus 2{\times} R(\Lambda_3{+}\Lambda_{10}) \oplus R(2\Lambda_1{+}\Lambda_{11}) \oplus 2{\times} R(\Lambda_2{+}\Lambda_{11}) \oplus R(\Lambda_1{+}2\Lambda_3) \oplus\ldots
\big)^{\ord{-1}} \nn\\
&\quad\quad  \oplus R(\Lambda_2)^{\ord{0}} \,, \CR
R_{\twelve}(\Lambda_1{+}\Lambda_{11}) &= \ldots \oplus \big( R(\Lambda_1{+}\Lambda_{10}) \oplus R(\Lambda_{11}) \oplus R(\Lambda_5) \oplus \ldots
\big)^{\ord{-1}}\oplus R(\Lambda_{10})^{\ord{0}} \,,\\
R_{\twelve}(\Lambda_{12}) &= \ldots \oplus \big( R(\Lambda_9)\oplus \ldots \big)^{\ord{-2}} \oplus R(\Lambda_{11})^{\ord{-1}}\,, \CR
R_{\twelve}(\Lambda_1{+}2\Lambda_4) &= \ldots \oplus \big(R(\Lambda_1{+}2\Lambda_3) \oplus R(\Lambda_3{+}\Lambda_4)\oplus R(\Lambda_2{+}\Lambda_3{+}\Lambda_{10})\oplus\ldots
\big)^{\ord{-1}} \oplus R(2\Lambda_3)^{\ord{0}} \, . \nonumber
 \end{align}
 The superscripts here refer to the eigenvalues under the adjoint action of  $\mf{gl}(1)$ when branching to $\mf{gl}(1)\oplus \mf{e}_{11}\subset \mf{e}_{12}$ and we shall refer to this as the $w_1$-degree.
We recognise at degree $w_1=1$ only $E_M\in \overline{R(\Lambda_1)}$ and at degree $w_1=0$ the generators $t^{\widehat{\alpha}} , t^\Lambda \in \cT_0({\mf e}_{11})$.\footnote{For ease of notation we use the same index $\Lambda$ for the reducible part $D_0$ of $\cT_0$ in \eqref{conjectured}, as we do for the generator $P^\Lambda \in \cT_2$, even though $D_0$   may turn out to do not include the entire bounded weight module $L(\Lambda_{10}) \subset \cT_2$. The proof given in this appendix does not rely on the assumption that $L(\Lambda_{10})\subset D_0$.} At degree $w_1=-1$ there are the generators $F^{M}\in R(\Lambda_1)$ as well as $F^{\tilde{I}} \in L(\Lambda_3)$ as given in~\eqref{eq:tildeI} and realising these latter in an algebra was one of the main reasons for going from $\cT(\mf{e}_{11})$ to $\cT(\mf{e}_{12})$. We define $L(\Lambda_3)$ inside the module $L_{\twelve}(\Lambda_3) $, such that it transforms indecomposably together with $\mf{e}_{11}$. This is indeed possible because one finds from  \eqref{eq:T0branch} that  $R(\Lambda_1) \otimes R(\Lambda_2) \ominus R(\Lambda_1 + \Lambda_2)$ is contained in  $R_{\twelve}(\Lambda_3)\subset L_{\twelve}(\Lambda_3)$.
We call the decomposable part $D_1 \subset L_{\twelve}( \Lambda_{1}+\Lambda_{11}) $. Note that it includes the representations in $\cT_1({\mf e}_{11}) \ominus  R(\Lambda_1)$ displayed in \eqref{eq:T1exp} and we shall therefore write $F^{\widetilde{M}}\in D_1 $. In summary, we deduce from~\eqref{eq:T0branch} that the branching of  $\cT_0(\mathfrak{e}_{12})$ under $\mf{e}_{11}$ gives 
\begin{align}
 \cT_0(\mathfrak{e}_{12}) = \dots  \oplus ( R(\Lambda_1) \oplus L(\Lambda_{3}) \oplus D_1)^\ord{-1}  \oplus \cT_0(\mf{e}_{11})^\ord{0} \oplus  \overline{R(\Lambda_1)}^\ord{1} \oplus \dots    
\end{align}
Note that the indecomposable structure induced from $ \mathfrak{e}_{12}\oleft R(\Lambda_3) $ is necessarily trivial in the category $\mathcal{O}$ of Kac--Moody representations, such that  $R(\Lambda_1) \oleft L(\Lambda_3) = R(\Lambda_1) \oplus L(\Lambda_3)$ as follows from~\cite{Kac}.

Restricting to $L(\Lambda_3)$, the algebra of $\cT_0(\mathfrak{e}_{12})$ can be written in BRST form as
\begin{align}
 \delta h_\alpha &=  \tfrac12 f^{\beta\gamma}{}_\alpha h_\beta h_\gamma - K^{\beta\tilde{\gamma}}{}_\alpha h_\beta h_{\tilde{\alpha}} - T_\alpha{}^M{}_N f_M e^N + C^{\tilde{I}}{}_{N\alpha } f_{\tilde{I}} e^N\; ,  \CR
\delta h_{\tilde{\alpha}} &= -T^{\beta\tilde{\gamma}}{}_{\tilde{\alpha}} h_\beta h_{\tilde{\gamma}} + C^{\tilde{I}}{}_{N\tilde{\alpha}} f_{\tilde{I}} e^N \; , \quad \delta h_{\Lambda} = -T^{\beta \Xi}{}_{\Lambda} h_\beta h_{\Xi} + \Pi^{\widetilde{M}}{}_{N\Lambda} f_{\widetilde{M}} e^N \; , \CR
\delta e^M &= T^{\widehat{\beta} M}{}_N h_{\widehat{\beta}} e^N \; , \\
\delta f_M &=- T^{\widehat{\beta} N}{}_M h_{\widehat{\beta}} f_N\; , \quad  \delta f_{\widetilde{M}} =- T^{\widehat{\beta} \widetilde{N}}{}_{\widetilde{M}} h_{\widehat{\beta}} f_{\widetilde{N}} \; ,  \CR
\delta f_{\tilde{I}} &= -T^{\widehat{\beta} \tilde{J}}{}_{\tilde{I}} h_{\widehat{\beta}} f_{\tilde{J}} \; , \nn
\end{align}
where we have omitted the terms bilinear in $h_{\tilde{\alpha}}, h_\Lambda$ and $f_{\tilde{I}}, f_{\widetilde{M}}$ and we write the structure constants as $C^{\tilde{I}}{}_{N\alpha } $ anticipating that we have identified $L(\Lambda_3)$ such that it agrees with the representation that appears in~\eqref{Father}. The structure of the algebra is constrained by  $\cT_0(\mathfrak{e}_{11})$ covariance and the  branching of $\mathfrak{e}_{12}$ modules under $\mathfrak{e}_{11}$, such that all the structure constants are fixed with  $C^{\tilde{I}}{}_{N\widehat{\alpha}} $ an invariant tensor under $\mf{e}_{11}$.

To see this we look at the Jacobi identity in $\cT_0(\mf{e}_{12})$ corresponding to the terms proportional to  $h_\alpha f_{\tilde{I}} e^N$ in $\delta^2 h_\alpha=0$. This gives
\begin{align}
 T^{\beta P}{}_N C^{\tilde{I}}{}_{P \alpha} - T^{\beta \tilde{I}}{}_{\tilde{J}} C^{\tilde{J}}{}_{N \alpha} - f^{\beta\gamma}{}_\alpha C^{\tilde{I}}{}_{N\gamma} + K^{\beta\tilde{\gamma}}{}_\alpha C^{\tilde{I}}{}_{N\tilde{\gamma}} = 0 \,.
\end{align}
From this identity we conclude that $C^{\tilde{I}}{}_{N \widehat{\alpha}}$ is an  $\mf{e}_{11}$-invariant tensor. In particular  $C^{\tI}{}_{N\ta}$ is a non-zero intertwiner that can be computed explicitly and the relation then gives some non-trivial components $C^{\tI}{}_{N\alpha}$, leading to a non-trivial tensor $C^{\tI}{}_{N\wa}$. This reasoning applies to all irreducible constituents of $L(\Lambda_3)$, giving in particular that $C^{\tilde{I}_\lambda}{}_{N \tilde{\alpha}}=c_\lambda \Pi^{\tilde{I}_\lambda}{}_{N \tilde{\alpha}}$ on any irreducible $\mf{e}_{11}$-representation $R(\lambda)$ inside $L(\Lambda_3)$.

We now consider the branching of $\cT_1(\mathfrak{e}_{12})$ under $\mathfrak{e}_{12}$
\be  \cT_1(\mathfrak{e}_{12}) = R_{\twelve}(\Lambda_1) \oplus  R_{\twelve}(\Lambda_2+\Lambda_{11})\oplus   R_{\twelve}(\Lambda_2+2\Lambda_4) \oplus   R_{\twelve}(\Lambda_1+\Lambda_{12}) \oplus\ldots \ee
that further branches under $\mathfrak{e}_{11}$ as
\begin{align}
R_{\twelve}(\Lambda_1) &= \dots \oplus  ( L(\Lambda_{10})  \oplus L(\Lambda_4) )^\ord{-\frac43} \oplus  R(\Lambda_1)^\ord{-\frac13} \oplus {\bf 1}^\ord{\frac23} \; , \CR
 R_{\twelve}(\Lambda_2+\Lambda_{11}) &= \dots \oplus  R(\Lambda_1+\Lambda_{10})^\ord{-\frac13} \; , \CR
  R_{\twelve}(\Lambda_2+2\Lambda_4) &= \dots \oplus  R(\Lambda_1+2\Lambda_3)^\ord{-\frac13}\; , \\
  R_{\twelve}(\Lambda_1+\Lambda_{12}) &= \dots\oplus  R(\Lambda_{11})^\ord{-\frac13}\nn
  \; , 
\end{align}
which shows that the $\mf{e}_{11}$-representation $R(\Lambda_1)$ arises together with all other components of $\cT_1(\mf{e}_{11})$ in $\cT_1(\mf{e}_{12})$ at $w_1=-\tfrac13$. It is possible that there is more than the module $\cT_1(\mf{e}_{11})$ at $w_1=-\tfrac13$, but this will not be relevant for the arguments of this appendix since we shall only be interested in the identities one can derive from the algebra $\cT(\mathfrak{e}_{12})$ for the submodule $R(\Lambda_1) \subset \cT_1(\mf{e}_{11})$, i.e.  $\widehat{M}$ restricted to $M$, where $\widehat{M}$ refers to all of $\cT_1(\mf{e}_{11})$.

Using the local super-algebra  of $\cT(\mathfrak{e}_{12})$ in \cite{Bossard:2017wxl} one deduces that the branching of $ \cT_{-1}(\mathfrak{e}_{12}) $ under $\mathfrak{e}_{11}$ gives\footnote{We assume here for simplicity that $D_0=L(\Lambda_{10})$ in~\eqref{conjectured}. In principle one should write different $L_\ell(\Lambda_k)$ with a label $\ell$ for different bounded weight modules extending the irreducible modules $R(\Lambda_k)$ and take into account that $\Lambda$ (or more precisely $\Lambda^\ell$) may be an index of different extensions $L_\ell(\Lambda_{10})$ of $R(\Lambda_{10})\oplus R(2\Lambda_3)$. This will not affect our conclusions in this appendix and we chose to avoid the label $\ell$ for brevity.}
\begin{multline} 
\label{eq:Tm1dec}
\cT_{-1}(\mathfrak{e}_{12}) = \dots \oplus L(\Lambda_3)^\ord{- \frac52} \oplus   ( \adjhhat \oplus L(\Lambda_{10})\oplus L(\Lambda_{10}) )^\ord{-\frac23}  \oplus  (  \cT_{-1}(\mf{e}_{11}) \oplus \overline{R(\Lambda_1)} )^\ord{\frac13} \\
\oplus    ( \overline{L(\Lambda_2)}\oplus  \overline{L(\Lambda_{10})} \oplus \overline{L(\Lambda_4)} )^\ord{\frac43} \oplus  \dots  
 \end{multline}
where\footnote{In this equation, we do not need to put parentheses since $L(\Lambda_2) \oleft  \mf{e}_{11} \oleft L(\Lambda_2)= L(\Lambda_2) \oleft \bigl(  \mf{e}_{11} \oleft L(\Lambda_2)\bigr) =\bigl( L(\Lambda_2) \oleft  \mf{e}_{11} \bigr) \oleft L(\Lambda_2) $. This module was already conjectured to exist in \cite{Bossard:2019ksx}, see footnote 16 there for its first levels in the $GL(3) \times E_8$ decomposition.} 
\begin{align}
 \adjhhat = L(\Lambda_2) \oleft  \mf{e}_{11} \oleft L(\Lambda_2)\;   \label{DoubleExt} 
 \end{align}
is a double extension of the adjoint representation $ \mf{e}_{11}$ that includes both indecomposable structures. The decomposition~\eqref{eq:Tm1dec} can be established by starting from the $GL(12)$ version of the local super-algebra of $\cT(\mf{e}_{12})$ and branching the $GL(12)$ representations to $GL(11)$ and then reassembling them into the structure of $\cT(\mf{e}_{11})$. In particular, there are two nine-forms of $GL(11)$ arising in the process from the field strength component $F_{mn}{}^p$ in the $GL(12)$ decomposition of $\cT_{-1}(\mf{e}_{12})$. The defining relations of $\cT(\mf{e}_{12})$ then implies the double indecomposability~\eqref{DoubleExt}.

The double extended module \eqref{DoubleExt} is defined such that $F^{\wwa} =( \acute{F}^{\tilde{\alpha}} ,F^\alpha,F^{\tilde{\alpha}})\in  L(\Lambda_2) \oleft  \mf{e}_{11} \oleft L(\Lambda_2) $ admits the non-covariant transformation
\begin{align}
 \Delta^\alpha F^{\tilde{\beta}} =  K^{\alpha \tb}{}_\gamma F^\gamma  + K^{\alpha \tb}{}_{\tg} \acute{F}^{\tg} \; , \qquad \Delta^\alpha F^{\beta} =  - \eta^{\alpha\iota} K_\iota{}^{ \tilde{\gamma}}{}_\delta \eta^{\beta\delta} \eta_{\tilde{\gamma}\tilde{\eta}}  \acute{F}^{\tilde{\eta}} \; , \qquad  \Delta^\alpha \acute{F}^{\tilde{\beta}} = 0 \; .  \label{DoubleExtend} 
\end{align}

Some of the generators of $\cT(\mf{e}_{12})$ in $\mf{e}_{11}$ language are summarised in Table~\hyperlink{tab:e12e11}{6}, where we also introduce names for the various elements of the representations discussed above.
\renewcommand{\arraystretch}{1.5}
\begin{table}[h!]
{\centering
\begin{tabular}{c|c|c|c}
& $w_1  + \frac{p_{\scalebox{0.45}{$12$}}}{3} = - 1 $  &  $w_1  + \frac{p_{\scalebox{0.45}{$12$}}}{3}= 0  $ &  $w_1  + \frac{p_{\scalebox{0.45}{$12$}}}{3}= 1$    \\
\hline\hline
$p_{\scalebox{0.5}{$12$}} = 1$ & $Q^\Lambda , Q^{\tL} $ & $ P^{\widehat{M}} $ & $p$\\
\hline
$ p_{\scalebox{0.5}{$12$}} = 0 $ & $ F^{{M}} , F^{\tilde{I}}, F^{\widetilde{M}}$ & $h,\, t^{\widehat{\alpha}}  , \, t^\Lambda  $ & $ E_M $\\
\hline
$ p_{\scalebox{0.5}{$12$}} = -1 $ & $ F^{\wwa} ,\, F^\Lambda ,\, \acute{F}^\Lambda $ & $ t_I, \, t_M $ & $ E_{\tilde{\alpha}},\, E_\Lambda,\, E_{\tL}  $\\
\end{tabular}
\\[3mm]}
{\small \hypertarget{tab:e12e11}{Table 6}: {\sl Decomposition of $\cT(\mf{e}_{12})$ generators under $\mathfrak{e}_{11}$, with $p_{\twelve}$ the level of $\cT_{p_{\twelve}}(\mf{e}_{12})$ and $w_1$ the $GL(1)$ weight under $E_{11}$. They are related to the degrees  $p = \frac{2p_{\twelve}}{3} - w_1$ and $q=-\frac{p_{\twelve}}{3} -w_1 $  of  \cite[Table 3]{Cederwall:2019bai}. $\wwatext$ is for ${\stackon[-10.5pt]{\stackon[-9.5pt]{\rm adj }{\widehat{\phantom{adj}}}}{\widehat{\phantom{adj}}}}$ and $\tL$ for $L(\Lambda_{4})$. The involution for $\cT(\mf{e}_{12})$ is centred around $p_{\twelve}=-\tfrac32$ and $q=\tfrac12$ but we shall not use this.  }}
\end{table}

\medskip

We now use the $\cT(\mf{e}_{12})$ machinery and ingredients just introduced to prove~\eqref{Father}, following \cite{Cederwall:2019bai}.  The starting point is the Jacobi identity involving the generators $F^{\wwa},\, E_M$ and $P^M$. Using the known commutators of $\cT(\mf{e}_{11})$ and using $\cT_0(\mf{e}_{11})$ covariance, one determines the following commutators within $\cT(\mf{e}_{12})$
\begin{align}
 [ h , p] &=- \tfrac23 p\; , \quad [ p , F^M ] = -P^M \; , \quad \{ p , F^{\widehat{\alpha}} \}  = t^{\widehat{\alpha}} \; , \quad \{ p , \acute{F}^{\tilde{\alpha}} \}  = 0 \; ,\quad  \{  p , t_M\} = - E_N\; , \CR
    {[}P^M , E_N {]} &= \delta^M_N p \; , \quad \{ P^M , t_N \} = \tfrac32 \delta^M_N h + T_\alpha{}^M{}_N t^\alpha \; ,   \CR
    {[} E_M, F^N {]} &= \tfrac32 \delta_M^N h - T_\alpha{}^N{}_M t^\alpha \; , \qquad {[} E_M , F^{\widetilde{N}} {]} = \Pi^{\widetilde{N}}{}_{M \Lambda} t^\Lambda\; , \CR
    \{ P^M , t_I \} &= - \Omega_{IJ} ( C^{JM}{}_{\widehat{\alpha}} t^{\widehat{\alpha}} + C^{JM}{}_\Lambda t^\Lambda ) \; , \qquad     {[} E_M , F^{\tilde{I}} {]} = C^{\tilde{I}}{}_{M \widehat{\alpha}} t^{\widehat{\alpha}}  \; . \label{FixedT12}
\end{align}
For $\{ P^M , t_N \} $, one can check using the $GL(12)$ local super-algebra that it does not produce the generator $t^{\tilde{\alpha}}\subset R(\Lambda_2)$. It follows by representation theory that this anti-commutator with $P^M \in R(\Lambda_1)$ cannot produce  generators in  $L(\Lambda_2)\oplus L(\Lambda_{10})$. It is consistent with the Jacobi identity 
\begin{align}
 [ F^M , \{ p , t_N \} ]  = \{ p , [ F^M,t_N]\} - \{ t_N , [p,F^M]\} 
 \end{align}
provided we define
\begin{align}
 \{ F^M , t_N\} = - 2 T_\alpha{}^M{}_N F^\alpha- 2  \eta_{\tilde{\alpha}\tilde{\beta}} T^{\tilde{\beta} P}{}_Q \eta^{QM} \eta_{PN} \acute{F}^{\tilde{\alpha}} 
 \end{align}
that can be checked to be $E_{11}$-covariant using  
\begin{align}
  \eta_{\alpha\beta} T^{\beta P}{}_Q \eta^{QM} \eta_{PN}  =  T_\alpha{}^M{}_N \; . 
\end{align}
Even though the two terms on the right-hand side are not individually $E_{11}$-covariant due to \eqref{DoubleExtend}, their combination is.

We make now the following ansatz for the remaining (anti-)commutators of $F^{\wwa}$ that we need for the Jacobi identity involving  $F^{\wwa}$, $E_M$ and $P^M$:
\begin{subequations}
\label{eq:CR1}
\begin{align}
[ E_M , F^{\wwa} ] &= \Omega^{IJ} \overline{C}_{IM}{}^{\wwa} t_J +  \delta^{\wwa}_{\widehat{\beta}}T^{\widehat{{\beta}}N}{}_M t_N \; , \\
\{ P^M , F^{\wwa} \} &=  \overline{C}_{\tilde{I}}{}^{M \wwa}  F^{\tilde{I}} + \delta^{\wwa}_{\widehat{\beta}}T^{\widehat{{\beta}}M}{}_N   F^N+C_{\widetilde{N}}{}^{M \wwa}  F^{\widetilde{N}}  \; ,
\end{align}
\end{subequations}
where  we used that there is a unique homomorphism from $\mf{e}_{11} \otimes R(\Lambda_1)\rightarrow R(\Lambda_1)$ to determine the structure coefficients multiplying respectively $t_N$ and $F^N$,\footnote{\label{R1edE1} The uniqueness of the homorphism can be proved in the $GL(11)$ level decomposition by solving the highest weight condition for the vector $X_n = c_0 h_n{}^m \partial_m + c_0^\prime h_m{}^m  \partial_n + c_1 A_{np_1p_2} \partial^{p_1p_2}  + \sum_{k=2}^\infty c^{\alpha_\dgr{k} M_{\dgr{-\frac32 -k}}}{}_n  A_{\alpha_\dgr{k}} \partial_{M_\dgr{-\frac32 -k}}$ for free coefficients $c^{\alpha M}{}_n$. One finds that $c_1= \frac12 c_0$ and $c_0^\prime = - \frac12 c_0$ and because the module $R(\Lambda_1)$ is generated from the highest weight vector $X_n$ we conclude that all the coefficients $c^{\alpha_\dgr{k} M_{\dgr{-\frac32 -k}}}{}_n $ are  determined as well as $c^{\alpha M}{}_n = c_0  T^{\alpha M}{}_n$ and the homomorphism is unique. The proof applies to all $\mf{e}_n$.} while $\overline{C}_{IM}{}^{\wwa}$, $\overline{C}_{\tilde{I}}{}^{M \wwa}$, and ${C}_{\widetilde{N}}{}^{M \wwa} $ are  defined on $\adjhhat$.  The latter satisfy the non-covariant transformations induced by the ones of $F^{\wwa} =( \acute{F}^{\tilde{\alpha}}, F^\alpha,F^{\tilde{\alpha}})\in  L(\Lambda_2) \oleft  \mf{e}_{11} \oleft L(\Lambda_2)$, see~\eqref{DoubleExtend}. The way they have been introduced above, $\overline{C}_{IM}{}^{\wwa}$ and $\overline{C}_{\tilde{I}}{}^{M \wwa}$ are a priori new objects and we shall argue in the final step of the proof how they are related to the tensors in~\eqref{Father}. The tensor ${C}_{\widetilde{N}}{}^{M \wwa} \ne 0$, but this will be irrelevant for us. The components of these objects are denoted by
\begin{align}
 \overline{C}_{IM}{}^{\wwa} &= (  \overline{C}_{IM}{}^{{\tilde{\alpha}}} , \overline{C}_{IM}{}^{{\alpha}} ,\acute{\overline{C}}_{IM}{}^{{\tilde{\alpha}}} )
\quad\text{and} \quad
\overline{C}_{\tilde{I}}{}^{M \wwa}  = (\overline{C}_{\tilde{I}}{}^{M \tilde{\alpha}} , \overline{C}_{\tilde{I}}{}^{M \alpha},\acute{\overline{C}}_{\tilde{I}}{}^{M \tilde{\alpha}}  ) 
\; . 
\end{align}
For the second terms~\eqref{eq:CR1} we note that the projection $ \delta^{\wwa}_{\widehat{\beta}}$ is consistent with the non-covariant\\[-2mm]transformation \eqref{DoubleExtend} because $\adjhat$ is the quotient of the module $\adjhhat$ by the submodule $L(\Lambda_2)$.

We now consider the Jacobi identity 
\begin{align} \label{JakobJacobi} 
 [ E_P , \{ P^M , F^{\wwa}\}] = \{ P^M , [ E_P , F^{\wwa}]\} + \{ F^{\wwa} , [ E_P , P^M]\} 
\end{align}
which reduces to the following expression along $t^{\widehat{\alpha}}$ when we use the commutators above:\footnote{The component along $t^\Lambda$ gives ${C}_{\widetilde{N}}{}^{M {\stackunder[-2.8pt]{\stackunder[-3.3pt]{ \scriptstyle \widehat{\phantom{\alpha}}}{\scriptstyle \widehat{\phantom{\alpha}}}}{ \scriptstyle \alpha}}} \Pi^{\widetilde{N}}{}_{P \Lambda}  =  \overline{C}_{IP}{}^{{\stackunder[-2.8pt]{\stackunder[-3.3pt]{ \scriptstyle \widehat{\phantom{\alpha}}}{\scriptstyle \widehat{\phantom{\alpha}}}}{ \scriptstyle \alpha}}} C^{IM}{}_{\Lambda} $, that we will not use, it implies ${C}_{\widetilde{N}}{}^{M {\stackunder[-2.8pt]{\stackunder[-3.3pt]{ \scriptstyle \widehat{\phantom{\alpha}}}{\scriptstyle \widehat{\phantom{\alpha}}}}{ \scriptstyle \alpha}}} \ne 0$.} 
\begin{align} 
\overline{C}_{\tilde{I}}{}^{M \wwa} C^{\tilde{I}}{}_{P \widehat{\beta}}  t^{\widehat{\beta}} -  \delta^{\wwa}_{\widehat{\gamma}}  T^{\widehat{\gamma}M}{}_R T_\beta{}^R{}_P t^\beta =    \overline{C}_{IP}{}^{\wwa} C^{IM}{}_{\widehat{\beta}} t^{\widehat{\beta}} +  \delta^{\wwa}_{\widehat{\gamma}}  T^{\widehat{\gamma} R}{}_P T_\beta{}^{M}{}_R t^\beta  - \delta_P^M  \delta^{\wwa}_{\widehat{\beta}} t^{\widehat{{\beta}}} \; . 
\end{align}
If one takes the Jacobi identity involving $\acute{F}^{\tilde{\alpha}}$, one has  $\delta^{\tilde{\alpha}}_{\widehat{\beta}}=0$ and this identity reduces to 
\begin{align}
\label{eq:almID6}
\acute{\overline{C}}_\tI{}^{M\ta} C^\tI{}_{P\wb} = \acute{\overline{C}}_{IP}{}^\ta C^{IM}{}_\wb\,,
\end{align}
which is already similar to~\eqref{eq:ID6} with $\acute{\overline{C}}_{\tilde{I}}{}^{M \ta}  =\Pi_\tI{}^{M\ta}$  the canonically normalised intertwiners. If one takes the Jacobi identity involving $F^{\alpha}$, one gets 
\begin{align}
\label{e12Equation}
 \overline{C}_{\tilde{I}}{}^{M \alpha} C^{\tilde{I}}{}_{P \widehat{\beta}}  t^{\widehat{\beta}} =    \overline{C}_{IP}{}^{\alpha} C^{IM}{}_{\widehat{\beta}} t^{\widehat{\beta}}  +  T^{\alpha M}{}_R T_\beta{}^R{}_P t^\beta+ T^{\alpha R}{}_P T_\beta{}^{M}{}_R t^\beta -\delta_P^M   t^{\alpha} \ .
 \end{align}
Replacing $t^{\widehat{\beta}}$ by the invariant tensors $T^{\widehat{\beta} N}{}_Q$ and contracting with derivatives $\partial_M\partial_N$ to enforce the section constraint one obtains
\begin{align}
\label{eq:almID5}
 \overline{C}_{\tilde{I}}{}^{M \alpha} C^{\tilde{I}}{}_{P \widehat{\beta}}  T^{\wb N}{}_Q \partial_M\partial_N &= \big( f^\alpha{}_{\beta\gamma} T^{\gamma M}{}_P T^{\beta N}{}_Q - 2\delta^M_{[P} T^{\alpha N}{}_{Q]} \big) \partial_M\partial_N
\end{align}
where we have used~\eqref{eq:THA1} to eliminate the first term on the right-hand side as well as~\eqref{eq:Rlamf} and the section condition~\eqref{eq:SC} to simplify the remaining terms into the form given. This consequence of the Jacobi identity is already very close to  the claimed identity \eqref{Father}, or equivalently~\eqref{eq:ID5}.

As the final step in the argument we must identify the relevant components of the tensors $\overline{C}_{IM}{}^{\wwa}$ and $\overline{C}_{\tilde{I}}{}^{M \wwa}$ with the unbarred ones that appear in the actual identities~\eqref{eq:ID5} and~\eqref{eq:ID6}. This is only possible for the components in $L(\Lambda_2) \oleft  \mf{e}_{11}\subset\adjhhat$ and the identification is 
\begin{align}
 \overline{C}_{IM}{}^{\wwa} &= (  \overline{C}_{IM}{}^{{\tilde{\alpha}}} , \overline{C}_{IM}{}^{{\alpha}} ,\acute{\overline{C}}_{IM}{}^{{\tilde{\alpha}}} ) = ( \overline{C}_{IM}{}^{{\tilde{\alpha}}} , \eta^{\alpha\beta} \eta_{IJ} \eta_{MN}   C^{JN}{}_{{\beta}} , \eta^{\tilde{\alpha}\tilde{\beta}} \eta_{IJ} \eta_{MN}   C^{JN}{}_{\tilde{\beta}}  ) \; , \CR
\overline{C}_{\tilde{I}}{}^{M \wwa}  &= (\overline{C}_{\tilde{I}}{}^{M \tilde{\alpha}} , \overline{C}_{\tilde{I}}{}^{M \alpha},\acute{\overline{C}}_{\tilde{I}}{}^{M \tilde{\alpha}}  ) = ( \overline{C}_{\tilde{I}}{}^{M \tilde{\alpha}}  , \eta^{\alpha\beta} \eta_{\tilde{I}\tilde{J}} \eta^{MN}   C^{\tilde{J}}{}_{N\beta} , \eta^{\tilde{\alpha}\tilde{\beta}} \eta_{\tilde{I}\tilde{J}} \eta^{MN}   C^{\tilde{J}}{}_{N\tilde{\beta}}  )  \; ,  \label{DoubleBarC} 
\end{align}
leaving $\overline{C}_{IM}{}^{{\tilde{\alpha}}}$ and $\overline{C}_{\tilde{I}}{}^{M \tilde{\alpha}}$ untouched as they have no correspondence with the tensors that arise in $E_{11}$ exceptional field theory.

To show this identification, we start from the component of \eqref{eq:almID6} along $\tilde{\beta}$. We replace $\acute{\overline{C}}_{IM}{}^{\tilde{\alpha}} $ on the left-hand side, and identify the right-hand side as the Clebsch--Gordan series for the tensor product $R(\Lambda_1) \otimes L(\Lambda_2)$
\begin{align}
\label{eq:CC=CCtt} 
 \eta^{\tilde{\alpha}\tilde{\gamma}} \eta_{IJ} \eta_{PQ}   C^{JQ}{}_{\tilde{\gamma}}  C^{IM}{}_{\tb} = \Pi_{\tilde{I}}{}^{M \tilde\alpha}  \eta^{\tilde{I}\tilde{J}} \eta_{PN} \eta _{\tb\tilde{\gamma}} \Pi_{\tilde{J}}{}^{N \tilde\gamma} \end{align}
 with the canonically normalised intertwiners $\Pi_\tI{}^{M\ta}$ corresponding to the tensor product~\eqref{eq:tildeI}
\begin{align}
\label{eq:tptI}
L(\Lambda_3) = R(\Lambda_1) \otimes L(\Lambda_2) \ominus R(\Lambda_1+\Lambda_2) \, , 
\end{align}
and the $K(E_{11})$-invariant tensors $\eta^{\tI\tJ}$ and $\eta_{\ta\tb}$ defined such that the equation holds. Of course we assume again that the bilinear form $\eta_{IJ}$ exists to make this ansatz. These bilinear forms  admit one free coefficient for each irreducible representation in $L(\Lambda_3)$ and $L(\Lambda_2)$, that can be chosen such that \eqref{eq:CC=CCtt} holds, provided the left-hand side vanishes in the irreducible representation $R(\Lambda_1 + \Lambda_2)$ not included in $L(\Lambda_3)$. That this choice of normalisation is possible can be verified explicitly for the first few $E_{11}$ representations in the tensor product by going through $GL(11)$ and $GL(3) \times E_8$ level decompositions, where we find explicitly that the left-hand side vanishes upon projection to the irreducible representation $R(\Lambda_1+\Lambda_2)$, see \eqref{FFGG11D} and \eqref{GG=FFE8}. This equation reproduces  \eqref{eq:almID6} along $\tilde{\beta}$ provided we identify 
\be 
\acute{\overline{C}}_{\tI}{}^{M \ta} =  \Pi_{\tI}{}^{M \ta}\; , \qquad C^{\tilde{I}}{}_{M\tilde{\alpha}} = \eta_{MN} \eta_{\tilde{\alpha}\tilde{\beta}} \eta^{\tilde{I}\tilde{J}} \Pi_{\tilde{J}}{}^{N \tilde{\beta}} \; . 
\ee
With the identification above, \eqref{eq:almID6} ensures that 
\begin{align}
\label{eq:CC=CCtw} 
\overline{C}_{IP}{}^{\tilde\alpha} C^{IQ}{}_{\wb} = \Pi_{\tilde{I}}{}^{Q\tilde\alpha} C^{\tilde{I}}{}_{P\wb} \,.
\end{align} 
Now taking \eqref{e12Equation} along $t^{\tilde{\beta}}$ gives exactly the conjugate of \eqref{eq:almID6}, which implies that 
\be 
{\overline{C}}_{IM}{}^{\alpha} =  \eta^{\alpha\beta} \eta_{IJ} \eta_{MN}   C^{JN}{}_{{\beta}} \; , \quad {\overline{C}}_{\tI}{}^{M \alpha} =  \eta^{\alpha\beta} \eta_{\tI\tJ} \eta^{MN}   C^{\tJ}{}_{{ N \beta}} \; . 
\ee
This concludes the proof that \eqref{DoubleBarC} is indeed the correct identification and we obtain therefore that \eqref{eq:almID5} is indeed Equation \eqref{Father} that we wanted to prove. Moreover, we checked in Appendices \ref{LowLevelFather} and \ref{LowLevelFatherE8} below that at low levels the tensors  \eqref{DoubleBarC}  satisfy  \eqref{eq:ID5}, which permits us to identify the relevant identities following from \eqref{Father} in eleven dimensions and in $E_8$ exceptional field theory. So we conclude that the claimed identities are satisfied in $E_{11}$ exceptional field theory.

\subsection{On the master identity}
\label{app:MID}

In this section, we shall investigate the central group-theoretic master identity~\eqref{NewMaster} that we recall here 
 \begin{align}
  \label{NewMasterXX}  \Omega_{IJ} C^{J M}{}_{\wa} T^{\wa N}{}_Q = \overline{C}_{I Q}{}^{\tilde{\beta}} \Pi_{\tilde{\beta}}{}^{MN} + \overline{\widetilde{ C}}{}_{I Q}{}^{\Lambda} \Pi_{\Lambda}{}^{MN}  + \overline{\widetilde{C}}{}_{I Q}{}^{\tL} \Pi_{\tL}{}^{MN}  \,.
  \end{align}
It is needed for gauge-invariance of the duality equation~\cite{Bossard:2019ksx} and of the Lagrangian~\eqref{eq:Lag}. In this equation $\overline{C}_{J Q}{}^{\tilde{\beta}}$ is the conjugate of the tensor hierarchy algebra structure constants $C^{IM}{}_{\tilde{\alpha}}$, whereas $ \overline{\widetilde{C}}{}_{J Q}{}^{\Lambda}$ and $ \overline{\widetilde{C}}{}_{J Q}{}^{\tL}$ are new tensors defined such that \eqref{NewMasterXX} holds. The tensor $\Pi_{\tilde{\alpha}}{}^{MN} $ is restricted to $R(\Lambda_2)$, while the tensors $\Pi_{\widetilde{\Lambda}}{}^{MN}$ and $\Pi_{\Lambda}{}^{MN}$ include a normalisation constant for each irreducible component contained in $L(\Lambda_4)$ and $L(\Lambda_{10})$, respectively. The same is true for the $K(E_{11})$-invariant metrics $\eta_{\widetilde{\Lambda}\widetilde{\Xi}}$ and $\eta_{\Lambda\Xi}$. These constants have to be fixed for each irreducible component such that~\eqref{NewMasterXX} holds. 

We will prove that this equation necessarily holds for some invariant tensor $ \overline{C}_{I Q}{}^{\tilde{\beta}} $ and the non-trivial proposition to prove is that 
\be  
\overline{C}_{I Q}{}^{\tilde{\beta}}  = \eta_{IJ} \eta_{QP} \eta^{\tilde{\beta}\tilde{\gamma}} C^{JP}{}_{\tilde{\beta}} 
\label{MasterSeed}
\ee
for $\tilde{\beta}$ restricted to the irreducible module $R(\Lambda_2)$. We will find that there is a unique homomorphism from $R_{-1} \otimes R(\Lambda_1) \rightarrow R(\Lambda_{2})$ where $R_{-1}$ is the irreducible component of $\cT_{-1}$ including the duality equation representations in supergravity. If $\cT_{-1}$ was completely reducible, we could always define the theory by restricting the index $I$ to the irreducible module $R_{-1}$, $\eta_{IJ}$ would then exist on $R_{-1}$, and the uniqueness of the homomorphism above would imply \eqref{NewMasterXX}. So the assumption of complete reducibility $\cT_{-1} = R_{-1} \oplus D_{-1}$ as an $\mf{e}_{11}$ module is a sufficient condition for all our identities to be satisfied. However, these identities could still be correct if $\cT_{-1}$ was not the direct sum of an irreducible module $R_{-1}$ and another module $D_{-1}$, so we will not assume it is the case. The general structure of an indecomposable module compatible with the existence of a $K(E_{11})$-invariant form is
\be 
J_{-1} \oleft R_{-1} \oleft \overline{J_{-1}} \subset \cT_{-1}\; , 
\ee
where $R_{-1}$ is irreducible and $J_{-1}$ is a possibly indecomposable module which decomposes as a vector space as the direct sum of a highest weight module $L$ and a lowest weight module $\overline{L}$. The existence of $\eta_{IJ}$ requires that $\overline{L}$ is the conjugate module to $L$, justifying the notation. The uniqueness of the homomorphism does not determine the structure constants $\overline{C}_{I Q}{}^{\tilde{\beta}} $ for $I$ valued in $\overline{J_{-1}} $, but we expect that we might still be able to redefine the corresponding structure constants to satisfy \eqref{MasterSeed}. In this Appendix we will only consider the module $R_{-1}$.

\vskip 3mm

Let us now describe the proof of \eqref{NewMasterXX} for $I$ valued in $R_{-1}$. We first consider the symmetric component in $MN$
 \be \label{NewMasterSym}  \Omega_{IJ} C^{J (M}{}_{\wa} T^{\wa N)}{}_Q =   \overline{ 
 \widetilde{C}}{}_{I Q}{}^{\Lambda} \Pi_{\Lambda}{}^{MN}  \; .   \ee
 This equation can be defined as a Jacobi identity for the tensor hierarchy algebra $\cT(\mf{e}_{11})$, see also~\cite{Bossard:2019ksx}.  The commutator of a generator of degree $2$ and a generator of degree $-1$ takes the form
\begin{align} \label{NewCtilde}
[ P^{\Lambda} , t_I ] = 2 \, \overline{\widetilde C}{}_{IM }{}^{\Lambda}  P^M 
\end{align}
for some $E_{11}$-invariant tensor $\overline{\widetilde{C}}_{IM}{}^{\Lambda} $. The Jacobi identity 
\be 2 [ P^{(M} , \{ P^{N)} , t_I \} ] =[  \{ P^M , P^N\} , t_I ]\ee
gives the identity for the structure constants
\begin{align}
\label{JacobiT2}  \Omega_{IJ} C^{J(M}{}_{\widehat{\alpha}} T^{\widehat{\alpha} N)}{}_Q   =  \overline{\widetilde{C}}{}_{IQ}{}^{\Lambda} \Pi_{\Lambda}{}^{MN} \; ,   
\end{align}
just like in \eqref{NewMasterSym}. Therefore this equation \eqref{NewMasterSym} holds for all indices $I$ in $\cT_{-1}$ provided we define $ \overline{\widetilde C}{}_{IM }{}^{\Lambda}$ from the tensor hierarchy algebra structure constants \eqref{NewCtilde}.

\medskip

Let us now consider the antisymmetric component in $MN$ of eq. \eqref{NewMasterXX}
 \be \label{NewMasterALT}  \Omega_{IJ} C^{J [M}{}_{\wa} T^{\wa |N]}{}_Q =  \overline{C}_{I Q}{}^{\tilde{\beta}} \Pi_{\tilde{\beta}}{}^{MN} + \overline{\widetilde{C}}{}_{I Q}{}^{\tL} \Pi_{\tL}{}^{MN} \; .  \ee
Written in this form, the right-hand side is simply the Clebsch--Gordan series for the tensor product $R(\Lambda_1) \ALT R(\Lambda_1)$ of the indices $MN$. The components $ \overline{\widetilde{C}}_{J Q}{}^{\tL}$ are defined by projection of the left-hand side. Its conjugate is given by
\begin{align}
\label{eq:cconj}
   \widetilde{C}^{IM}{}_{\tL} = \eta^{IJ} \eta^{MN} \eta_{\tL\tXi} \overline{\widetilde{C}}_{J N}{}^{\widetilde{\Xi}}\,.
\end{align}
The only non-trivial equation that remains to check then is that $ \overline{C}_{J Q}{}^{\tilde{\beta}} $ is indeed the conjugate \eqref{MasterSeed} of the tensor hierarchy structure constants $C^{IM}{}_{\tilde{\alpha}}$ on $R(\Lambda_2)$.

To show this for $I$ valued in $R_{-1}$ we prove that there is a unique homomorphism from $R_{-1} \otimes R(\Lambda_1) \rightarrow R(\Lambda_{2})$ determined by the map to the lowest weight vector in $R(\Lambda_{2})$. This lowest weight vector must be annihilated by $\delta_f$ in the $GL(11)$ decomposition, similar to above. Writing the ansatz for the lowest weight vector as\footnote{Note that for the highest weight module $R(\Lambda_2)$, the set of components is generated from the lowest weight component $X_{n_1n_2}$.}
\begin{align}
 X_{n_1n_2}  = \partial_p F_{n_1n_2}{}^p + 2 c_0 \partial_{[n_1} F_{n_2] p}{}^p + \frac{c_1}{2} \partial^{p_1p_2} F_{n_1n_2p_1p_2} + \dots \,,
 \end{align}
the lowest weight condition is  
\begin{align}
 \delta_f X_{n_1n_2}  = \frac{1}{2} ( c_1 - 1) f^{p_1p_2p_3} \partial_{p_1} F_{n_1n_2p_2p_3} +  \frac{1}{9} ( c_0-1) \partial_{[n_1} F_{n_2]p_1p_2p_3}  + \dots  = 0 
\end{align}
which can only vanish if $c_0=c_1=1$, so that 
\begin{align}
 X_{n_1n_2}  = \partial_p F_{n_1n_2}{}^p + 2 \partial_{[n_1} F_{n_2]p}{}^p + \frac{1}{2} \partial^{p_1p_2} F_{n_1n_2p_1p_2} + \dots 
\end{align}
Just like the coefficient $c_1$ is uniquely fixed by the coefficients $c_0$ and $1$, all the coefficients $c_n$ associated to the terms involving the derivative of weight $- \frac{3}{2}-n$ are uniquely determined by the equation $ \delta_f X_{n_1n_2} =0$ because the module $R(\Lambda_2)$ is generated from the lowest weight vector. We conclude that the homomorphism is unique so that 
\begin{align}
 \overline{C}_{IP}{}^{\tilde{\alpha}}  = \eta_{IJ} \eta_{PQ} \eta^{\tilde{\alpha}\tilde{\beta}} C^{JQ}{}_{\tilde{\beta}}\; , 
 \end{align}
on $R(\Lambda_2)$. We conclude that \eqref{NewMasterALT} is indeed satisfied on the whole irreducible component $R(\Lambda_2)$. Since the $L(\Lambda_4)$ components work by construction due to~\eqref{eq:cconj} we thus have shown~\eqref{NewMasterALT} on the whole antisymmetric part in $[MN]$ for $I$ restricted to $R_{-1}$.

\medskip

\medskip

Now that we have proved \eqref{NewMasterXX} for $I$ restricted to $R_{-1}\subset \cT_{-1}$, we would like to describe in more detail this identity. In particular, we shall find that all structure constants $\overline{\widetilde{C}}{}_{I P}{}^{\hL}$ are non-zero, so we must consider the complete module $L(\Lambda_{10})\oplus L(\Lambda_4)$ and not a submodule. We will moreover discuss the possibility that $\widetilde{C}^{I M}{}_{\Lambda}$ is the same tensor as the tensor hierarchy algebra structure constants $C^{I M}{}_{\Lambda}$ appearing in the commutator $[P^M,\bar{t}_\Lambda]= C^{IM}{}_\Lambda t_I$ for $\Lambda$ valued in $\cT_{-2}$. This assumption is not necessary for the consistency of the theory, but would make the relation to the tensor hierarchy algebra more direct and would therefore be aesthetically pleasing.

We shall start by proving the complete module $L(\Lambda_{10})\oplus L(\Lambda_4)$ is required. For this we can use the consistency of~\eqref{NewMasterXX} with  \eqref{e12Equation}, which gives
\bea
\label{eq:CC0}
&& \overline{C}_{\tilde{I}}{}^{M \alpha} C^{\tilde{I}}{}_{P \widehat{\beta}}  T^{\widehat{\beta} N}{}_Q \bar{\partial}^P \otimes \bar{\partial}^Q    \nn\\
 &=&   \Bigl(  \overline{C}_{IP}{}^{\alpha} C^{IM}{}_{\widehat{\beta}}  T^{\widehat{\beta} N}{}_Q  +  T^{\alpha M}{}_R T_\beta{}^R{}_P  T^{{\beta} N}{}_Q +  T_\beta{}^{M}{}_R T^{\alpha R}{}_P T^{{\beta} N}{}_Q \ -\delta_P^M  T^{{\alpha} N}{}_Q \Bigr)  \bar{\partial}^P \otimes \bar{\partial}^Q \CR
 &=& -   \Omega^{IJ} \overline{C}_{IP}{}^{\alpha} \bigl( \overline{C}_{J Q}{}^{\tilde{\beta}} \Pi_{\tilde{\beta}}{}^{MN} + \overline{\widetilde{C}}{}_{J Q}{}^{\Lambda} \Pi_{\Lambda}{}^{MN}  + \overline{\widetilde{C}}{}_{J Q}{}^{\tL} \Pi_{\tL}{}^{MN} \bigr)  \bar{\partial}^P \otimes \bar{\partial}^Q \CR
 && \qquad  + \bigl( - f^{\alpha\beta}{}_\gamma T^{\gamma M}{}_P T_\beta{}^N{}_Q + 2 \delta^{N}_{[P} T^{\alpha M}{}_{Q]} - 2 \delta_P^{[M} T^{\alpha N]}{}_Q  \bigr) \bar{\partial}^P \otimes \bar{\partial}^Q\; \CR
 &=&0 \,,
\eea 
because the first line vanishes according to \eqref{eq:ID9}.  One can check the consistency of~\eqref{eq:CC0}
\bea
 &&    \Omega^{IJ} \overline{C}_{IP}{}^{\alpha} \bigl( \overline{C}_{J Q}{}^{\tilde{\beta}} \Pi_{\tilde{\beta}}{}^{MN} + \overline{\widetilde{C}}{}_{J Q}{}^{\Lambda} \Pi_{\Lambda}{}^{MN}  + \overline{\widetilde{C}}{}_{J Q}{}^{\tL} \Pi_{\tL}{}^{MN} \bigr)  \bar{\partial}^P \otimes \bar{\partial}^Q \CR
 &=&  \bigl(  -f^{\alpha\beta}{}_\gamma T^{\gamma M}{}_P T_\beta{}^N{}_Q +2 \delta^{N}_{[P} T^{\alpha M}{}_{Q]} - 2 \delta_P^{[M} T^{\alpha N]}{}_Q  \bigr) \bar{\partial}^P \otimes \bar{\partial}^Q\; , \label{ConsistentlyNewMaster}
 \eea
 by projecting $MN$ on both sides onto possible irreducible representations. One finds that when projecting $MN$ to $R(2\Lambda_1)$, both sides vanish consistently. The symmetric component of the second line in $MN$ is antisymmetric in $PQ$,  which implies that the symmetric component in both $MN$ and $PQ$ of the first line vanishes, i.e.
 \be  
 \Omega_{IJ} C{}^{IP}{}_{\alpha}  \widetilde{{C}}{}^{J Q}{}_{\Lambda}   \partial_P \partial_Q =  0 \; , \label{asforTHA} 
\ee
after conjugation, with by definition 
\be  
\widetilde{{C}}{}^{JQ}{}_{\Lambda}  = \eta^{JI} \eta^{QM} \eta_{\Lambda\Xi} \overline{\widetilde{C}}{}_{IM}{}^\Xi\; . 
\ee
Note that \eqref{asforTHA} would be satisfied if $\widetilde{C}^{IM}{}_\Lambda$ was identified with  the tensor hierarchy algebra structure constant $C^{IM}{}_\Lambda$ 
\be 
[ P^M , \bar t_{\Lambda^\prime} ] = C^{IM}{}_{\Lambda^\prime} t_I \; , \label{D0structureCons}  
\ee
because of the Jacobi identities. Here we defined  $\Lambda^\prime$ the index of $D_0\subset \cT_{0}$, that we identify with $\Lambda$ in the common submodule of $D_0$ and $L(\Lambda_{10})$. We shall give evidence that $L(\Lambda_{10}) \subset D_0$ such that we could identify $\Lambda^\prime = \Lambda$ and that the structure constants  are equal 
\be 
\widetilde{{C}}{}^{IM}{}_{\Lambda}  = C^{IM}{}_{\Lambda}  
\label{eq:2prove} 
\ee
As explained before, this identification is not essential for the consistency of the theory, but it would make the algebraic structure more pleasing and it might be useful when considering additional gauge symmetries of the theory.\footnote{One could wonder if the additional gauge parameter in $\cT_{-3}\ominus \overline{R(\Lambda_1)}$ could also define gauge symmetries of the theory, and in this case it would be useful to have the identification such that $(C^{IM}{}_{\wa} T^{\wa N}{}_{\tilde{P}} + C^{IM}{}_{\Lambda} T^{\Lambda N}{}_{\tilde{P}}) \partial_M \partial_N \xi^{\tilde{P}} = 0 $.} For ease of notations we drop the tilde and write $C^{IM}{}_{\Lambda}$ and $C^{IM}{}_{\tL}$ for  $\widetilde{C}^{IM}{}_{\Lambda}$ and $\widetilde{C}^{IM}{}_{\tL}$ in the body of the paper.

Projecting $MN$ to $R(\Lambda_2)$ one finds that the first line in \eqref{ConsistentlyNewMaster} equals the second  using the identity
\be
 \Pi_{\tilde{\alpha}}{}^{PQ} \Pi^{\tilde{\alpha}}{}_{MN} \partial_P \otimes \partial_Q = - 2 \partial_{[M} \otimes \partial_{N]} \; . 
 \label{A66}  
 \ee
However, if one takes $MN$ in another irreducible component $R(\lambda)$ of $R(\Lambda_1)\ALT R(\Lambda_1)$, one can check that it satisfies 
\be 
T_{\beta}{}^{M}{}_P T^{\beta N}{}_Q X^{[PQ]}_\lambda =  - ( \tfrac32 + n_\lambda ) X^{MN}_\lambda 
\ee
which corresponds to the `cross-term' in the action of the quadratic Casimir on the tensor product such that\footnote{With this we mean that $\kappa_{\alpha\beta}t^\alpha v \otimes t^\beta w = \frac12 \kappa_{\alpha\beta} t^\alpha  t^\beta (v\otimes w) - \frac12 [\kappa_{\alpha\beta} t^\alpha t^\beta v]\otimes w -\frac12 v\otimes [\kappa_{\alpha\beta} t^\alpha t^\beta w]$.} 
\be 
n_\lambda = - \tfrac12 (\lambda,\lambda) - (\varrho,\lambda) +  (\Lambda_1,\Lambda_1) + 2 (\Lambda_1,\rho) - \tfrac32  =  - \tfrac12 (\lambda,\lambda) - (\varrho,\lambda)-30 
\ee 
and  $n_\lambda$ is a strictly positive integer for $X^{MN}_\lambda $ in an irreducible submodule of 
\be R(\lambda) \subset L(\Lambda_4) = R(\Lambda_1)\ALT R(\Lambda_1) \ominus R(\Lambda_2) \; . \ee
For example $n_{\Lambda_4} = 36$. 
To prove the positivity of $n_\lambda$, one uses that $\lambda = n_i \Lambda_i$ is a dominant weight with $n_i \in \mathds{N}$ and also $\lambda = \Lambda_2 - \alpha$ with $\alpha $ positive in the root lattice. Therefore one gets
\be 
n_\lambda = 
n_{\Lambda_2-\alpha} = ( \alpha , \tfrac12( \lambda+\Lambda_2) + \varrho)>0 \; . 
\ee
 If one takes $[MN]$ in $R(\lambda)$ and $PQ$ symmetric one obtains then
\be  
 \Omega^{IJ} \overline{C}_{IP}{}^{\alpha} \overline{\widetilde{C}}{}_{J Q}{}^{\tL_\lambda} \Pi_{\tL_\lambda}{}^{MN}   \bar{\partial}^P  \bar{\partial}^Q  = - n_\lambda T^{\alpha [ M}{}_{P} \delta_Q^{N]_\lambda} \bar{\partial}^P  \bar{\partial}^Q \ne 0 
 \ee 
where $[MN]_\lambda$ denotes the projection to $R(\lambda)$. It follows that for all $R(\lambda) \in L(\Lambda_4)$ the left-hand side does not vanish and $\widetilde{C}{}^{J Q}{}_{\tL_\lambda} $ is not part of the $\cT(\mf{e}_{11})$ structure constants $C^{J Q}{}_{\Lambda^\prime}$ \eqref{D0structureCons}. This is why we have to introduce the additional constrained field $\zeta_M{}^{\tL}$ valued in $L(\Lambda_4)$, which was not found to be necessary for gauge-invariance in \cite{Bossard:2019ksx} because it does not contribute at low-levels that were analysed.

One can use the same argument to analyse the component symmetric in  $MN$ and antisymmetric in $PQ$. One has in this case 
\be 
T_{\beta}{}^{M}{}_P T^{\beta N}{}_Q X^{(PQ)}_\lambda =  ( \tfrac12 - n_\lambda ) X^{MN}_\lambda 
\ee
where
\be 
n_\lambda = - \tfrac12 (\lambda,\lambda) - (\varrho,\lambda) +  (\Lambda_1,\Lambda_1) + 2 (\Lambda_1,\rho) + \tfrac12  =  - \tfrac12 (\lambda,\lambda) - (\varrho,\lambda)-28 
\ee 
and  $n_\lambda$ is a strictly positive integer for $X^{MN}_\lambda $ in an irreducible submodule of $R(\lambda) \subset L(\Lambda_{10}) =  R(\Lambda_1)\SYM R(\Lambda_1) \ominus R(2\Lambda_1)$. One obtains then 
\be 
 \Omega^{IJ} \overline{C}_{IP}{}^{\alpha}  \overline{\widetilde{C}}{}_{J Q}{}^{\Lambda_\lambda} \Pi_{\Lambda_\lambda}{}^{MN}   \bar{\partial}^P \otimes \bar{\partial}^Q  = - n_\lambda T^{\alpha ( M}{}_{[P} \delta_{Q]}^{N)_\lambda} \bar{\partial}^P \otimes  \bar{\partial}^Q \ne 0 \; , 
 \ee
from which we find that $ \overline{\widetilde{C}}_{J Q}{}^{\Lambda_\lambda}$ and $\Pi_{\Lambda_\lambda}{}^{MN} $ do not vanish for any $R(\lambda) \subset L(\Lambda_{10})$. This justifies the definition of the constrained fields $\zeta_M{}^\Lambda$ in $ L(\Lambda_{10})$. 

\medskip

To summarise, we have obtained that $\Lambda$ and $\tilde{\Lambda}$ must range over the complete modules $L(\Lambda_{10})$ and $L(\Lambda_4)$ and that $\zeta_M{}^{\tilde{\Lambda}}$ is not a field valued in $\cT_0$ whereas it seems plausible that $\zeta_M{}^\Lambda$ would be defined in $D_0\subset \cT_0$. Note that we have proved that $R(\Lambda_{10})\oplus R(2\Lambda_3)\subset D_0$ and we will show below that one can identify the structure coefficients as in \eqref{eq:2prove} for $\Lambda$ valued in $R(\Lambda_{10})\oplus R(2\Lambda_3)$. 

\medskip

Let us first consider the irreducible components $R(\Lambda_{10})$. The tensors $\widetilde{C}^{IP}{}_{\Lambda} $ and $C^{IP}{}_{\Lambda}$ define $E_{11}$-homomorphisms from $R_{-1} \otimes R(\Lambda_1) \rightarrow R(\Lambda_{10})$ and are therefore uniquely determined by the map to the lowest weight vector in $R(\Lambda_{10})$. We recall from~\eqref{eq:Tminus1} that $R_{-1}$ denotes the irreducible submodule of $\cT_{-1}$. Written in $GL(11)$ level decomposition, this lowest weight vector must be annihilated by the lowering operators $\delta_f$ with negative $GL(11)$ level. By $GL(11)$ grading, the general ansatz for the lowest weight vector is
\begin{align}
 Y^m = \partial_n F^{n,m} + c_0 \partial_n F_p{}^{mnp}  + c_1 \partial^{n_1n_2} F_{n_1n_2}{}^m + c_1^\prime \partial^{mn} F_{np}{}^p + \dots 
\end{align}
and it must satisfy the condition under $GL(11)$ level $-1$
\begin{align}
 \delta_f Y^m &= \Bigl( \frac{1}{4} - \frac{c_0}{8} \Bigr) f^{p_1p_2m}  \partial_n F_{p_1p_2}{}^n + \Bigl( \frac{1}{4} + \frac{c_0}{8} + c_1 \Bigr) f^{p_1p_2n}  \partial_n F_{p_1p_2}{}^m + ( c_0-c_1^\prime) f^{mnp} \partial_n F_{pq}{}^q + \dots  \nn\\
& \stackrel{!}{=} 0 \,.
 \end{align}

This implies by linear independence that
\begin{align}
 c_0 = 2\; , \quad c_1 = - \frac12 \; , \quad c_1^\prime = 2 \,,
\end{align}
leading to 
\be Y^m = \partial_n F^{n,m} + 2 \partial_n F_p{}^{mnp} - \frac12 \partial^{n_1n_2} F_{n_1n_2}{}^m + 2 \partial^{mn} F_{np}{}^p + \dots \ee
Just like the coefficients $c_1$ and $c_1^\prime$ are uniquely fixed by the coefficients $c_0$ and $1$, all the coefficients $c_n$ associated to terms in the ansatz involving the derivative of weight $- \frac{3}{2}-n$ are uniquely determined by the equation $ \delta_f Y^m =0$ because the module $R(\Lambda_1)$ is generated from the lowest weight vector. We conclude that the homomorphism is unique (up to an overall factor) so that we have indeed $\widetilde{C}^{I M}{}_{\Lambda}  = {C}^{I M}{}_{\Lambda} $ for this irreducible component.

 The same argument allows us to prove that there is a unique $E_{11}$-homomorphism from $R_{-1} \otimes R(\Lambda_1) \rightarrow R(2\Lambda_{3})$, as we show in \eqref{Ext2L3}. This extends the proof of    \eqref{eq:2prove}  to the reducible submodule $R(\Lambda_{10})\oplus R(2\Lambda_3) \subset L(\Lambda_{10})$. 

 We moreover argue in Appendix~\ref{Casimir} using a Casimir homomorphism that we have $L(\Lambda_{10})\subset \mathcal{T}_0(\mf{e}_{11})$, so that we can consistently identify the index $\Lambda^\prime$ in \eqref{D0structureCons} with $\Lambda$ of $L(\Lambda_{10})$ on both sides of~\eqref{eq:2prove}. Moreover we have seen from \eqref{asforTHA} that $\widetilde{C}^{IM}{}_{\Lambda}  = {C}^{IM}{}_{\Lambda} $ is compatible with the Jacobi identity. However, unlike for the $R(\Lambda_{10})\oplus R(2\Lambda_3)$ component, we do not have a full proof, and this identification is not necessary for the consistency of the theory.

\medskip

\subsection{\texorpdfstring{Proof of $\Omega_{IJ} C^{IM}{}_\wa C^{JN}{}_\wb$ identities}{Proof of Omega CC identities}}
\label{app:Moth}

In this section, we prove the identities~\eqref{eq:ID3},~\eqref{eq:ID7},~\eqref{eq:ID8},~\eqref{eq:ID2} and~\eqref{eq:ID4}, see the summary in Table~\ref{tab:ids}.  A first observation for all these identities is that by \eqref{eq:sympart} it is sufficient to consider the parts of these identities where $M$ and $N$ are antisymmetrised, which means we only have to consider the antisymmetric part of the section constraint~\eqref{eq:SC}.

The first three identities are simple consequences of tensor products of $E_{11}$ highest weight representations when we look at their structure in the $M$ and $N$ indices. As the identities are to be valid only on section, we have to make sure that the tensor products do not contain the representation $R(\Lambda_2)$ that is the only non-trivial one on the antisymmetric section according to~\eqref{eq:L1tens}. For this we look at the structure of the lower indices.

For the first identity
\begin{align}
\label{eq:ID3app}
\Omega_{IJ} C^{IM}{}_\ta C^{JN}{}_\tb \partial_M\otimes \partial_N =0
\end{align}
we therefore need to consider the tensor product $L(\Lambda_2)\otimes L(\Lambda_2)$. A quick check of the dominant weights that can occur shows that it is sufficient to consider the part $R(\Lambda_2)\otimes R(\Lambda_2)$ as all other products can never contain the non-trivial section representations. The tensor product is\footnote{Such tensor products can be computed using the character formulas for highest modules. Any product of two integrable highest weight representations is completely reducible as an infinite sum of integrable highest weight representations and the terms can be (partly) ordered by the height of their highest weights~\cite[\S10.7]{Kac}.}
\begin{align} 
R(\Lambda_2) \otimes R(\Lambda_2) = R(2\Lambda_2) \oplus R(\Lambda_4) \oplus  R(2\Lambda_1 + \Lambda_{10})  \oplus  R(\Lambda_1 + \Lambda_3) \oplus R(\Lambda_2 + \Lambda_{10}) \oplus \dots 
\end{align}
showing that it includes neither $R(2\Lambda_1)$ nor $R(\Lambda_2)$. From this we can conclude based on $E_{11}$ representation theory that~\eqref{eq:ID3app} holds. 

A similar calculation of the tensor products $R(\Lambda_2) \otimes R(\Lambda_{10}) $ and  $R(\Lambda_{10}) \otimes R(\Lambda_{10}) $ shows by the same argument that~\eqref{eq:ID7} and~\eqref{eq:ID8} hold. Again the arguments can be extended from $R(\Lambda_2)$ and $R(\Lambda_{10})$ to $L(\Lambda_4)$ and $L(\Lambda_{10})$, giving also~\eqref{eq:ID7} and~\eqref{eq:ID8} in full generality.

We now turn to proving~\eqref{eq:ID2}. A first observation is that the combination
\begin{align}
\label{eq:Omata}
\Omega_{IJ} C^{IM}{}_{\tilde{\alpha}} C^{JN}{}_{\beta} \partial_M \otimes \partial_N
\end{align}
transforms covariantly under $E_{11}$ even though it contains the non-covariant object $C^{JN}{}_\beta$. Its non-covariance~\eqref{eq:DeltaC} is indeed contracted in such a way as to yield~\eqref{eq:ID3app} and thus disappears. 

Since the symmetric part $(MN)$ vanishes according to \eqref{eq:sympart}, we can view~\eqref{eq:Omata} as an $\mf{e}_{11}$-homomorphism  $\mathfrak{e}_{11} \otimes R(\Lambda_2)\to R(\Lambda_2)$, where the target $R(\Lambda_2)$ is the only non-trivial component of $[MN]$ on section. The representation matrices $T_\beta{}^{\tb}{}_\ta$ are similarly the components of a homomorphism from $\mathfrak{e}_{11} \otimes R(\Lambda_2)$ to $R(\Lambda_2)$ and we shall prove that this homomorphism is unique (up to a multiplicative constant) which will then imply~\eqref{eq:ID2} after fixing the constant on one component.
 
 Any homomorphism $\mathfrak{e}_{11} \otimes R(\Lambda_2)\to R(\Lambda_2)$ defines a highest weight representative in the target for which we make the following ansatz in the $GL(11)$ level decomposition
 \begin{align}
  X^\prime_{n_1\dots n_9} &= c_0 9 h_{[n_1}{}^p  X_{n_2\dots n_9]p} + c_0^\prime h_p{}^p X_{n_1\dots n_9} + c_1 \bar A^{p_1p_2p_3} X_{n_1\dots n_9p_1,p_2p_3}\nn\\
  &\hspace{10mm} + c_1^\prime \bar A^{p_1p_2p_3} X_{n_1\dots n_9p_1p_2,p_3} + \dots \nn\\
 &= \sum_{k=0}^\infty c_k ( \phi^\ord{-k} , X^\ord{3+k} ) \,.
 \end{align}
 To be highest weight, the $\delta_e$ variation \eqref{adjoint} must cancel $X^\prime_{n_1\dots n_9}$. Checking the terms in $\bar A^{p_1p_2p_3} X_{n_1\dots n_9}$ one finds 
\begin{align}
 c_0^\prime = 0 \; , \qquad c_1 = - \frac{c_0}{2}\; , \qquad  c_1^\prime = c_0 \; . 
\end{align}
To show that it is the unique solution, one uses the general variation 
\begin{align}
  \delta_e  X^\prime_{n_1\dots n_9} = \sum_{k=0}^\infty \Bigl( c_k ( \delta_e \phi^\ord{-k} , X^\ord{3+k} )  + c_{k+1}  (  \phi^\ord{-k-1} , \delta_e X^\ord{4+k} )  \Bigr) = 0  
\end{align}
which implies by the $GL(11)$ grading 
\begin{align}  
c_k ( \delta_e \phi^\ord{-k} , X^\ord{3+k} )  + c_{k+1}  (  \phi^\ord{-k-1} , \delta_e X^\ord{4+k} ) = 0 
\end{align}
for all $k$. But the module $R(\Lambda_2)$ is generated from the highest weight, so 
\begin{align}
 c_{k+1}  (  \phi^\ord{-k-1} , \delta_e X^\ord{4+k} ) = 0 
\end{align}
for $k\ge 0$  if and only if $c_{k+1}(  \phi^\ord{-k-1} ,  X^\ord{4+k} ) =0$. We conclude that there is no solution with $c_0 = 0$, and since the solution with $c_0\ne 0$ is unique up to a multiplicative constant, the homomorphism is unique up to a multiplicative constant. Fixing this constant by using the explicit expressions for the lowest components of tensors in $GL(11)$ decomposition proves \eqref{eq:ID2}.

\medskip

We now prove~\eqref{eq:ID4}. Since  \eqref{eq:ID2} is true, we deduce that 
\begin{align}
\label{eq:Omaa}
\Bigl(  \Pi_\ta{}^{MN} K_{(\alpha}{}^\ta{}_{\beta)} + \frac{1}{2} \Omega_{IJ} C^{IM}{}_{(\alpha} C^{JN}{}_{\beta)} \Bigr) \partial_M \otimes \partial_N 
\end{align}
is an invariant $E_{11}$ tensor despite appearances. The reason is that its non-covariant $E_{11}$ variation is proportional to~\eqref{eq:ID2} which was just shown to vanish. Therefore,~\eqref{eq:Omaa} defines a homomorphism from $(\mathfrak{e}_{11} \otimes \mathfrak{e}_{11} )_{\rm sym}$ to $R(\Lambda_2)$. We are now going to prove that there is no non-trivial such homomorphism. 

The existence of any such homomorphism would imply that one can define a highest weight vector of $R(\Lambda_2)$ for which we make the ansatz
\begin{align}
X^\prime_{n_1\dots n_9} &= c_0 A_{[n_1n_2n_3} A_{n_4\dots n_9]} + c_1 h_{[n_1}{}^p h_{n_2\dots n_9],p}  + c_2 \bar A^{p_1p_2p_3} A_{n_1\dots n_9,p_1p_2p_3}+ \dots \CR
 &= \sum_{k=0}^\infty c_k ( \phi^\ord{1-k} , \phi^\ord{2+k} ) \label{XfromPhiPhi}
\end{align}
in $GL(11)$ level decomposition. The $E_{11}$ variation of the ansatz is 
\begin{align}
\delta X^\prime = \sum_{k=0}^\infty \Bigl( c_k ( \delta_e \phi^\ord{1-k} , \phi^\ord{2+k} )+c_{k+1} ( \phi^\ord{-k} ,\delta_e  \phi^\ord{3+k} ) \Bigr) 
\end{align}
which implies by $GL(11)$ grading that 
\begin{align}
c_k ( \delta_e \phi^\ord{1-k} , \phi^\ord{2+k} )+c_{k+1} ( \phi^\ord{-k} ,\delta_e  \phi^\ord{3+k} )= 0
\end{align}
for all $k\ge 0$ and $c_0( \phi^\ord{1}, \delta_e \phi^\ord{2})= 0$. Once again, $\mathfrak{e}_{11}$ is  generated by the local subalgebra, and therefore 
\begin{align}
c_{k+1} ( \phi^\ord{-k} ,\delta_e  \phi^\ord{3+k} )= 0 \quad \Rightarrow  \quad c_{k+1} ( \phi^\ord{-k} , \phi^\ord{3+k} ) = 0 
\end{align}
for all $k\ge 0$, so $c_0$ must be different from zero, and then all the $c_{k}$ are uniquely determined by $c_0$. In order to fix $c_0$ we look at the terms in $e h_1{}^1 A_6$ obtained by varying the ansatz
\begin{align}
\delta X^\prime = - ( 3 c_0 + 56 c_1) e_{p[n_1n_2} h_{n_3}{}^p A_{n_4\dots n_9]} + 28 c_1 h_{[n_1}{}^p ( \tfrac23 e_{n_2n_3n_4} A_{n_5 \dots n_9]p} + \tfrac13 e_{|p|n_2n_3}  A_{n_4 \dots n_9]}) + \dots  
\end{align}
This expression can only vanish for zero coefficients which implies $c_0=0$. This concludes the proof that there is no non-trivial homomorphism from $(\mathfrak{e}_{11} \otimes \mathfrak{e}_{11} )_{\rm sym}$ to $R(\Lambda_2)$, and therefore that  \eqref{eq:ID4} is true.

\subsection{\texorpdfstring{Casimir and homomorphic image of $L(\Lambda_{10})$ in $\cT_0$}{Casimir and homomorphic image of L10 in T0}}
\label{Casimir}

In this appendix, we study the $\cT_0$-homomorphism $\omega_1 : \cT_2 \rightarrow \cT_0$ in order to provide evidence that $L(\Lambda_{10}) \subset \cT_0$. This question arises in view of the constrained fields $\zeta_M{}^\Lambda$ appearing in the field strength~\eqref{eq:FStemp}. 
We define $\zeta_M{}^\Lambda$ as being valued in $L(\Lambda_{10})\subset \cT_2$ in this paper, but as we explained in Appendix \ref{app:MID}, it would be more satisfactory, but not essential, to also be able to identify them in $\cT_0$.
One can check at low levels that the components $R(\Lambda_{10}) \oplus R(2\Lambda_3)  \subset L(\Lambda_{10})$ and the decomposable part $D_0$ of $\cT_0$ are indeed the same. This is done explicitly in $GL(3)\times E_8$ decomposition in Appendix~\ref{app:e8tha} below, see  \eqref{TpE11Decompose}. 

The $\cT_0$-homomorphism $\omega_1 : \cT_2 \rightarrow \cT_0$ is defined  from the quadratic Casimir $\omega$ for $\cT(\mathfrak{e}_{11})$~\cite{Bossard:2017wxl}.  To show $L(\Lambda_{10})\subset \cT_0$ we would require this homomorphism to be injective as a $\cT_0$-homomorphism on $L(\Lambda_{10})\subset \cT_2$.  
We check explicitly that $R(\Lambda_{10}) \subset \cT_2$ is not in the kernel of $\omega_1$. It is plausible that the $\cT_0$ module generated from $R(\Lambda_{10}) \oplus R(2\Lambda_3) $ is the whole $L(\Lambda_{10})$, in which case we would indeed have that $\omega_1$ is injective on $L(\Lambda_{10})$, establishing the isomorphism. 

The Casimir $\omega$ defines a $\cT(\mf{e}_{11})$-homomorphism 
that  
acts on the category $\mathcal{O}$ of $\cT(\mathfrak{e}_{11})$-modules. 
We will not use such modules, but they could be defined as highest weight modules for example. 
The quadratic Casimir $\omega$ decomposes in the sum of $\cT_0$-invariant quadratic operators $\omega_p$ according to~\eqref{eq:Tdec}.  $\omega_p$ can also be defined on some of the modules of the category $\mathcal{O}$ of $\cT_0$-modules that are not inside modules of the category $\mathcal{O}$ of  $\cT(\mathfrak{e}_{11})$-modules. In particular, one can act with $\omega_q$  on $\cT_p$ for $p> q$. 

Recall the standard quadratic Casimir of $\mathfrak{e}_{11}$ in the $GL(11)$ decomposition 
\begin{align}
 c = \tfrac12 K^p{}_q K^q{}_p   - \tfrac1{18} K^p{}_p K^q{}_q +\tfrac{13}{3} K^p{}_p + \tfrac16 F_{p_1p_2p_3} E^{p_1p_2p_3} 
+ \dots   
\end{align}
The second component $\omega_2$ can be defined in the same way  by pairing $\cT_0$ with its conjugate:
\begin{align}
\omega_{2} &= \tfrac12 K^p{}_q \bar K^q{}_p +  \tfrac12 \bar K^p{}_q  K^q{}_p  - \tfrac1{18} K^p{}_p \bar K^q{}_q  - \tfrac1{18} \bar K^p{}_p K^q{}_q + \tfrac{26}{3} \bar K^p{}_p \nn\\
&\quad + \tfrac16 F_{p_1p_2p_3} \bar E^{p_1p_2p_3} 
+  \tfrac16 \bar F_{p_1p_2p_3} E^{p_1p_2p_3} + \dots   
\end{align}
One defines similarly $\omega_1$ as
\begin{multline}  
\omega_1 = - \tfrac{35}{3} \bar K^n{}_n + \tfrac1{24} t^{n_1n_2n_3n_4} \bar t_{n_1n_2n_3n_4} - \tfrac12 t^{n_1n_2}{}_m \bar t_{n_1n_2}{}^m + \tfrac19 t^{np}{}_p \bar t_{nq}{}^q\\  - \tfrac12 t_{m,n} \bar t^{m,n} - \tfrac16 t^m{}_{n_1n_2n_3} \bar t_m{}^{n_1n_2n_3} + \tfrac{1}{16} t^p{}_{n_1n_2p} \bar t_q{}^{n_1n_2q} + \dots
 \end{multline} 
which is a normal-ordered form of $\frac12 \Omega^{IJ} t_I t_J$ where all the positive $GL(11)$ weight generators are on the right.\footnote{The normal ordering is responsible for the term $\bar{K}^n{}_n$.}  We have written them as $\bar t^I = \Omega^{IJ} t_J$ to simplify the expressions. They are defined such that the field strength as an element of $\cT_{-1}$ reads 
\begin{multline} 
F = \dots + \tfrac{1}{2\cdot 9!} \varepsilon^{n_1\dots n_9p_1p_2} F_{n_1\dots n_9;m} \bar t_{p_1p_2}{}^m + \tfrac1{4!\cdot 7!} \varepsilon^{n_1\dots n_7p_1p_2p_3p_4} F_{n_1\dots n_7} \bar t_{p_1p_2p_3p_4} \\
+ \tfrac{1}{4!} F_{n_1n_2n_3n_4} t^{n_1n_2n_3n_4} + \tfrac12 F_{n_1n_2}{}^m t_{n_1n_2}{}^m +\tfrac16 F_m{}^{n_1n_2n_3} t^m{}_{n_1n_2n_3} + F^{m,n} t_{m,n} + \dots 
\end{multline}
One can prove that $\omega_1$ and $\omega_2$  commute with $\mathfrak{e}_{11}$ by showing that they commute with $E^{n_1n_2n_3}$ and $F_{n_1n_2n_3}$ and we give the relevant commutators at the end of this section.  All the terms in the ellipses are determined by the invariant symplectic form on $\cT(\mathfrak{e}_{11})$ and are automatically invariant because their invariance does not require to modify the ordering of the generators. 

For  $p>2$, $\cT_{-p}$ admits a lowest weight and one can define the corresponding Casimir by the pairing $\omega_p = P^\ord{p-2} \cdot  \bar P^\ord{-p}$. Its $\cT_0$ invariance does not require reordering, so the linear terms in $\bar K^n{}_n$ only appear in $\omega_1$ and $\omega_2$.  Note that for the total Casimir $\omega = \sum_{p=1}^\infty \omega_p$, these linear terms sum up to $-3 \bar K^n{}_n$.

One checks that  $\omega$ is indeed a Casimir of $\cT(\mathfrak{e}_{11})$ by proving it commutes with $\cT_1$ and $\cT_{-1}$. We use $GL(11)$ level decomposition to show
\begin{align}
 [ P^m , \omega ] = - 3 t^{mn}{}_p + \tfrac16 [ F_{n_1n_2n_3} , t^{mn_1n_2n_3}] + \tfrac12 [ K^p{}_q , t^{mq}{}_q] - \tfrac1{18} [ K^p{}_p , t^{mq}{}_q] - \tfrac12 [ \bar K^m{}_n , P^n ] = 0 \,,
 \end{align}
where we have only written the commutators that appear in putting all the generators with the same ordering. Because $[ \bar t_{p_1p_2p_3p_4} , \bar K^m{}_n ] = 0 $ we do not need to check the ordering ambiguity to obtain that $[ \bar t_{p_1p_2p_3p_4},\omega ] = 0 $. Using commutation with $
\cT_0$ we generate the whole local algebra \cite[\S B.3.1]{Bossard:2017wxl}, $\cT(\mathfrak{e}_{11})$, which concludes the proof that $\omega$ is indeed a Casimir of  $\mathcal{T}(\mathfrak{e}_{11})$.

Because $\omega$ is invariant under $\mathcal{T}(\mathfrak{e}_{11})$, it follows that $\omega_p$ are invariant under $\cT_0$ and not only under $\mathfrak{e}_{11}$. This is important because it means that $\omega_1 \cT_2 \subset \cT_0$ is an ideal of $\cT_0$, and so the quotient $\cT_0 / \omega_1 \cT_2$ is an algebra. Since $\omega_1 P_m = 356 F_m \in \cT_0$ in $GL(11)$ level decomposition (see~\eqref{eq:om1Pm}), $R(\Lambda_{10})$ is not in the kernel of $\omega_1$. In particular, $L(\Lambda_{10})$ is a subalgebra $\cT_0$ that admits a highest weight $\Lambda_{10}$. It follows that its action on $\cT_p$ for $p\ge 1$ and it own commutation relations are consistent with highest weight representation theory of $\mathfrak{e}_{11}$. So in particular it is a solvable algebra and its representation matrices are $\mathfrak{e}_{11}$ highest weight module intertwiners. 

\medskip

{\allowdisplaybreaks
To prove the results of this section, we had to compute some commutation relations of $\cT(\mathfrak{e}_{11})$ that we now display. The representation $\cT_{-1}$ of $\mathfrak{e}_{11}$ is defined through the commutators 
\begin{subequations}
\begin{align}
 [ E^{n_1n_2n_3} , \bar t_{p_1p_2p_3p_4}] &= - 36 \delta_{[p_1p_2}^{[n_1n_2} \bar t_{p_3p_4]}{}^{n_3]} - 8 \delta_{[p_1p_2p_3}^{n_1n_2n_3} \bar t_{p_4]q}{}^q \,,\\
 [ E^{n_1n_2n_3} ,  t^{p_1p_2p_3p_4} ] &= - \tfrac{1}{24} \varepsilon^{n_1n_2n_3p_1\dots p_4q_1\dots q_4} \bar t_{q_1q_2q_3q_4} \,,\\
 [ E^{n_1n_2n_3} ,   t^{p_1p_2}{}_m ] &= -3 \delta_m^{[n_1} t^{n_2n_3]p_1p_2} \,,\\
[ E^{n_1n_2n_3} ,  t^m{}_{p_1p_2p_3} ] &=18 \delta_{[p_1p_2}^{[n_1n_2} t^{n_3]m}{}_{p_3]} + 2 \delta^{n_1n_2n_3}_{p_1p_2p_3} t^{mq}{}_q \,,\\
[ E^{n_1n_2n_3} ,   t_{m,p} ] &= 6 \delta^{[n_1}_{(m} t^{p_2p_3]}{}_{n)} \,,\\
[ F_{n_1n_2n_3} , \bar t_{p_1p_2}{}^m ] &= - 3\delta^m_{[n_1} \bar t_{n_2n_3]p_1p_2}  \,,\\
 [ F_{n_1n_2n_3} ,  \bar t_{p_1p_2p_3p_4} ] &=  \tfrac{1}{24} \varepsilon_{n_1n_2n_3p_1\dots p_4q_1\dots q_4} t^{q_1q_2q_3q_4} \,,\\
 [ F_{n_1n_2n_3} ,  t^{p_1p_2p_3p_4} ] &=- 36 \delta^{[p_1p_2}_{[n_1n_2}  t^{p_3p_4]}{}_{n_3]} - 8 \delta^{[p_1p_2p_3}_{n_1n_2n_3}  t^{p_4]q}{}_q \,,\\
 [ F_{n_1n_2n_3} ,   t^{p_1p_2}{}_m ] &= 6 \delta^{p_1p_2}_{[n_1n_2} t_{n_3],m} + 6 \delta^{[p_1}_{[n_1} t^{p_2]}{}_{n_2n_3]m} - \tfrac32 \delta^{p_1p_2}_{[n_1n_2} t^q{}_{n_3]mq} \,.
 \end{align}
 \end{subequations}
The commutators of  $\cT_{-1}$ with itself are
\begin{subequations}
\begin{align}
 \{ t^{n_1n_2n_3n_4} , t^{p_1p_2p_3p_4}\} &= \tfrac16 \varepsilon^{n_1n_2n_3n_4p_1p_2p_3p_4q_1q_2q_3} \bar F_{q_1q_2q_3} \,\\
\{ \bar  t_{n_1n_2n_3n_4} , \bar t_{p_1p_2p_3p_4}\} &= - \tfrac16 \varepsilon_{n_1n_2n_3n_4p_1p_2p_3p_4q_1q_2q_3} \bar E^{q_1q_2q_3}  \,\\
\{ \bar  t_{n_1n_2n_3n_4} ,  t^{p_1p_2p_3p_4}\} &= 96 \delta_{[n_1n_2n_3}^{[p_1p_2p_3} \bar K^{p_4]}{}_{n
_4]} - 8 \delta_{n_1n_2n_3n_4}^{p_1p_2p_3p_4} \bar K^m{}_m   \,.
\end{align}
\end{subequations}
These relations allows checking the invariance of $\omega_1$ under $\mathfrak{e}_{11}$. 

We also need the commutators involving $P_m$ and $P_{n_1n_2n_3n_4}$ of $R(\Lambda_{10}) \subset \cT_2$ 
\begin{subequations}
\begin{align}
 [ F_{n_1n_2n_3} , P_m ] &= P_{mn_1n_2n_3} \; , \qquad   [ E^{n_1n_2n_3} , P_{p_1p_2p_3p_4} ] = - 24 \delta^{n_1n_2n_3}_{[p_1p_2p_3} P_{p_4]} \,,\\
 [ t^{n_1n_2n_3n_4} , P_m] &= \tfrac{1}{7!} \varepsilon^{n_1n_2n_3n_4p_1\dots p_7} ( - P_{p_1\dots p_7,m} + 5 P_{p_1\dots p_7m}) \,\\
 [ \bar t_{n_1n_2n_3n_4} , P_m ] &= P_{n_1n_2n_3n_4m} \; , \qquad  [ \bar t_{p_1p_2}{}^m , P_{n} ]  = 4 \delta^m_{[p_1} P_{p_2]n} - \delta^m_n P_{p_1p_2}  \,,\\
 [ \bar t_{p_1p_2}{}^m , P_{n_1n_2n_3n_4} ] &= - 4\delta^m_{[n_1} P_{n_2n_3n_4]p_1p_2}  + 4 \delta^m_{[p_1} P_{p_2] n_1n_2n_3n_4} \,,\\
 [ \bar t_m{}^{p_1p_2p_3}  , P_n  ] &=  12 \delta_{nm}^{[p_1p_2} P^{p_3]}\; , \qquad   [ \bar t^{m,p}  , P_n  ] = - 6 \delta_n^{(m} P^{n)}
 \end{align}
 \end{subequations}
and 
\begin{align}
[ \bar K^p{}_n , P_m ] - 4 \delta^p_{(m} F_{n)} + \dots \,,
\end{align}
where $F_n$ is the generator of $\cT_0$ in $R(\Lambda_{10})$ and we omit the other terms. The commutators with the $\cT_1$ module are the ones written in \cite{Bossard:2017wxl} and we do not display them here. We have the following anti-commutator of $\{ \cT_{-1} , \cT_1 \} \subset \cT_0$ 
\begin{subequations}
\begin{align}
\{ \bar t_{n_1n_2n_3n_4},P^m \} &= 4 \delta_{[p_1}^m F_{p_2p_3p_4]}\; , \qquad \{ \bar t_{p_1p_2}{}^n , P^m\} = - 4 \delta^{[m}_{[p_1} K^{n]}{}_{p_2]} \,,\\
 \{ \bar t_{n_1n_2n_3n_4},P_{p_1p_2}  \} &= - F_{n_1n_2n_3n_4p_1p_2} \; , \\
    \{ \bar t_{p_1p_2}{}^m , P_{n_1n_2}\} &= 2 \delta^{m}_{[n_1} F_{n_2]p_1p_2} - 2 \delta^m_{[p_1} F_{p_2] n_1n_2} \,,\\
 \{ t^{n_1n_2n_3n_4} , P^m\} &= - \tfrac1{6!} \varepsilon^{mn_1n_2n_3n_4p_1\dots p_6} F_{p_1\dots p_6} \,,\\
 \{ t^{n_1n_2n_3n_4} , P_{p_1p_2}\} &= - \tfrac1{7!} \delta_{[p_1|}^{[n_1} \varepsilon^{n_2n_3n_4]q_1\dots q_8} F_{q_1\dots q_8,|p_2]} + 12 \delta_{p_1p_2}^{[n_1n_2} \widehat{F}^{n_3n_4]} \,,\\
 \{ t^{p_1p_2}{}_m , P^n \} &=  \tfrac1{8!}  \varepsilon^{np_1p_2q_1\dots q_8} F_{q_1\dots q_8,m} + 3 \delta^{[n}_{m} \widehat{F}^{p_1p_2]} \,,
 \end{align}
 \end{subequations}
where $\widehat{F}^{n_1n_2}$ is the highest weight generator in $R(\Lambda_2)$. After this level it becomes more difficult to obtain all terms, but since we know that $\omega_1 P_n$, if non-zero, must give $F_n \in R(\Lambda_{10})$, we shall only write the generators in $R(\Lambda_{10})$. Obtaining other generators would be inconsistent with the property that $\omega_1$ is a homomorphism. We determine these terms by consistency with the property that  $L(\Lambda_{10})$ is an $\mathfrak{e}_{11}$-submodule of $\cT_0$, and in particular $[ \mathfrak{e}_{11} ,L(\Lambda_{10})] \subset L(\Lambda_{10})$.  We use ellipses for the generator in $\widehat{\rm adj}$ with
\begin{subequations}
\begin{align}
 \{ t^m{}_{p_1p_2p_3} , P^n \} &= - 60 \delta^{mn}_{[p_1p_2} F_{p_3]} + \dots \,, \\
  \{ t_{m,p} , P^n \} &= - \tfrac12 \delta^n_{(m} F_{p)} + \dots \,,\\
\{ t^{p_1p_2p_3p_4} , P_{n_1n_2n_3n_4n_5} \} &= 60 \delta^{p_1p_2p_3p_4}_{[n_1n_2n_3n_4} F_{n_5]} + \dots \; , \\
 \{ t^{p_1p_2}{}_m , P_{n_1n_2}\} &= \delta_{n_1n_2}^{p_1p_2} F_m - 4 \delta^{p_1p_2}_{m[n_1} F_{n_2]} 
\end{align}
\end{subequations}
and
\begin{subequations}
\begin{align}
\{ t^{n_1n_2}{}_m , P_{q_1\dots q_5} \} &= - 10 \delta^{n_1n_2}_{[q_1q_2} F_{q_3q_4q_5]m} - 10  \delta^{n_1n_2}_{m[q_1} F_{q_2q_3q_4q_5]} + \dots \,,\\
 \{ t^{n_1n_2n_3n_4} , P_{p_1\dots p_7,m} \} &= 210 \delta^{n_1\dots n_4}_{[ p_1\dots p_4} F_{p_5p_6p_7]m} + 210  \delta^{n_1\dots n_4}_{m[ p_1p_2p_3} F_{p_4p_5p_6p_7]} + \dots \,,\\
 \{ t^{n_1n_2n_3n_4} , P_{p_1\dots p_8}\} &= - 420 \delta^{n_1n_2n_3n_4}_{[p_1\dots p_4} F_{p_5p_6p_7p_8]}  + \dots 
 \end{align}
 \end{subequations}
The latter are needed to compute $[ \bar K^n{}_p , P_m]$. With this, one has all the information to compute 
\begin{align}
\label{eq:om1Pm}
 \omega_1 P_m &= 280 F_m + \tfrac{1}{4!} \{ t^{n_1n_2n_3n_4}, P_{mp_1p_2p_3p_4}\} + \tfrac12 \{ t^{n_1n_2}{}_m , P_{n_1n_2}\} + \tfrac13 \{ t^{np}{}_p , P_{mn}\} \CR
& \hspace{40mm} + 3 \{ t_{m,n} , P^n\} - \tfrac14 \{ t^q{}_{mnq} , P^n\} \CR
&= ( 280  + 105 + \tfrac52 - \tfrac{63}{2}) F_m = 356 F_m \,,
\end{align}
 which completes the proof that $R(\Lambda_{10})$ is not in the kernel of $\omega_1$. 
We do not have a proof that $\omega_1$ is injective on all of  $L(\Lambda_{10})\subset \cT_2$, however $\cT_0$-equivariance may imply that it is.
}


\subsection{\texorpdfstring{$\cT(\mathfrak{e}_{11})$ under $\mf{gl}(3)\oplus \mf{e}_8$}{T(e11) under gl(3)+e8}}
\label{app:e8tha}

The local super-algebra of $\cT(\mathfrak{e}_{11})$ associated to $\mf{gl}(3)\oplus \mf{e}_8$ has been defined in \cite{Bossard:2019ksx}. Here we shall compute the next level of superfields to obtain more information about the irreducible modules that appear in $\cT_p$. For this we recall that $\cT$ admits a $\mathds{Z}^2$ grading 
\be 
\cT(\mathfrak{e}_{11})  = \sum_{p\in \mathds{Z}} \cT_{p} = \sum_{p\in \mathds{Z}} \sum_{q\in \mathds{Z}} \cT_{p,q} = \sum_{q\in \mathds{Z}} \mathcal{S}_{q} \,,
\ee
where $q- \frac12 p$ is the weight under $GL(1) \subset GL(3) \times E_8 \subset E_{11}$.  $q$ is not consistent with the $\mathds{Z}_2$ grading of the super-algebra, but each component $\mathcal{S}_q $ at fixed $q$ is finite-dimensional. 

The algebra can be defined using a BRST complex, with superfields of fixed $q$ degree that expand as superfields of Grassmann coordinates $\vartheta_\mu$ of $p$-degree $1$. The components of degree $q=0$ are parametrised by a bosonic vector superfield  $V_\mu(\vartheta)$ generating the reparametrisation in three Grassmann variables $\vartheta_\mu$, and scalar fermionic superfield $\Phi^A(\vartheta)$ in $\mathfrak{e}_8$. We use $\iota^\mu = \frac{\partial\; }{\partial \vartheta_\mu}$. The components of degree $q=1$ are parametrised by the fermionic superfield $\psi_\mu^A$ and the bosonic superfield $T^{AB}$ in the ${\bf 3875}\oplus {\bf 1}$. The components of degree $q=-1$ are parametrised by the bosonic superfield $S^A$ and the fermionic superfield $\Theta^\mu$. The BRST differential is then 
\begin{subequations}
\begin{align}
\delta V_\mu &= V_\nu \iota^\nu V_\mu + \psi^A_\mu S_A \ , \\
\delta \Phi^A &= \tfrac12 f_{BC}{}^A \Phi^B \Phi^C + V_\mu \iota^\mu \Phi^A + T^{AB} S_B + f_{BC}{}^A \bigl(  \psi_\mu^B \iota^\mu S^C -\tfrac12 \iota^\mu \psi_\mu^B S^C \bigr) + \psi_\mu^A \Theta^\mu  \ , \\
\delta S^A &= V_\mu \iota^\mu S^A + \iota^\mu V_\mu S^A + f_{BC}{}^A \Phi^B S^C  , \\
\delta \Theta^\mu &= V_\nu \iota^\nu \Theta^\mu - \iota^\mu V_\nu \Theta^\nu + \iota^\nu V_\nu \Theta^\mu - \iota^\mu \Phi^A S_A  , \\
\delta \psi_\mu^A &= V_\nu \iota^\nu \psi_\mu^A + \iota^\nu V_\mu \psi_\nu^A - \iota^\nu V_\nu \psi_\mu^A + f_{BC}{}^A \Phi^B \psi_\mu^C  , \\
\delta T^{AB} &= V_\nu \iota^\nu T^{AB}   - \iota^\nu V_\nu T^{AB}  + 2 \Phi^C f_{CD}{}^{(A} T^{B)D} - 2 \iota^\mu \Phi^{(A} \psi_\mu^{B)}{-} f^{E(A}{}_C f_{ED}{}^{B)} \iota^\mu \Phi^C \psi^D_\mu  \, .
\end{align}
\end{subequations} 
One checks indeed that $\delta^2 = 0$ on $V_\mu$ and $\Phi^A$ and vanishes up to terms quadratic in the degree $q=\pm1$ on the components of degree $q=\pm 1$ respectively, showing that this defines a local super-algebra in the sense of Kac~\cite{Kac2}. The tensor hierarchy algebra is defined as the quotient of the super-algebra freely generated from this local super-algebra by its maximal ideal. The super-algebra associated to $V_\mu$ is the Kac super-algebra $W(3)$ of super-diffeormorphisms in three dimensions, while $\phi_A$ is the super-algebra $G_3(\mathfrak{e}_8)$ of functions of three Grassmann variables in $\mathfrak{e}_8$, and 
\be \mathcal{S}_{0} =  \sum_{p=-3}^1 \mathcal{T}_{p,0} = W(3)  \oright  G_3(\mathfrak{e}_8) \; . \ee

Now we want to determine the superfields that appear at $q=2$. We know from $\mathfrak{e}_{11}$ that at level $p=0$ we must have $\psi_{\mu\nu}^{AB}$ in the ${\bf 1}\oplus {\bf 3875}$ and $\psi_{\mu,\nu}^A$ in the ${\bf 248}$. Checking the generators that they include at $p=-2$, one obtains $\psi_{\mu_1\mu_2}^{\nu_1\nu_2 AB}$ that gives one $(2,1)$-form and one three-form whereas  $\psi_{\mu,\nu}^{\sigma_1\sigma_2A}$ gives one symmetric $(1,1,1)$ tensor and one $(2,1)$-form. Together $\cT_0 \subset \mathfrak{e}_{11} \oleft R(\Lambda_2)\oplus R(\Lambda_{10})$ at level 3 include these representation and more\footnote{Here we use the convention that $\Lambda_1$ of $E_8$ is the ${\bf 248}$, $\Lambda_2$ the ${\bf 30\, 380}$, $\Lambda_7$ the ${\bf 3875}$ and $\Lambda_8$ the ${\bf 147\, 250}$.}
\begin{align}
 \mathfrak{e}_{11}|_3 &= (1,1,1;\Lambda_1) \oplus (2,1;\Lambda_1) \oplus (3;\Lambda_1)  \oplus (2,1;\Lambda_7)  \oplus (3;\Lambda_7) \oplus (2,1;\Lambda_2) \oplus (3;\Lambda_8) \oplus (2,1;0) \CR
R(\Lambda_2)|_3 &= (2,1;\Lambda_1) \oplus (3;\Lambda_1)  \oplus (2,1;\Lambda_7)  \oplus (3;\Lambda_7) \oplus (3;\Lambda_2) \oplus (3;\Lambda_8) \oplus (2,1;0) \oplus (3;0) \CR
R(\Lambda_{10})|_3 &= (3;\Lambda_7) \label{Level3E8}
 \end{align}
so we need new superfields for the remaining components 
\be  
(2,1;\Lambda_1) \oplus 2\times  (3;\Lambda_1) \oplus (2,1;\Lambda_7)  \oplus 2\times (3;\Lambda_7)  \oplus  (2,1;\Lambda_2)\oplus  (3;\Lambda_2) \oplus (3;\Lambda_8) \oplus (2,1;0) 
\ee
To have a $(2,1)$ form one must either have a new superfield $T_\mu$ at $p=-1$, or a new superfield $\lambda_{\mu}{}^\nu$ at $p=-2$. One checks, however, that the second would lead to $(2,2,1)$ forms at level $-3$ that are incompatible with the level decomposition of $\cT_{-3}$. We must therefore introduce the superfields $T_\mu^{[AB]}$ in the antisymmetric tensor product of two ${\bf 248}$ and $T_\mu^{AB}$ in ${\bf 1}\oplus {\bf 3875}$ at $p=-1$. However, $T_\mu^{AB}$ includes one more $E_8$ singlet and we conclude that $\cT_0 \supset \mathfrak{e}_{11} \oleft R(\Lambda_2)\oplus R(\Lambda_{10}) \oplus R(2\Lambda_3)$. We must moreover include the superfields $\lambda^{A_8}$ in $R(\Lambda_8)$ of $E_8$, $\acute{\lambda}^{AB}$ in the ${\bf 3875}$ and $\lambda^A$ in the ${\bf 248}$ at $p=-2$ to get all the irreducible representations. To remove the $R(2\Lambda_3)$ representation one would need to constrain the superfield $T_\mu = \frac{1}{248} \kappa_{AB} T^{AB}_\mu$ to $\iota^\mu T_\mu = 0 $. But the $W(3)$ representation 
\be
 \delta T_\mu  = V_\nu \iota^\nu T_\mu + \iota^\nu V_\mu T_\nu -2 \iota^\nu V_\nu T_\mu 
\ee
is incompatible with this constraint. (One would need weight $+1$ and not $-2$.) There cannot be any other superfield starting at $p=-3$ because it would contribute to $\cT_4 \supset R(\Lambda_9)$ that only starts at $q=3$. Together this set of superfields implies
\bea 
\cT_0 &\supset& \mathfrak{e}_{11} \oleft  R(\Lambda_2)\oplus R(\Lambda_{10}) \oplus R(2\Lambda_3) \CR
\cT_1 &\supset&R(\Lambda_1) \oplus R(\Lambda_1 + \Lambda_{10}) \oplus R(\Lambda_{11}) \oplus R(\Lambda_1+2\Lambda_3) \CR 
\cT_2 &\supset&R(\Lambda_{10})\oplus R(2\Lambda_3)  \oplus R(\Lambda_2 + \Lambda_{10}) \oplus R(\Lambda_1+\Lambda_{11})   \oplus R(\Lambda_2+2\Lambda_3) \CR
\cT_3 &\supset&R(\Lambda_{11})\oplus  R(\Lambda_3 + \Lambda_{10}) \oplus R(\Lambda_3+\Lambda_{4})   \oplus R(\Lambda_2+\Lambda_{11}) \label{TpE11Decompose} 
\eea
This consistency check shows that at least these representations are in $\cT(\mathfrak{e}_{11})$, but one may in principle have more already at $q=2$. We shall now check that there is no extra superfield. We shall see moreover that the only indecomposable sum is the  $ \mathfrak{e}_{11} \oleft  R(\Lambda_2)$ at this order.

\medskip

At the next order one computes the following BRST variations 
\bea \delta \psi_\mu^A \hspace{-0.7mm} &=& \hspace{-0.7mm}  V_\nu \iota^\nu \psi_\mu^A + \iota^\nu V_\mu \psi_\nu^A - \iota^\nu V_\nu \psi_\mu^A + f_{BC}{}^A \Phi^B \psi_\mu^C \CR
&& - T_\mu^{A;B} S_B + \psi_{\mu,\nu}^A \Theta^\nu + \tfrac12 f_{BC}{}^A  \psi_{\mu,\nu}^B \iota^\nu S^C + \psi_{\mu\nu}^{AB} \iota^\nu S_B  \ , \CR
\delta \psi_{\mu\nu}^{AB} \hspace{-0.7mm} &=& \hspace{-0.7mm} V_\sigma \iota^\sigma \psi_{\mu\nu}^{AB} \hspace{-0.7mm} - \hspace{-0.7mm} 2 \iota^\sigma V_{[\mu} \psi_{\nu]\sigma}^{AB}\hspace{-0.7mm}  - \hspace{-0.7mm} 2 \iota^\sigma V_\sigma \psi_{\mu\nu}^{AB} \hspace{-0.7mm} - \hspace{-0.7mm} 2 \Phi^C f_{CD}{}^{(A} \psi_{\mu\nu}^{B)D}  \hspace{-0.7mm} + \hspace{-0.7mm} 2 \psi_\mu^{(A} \psi_\nu^{B)}\hspace{-0.7mm}  + \hspace{-0.7mm} f_{CE}{}^{(A} f^{B)E}{}_D \psi_\mu^C \psi_\mu^D \; , \CR
\delta \psi_{\mu,\nu}^{A} \hspace{-0.7mm} &=& \hspace{-0.7mm}  V_\sigma \iota^\sigma \psi_{\mu,\nu}^{A} + 2 \iota^\sigma V_{(\mu} \psi_{\nu),\sigma}^{A} - 2 \iota^\sigma V_\sigma \psi_{\mu,\nu}^{A} +  f_{BC}{}^{A} \Phi^B \psi_{\mu,\nu}^{C}  +f_{BD}{}^A \psi_\mu^B \psi_\nu^C  \; , \CR
\delta T^{A;B}_\mu &=& V_\nu \iota^\nu T_\mu^{A;B} + \iota^\nu V_\mu T_\nu^{A;B} - 2 \iota^\nu V_\nu T_\mu^{A;B} + f_{CD}{}^A \Phi^C T_\mu^{D;B} +  f_{CD}{}^B \Phi^C T_\mu^{A;D} \CR
&& +( \psi_{\mu\nu}^{AB}- \tfrac{1}{2} f^{AB}{}_C \psi_{\mu,\nu}^C)  \iota^\nu \iota^\sigma V_\sigma - \psi_{\mu,\nu}^A \iota^\nu \Phi^B - \tfrac12 f_{CE}{}^{A} f^{BE}{}_D \psi_{\mu,\nu}^C \iota^\nu \Phi^D + f_{CD}{}^B \psi_{\mu\nu}^{AC} \iota^\nu \Phi^D \CR
&&  + \psi_\mu^A \iota^\nu \psi_\nu^B + \tfrac12 f_{CE}{}^A f^{BE}{}_D \psi_\mu^C \iota^\nu \psi_\nu^D - 2 \psi_\nu^{[A} \iota^\nu \psi_\mu^{B]} + f_{CD}{}^A T^{BC} \psi_\mu^D \; , \label{BRSTextended}
\eea
where $T_\mu^{A;B} = T_\mu^{[AB]} + T_\mu^{AB} $ is in $R(\Lambda_1) \otimes R(\Lambda_1) \ominus R(2\Lambda_1)$ of $E_8$. One can compute similarly the terms bilinear in $(\Theta^\mu , S^A)$ and $( \psi_{\mu\nu}^{AB}, \psi_{\mu,\nu}^{A},T_\mu^{A;B}, \lambda^{AB;C})$ in $\delta T^{AB}$  by requiring nilpotency of $\delta$.  But the only terms involving $\lambda^{AB;C}$ is of the form 
\be
 \delta T^{AB} =- \lambda^{AB;C} S_C + \dots
  \ee
and so all the remaining superfields must be included in $\lambda^{AB;C}$. The general decomposition of $\lambda^{AB;C}$ is
\be 
\lambda^{AB;C} = \Pi_{A_8}{}^{AB;C} \lambda^{A_8} +  f^{C(A}{}_D \acute{\lambda}^{B)D} + 2 P^{AB,CD} \lambda_C + \frac14  \kappa^{AB} \tL^C  \,,
\ee
where we used the property that there is no $E_{11}$ highest weight representation that would start at $w_3=3$ with an $E_8$ representation $R(\Lambda_2)$ or $R(\Lambda_1+\Lambda_7)$ to disregard these possibilities. We have introduced two adjoint superfields to allow a priori for the need of an extra superfield $\tL^A$. To check the variation of $\lambda^{AB;C}$, we shall only consider monomial in $T^{AB} \psi_\mu^C S^D$ in $\delta^2 T^{AB}$.  We use that 
\be 
\kappa_{BC}  \lambda^{AB;C} = \frac{125}{4} \lambda^A + \frac14   \tL^A \; , \qquad \kappa_{AB} \lambda^{AB;C} = 62 \tL^C 
\ee
to project in the same way the terms in $\delta^2 T^{AB}$, after having removed all we could from the variation of bilinear in  $T^{A;B}_\mu S^C$. 
One obtains 
\be
 \delta \lambda^A  = - \delta \tL^A = T^{AB} \iota^\mu \psi_{\mu B} + \frac{1}{31} \kappa_{BC}  \iota^\mu T^{BC}  \psi_\mu^A \; .  
 \ee
This is the only term that could require a new adjoint field and we conclude therefore that $\tL^A  = - \lambda^A$ and there is no more irreducible representation at degree $q=2$. 

Using \eqref{BRSTextended} one moreover checks the $E_{11}$ variation of the scalar superfields and one obtains that 
\be 
E_A^\sigma ( \iota^\mu\iota^\nu \psi_{\mu\nu}  + 4 \iota^\mu T_\mu ) = 0 \; , \label{Reducibility} 
\ee
so the module $R(\Lambda_k + 2\Lambda_3)\subset \cT_k$ do not mix with lower level module and one can write a direct sum, which proves that one has only direct sum in \eqref{TpE11Decompose}.

This provides a non-trivial consistency check that $L(\Lambda_{10}) \subset  \cT_0$  with 
\be 
L(\Lambda_{10}) = R(\Lambda_{10}) \oplus R(2\Lambda_2) \oplus \dots \; . 
\ee
We have also obtained that there is no $R(\Lambda_4)$ in $\cT_0$, consistently with the property that it belongs to $L(\Lambda_4)$.  In terms of the decomposition of $\cT_2$ given in~\eqref{eq:THA2}, we have checked that $R(\Lambda_1+\Lambda_{11})$ is not part of $L(\Lambda_{10})$ but is part of $\cT_2$ so that $D_2\neq 0$. This is not in contradiction with our claim of a homomorphic image of $L(\Lambda_{10})$ in $\cT_0$.\footnote{We thank M. Cederwall for discussions related to this point.}


\section{Representation extensions and cohomology}
\label{app:ext}

In this appendix, we take a slightly more formal look at the indecomposable structure of $\adjhat$ as described in Section~\ref{sec:THA}. In particular, we spell out the sense in which the cocycle $K^{\alpha\ta}{}_{\tb}$ is an element of degree one in the Chevalley--Eilenberg cohomology of $\mf{e}_{11}$ in a certain module.  We shall return to the conjecture that $L(\Lambda_2)=R(\Lambda_2)$ in this context at the end.
 
Generally, an extension of the adjoint of $\mf{e}_{11}$ by a module $R$ to an indecomposable $\mf{e}_{11}$-representation $\mf{e}_{11} \oleft R$ is determined by a linear map\footnote{$K$ is only a linear map, not an $\mf{e}_{11}$ homomorphism. The target $\text{Hom}(R,\mf{e}_{11})$ is the space of $\mf{e}_{11}$-homomorphisms between the modules, however.} $K: \mf{e}_{11} \to \text{Hom}(R,\mf{e}_{11})$ such that the action of $t^\alpha\in \mf{e}_{11}$ on $\mf{e}_{11} \oleft R$ is defined by
\begin{align}
t^\alpha \cdot \Big( \phi_\beta t^\beta + X_\tb t^\tb \Big)  
&=  \phi_\beta f^{\alpha\beta}{}_\gamma - K(t^\alpha) \big(X_\tb t^\tb\big) - X_\tb T^{\alpha \tb}{}_\tg t^\tg\nn\\
&= \Big( \phi_\beta f^{\alpha\beta}{}_\gamma  - X_\tb K^{\alpha \tb}{}_\gamma \Big) t^\gamma
- X_\tb T^{\alpha \tb}{}_\tg t^\tg
\end{align}
where we denote the basis elements of $\mf{e}_{11}$ by  $t^\gamma$ and those of $R$ by $t^\tg$ and have used arbitrary coefficients. The notation is motivated by $E_{11}$ exceptional field theory and~\eqref{eq:T0CR}. The matrix components of the map $K(t^\alpha)\in \text{Hom}(R,\mf{e}_{11})$ are written as $K^{\alpha\ta}{}_\gamma$ and these off-diagonal terms go beyond the $\mf{e}_{11}$ action on the direct sum of the two modules.
For this action to define a representation we must have
\begin{align}
2 t^{[\alpha}\cdot t^{\beta]} \cdot  \Big( \phi_\gamma t^\gamma + X_\tg t^\tg \Big)  &= t^\alpha \cdot \left[\Big( \phi_\gamma f^{\beta\gamma}{}_\delta  - X_\tg K^{\beta \tg}{}_\delta \Big) t^\delta
- X_\tg T^{\beta \tg}{}_\td t^\td\right] - (\alpha\leftrightarrow\beta)\nn\\
&=  \Big( \phi_\gamma  f^{\beta\gamma}{}_\delta f^{\alpha\delta}{}_\epsilon - X_\tg f^{\alpha\delta}{}_\epsilon K^{\beta \tg}{}_\delta  +X_\tg T^{\beta\tg}{}_\td K^{\alpha \td}{}_\epsilon\Big) t^\epsilon \nn\\
&\hspace{10mm} + X_\tg T^{\beta \tg}{}_\td T^{\alpha \td}{}_{\tilde{\epsilon}} t^{\tilde{\epsilon}}- (\alpha\leftrightarrow\beta)\nn\\
&\stackrel{!}{=} \Big( \phi_\gamma  f^{\alpha\beta}{}_\delta f^{\delta\gamma}{}_\epsilon - X_\tg f^{\alpha\beta}{}_\delta K^{\delta \tg}{}_\epsilon \Big) t^\epsilon - X_\tg f^{\alpha\beta}{}_\delta T^{\delta \tg}{}_{\tilde\epsilon} t^{\tilde\epsilon}
\end{align}
for all $\phi_\gamma$ and $X_\tg$. After renaming dummy indices, this leads to the $\mf{e}_{11}$ Jacobi identity, the representation property
\begin{align}
T^{\alpha \ta}{}_\tb T^{\beta \tb}{}_\tg - T^{\beta \ta}{}_\tb T^{\alpha \tb}{}_\tg = f^{\alpha\beta}{}_\gamma T^{\gamma \ta}{}_\tg ,
\end{align}
on the module $R$ and the identity
\begin{align}
\label{eq:1coc}
-f^{\alpha \delta}{}_\gamma K^{\beta\ta}{}_\delta + f^{\beta \delta}{}_\gamma K^{\alpha\ta}{}_\delta + T^{\beta\ta}{}_\tb K^{\alpha \tb}{}_\gamma  - T^{\alpha\ta}{}_\tb K^{\beta\tb}{}_\gamma  = - f^{\alpha\beta}{}_\delta K^{\delta \ta}{}_\gamma
\end{align}
that was already given in~\eqref{eq:Ktrm}. The linear map $K$ is only defined up to a choice of  basis, such that  
\begin{align}
\label{eq:exact}
 K^{\prime \alpha\ta}{}_\beta = K^{\alpha\ta}{}_\beta + T^{\alpha \ta}{}_\tb K^{\tb}{}_\beta +f^{\alpha\gamma}{}_\beta K^\ta{}_\gamma\; ,
 \end{align}
in the basis $t^{\prime \ta} =  t^\ta - K^{\ta}{}_\beta t^\beta$ with 
\be 
\phi_\alpha t^\alpha + X_\ta t^\ta = ( \phi_\alpha +K^{\tb}{}_\alpha X_\tb ) t^\alpha + X_\ta ( t^\ta - K^{\ta}{}_\beta t^\beta )  =  \phi^\prime_\alpha  t^\alpha + X_\ta  t^{\prime \ta}  \; . 
\ee 
This can be reformulated in the language of Lie algebra cohomology~\cite{Knapp:1988} as follows. Rewriting $\text{Hom}(R,\mf{e}_{11})\cong \overline{R}\otimes \mf{e}_{11}$, where $\overline{R}$ is the dual representation to $R$, gives us an interpretation of $K$ as a map from $\mf{e}_{11}$ to the $\mf{e}_{11}$-module $M=\overline{R}\otimes \mf{e}_{11}$ and therefore as an element of degree one in the Chevalley--Eilenberg complex $H( \Lambda^\bullet(\mf{e}_{11}), M)$. The nilpotent differential on this complex is defined for an element $f$ of degree $n$ by (hat means omission of the corresponding argument)
\begin{align}
df(x_1,\ldots, x_{n+1}) &= \sum_{i=1}^{n+1} (-1)^{i+1} x_i\cdot f(x_1,\ldots, \hat{x}_i,\ldots ,x_{n+1})\nn\\
&\hspace{10mm} + \sum_{i<j} (-1)^{i+j} f([x_i,x_j], x_1,\ldots ,\hat{x}_i,\ldots, \hat{x}_j,\ldots, x_{n+1})\; , 
\end{align}
with $\cdot$ denoting the action on the module $ \overline{R} \otimes \mf{e}_{11}$. The cocycle condition for an element $K$ of degree one means therefore
\begin{align}\label{Cocycledclosed}
dK(x_1 ,x_2) = x_1 \cdot K(x_2) -x_2  \cdot K(x_1)  - K([x_1,x_2]) \stackrel{!}{=} 0\,.
\end{align}
Writing this equation in a basis with $K(t^\alpha) = K^{\alpha\ta}{}_\beta \bar{t}_\ta \otimes t^\beta$ leads to
\begin{align}
K^{\beta \ta}{}_\delta \big(T^{\alpha \tb}{}_\ta \bar{t}_\tb \otimes t^\delta + f^{\alpha\delta}{}_\gamma  \bar{t}_\ta \otimes t^\gamma\big) 
- K^{\alpha \ta}{}_\delta \big(T^{\beta \tb}{}_\ta \bar{t}_\tb \otimes t^\delta + f^{\beta\delta}{}_\gamma  \bar{t}_\ta \otimes t^\gamma\big) 
- f^{\alpha\beta}{}_\delta K^{\delta \ta}{}_\gamma \bar{t}_\ta \otimes t^\gamma = 0\,,
\end{align}
which is equivalent to~\eqref{eq:1coc}. 

An exact cocycle of degree one in the Chevalley--Eilenberg complex is the differential of a degree zero element $\lambda\in M $, viewed as a linear map from the one-dimensional vector space $\Lambda^0(\mf{e}_{11})\cong \mathds{R}$ to $M=\overline{R}\otimes \mf{e}_{11}$, 
\be  d \lambda(x)  = x \cdot \lambda \; . \ee
Writing this element as $\lambda = K^{\ta}{}_\beta \bar{t}_\ta\otimes t^\beta$, an exact cocycle expands in components as
\begin{align}
d\lambda(t^\alpha) = t^\alpha \cdot K^\ta{}_\beta \bar{t}_\ta\otimes t^\beta = K^\ta{}_\beta T^{\alpha \tb}{}_\ta \bar{t}_\tb \otimes t^\beta + f^{\alpha\beta}{}_\gamma K^\ta{}_\beta \bar{t}_\ta \otimes t^\gamma\,,
\end{align}
which means we have to identify
\begin{align}
K^{\alpha\ta}{}_\beta \sim K^{\alpha\ta}{}_\beta + T^{\alpha \ta}{}_\tb K^{\tb}{}_\beta +f^{\alpha\gamma}{}_\beta K^\ta{}_\gamma\,,
\label{exact}
\end{align}
in agreement with~\eqref{eq:exact}.

The existence of the tensor hierarchy algebra $\cT(\mf{e}_{11})$ gives the existence of a non-trivial cocycle for the case of $R=R(\Lambda_2)$. We see directly that condition \eqref{Cocycledclosed} is satisfied. To check that there is no trivialisation $K^\ta{}_\alpha$ one can resort to level decomposition: The structure constants are invariant under the Levi subgroup $L\subset E_{11}$ associated to a level decomposition, and one finds that the irreducible representations of $L$ in $R(\Lambda_2)$ do not all lie in $\mf{e}_{11}$. Therefore no $K^\ta{}_\alpha$ can trivialise the cocycle, given that by Schur's lemma there are no non-trivial homomorphisms between non-isomorphic irreducible representations of $L$. 

It would be interesting to investigate which representations can be added indecomposably to the adjoint $\mf{e}_{11}$. There is a non-trivial cocycle  mixing $\mf{e}_{11}$ with $R(n\Lambda_2)$ for all $n\ge 1$. They are direct generalisations of the Virasoro extensions $L_n$ of affine Lie algebras. To prove the existence of these cocycles, one defines  an algebra $\mathfrak{V}_n(\mf{e}_{11}) \supset \mf{e}_{11} \oleft R(n\Lambda_2)$ extending $\mf{e}_{11}$ for any $n\ge 1$ using the local algebra construction~\cite{Kac2}. The local algebra  of $\mathfrak{V}_n(\mf{e}_{11})$ is defined in the branching under  $ \mf{e}_9\oplus \mf{sl}(2)$ such that the grading is the $\mf{e}_9$ central charge, with 
\be  \mf{V}_n(\mf{e}_{11}) \supset  \bigl(  {\bf 2}\otimes \overline{R(\Lambda_0)_{\scalebox{0.6}{$- \frac32 $}}} \bigr)^\ord{-1}   \oplus \bigl( \mathfrak{e}_9\oleft  \langle L_{n} \rangle  \oplus \mathfrak{sl}(2) \bigr)^\ord{0} \oplus \bigl( {\bf 2}\otimes R(\Lambda_0)_{\scalebox{0.6}{$- \frac32 $}}\oplus {\bf 2}\otimes R(\Lambda_0)_{\scalebox{0.6}{$n- \frac32 $}} \bigr)^\ord{1} \; , \label{VirE11}  \ee
and
\bea
\mf{e}_{11} &\supset&   \bigl(  {\bf 2}\otimes \overline{R(\Lambda_0)_{\scalebox{0.6}{$- \frac32 $}}} \bigr)^\ord{-1}   \oplus \bigl( \mathfrak{e}_9  \oplus \mathfrak{sl}(2) \bigr)^\ord{0} \oplus \bigl( {\bf 2}\otimes R(\Lambda_0)_{\scalebox{0.6}{$- \frac32 $}} \bigr)^\ord{1} \; , \CR
R(n\Lambda_2) &\supset&  \langle L_{n} \rangle^\ord{0} \oplus \bigl( {\bf 2}\otimes R(\Lambda_0)_{\scalebox{0.6}{$n- \frac32 $}} \bigr)^\ord{1} \; , 
\eea
such that the subscripts $w=-\frac{3}{2}$ or $n-\frac{3}{2}$ are the eigenvalues of the ${\mf e}_9$ derivation ${\rm d} = L_0 + w$ on the highest weight vector. One moreover checks that the commutation relations are uniquely fixed by solving Jacobi identities.\footnote{One can combine all modules with $n \ge 1$ to obtain an algebra  $\mathfrak{V}_+(\mf{e}_{11}) \supset \mf{e}_{11} \oleft \bigl(  \bigoplus_{n\ge 1} R(n\Lambda_2)\bigr) $. Performing a level decomposition of the latter with respect to the grading element $H_{\Lambda_2}$, the degree zero subalgebra is  $ \mathfrak{e}_9\oleft  \bigl( \bigoplus_{m>0} \langle L_{m} \rangle \bigr) \oplus \mathfrak{sl}(2) $  where the Virasoro generators $L_m$ with $m\ge 0$ satisfy the Virasoro algebra $[ L_m , L_n ] = (m-n) L_{m+n} $.} Preliminary checks suggests that there might be many more possibilities, and we have evidence for a non-trivial cocycle mixing $\mf{e}_{11}$ with $R(2\Lambda_{10})$.\footnote{We have checked that it is consistent with the positive Borel subalgebra $\mf{b}_+\subset \mf{e}_{11}$. For discussions of Lie algebra cohomology for Kac--Moody algebras see~\cite{Morita,Farnsteiner}.} 

In the context of the tensor hierarchy algebra, our conjecture is that for $\adjhat$ we have $L(\Lambda_2)=R(\Lambda_2)$. From the analysis of the tensor hierarchy algebra in $GL(3)\times E_8$ we can show that there is no $R(2\Lambda_2)$ inside $\adjhat$, suggesting that there is no $R(n\Lambda_2)$. The structure of the Virasoro algebra indeed requires that if $R(n\Lambda_2)$ is present, all $R(m\Lambda_2)$ for $m>n$ must also be. It would be rather disturbing to have such an infinite sequence with a gap between $R(\Lambda_2)$ and the first occurring $R(n\Lambda_2)$. Moreover, we have partly checked the local algebra construction of $\cT(\mf{e}_{11})$ based on the degree zero component $\mf{V}_1(\mf{e}_{11})$, which would prove that none of the $R(n\Lambda_2)$ with $n\ge 2$ appear in $\cT_0$. Nevertheless, the statement $L(\Lambda_2)=R(\Lambda_2)$ remains conjectural.

\section{\texorpdfstring{$GL(11)$ formulae}{GL(11) formulae}}
\label{app:GL11}

In this appendix we collect some relevant reference expressions for the $GL(11)$ level decompositions of the various objects of $E_{11}$ exceptional field theory. The $GL(11)$ level decomposition of $E_{11}$ was originally studied in~\cite{West:2002jj}, see also~\cite{Nicolai:2003fw,Kleinschmidt:2003mf}. The extension to the tensor hierarchy algebra $\cT(\mf{e}_{11})$ and in particular the field strength representation was given in~\cite{Bossard:2017wxl,Bossard:2019ksx}.
$E_{11}$ is generated by the level $\pm 1$ generators $E^{n_1n_2n_3}$ and $F_{n_1n_2n_3}$, respectively.

\subsection{\texorpdfstring{$E_{11}$ transformations}{E11 transformations}}
\label{GL11E11tr}

In order to describe the adjoint of $\mf{e}_{11}$, we introduce formal fields that are not projected to the coset component, with a $+$ superscript for positive or null $GL(11)$ level and a $-$ superscript for negative level. More specifically we need to describe the indecomposable representation $\adjhat$, so we define this formal field in $\cT_{-2}$ as explained in~\cite[Eq.~(4.30)]{Bossard:2017wxl} 
\begin{align}
{\phi}^\wa \bar{t}_\wa &=   \ldots + \frac{1}{8!} h_-^{n_1\ldots n_8,m} \bar{F}_{n_1\ldots n_8,m} + \frac1{6!} A_-^{n_1\ldots n_6}\bar{F}_{n_1\ldots n_6} + \frac1{3!} A_-^{n_1n_2n_3} \bar{F}_{n_1n_2n_3} + h^+_n{}^m \bar{K}^n{}_m \nn\\
&\quad + \frac{1}{3!} A^+_{n_1n_2n_3} \bar{E}^{n_1n_2n_3} + \frac1{6!} A^+_{n_1\ldots n_6} \bar{E}^{n_1\ldots n_6} + \frac1{8!} h^+_{n_1\ldots n_8,m} \bar{E}^{n_1\ldots n_8,m} \\
&\quad + \frac{1}{8!} X_{n_1\ldots n_9} \bar{E}^{n_1\ldots n_9} +\frac{1}{9!} X_{n_1\ldots n_{10},p_1p_2} \bar{E}^{n_1\ldots n_{10},p_1p_2} +\frac{1}{9!} X_{n_1\ldots n_{11},m} \bar{E}^{n_1\ldots n_{11},m} + \ldots \,.
\nn 
\end{align}
Under the $E_{11}$ generator
\begin{align}
\label{eq:E11elt}
\Lambda = \frac1{3!} e_{n_1n_2n_3} E^{n_1n_2n_3} + \frac1{3!} f^{n_1n_2n_3} F_{n_1n_2n_3}\; , 
\end{align}
these components transform in our conventions as~\cite{Bossard:2017wxl}
\begin{subequations}
\label{adjoint}
\begin{align} 
\delta_\Lambda {h}_-^{n_1\cdots n_8,m} &= -56 f^{\lsharp n_1n_2n_3}A_-^{n_4\cdots n_8,m\rsharp} + \dots    \ ,
\\
\delta_\Lambda {A}_-^{n_1\cdots n_6} &= -20 f^{[n_1n_2n_3} A_-^{n_4n_5n_6]} 
+\frac{1}{2} e_{p_1p_2q} h_-^{n_1\cdots n_6p_1p_2,q} \ ,
\\
\delta_\Lambda A_-^{n_1n_2n_3} &= \frac{1}{6} e_{p_1p_2p_3} 
  A_-^{n_1n_2n_3p_1p_2p_3} + 3 f^{p[n_1n_2} h^+_p{}^{n_3]} \,,
\\
\delta_\Lambda h^+_n{}^m &=\frac{1}{2}e_{np_1p_2} A^{mp_1p_2}_- - \frac{1}{2} f^{mp_1p_2} A^+_{np_1p_2}
\nn\\ 
&\quad\, - \frac{1}{18} \delta_n^m \scal{e_{p_1p_2p_3}  A_-^{p_1p_2p_3} - f^{p_1p_2p_3} A^+_{p_1p_2p_3}}\,, 
\\
\delta_\Lambda A^+_{n_1n_2n_3} &= -\frac{1}{6} f^{p_1p_2p_3} A^+_{n_1n_2n_3p_1p_2p_3} 
- 3 e_{p[n_1n_2} h^+_{n_3]}{}^{p}\,,
\\
\delta_\Lambda A^+_{n_1\cdots n_6} &= 20 e_{[n_1n_2n_3} A^+_{n_4n_5n_6]} 
-\frac{1}{2} f^{p_1p_2q} h^+_{n_1\cdots n_6p_1p_2,q}  \,,
\\
\delta_\Lambda h^+_{n_1\cdots n_8,m} &= 56 e_{\lsharp n_1n_2n_3} A^+_{n_4\cdots n_8,m\rsharp} + \dots\,, 
\\
\label{eq:deltaX}
\delta_\Lambda X_{n_1\cdots n_9} &=-  \tfrac12 f^{p_1p_2p_3}  X_{n_1\ldots n_9p_1,p_2p_3}+ f^{p_1p_2p_3}  X_{n_1\ldots n_9p_1p_2,p_3} - 28 e_{[n_1n_2n_3} A^+_{n_4\cdots n_9]} + \dots\ ,
\end{align}
\end{subequations}
where $\lsharp n_1\dots n_k,m\rsharp$ denotes  the projection to the irreducible  representation of Young symmetry $(k,1)$.

The derivatives $\partial_M$ belong to the $\overline{R(\Lambda_1)}$ module, see Section~\ref{sec:coord}. Their level decomposition was first analysed in~\cite{West:2003fc}, see also~\cite{Kleinschmidt:2003jf}. At lowest $GL(11)$ levels they are
\begin{align}
\partial_M = ( \partial_m ; \partial^{n_1n_2}; \partial^{n_1\ldots n_5}; \partial^{n_1\dots n_7,m}, \partial^{n_1\ldots n_8};\ldots )\,.
\end{align}
Under the $E_{11}$ generator
\begin{multline} \Upsilon = \frac{1}{8!} e^{n_1\dots n_8,m} E^{n_1\dots n_8,m}+ \frac{1}{6!} e_{n_1\dots n_6} E^{n_1\dots n_6}  +  \frac1{3!} e_{n_1n_2n_3} E^{n_1n_2n_3} \\
+ \frac1{3!} f^{n_1n_2n_3} F_{n_1n_2n_3} + \frac{1}{6!} f^{n_1\dots n_6} F_{n_1\dots n_6} + \frac{1}{8!} f^{n_1\dots n_8,m} F_{n_1\dots n_8,m} \end{multline}
they transform in our conventions as~\cite{Bossard:2017wxl}
\begin{align}
\delta_\Upsilon \partial_m &= \tfrac12 e_{mn_1n_2} \partial^{n_1n_2}  + \tfrac1{5!} e_{mn_1\dots n_5} \partial^{n_1\dots n_5} + \tfrac{1}{7!} e_{mn_1\dots n_7,p} \partial^{n_1\dots n_7,p} + \tfrac{3}{7!} e_{mn_1\dots n_7p} \partial^{n_1\dots n_7p} 
\nn\\
\delta_\Upsilon \partial^{n_1n_2} &= f^{n_1n_2p} \partial_p + \tfrac1{6} e_{p_1p_2p_3} \partial^{n_1n_2p_1p_2p_3} + \tfrac{1}{5!} e_{p_1\dots p_5q} \partial^{n_1n_2p_1\dots p_5,q} + \tfrac{1}{3\cdot 5!} e_{p_1\dots p_6} \partial^{n_1n_2p_1\dots p_6}  
\nn\\
\delta_\Upsilon \partial^{n_1\dots n_5} &= 10 f^{[n_1n_2n_3} \partial^{n_4n_5]}  - f^{n_1\dots n_5p} \partial_p + \tfrac12 e_{p_1p_2q} \partial^{n_1\dots n_5p_1p_2,q} + \tfrac16 e_{p_1p_2p_3} \partial^{n_1\dots n_5p_1p_2p_3} 
\nn\\
\delta_\Upsilon \partial^{n_1\dots n_7,m} &= 35 f^{\langle n_1n_2n_3} \partial^{n_4\dots n_7,m\rangle }- 7 f^{\langle n_1\dots n_6} \partial^{n_7,m\rangle} + f^{p\langle n_1\dots n_7,m\rangle } \partial_p  
\nn\\
\delta_\Upsilon \partial^{n_1\dots n_8} &= -7  f^{[n_1n_2n_3} \partial^{n_4\dots n_8]}+ 7 f^{[n_1\dots n_6} \partial^{n_7n_8]} - \tfrac38  f^{n_1\dots n_8,p} \partial_p  \; ,
\label{eq:L1mod}
\end{align}
while they transform under $\mf{gl}(11)$ as tensors of density weight $\frac12$. For instance, one has the $\mf{gl}(11)$ transformation with $h\cdot K = h_m{}^n K^m{}_n$
\begin{align}
\label{eq:L1wt}
\delta_{h \cdot  K}  \partial_m = h_m{}^p \partial_p -\frac12 h_p{}^p \partial_m + \ldots\, , \qquad 
\delta_{h \cdot  K}  \partial^{n_1n_2} =  2 h_p{}^{[n_1} \partial^{n_2]p} -\frac12 h_p{}^p \partial^{n_1n_2} + \ldots\,,
\end{align}
and similarly for the lower level components. We need these components to write the potential terms \eqref{eq:Lpot1} and \eqref{tpot} in $GL(11)$ level decomposition, this is why we give them explicitly.

The $E_{11}$ transformation rules of the components of the field strengths $F^I$ in $\cT_{-1}$ under~\eqref{eq:E11elt} are given by~\cite{Bossard:2017wxl,Bossard:2019ksx} (see for example \cite[Table 3]{Bossard:2017wxl})\footnote{The last two field strengths transformations were not given in eq. (4.37) of \cite{Bossard:2019ksx}, and here we choose to represent $\cF_{n_1\dots n_9;m} = \cF_{n_1\dots n_9,m} + \cF_{mn_1\dots n_9}$ in the reducible representation.}
\begin{subequations}
\label{evF}
\begin{align}
\delta_\Lambda \cF^{m,n} &= \frac{1}{2} f^{p_1p_2(m} \cF_{p_1p_2}{}^{n)}  - \frac{1}{6} e_{p_1p_2p_3} \cF^{p_1p_2p_3(m,n)}\ , 
\\
\delta_\Lambda \cF_{m}{}^{n_1n_2n_3} &= -3 f^{p[n_1n_2} \cF_{mp}{}^{n_3]} + \frac{3}{4} f^{p_1p_2[n_1} \delta_m^{n_2} \cF_{p_1p_2}{}^{n_3]} 
- \frac{1}{6}e_{p_1p_2p_3} \cF_m{}^{n_1n_2n_3p_1p_2p_3}
\nn\\
&\quad  - e_{mpq} \cF^{n_1n_2n_3p,q} + \frac{3}{8} \delta_m^{[n_1} e_{p_1p_2q} \cF^{n_2n_3]p_1p_2,q} \ ,
\\
\delta_\Lambda \cF_{n_1n_2}{}^m &= e_{p_1p_2[n_1} \cF_{n_2]}{}^{mp_1p_2} - \frac{1}{9} e_{p_1p_2p_3}\delta^m_{[n_1} \cF_{n_2]}{}^{p_1p_2p_3} +  e_{pn_1n_2} \cF^{m,p} 
\nn\\
&\quad- \frac{1}{2} f^{mp_1p_2} \cF_{n_1n_2p_1p_2} - \frac{1}{9} f^{p_1p_2p_3} \delta^m_{[n_1} \cF_{n_2]p_1p_2p_3} \ ,
\\
\delta_\Lambda \cF_{n_1n_2n_3n_4} &= - 6 e_{p[n_1n_2} \cF_{n_3n_4]}{}^p - \frac{1}{6} f^{p_1p_2p_3} \cF_{n_1n_2n_3n_4p_1p_2p_3} \ ,
\\
\delta_\Lambda \cF_{n_1\ldots n_7}& = -35 e_{[n_1n_2n_3} \cF_{n_4...n_7]} - \frac{1}{2} f^{p_1p_2p_3} \cF_{n_1\dots n_7p_1p_2;p_3} \ ,
\label{evF3}
\\
\delta_\Lambda \cF_{n_1\ldots n_{10}} &= 4 e_{[n_1n_2n_3} \cF_{n_4\ldots n_{10}]}+\frac1{18} f^{p_1p_2p_3} \cF_{p_1[n_1\dots n_9,n_{10}]p_2p_3} \ ,
\\
\delta_\Lambda \cF_{n_1\ldots n_9;m} &= -28 e_{[n_1n_2n_3} \cF_{n_4\ldots n_9]m} -24  e_{m[n_1n_2} \cF_{n_3\ldots n_9]}
\label{evF11} \\
& - \frac12 f^{p_1p_2p_3} \cF_{n_1\dots n_9p_1;p_2p_3m} 
+\frac1{18} f^{p_1p_2p_3} \cF_{n_1\dots n_9m;p_1p_2p_3}  +f^{p_1p_2p_3} \cF_{n_1\dots n_9p_1p_2;p_3,m}  \ , \nn
\\
\delta_\Lambda \cF_{n_1\ldots n_{10};p_1p_2p_3} &=15 e_{p_1p_2][n_1} \cF_{n_2\dots n_{10}];[p_3} - \frac{135}{2} e_{p_1][n_1n_2} \cF_{n_3\dots n_{10}][p_2;p_3}   + f^{p_1p_2p_3} (\dots ) \; , 
\\
\delta_\Lambda \cF_{n_1\ldots n_{11};p,q} &= \frac{55}{2}  e_{p)[n_1n_2} \cF_{n_3\dots n_{11}];(q}  + f^{p_1p_2p_3} (\dots )\; , 
\label{evF22}
\end{align}
\end{subequations}
where we use the notation that two sets of indices separated by a semi-column are in the tensor product of the two respective antisymmetric tensor representations, while two  sets of indices separated by a comma are in the biggest irreducible component of the same tensor product, excluding further antisymmetrisation. For example, $\cF^{m,n}$ is a symmetric tensor while $ \cF_{n_1\ldots n_9;m} $ includes both the hooked irreducible representation $(9,1)$, written as $\cF_{n_1\ldots n_9,m}$,  and the antisymmetric rank $10$ tensor, written as $\cF_{n_1\ldots n_{10}}$.

We also need to describe the module $L(\Lambda_3)$, which we write with an auxiliary field strength $G^{\tI}$. The first components of $G^{\MLthree} \in R(\Lambda_3)\subset L(\Lambda_3)$ are
\be G^{\MLthree} = ( G_{n_1\dots n_8}; G_{n_1\dots n_9,p_1p_2}; \dots )\; , \ee
and they transform under~\eqref{eq:E11elt} as 
\be 
\delta G_{n_1\dots n_8} = - \frac12 f^{p_1p_2p_3} G_{n_1\dots n_8p_1;p_2p_3} \; , \qquad \delta G_{n_1\dots n_9;p_1p_2} = - 9 e_{p_1p_2[n_1} G_{n_2\dots n_9]} + f^{p_1p_2p_3} (\dots ) \; . 
\ee

\subsection{Field strengths and gauge transformations} 

\label{sec:GL11adj}

An important r\^ole is played by the tensor $C^{IM}{}_{\wa}$ arising from the tensor hierarchy algebra since it defines the field strengths out of the currents according to~\eqref{eq:FS}. {\allowdisplaybreaks 
To display the components  $C^{IM}{}_{\wa}$ it is convenient to write the linearised field strength 
\be 
F^I_\lin = C^{IM}{}_\alpha \partial_M \phi^\alpha+C^{IM}{}_{\ta} \partial_M X^\ta+C^{IM}{}_{{\Lambda}} \partial_M Y^{{\Lambda}}  +C^{IM}{}_{\tL} \partial_M Y^{\tL} \; , 
 \ee
 in terms of the field ${\phi}^\wa = (\phi^\alpha , X^{\tilde{\alpha}})$ and $Y^{\widehat{\Lambda}}$, where we use $\chi_M{}^{\tilde{\alpha}} = \partial_M X^{\tilde{\alpha}}$ and $\zeta_M{}^{\widehat{\Lambda}} = \partial_M Y^{\widehat{\Lambda}}$ to simplify notations. These linearised field strength $F^I_\lin $ are given by~\cite{Bossard:2017wxl,Bossard:2019ksx}\footnote{The last two field strengths, the $\partial^{n_1...n_5}$ terms in $\Fs{1}_{m_1\ldots m_7}^\lin $  and the $\partial^{pq}$ terms in $\Fs{3}_{n_1\ldots n_9;m}^\lin$  were not given in Eq. (4.36) of~\cite{Bossard:2017wxl}. Note that there is a typo in Eq.~(4.36g) of~\cite{Bossard:2017wxl} where the correct coefficient of the last term should be $\frac{48}{9!}$ instead $\frac{1}{12\cdot 9!}$. Note that the tensor $C^{IM}{}_\Lambda$ defined in this paper agrees with the one defined in~\cite{Bossard:2017wxl} for $\Lambda$ valued in $R(\Lambda_{10})$ which is all that enters here.}
 \begin{subequations}
\label{F}
\begin{align}
\Fs{-5}^{n_1,n_2}_\lin &=  \partial^{q(n_1} h_q{}^{n_2)} + \frac{1}{6!} \partial^{p_1p_2p_3p_4p_5p_6(n_1,n_2)} A_{p_1p_2p_3p_4p_5p_6}  +\dots\ , 
\\
\Fs{-5}_m{}^{n_1n_2n_3}_\lin &= - \eta^{n_1p_1} \eta^{n_2p_2} \eta^{n_3p_3} \partial_m A_{p_1p_2p_3} + 3 \partial^{[n_1n_2} h_m{}^{n_3]} + \frac{1}{2}\partial^{n_1n_2n_3p_1p_2} A_{mp_1p_2} 
\nn\\
 &\quad + \frac{1}{4!}    \partial^{n_1n_2n_3p_1p_2p_3p_4,q} A_{mp_1p_2p_3p_4q} - \frac{1}{5!} \partial^{n_1n_2n_3p_1p_2p_3p_4p_5} A_{mp_1p_2p_3p_4p_5}
\nn\\
&\quad + \frac{3}{2} \delta_m^{[n_1} \Bigl( \partial^{n_2|q} h_q{}^{n_3]} 
- \frac{1}{6} \partial^{n_2n_3]p_1p_2p_3} A_{p_1p_2p_3} - \frac{3}{2\cdot 5!}  \partial^{n_2n_3]p_1\dots p_5,q} A_{p_1\dots p_5q}\nn\\
& \hspace{40mm}  
+\frac{1}{6!}  \partial^{n_2n_3]p_1\dots p_6} A_{p_1\dots p_6} \Bigr) +\dots \ , 
\\
\Fs{-3}^\lin_{n_1n_2}{}^m &= 2 \partial_{[n_1} h_{n_2]}{}^m + \partial^{mp} A_{n_1n_2p} + \frac13\delta_{[n_1}^m  \partial^{p_1p_2} A_{n_2]p_1p_2} +\ldots\ ,
\\
\Fs{-1}_{m_1\ldots m_4}^\lin &= 4 \partial_{[m_1} A_{m_2m_3m_4]} - \frac12 \partial^{n_1n_2} A_{m_1\ldots m_4 n_1n_2} - \frac1{24} \partial^{n_1\ldots n_5} h_{m_1\ldots m_4n_1\ldots n_4,n_5}
\nn\\
&\quad + \frac1{5!} \partial^{n_1\ldots n_5} X_{m_1\ldots m_4n_1\ldots n_5} +\ldots \ ,
\\
\label{eq:F7imp}
\Fs{1}_{m_1\ldots m_7}^\lin &= 7 \partial_{[m_1} A_{m_2\ldots m_7]} + \partial^{n_1n_2} h_{m_1\ldots m_7n_1,n_2} - \frac12\partial^{n_1n_2} X_{m_1\ldots m_7n_1n_2} 
\nn\\
&\quad -\frac1{12} \partial^{n_1\ldots n_5} X_{m_1\ldots m_7n_1n_2n_3,n_4n_5} 
 - \frac1{24} \partial^{n_1\ldots n_5} X_{m_1\ldots m_7n_1\ldots n_4,n_5}   
 \nn\\
 &\quad - \frac1{24} \partial^{n_1\ldots n_5} Y_{m_1\ldots m_7n_1\ldots n_4,n_5} +\ldots \ ,
\\
\Fs{3}_{n_1\ldots n_9;m}^\lin &= 9 \partial_{[n_1} h_{n_2\ldots n_9],m}  +\partial_m X_{n_1\ldots n_9} + \frac12 \partial^{p_1p_2} A_{n_1\dots n_9,mp_1p_2}
\\*
& \quad +\frac{9}{10} \left(\partial^{pq} X_{p n_1\ldots n_9,m q}+\partial^{pq} X_{pm[ n_1\ldots n_8,n_9] q}\right) + \frac1{15}\partial^{pq} X_{n_1 \ldots n_{9}m ,pq} 
\nn\\*
&\quad 
 + \frac{27}{20} \left(\partial^{pq} X_{pq n_1\ldots n_9, m} +\partial^{pq} X_{pqm [n_1\ldots n_8,n_9]}\right)+\frac1{30} \partial^{pq} X_{n_1 \ldots n_{9}m p,q}
 \nn\\
&\quad+ \frac{9}{20}  \left(\partial^{pq} Y_{pq n_1\ldots n_9, m} +\partial^{pq} Y_{pqm [n_1\ldots n_8,n_9]}\right)+ \frac7{30} \partial^{pq} Y_{n_1 \ldots n_{9}m p,q} +\ldots \ , \nn
\\
\Fs{5}_{m_1\ldots m_{10};n_1n_2n_3}^\lin  & = 10 \partial_{[m_1} A_{m_2\dots m_{10}],n_1n_2n_3} + 3 \partial_{[n_1|} X_{m_1\dots m_{10},|n_2n_3]}\nn\\*
& \quad + \frac32   \partial_{[n_1|} X_{m_1\dots m_{10}|n_2,n_3]} + \frac32   \partial_{[n_1|} Y_{m_1\dots m_{10}|n_2,n_3]}  +  \dots \; , 
\\*
\Fs{5}_{m_1\ldots m_{11};n,p}^\lin  &= 11 \partial_{[m_1} B_{m_2\dots m_{11}],n,p} + 2 \partial_{(m|} C_{n_1\dots n_{11},|p)} \nn\\
& \quad + \frac52 \partial_{(n|} X_{m_1\dots m_{11},|p)} +   \frac12 \partial_{(n|} Y_{m_1\dots m_{11},|p)}+ \dots\; . 
\end{align}
\end{subequations}
Note that compare to~\cite{Bossard:2017wxl,Bossard:2019ksx} we find convenient to use the reducible field strength $\Fs{3}_{n_1\ldots n_9;m}^\lin = \Fs{3}_{n_1\ldots n_9,m}^\lin+\Fs{3}_{mn_1\ldots n_9}^\lin$, instead of its irreducible components. 
} 

We now write out the linearised gauge transformations of the coset potentials using the coset element~\eqref{eq:U11} together with the linearised metric  $g_{mn} = \eta_{mn} + \eta_{p(m} h_{n)}{}^p$. We also present the corresponding formulas for the lowest level constrained fields~\eqref{eq:L211} and~\eqref{eq:L1011}. According to~\eqref{eq:chiGT}, these gauge variations contain the tensors $T^{\alpha\ta}{}_\tb$, $\Pi^\ta{}_{MN}$ and $\Pi^\Lambda{}_{MN}$
\begin{subequations}
\label{xt}
\begin{align}
\delta_\xi h_{n}{}^m &=  \partial_n \xi^m -\partial^{mp} \lambda_{np} + \partial^m \xi_n - \partial_{np} \lambda^{mp} +\frac13\delta_n^m \partial_{pq} \lambda^{pq} + \ldots\ ,
\label{xt1}\\
\delta_\xi A_{n_1n_2n_3} &= 3\partial_{[n_1} \lambda_{n_2n_3]} + \frac12 \partial^{p_1p_2} \lambda_{n_1n_2n_3p_1p_2}  +  3 \partial_{[n_1n_2} \xi_{n_3]} +\ldots\ ,
\label{xt2}\w2
\delta_\xi  A_{n_1\cdots n_6} &=  6 \partial_{[n_1} \lambda_{n_2\cdots n_6]} 
- \partial^{p_1p_2} \xi_{n_1\cdots n_6p_1,p_2} 
+  \partial^{p_1p_2} \lambda_{n_1\cdots n_6p_1p_2}+\cdots \ ,
\label{xt22}\w2
\delta_\xi h_{n_1\cdots n_8,m} &= 8 \partial_{[n_1} \xi_{n_2\cdots n_8],m}
+24 \partial_{\!\lsharp \! n_1} \lambda_{n_2\cdots n_8 ,m\!\rsharp}  +\cdots \ ,
\label{xt23}\w2
\delta_\xi \chi_{M;n_1\ldots n_9} &=    24 \partial_M \partial_{[n_1} \lambda_{n_2\cdots n_9]}     - \varepsilon_{n_1\ldots n_9pq}  \partial_M \partial^{p} \xi^q +\ldots\ ,
\label{xt3}\w2
\delta_\xi \chi_{M;\, m_1\ldots m_{\scalebox{0.6}{$10$}},n_1n_2} &=  \varepsilon_{m_1\ldots m_{\scalebox{0.6}{$10$}}p} \partial_M \bigl(  \partial^p \lambda_{n_1n_2} + \tfrac15 \delta^p_{[n_1} \partial^q \lambda_{n_2]q} - \partial_{n_1n_2} \xi^p - \tfrac15 \delta^p_{[n_1} \partial_{n_2]q} \xi^q \bigr) +\ldots\ , 
\label{xt4}\\
\delta_\xi \chi_{M;\, m_1\ldots m_{\scalebox{0.6}{$11$}},n} &=  \frac1{10} \varepsilon_{m_1\ldots m_{\scalebox{0.6}{$11$}} } \partial_M ( \partial^p \lambda_{np} -  \partial_{np} \xi^p ) +\ldots \ ,
\label{xt5}\\
\delta_\xi \zeta_{M;\, m_1\ldots m_{\scalebox{0.6}{$11$}},n} &=  -\frac1{2} \varepsilon_{m_1\ldots m_{\scalebox{0.6}{$11$}} } \partial_M (  \partial^p \lambda_{np} + \partial_{np} \xi^p ) +\ldots\,,
\label{xt6}
\end{align}
\end{subequations}
where indices are lowered and raised with the Minkowski metric $\eta_{mn}$, and we distinguished $\xi_{n_1\dots n_7,m}$ from $\lambda_{n_1\dots n_8}$ at level $\frac92$.

We finally write the first components of $G^{\tI}$ appearing in~\eqref{eq:FM3} that defines the tensor $C^{\tI}{}_{M\wa}$ through $G^\tI_\lin = C^{\tI}{}_{M\wa} \eta^{MN} \partial_N \phi^\wa$.\footnote{Recall that there is no direct use of this field strength $G^{\tI} = C^{\tI}{}_{M\wa} \cM^{MN} J_N{}^\wa$ in the theory, but it is simply convenient to use $G^\tI_\lin$ to write the components of the tensor $C^{\tI}{}_{M\wa}$.}
The components of $G^{\MLthree}\in R(\Lambda_3)$ are given by~\cite{Bossard:2017wxl}
\begin{subequations}
\begin{align} 
G^\lin_{n_1\dots n_8} &= \bar  \partial^q ( h_{n_1\dots n_8,q}  +X_{ n_1\dots n_8q} )- 28 \bar \partial_{[n_1n_2} A_{n_3\dots n_8]} - 56 \bar \partial_{[n_1\dots n_5} A_{n_6n_7n_8] } \\ &\qquad  +  8  \bar \partial_{[n_1\dots n_7|,q} h_{n_8]}{}^q - 24 \bar \partial_{q[n_1\dots n_7} h_{n_8]}{}^q  \nn \nn\\
G^\lin_{n_1\dots n_9;p_1p_2} &= \bar \partial_{p_1p_2} X_{n_1\dots n_9} - \bar \partial^m X_{n_1\dots n_9 m,p_1p_2} + 2 \bar \partial^m X_{n_1\dots n_9 m[p_1,p_2]} 
+ \dots  
\end{align}
\end{subequations}
while the highest weight components in $R(\Lambda_1 + \Lambda_{10})\oplus R(\Lambda_{11})$ are combined in the reducible tensor
\be 
\tilde{G}^{ \lin}_{n_1\dots n_{10};m} = 10  \bar \partial_{m[n_1} X_{n_2\dots n_{10}]} -  \bar \partial^p X_{n_1\dots n_{10},mp} - 2\bar  \partial^p X_{n_1\dots n_{10}[p,m]} +\dots \; ,
\ee 
where we only give the components $C^{\tI}{}_{M\tilde{\alpha}} \eta^{MN} \partial_N X^{\ta}$ at level $\frac{11}{2}$.

\subsection{Bilinear forms}
\label{sec:bilGL11}

One defines the following invariant bilinear forms. The Killing--Cartan form expands as 
\begin{multline} 
\kappa^{\alpha\beta} \Phi^+_\alpha \Phi^+_\beta = h^+_m{}^n h^+_n{}^m - \tfrac12 h^+_m{}^m h^+_n{}^n + \frac{1}{3}  A^+_{n_1n_2n_3} A_-^{n_1n_2n_3}\\ +  \frac{2}{6!}  A^+_{n_1\dots n_6} A_-^{n_1\dots n_6} + \frac{2}{8!}  h^+_{n_1\dots n_8,m} h_-^{n_1\dots n_8,m} + \dots  \label{KCG11}
\end{multline}
One can also check that the $K(E_{11})$ invariant bilinear forms on $R(\Lambda_1)$ and $\mathcal{T}_{-1}$ respectively expand as
\begin{multline} 
\eta^{MN} \partial_M \partial_N  =\eta^{mn} \partial_m \partial_n +\frac12  \eta_{n_1p_1} \eta_{n_2p_2} \partial^{n_1n_2} \partial^{p_1p_2} + \frac1{5!}  \eta_{n_1p_1} \cdots \eta_{n_5p_5} \partial^{n_1\dots n_5} \partial^{p_1\dots p_5}   \\
+  \frac1{7!}  \eta_{n_1p_1} \cdots \eta_{n_7p_7}\eta_{mq}  \partial^{n_1\dots n_7,m} \partial^{p_1\dots p_7,q}+\frac1{7!}  \eta_{n_1p_1} \cdots \eta_{n_8p_8} \partial^{n_1\dots n_8} \partial^{p_1\dots p_8} + \dots  \; , 
\end{multline}
and 
\begin{align} 
\label{eq:LLT0}
\eta_{IJ} F^I F^J  &=\dots - \frac{1}{3 \cdot 10!} \cF_{n_1\dots n_{10};p_1p_2p_3} \cF^{n_1\dots n_{10};p_1p_2p_3}  - \frac1{2\cdot 9!} \cF_{n_1\dots n_{10};p_1p_2p_3} \cF^{p_1n_1\dots n_9;n_{10}p_2p_3} \nn\\
&\quad  + \frac1{11!} \cF_{n_1\dots n_{11};m,p} \cF^{n_1\dots n_{11};m,p}   + \frac{1}{8!} \cF_{n_1\dots n_9;m} \cF^{mn_1\dots n_8;n_9}+ \frac1{7!} \cF_{n_1\dots n_7 }  \cF^{n_1\dots n_7}  
\CR
& \quad + \frac{1}{4!} \cF_{n_1\cdots n_4}  \cF^{n_1\cdots n_4}   +\frac{1}{2} \cF_{n_1n_2}{}^m \cF^{n_1n_2}{}_m -\cF_{np}{}^p \cF^{nq}{}_q\CR
&\qquad +\frac{4}{6} \cF^{n_4}{}_{[n_1n_2n_3} \cF_{n_4]}{}^{n_1n_2n_3} + \cF_{m,n} \cF^{m,n} + \dots \ .
\end{align}
We have the $E_{11}$ invariant symplectic form 
 \begin{multline}
 \Omega_{IJ} F^I F^{\prime J} = \frac1{4!7!}  
 \varepsilon^{n_1\dots n_{11}} \Bigl( \tfrac16 \cF_{n_1n_2}{}^m \cF^\prime_{n_3\dots n_{11};m} - \tfrac13 \cF_{n_1m}{}^m \cF^\prime_{n_2\dots n_{10};n_{11}} - \cF_{n_1n_2n_3n_4} \cF^\prime_{n_5\dots n_{11}}  \\
  - \tfrac{1}{180} \cF_{n_1}{}^{p_1p_2p_3} \cF^\prime_{n_2\dots n_{11};p_1p_2p_3} + \tfrac{1}{60} \cF_q{}^{p_1p_2q} \cF^\prime_{n_1\dots n_{10};n_{11}p_1p_2} - \tfrac{1}{165} \cF^{m,p} \cF^\prime_{n_1\dots n_{11};m,p}+ \dots  \Bigr) \ ,
 \end{multline}
 where we included in the ellipses the terms with $\cF$ and $\cF^\prime$ exchanged that can be deduced by antisymmetry of the symplectic form. 

One compute that~\eqref{eq:CC=CCtw} is satisfied at this level provided the $K(E_{11})$ invariant bilinear form $\eta_{\tI\tJ}$ expands as
\bea \label{FFGG11D} 
\eta_{\tI \tJ} G^{\tI} G^{\tJ} &=& \frac{1}{8!} G_{n_1\dots n_8}  G^{n_1\dots n_8} + \frac{1}{2\cdot 9!} G_{n_1\dots n_9;p_1p_2}  G^{n_1\dots n_9;p_1p_2} 
\CR
&& \quad - \frac{1}{10!} \tilde{G}_{n_1\dots n_{10};m} \tilde{G}^{ n_1\dots n_{10};m} + \frac{11}{4 \cdot 10! } \tilde{G}_{[n_1\dots n_{10};m]} \tilde{G}^{ n_1\dots n_{10};m} \; .  
\eea
This exhibits in particular that the module $L(\Lambda_3)$ must indeed include  $R(\Lambda_1 + \Lambda_{10})\oplus R(\Lambda_{11})$. We note also that the coefficients of the irreducible components of the bilinear form $\eta_{\tI\tJ}$ are not all positive, it is positive for $R(\Lambda_3)$, negative for $R(\Lambda_1 + \Lambda_{10})$ and positive for $R(\Lambda_{11})$. 

\subsection{Closure of gauge transformations on the dual graviton}
\label{LowLevelFather}

Here, we shall verify the first non-trivial component of the identity~\eqref{eq:ID5} in $GL(11)$ level decomposition and show how it relates to the closure of gauge transformations on the dual graviton.
The first non-trivial component of this identity in the $GL(11)$ decomposition is when $\tilde{I}$ is at level $\frac{9}{2}$ and it corresponds to an eight-form. Let us write it as it appears in the closure of the algebra of generalised diffeomorphisms 
\be \label{Tocheck}h_\alpha C^{\tilde{I}}{}_{P \widehat{\beta}} T^{\widehat{\beta} M}{}_Q \bar C_{\tilde{I}}{}^{N\alpha}  \xi^P \partial_M  \partial_N  \xi^Q =  h_{\alpha} \Bigl( f^{\alpha\beta}{}_\gamma T^{\gamma  M}{}_P T_\beta{}^{N }{}_Q -2 \delta^{ M}_{[P} T^{\alpha N}{}_{Q]} \Bigr)  \xi^P \partial_M  \partial_N \xi^Q\; , \ee
in which case the level $\frac{9}{2}$ component $\tI$ of $\Sigma_M{}^{\tilde{I}}$ corresponds to the St\"{u}ckelberg gauge parameter $\Sigma_{m,n_1\dots n_8}$ that allows us to reabsorb the dual graviton field in a redefinition of the constrained  field $\chi_{m;n_1\dots n_9}$. 

For the left-hand side we observe that 
\begin{align} 
 C^{\tilde{I}}{}_{P \widehat{\beta}} T^{\widehat{\beta} M}{}_Q   \xi^P \partial_M   \xi^Q  =  C^{\tilde{I}}{}_{P \widehat{\beta}}   \xi^P \delta^+_\xi h^{\widehat{\alpha}} \,,
\end{align}
where we have introduced the notation 
\be  
\delta^+_\xi h^{\widehat{\alpha}}   = T^{\widehat{\alpha} M}{}_N \partial_M \xi^N \; , \label{Xiplus} 
\ee
so for the eight-form component one gets 
\bea
&&  \xi^q  \delta^+_\xi( h_{n_1\dots n_8,q}  +X_{ n_1\dots n_8q} )- 28 \lambda_{[n_1n_2}  \delta^+_\xi A_{n_3\dots n_8]} - 56 \lambda_{[n_1\dots n_5}  \delta^+_\xi A_{n_6n_7n_8] } \CR &&\qquad  +  8  \xi_{[n_1\dots n_7|,q}  \delta^+_\xi h_{n_8]}{}^q - 24 \lambda_{q[n_1\dots n_7}  \delta^+_\xi h_{n_8]}{}^q  \CR
&=& 8 \bigl( \xi^q \partial_{[n_1} ( \xi_{n_2\dots n_8],q} + 3 \lambda_{n_2\dots n_8]q})-21 \lambda_{[n_1n_2} \partial_{n_3} \lambda_{n_4\dots n_8]}-21 \lambda_{n_1\dots n_5} \partial_{n_6} \lambda_{n_7n_8} \CR
&& \qquad +( \xi_{[n_1\dots n_7|,q} + 3 \lambda_{[n_1\dots n_7|q}) \partial_{n_8]} \xi^q  \; . 
\eea
Using then the eight-form component of $ \bar C_{\tI}{}^{N\alpha}  \partial_N h_\alpha$
\be 
 \partial_m \bar h^{n_1\dots n_8,m} 
 \ee
one obtains for the left-hand side of \eqref{Tocheck}
\begin{multline}
 \label{lefteight}
  L^{n_1\dots n_8} =  \frac{1}{7!}  \bar h^{n_1\dots n_8,m}  \bigl( \xi^q \partial_m\partial_{[n_1} ( \xi_{n_2\dots n_8],q} + 3 \lambda_{n_2\dots n_8]q})+( \xi_{[n_1\dots n_7|,q} + 3 \lambda_{[n_1\dots n_7|q}) \partial_{n_8]} \partial_m \xi^q\\
-21 \lambda_{[n_1n_2|} \partial_m \partial_{n_3} \lambda_{n_4\dots n_8]}-21 \lambda_{n_1\dots n_5|} \partial_m  \partial_{n_6} \lambda_{n_7n_8} \Bigr)  \ . 
\end{multline}

For the right-hand side of \eqref{Tocheck} one first identifies
\be 
f^{\alpha\beta}{}_\gamma h_\alpha T^{\gamma M}{}_P \xi^P T_{\beta}{}^N{}_Q \xi^Q =- \delta_{\mathfrak{e}_{11}}( \delta_{\mathfrak{e}_{11}}({T_\beta}) h) \xi^M \times \delta_{\mathfrak{e}_{11}}({T^\beta}) \xi^N 
\ee
and we expand this component for $M=m$ and $N=n$
\begin{align}
 &\quad  \delta_{\mathfrak{e}_{11}}( \delta_{\mathfrak{e}_{11}}({T_\beta}) h) \xi^m \times \delta_{\mathfrak{e}_{11}}({T^\beta}) \xi^n \nn\\
 &= \delta_{\mathfrak{e}_{11}}(  \delta_{\mathfrak{e}_{11}}({K^q{}_p}) h) \xi^m \times  \delta_{\mathfrak{e}_{11}}({K^p{}_q}) \xi^n- \tfrac19 \delta_{\mathfrak{e}_{11}}(  \delta_{\mathfrak{e}_{11}}({K^p{}_p} ) h) \xi^m \times  \delta_{\mathfrak{e}_{11}}({K^q{}_q} )\xi^n \CR
 & \hspace{10mm}  +\frac{1}{6} \delta_{\mathfrak{e}_{11}}(  \delta_{\mathfrak{e}_{11}}({F^{p_1p_2p_3} })h) \xi^m \times  \delta_{\mathfrak{e}_{11}}({E_{p_1p_2p_3} }  )\xi^n + \dots  \CR
 &= \delta_{\mathfrak{e}_{11}}(  \delta_{\mathfrak{e}_{11}}({K^n{}_q}) h) \xi^m \times   \xi^q   +\frac{1}{2} \delta_{\mathfrak{e}_{11}}(  \delta_{\mathfrak{e}_{11}}({F^{nq_1q_2} })h) \xi^m \times  \lambda_{q_1q_2}  + \dots \CR
 &= -  \frac{1}{7!} \Bigl( \delta_{\mathfrak{e}_{11}}({K^n{}_q})  \bar h^{mp_1\dots p_7,r} ( \xi_{p_1\dots p_7,r} + 3 \lambda_{p_1\dots p_7r}) \times   \xi^q   \nn\\
 &\hspace{50mm}+ 21 \delta_{\mathfrak{e}_{11}}({F^{nq_1q_2} }) \bar A^{mp_1\dots p_5} \lambda_{p_1\dots p_5} \times \lambda_{q_1q_2} + \dots \Bigr) \CR
 &=  -  \frac{1}{7!} \Bigl(   \bar h^{np_1\dots p_7,r} ( \xi_{p_1\dots p_7,r} + 3 \lambda_{p_1\dots p_7r}) \times   \xi^m + 7  \bar h^{mnp_2\dots p_7,r} ( \xi_{p_1\dots p_7,r} + 3 \lambda_{p_1\dots p_7r}) \times   \xi^{p_1} \CR
 &\quad +    \bar h^{mp_1\dots p_7,n} ( \xi_{p_1\dots p_7,r} + 3 \lambda_{p_1\dots p_7r}) \times   \xi^r    -3\times  21  \bar h^{mp_1\dots p_5[q_1q_2,n]} \lambda_{p_1\dots p_5} \times \lambda_{q_1q_2} + \dots \Bigr)  
 \end{align}
Then one uses that 
\be 
h_{\alpha} T^{\alpha N}{}_Q  \xi^{Q} = \delta_{\mathfrak{e}_{11}}(h) \xi^N 
\ee
to write 
\be 
\delta_{\mathfrak{e}_{11}}(h) \xi^n =  -  \frac{1}{7!}  \bar h^{np_1\dots p_7,r} ( \xi_{p_1\dots p_7,r} + 3 \lambda_{p_1\dots p_7r}) \; ,
\ee
for the second term. Combining the two terms in the right-hand side \eqref{Tocheck} one finds indeed the left-hand side \eqref{lefteight} as computed above. 

This shows in particular from \eqref{lefteight} that the 3-form and 6-form potential gauge transformations do not close on themselves on the dual graviton field at the non-linear level, but they close up to a St\"{u}ckelberg gauge transformation of parameter 
\be \Sigma_{n_1\dots n_8,m} =  -21 \lambda_{[n_1n_2|} \partial_m \partial_{n_3} \lambda_{n_4\dots n_8]}-21 \lambda_{n_1\dots n_5|} \partial_m  \partial_{n_6} \lambda_{n_7n_8}  \; , \ee
 that does not affects the dual graviton field strength. The same is true for the commutator of a dual-graviton gauge transformation and a diffeomorphism, with 
\be \Sigma_{n_1\dots n_8,m} =    \xi^q \partial_m\partial_{[n_1} ( \xi_{n_2\dots n_8],q} + 3 \lambda_{n_2\dots n_8]q})+( \xi_{[n_1\dots n_7|,q} + 3 \lambda_{[n_1\dots n_7|q}) \partial_{n_8]} \partial_m \xi^q \; , \ee
similarly as in the vielbein formulation \cite{Boulanger:2008nd}.

 \subsection{On the cancelations in the pseudo-Lagrangian}
 \label{CancelO3}
 
 In this appendix we illustrate the general proof of Section \ref{sec:extall} by showing explicitly how the cancelation \eqref{OfromL} occurs for $k=3$. The case $k=3$ is particularly relevant to construct the dual graviton Lagrangian \eqref{DualGravityL} in Section \ref{DualGravitonIn11D}.
 
 We expand the terms in \eq{OfromL} for $k=3$, using~\eqref{eq:LLT0}. From the kinetic term we find
\be
\sum_{k=0,3} {\mathcal L}_{\rm kin} \big|_{k} = \frac14 \sqrt{-g} \Big( \frac{1}{2} \cF_{n_1n_2}{}^m \cF^{n_1n_2}{}_m -\cF_{nm}{}^m \cF^{np}{}_p - \frac{1}{8!} \cF_{n_1\dots n_{8}p;q} \cF^{n_1\cdot n_8q;p}  \Big)\ .
\ee
The potential term~\eqref{p1} gives 
\bea 
\mathcal{L}_{\text{pot}_1} \big|_{k=3} &=&  - \frac12    m_{\alpha_\dgr{3}\beta_{\dgr{3}}} m^{mn} \tilde{\mathcal{J}}_m{}^{\alpha_\dgr{3}} \tilde{\mathcal{J}}_n{}^{\beta_\dgr{3}} + \frac12  m_{\alpha_\dgr{3}\gamma_\dgr{3}}  T^{\gamma_\dgr{3} n}{}_{Q} T_{\beta_\dgr{3}}{}^Q{}_p m^{pm}   \tilde{\mathcal{J}}_m{}^{\alpha_\dgr{3}} \tilde{\mathcal{J}}_n{}^{\beta_\dgr{3}}   \CR
&=& -\frac{1}{2\cdot 8!} \sqrt{-g} g^{mn}  \cJ_{m;p_1\ldots p_8,q} \cJ_{n;}{}^{p_1\ldots p_8,q} \CR
&& \hspace{30mm} + \frac1{2\cdot 8!} \sqrt{-g}  g^{mn}  \Big( \cJ_{q;mp_1\ldots p_7,r} \cJ_{n;}{}^{qp_1\ldots p_7,r}  - \cJ_{q;mp_1\ldots p_7,r} \cJ_{n;}{}^{p_1\ldots p_7r,q} \Big)\CR
&=& \sqrt{-g} \Bigl(  -\frac{9}{2\cdot 8!} \cJ_{[n_1;n_2\ldots n_9],q} \cJ^{n_1;n_2\ldots n_9,q}
+ \frac{1}{2\cdot 8!} \cJ_{q;n_1\ldots n_8,p} \cJ^{p;n_1\ldots n_8,q} \Bigr) \ .
\eea
Finally, using~\eqref{FFGG11D}, \eqref{eq:Pi11} and \eqref{topoKT} we compute
\bea
\mathcal{L}_{\text{pot}_2}\big|_{k=3} \hspace{-1mm} &=&\hspace{-1mm}  -\frac{ \sqrt{-g}}{2\cdot 8!}\Big(  \cJ_{p;n_1\ldots n_8,q} \cJ^{q;n_1\ldots n_8,p}  + 2 \cJ_{[n_1;n_2\ldots n_9],p}\chi^{p;n_1\ldots n_9} +  \chi_{p;n_1\ldots n_9} \chi^{n_1;n_2\ldots n_9 p}\Big)\ ,\nn
\w2
\mathcal{L}_{\text{top}}\big|_{k=0,3}  \hspace{-1mm} &=&\hspace{-1mm}- \frac1{4\cdot 9!} \varepsilon^{n_1\ldots n_{11}}  \cF_{n_1n_2}{}^s \big(\cF_{n_3\ldots n_{11},s} + 9 \cF_{n_3\ldots n_{11}s}\big) \ .
\eea
Combining the results for $k=3$ given above allows us to check that 
\be
2 \cL_{\rm kin} \big|_{k=3} + \mathcal{L}_{\text{pot}_1} \big|_{k=3}  + \mathcal{L}_{\text{pot}_2}\big|_{k=3} = 0 \; , 
\ee
as proved for all $k\ge 3$ in Section \ref{sec:extall},  such that the sum of all terms gives $\mathcal{O}_3$ as in \eqref{OfromL}
\bea  
&&  \cL_{\rm kin} \big|_{k=0} + \cL_{\rm kin} \big|_{k=3} + \mathcal{L}_{\text{pot}_1} \big|_{k=3}  + \mathcal{L}_{\text{pot}_2}\big|_{k=3} +\mathcal{L}_{\text{top}}\big|_{k=0,3} \CR
&=& \sqrt{-g} \Big(\frac{1}{8} \cF_{n_1n_2}{}^m \cF^{n_1n_2}{}_m -\frac14 \cF_{nm}{}^m \cF^{np}{}_p- \frac{1}{4\cdot 8!} \cF_{n_1\dots n_{8}p;q} \cF^{n_1\cdot n_8q;p}   \Big)
\label{eq:L113} \CR
&& \hspace{20mm}- \frac1{4\cdot 8!} \varepsilon^{n_1\ldots n_{11}}\cF_{n_1n_2}{}^p \cF_{pn_3\ldots n_{10};n_{11}}   \CR
&=&\mathcal{O}_3 \ . 
\eea

\section{\texorpdfstring{$GL(3)\times E_8$ formulae}{GL(3)xE8 formulae}}
\label{app:E8}

Many details for the $E_8$ decomposition and the tensors appearing in the tensor hierarchy can be found in~\cite{Bossard:2019ksx}. We recall the salient features here and refer to Table~\ref{tab:e8dec} for a summary of the representations. We use the conventions and the explicit projectors onto $E_8$ irreducible representations of~\cite{Koepsell:1999uj}.\footnote{We here use capital latin letters $A,B,C$ for the adjoint instead of lower case.}

\subsection{\texorpdfstring{$E_{11}$ transformations}{E11 transformations}}

In order to describe the adjoint of $\mf{e}_{11}$ in this basis, we introduce formal fields that are not projected to the coset component, with a $+$ superscript for positive or null level and a $-$ superscript for negative level. These transform under the adjoint action of the elementary level $\pm 1$ $\mf{e}_{11}$ element 
\be \Lambda = f^\mu_A F^A_\mu+e_\mu^A E^\mu_A   \; , \ee
as
\begin{subequations}
\label{eq:adjointE8} 
\begin{align}
\delta_\Lambda h^+_\mu{}^\nu &= e_\mu^A A_A^{-\nu} - f^\nu_A A^+_\mu{}^A - \delta_\mu^\nu \bigl( e_\sigma^A  A_A^{-\sigma} - f^\sigma_A A^{+A}_\sigma \bigr) \,,\\ 
 \delta_\Lambda \Phi^+_A &= f_{AB}{}^C e_\mu^B A^{-\mu}_C - f_{AB}{}^C f^\mu_C A_\mu^{+B} \,,\\
  \delta_\Lambda A_\mu^{+A} &= - e_\nu^A h^+_\mu{}^\nu + f^{CA}{}_B e_\mu^B \Phi^+_C - f^\nu_B B^+_{\mu\nu}{}^{AB} - f^{AB}{}_C f^\nu_B h^+_{\mu,\nu}{}^C \,, \\
    \delta_\Lambda B_{\mu\nu}^{+AB}  &= 28 P^{AB}{}_{CD}  e_{[\mu}^C A^+_{\nu]}{}^D + \tfrac12 \kappa^{AB} \kappa_{CD} e_{[\mu}^C A^+_{\nu]}{}^D  + f^\sigma_C (\dots)   \,, \\
        \delta_\Lambda h_{\mu,\nu}^{+A}  &= -  f_{BC}{}^A e_{(\mu}^B A^+_{\nu)}{}^C + f^\sigma_B (\dots ) \,.
\end{align}
\end{subequations}
To write the relevant structure constant components of $T^{\alpha \ta}{}_\tb$ and $K^{\alpha \ta}{}_\beta$ we introduce a more general element 
\be \Upsilon =f^\mu_A F^A_\mu+ k_\mu{}^\nu K^{\mu}{}_\nu + k_A K^A + e_\mu^A E^\mu_A  + \frac{1}{28} \acute{e}_{\mu\nu}^{AB} \acute{E}_{AB}^{\mu\nu} + 2 e_{\mu\nu} E^{\mu\nu} + e_{\mu,\nu}^A E_A^{\mu,\nu}  \; ,  \ee
and the formal field $X^\ta$ in $L(\Lambda_2)$  with the indecomposable transformation 
\bea \label{deltaX} \delta_\Upsilon X_\mu &=& k_\mu{}^\nu X_\nu + f^\nu_A ( X_{\mu\nu}^A + h_{\mu,\nu}^A) + e_\mu^A \Phi_A - 2 e_{\mu\nu}^A A^{-\nu}_A \; , \CR
\delta_\Upsilon X_{\mu\nu}^A &=&  - 2 k_{[\mu}{}^\sigma X_{\nu]\sigma}^A - f^{CA}{}_B k_C X_{\mu\nu}^B - 2 e^A_{[\mu} X_{\nu]} - f_{BC}{}^A e_{[\mu}^B A_{\nu]}^C - e_{\mu\nu}^{AB} \Phi_B - 2 e_{\sigma,[\mu}^A h_{\nu]}{}^\sigma  \CR
&& + f_B^\sigma ( X_{\mu\nu\sigma}^{AB} + \dots ) \; ,  \CR
\delta_\Upsilon X_{\mu\nu\sigma}^{AB} &=&   3 k_{[\mu}{}^\rho X_{\nu\sigma]\rho}^{AB} +2 f^{C[A}{}_D k_C X_{\mu\nu\sigma}^{B]D} - 6 e_{[\mu}^{[A} X_{\nu\sigma]}^{B]} +  f_{CD}{}^{[A} e_{[\mu}^{|C|} B
_{\nu\sigma]}^{B]D} - 2 f_{CD}{}^{[A} e_{[\mu\nu}^{B]C} A_{\sigma]}^D  \CR
&& \qquad + f^{AB}{}_C  \bigl( 2 e_{[\mu}^C B_{\nu\sigma]} + \tfrac12 \kappa_{DE} e_{[\mu\nu}^{CD} A_{\sigma]}^E \bigr) + f_C^\rho (\dots )  \; ,  \eea
where $X_{\mu\nu\sigma}^{AB}$ is a reducible antisymmetric tensor in $AB$. It is not the only level 3 component (see Eq. \eqref{Level3E8} 
 for the exhaustive list), but the others structure coefficients do not appear in the topological term. 

The module $R(\Lambda_1)$ of derivatives transforms under $\mf{e}_{11}$ as
 \begin{subequations}
 \begin{align}
 \delta_\Lambda \partial_\mu &= e_\mu^A \partial_A \,,\\
  \delta_\Lambda \partial_A &= f_A^\mu \partial_\mu + e_\mu^B \partial^\mu_{AB} + f_{AB}{}^C e_\mu^B \partial^\mu_C \,,\\
  \delta_\Lambda \partial^\mu_{AB} &= 14 P^{CD}{}_{AB}   f^\mu_C \partial_D + \tfrac14 \kappa_{AB} \kappa^{CD} f^{\mu}_C \partial_D  + e_\nu^C (\dots)  \,, \\
  \delta_\Lambda \partial^\mu_A &= \tfrac12 f^{BC}{}_A f_B^\mu \partial_C + e_\nu^B (\dots ) \ ,
 \end{align}
 \end{subequations}
which determines the structure constants $T^{\alpha M}{}_N$. More specifically, we need  the transformation $\delta_\Phi \partial_M = - \Phi_\alpha T^{\alpha N}{}_M \partial_N$ for the coset element $\Phi_\alpha$ in order to write the potential terms \eqref{eq:Lpot1} and \eqref{tpot}.  One computes that  
 \bea \delta_\Phi^{\; 2} \partial_\mu \hspace{-0.5mm} &=& \hspace{-0.5mm} \bigl( ( h_\mu{}^\sigma  - \tfrac12 \delta_\mu^\sigma h_\rho{}^\rho) ( h_\sigma{}^\nu- \tfrac12 \delta_\sigma^\nu h_\lambda{}^\lambda)   + A_\mu^A \bar A_A^\nu + \tfrac1{14} \acute{B}_{\mu\sigma}^{AB} \acute{\bar B}_{AB}^{\nu\sigma} + 4 B_{\mu\sigma} \bar B^{\nu\sigma} + 2 h_{\mu\sigma}^A \bar h_A^{\nu\sigma} \bigr) \partial_\nu  \nn\\
&& \hspace{10mm} + \bigl( ( h_\mu{}^\nu  -   \tfrac12 \delta_\mu^\nu h_\sigma{}^\sigma) A_\nu^A - \tfrac12 A_\mu^A h_\nu{}^\nu -f^{AB}{}_C  A_\mu^C \Phi_B  +  B_{\mu\nu}^{AB} \bar A^\nu_B +f^{AB}{}_C h_{\mu\nu}^C \bar A^\nu_B \bigr) \partial_A  \nn\\
\delta_\Phi^{\; 2} \partial_A \hspace{-0.5mm} &=& \hspace{-0.5mm} \bigl( f^{CE}{}_A f^{DB}{}_E \Phi_C \Phi_D - f^{CB}{}_A \Phi_C h_\mu{}^\mu + \tfrac14 h_\mu{}^\mu h_\nu{}^\nu \\
&& \hspace{20mm} + \bar A_A^\mu A_\mu^B + A_\mu^B \bar A_A^\mu+ \delta_A^B A_\mu^C \bar A^\mu_C - f^{ED}{}_C f_{AE}{}^B A_\mu^C \bar A_D^\mu   \nn\\
&&\hspace{3mm} + f_{EA}{}^B f^{EC}{}_D ( \tfrac1{14} \acute{B}_{\mu\nu}^{DF} \acute{\bar{B}}_{CF}^{\mu\nu} + h^D_{\mu,\nu} \bar h^{\mu,\nu}_C)   + ( \tfrac{1}{14}   \acute{B}_{\mu\nu}^{CD} \acute{\bar{B}}_{CD}^{\mu\nu}  + 4  {B}_{\mu\nu} {\bar{B}}^{\mu\nu}  + 2 h_{\mu,\nu}^C \bar h_C^{\mu,\nu}) \delta_A^B  \bigr) \partial_B  \nn\\
&& \hspace{10mm} + \bigl( f^{CB}{}_A \Phi_C \bar A_B^\mu + \bar A_A^\nu ( h_\nu{}^\mu - \tfrac12 \delta_\nu^\mu h_\sigma{}^\sigma) - \tfrac12 h_\nu{}^\nu \bar A_A^\mu  + A_\nu^B \bar B^{\mu\nu}_{AB} - f_{AB}{}^C A_\nu^B \bar h_C^{\mu\nu} \bigr) \partial_\mu  \nn \eea
 where we keep track of the ordering of the fields to recall on which term the derivative acts in the potential terms.  Here we used the notation that 
 \be \bar A^\mu_A = \delta_{AB} \eta^{\mu\nu} A_\nu^B \; , \quad \bar B^{\mu\nu}_{AB} = \delta_{AC} \delta_{BD}  \eta^{\mu\sigma}\eta^{\nu\rho} B_{\sigma\rho}^{CD} \; ,  \quad  \bar h^{\mu,\nu}_{A} = \delta_{AB} \eta^{\mu\sigma}\eta^{\nu\rho} h_{\sigma,\rho}^{B}\; .   \ee

 The dual module of gauge parameters can be obtained by invariance of the bilinear form
 \begin{align}
  \xi^M \partial_M  = \xi^\mu \partial_\mu    +  \lambda^A \partial_A  + \tfrac1{14} \acute{\lambda}_{\mu}^{AB} \acute{\partial}_{AB}^{\mu}  +4 \lambda_{\mu} \partial^\mu + 2 \xi_\mu^{A} \partial_A^\mu\; . 
  \end{align}
  The transformations are simply obtained by using that the two modules induce the same $K(E_{11})$ representations. 
  
 The field strength $F^I$ in the representation $\mathcal{T}_{-1}$ decomposes into components that transform as
\begin{subequations}
 \begin{align}
   \delta_\Lambda F_{\mu A} &= e_\mu^B F_{AB} + e_\nu^B f_{AB}{}^C F_{\mu }{}^\nu_C - f_{AB}{}^C f_C^\nu F_{\mu\nu}^B + f^\nu_A F_{\nu;\mu}  \,,\\
 \delta_\Lambda F_{\mu\nu}{}^\sigma &= - 2 e_{[\mu}^A  F_{\nu]}{}^\sigma_A + 2 e_\rho^A \delta^\sigma_{[\mu} F_{\nu]}{}^\rho_A - f_A^\sigma F_{\mu\nu}^A - 2 f_A^\rho \delta^\sigma_{[\mu} F_{\nu]\rho}^A \,,\\
 \delta_\Lambda F_{\mu\nu}^A &= - e_\sigma^A F_{\mu\nu}{}^\sigma + 2 f^{AB}{}_C e_{[\mu}^C F_{\nu]B}  + f^B_\mu ( \dots) \,,\\
 \delta_\Lambda F_{\mu;\nu} &=  e_\nu^A  F_{\mu A} + f^\sigma_A ( \dots ) \,,\\
 \delta_\Lambda F_{AB} &= \bigl( 14 P^{CD}{}_{AB} + \tfrac14 \kappa_{AB} \kappa^{CD} \bigr) f_C^\mu F_{\mu D} +  e_\mu^C  (\dots )  \,,\\ 
 \delta_\Lambda F_{\mu}{}^\nu_A  &= f_A^\sigma F_{\mu\sigma}{}^\nu +f^{BC}{}_A \bigl( f_B^\nu F_{\mu C} - \tfrac12 \delta_\mu^\nu f_B^\sigma F_{\sigma C} \bigr)  +e_\sigma^B \bigl(  \dots  \bigr)   \ , \qquad  
\end{align}
\end{subequations}
and the auxiliary field strength $G^{\tilde{I}}$ in ${L(\Lambda_3)}$, have components in $R(\Lambda_3)$ transforming as
\begin{subequations}
\begin{align}
 \delta_\Lambda G &= - f_A^\mu G_\mu^A \; , \qquad  \delta G_\mu^A = - e_\mu^A G + f^\nu_B ( G_{\mu\nu}^{A;B} +G_{\mu,\nu}^{A;B} ) \; , \\
 \delta_\Lambda G_{\mu\nu}^{A;B} &= e_{[\mu}^A G_{\nu]}^B - 3  e_{[\mu}^B G_{\nu]}^A - f_{CE}{}^B f^{AE}{}_D e_{[\mu}^C G_{\nu]}^D + f^\sigma_C (\dots)   \; , \\
  \delta_\Lambda G_{\mu,\nu}^{A;B} &= 2 e_{(\mu}^{(A} G_{\nu)}^{B)} +f_{CE}{}^B f^{AE}{}_D e_{(\mu}^C G_{\nu)}^D  + f^\sigma_C (\dots) \; , 
\end{align}
\end{subequations}
 where $G_{\mu\nu}^{A;B} $ belongs to the ${\bf 1}\oplus {\bf 3875}\oplus{\bf 248} \oplus {\bf 30380}$, with
\begin{align}
 G_{\mu\nu}^{A;B} =  G_{\mu\nu}^{[A;B]} + \acute{ G}_{\mu\nu}^{AB}  + \kappa^{AB} G_{\mu\nu} \,,\end{align}
while $G_{\mu,\nu}^{A;B} $ belongs to the ${\bf 1}\oplus {\bf 3875}\oplus{\bf 248}$ with
\begin{align}
G_{\mu,\nu}^{A;B} =   \acute{ G}_{\mu,\nu}^{AB}  + \kappa^{AB} G_{\mu,\nu} + f^{AB}{}_C G_{\mu,\nu}^C\; .  
\end{align}

\subsection{Gauge transformations and field strengths} 

 In this section we give the linearised gauge transformations and field strengths. We introduce the notation that bar fields or derivative are understood to be conjugated with the background $\cM$ matrix as
 \be \bar \partial^\mu = \eta^{\mu\nu} \partial_\nu\; , \quad \bar \partial^A = \delta^{AB} \partial_B\; , \quad \bar \partial_\mu^{AB} = \delta^{AC} \delta^{BD} \eta_{\mu\nu} \partial^\nu_{CD} \; , \quad \bar \partial_\mu = \eta_{\mu\nu} \bar \partial^\nu \; , \ee
 and in particular the level $ \frac52$ derivative $\bar \partial_\mu$  should not be confused with the level $-\frac12 $ derivative $\partial_\mu$. 
 
 Using this notation, the linearised gauge transformations of the $\mf{e}_{11}$ fields  read
  \begin{subequations}
   \label{GaugeStrucCoeff} 
\begin{align} 
  \delta_\xi  h_\mu{}^\nu &= \partial_\mu \xi^\nu - \tfrac1{14} \acute{\partial}^\nu_{AB} \acute{\lambda}_\mu^{AB} - 4 \partial^\nu \lambda_\mu + 2 \partial^\nu_A \xi_\mu^A + 2 \delta_\mu^\nu \Bigl( \partial_A \lambda^A+ \tfrac17  \acute{\partial}^\sigma_{AB} \acute{\lambda}_\sigma^{AB}+8 \partial^\sigma \lambda_\sigma   \Bigr) \nn\\
  &\quad + \bar{\partial}^\nu \bar{\xi}_\mu - \tfrac1{14} \acute{\bar{\partial}}_\mu^{AB} \acute{\bar{\lambda}}^\nu_{AB} - 4 \bar{\partial}_\mu \bar{\lambda}^\nu + 2 \bar{\partial}_\mu^A \bar{\xi}^\nu_A  + \dots \ ,\\
  \delta_\xi \Phi_A &= ( f_{AB}{}^C - \delta_{AD}  f^{DC}{}_B) \bigl( - \partial_C \lambda^B- \tfrac17  \partial_{CD}^\mu \lambda_\mu^{BD} + 2 \partial_C^\mu \xi_\mu^B  \bigr) + \dots \\
  \delta_\xi  A_\mu{}^A &= \partial_\mu \lambda^A + \partial_B \lambda^{AB}_\mu - f^{AB}{}_C \partial_B \xi_\mu^C+ \bar \partial^A \bar \xi_\mu  + \bar{\partial}^{AB}_\mu \bar\lambda_B -f^{AB}{}_C \bar{\partial}^C_\mu \bar\lambda_B  + \dots \\
  \delta_\xi B_{\mu\nu}^{AB} &= 2 \partial_{[\mu} \lambda_{\nu]}^{AB}   -2\bar{\partial}^{AB}_{[\mu}  \bar{\xi}_{\nu]}+ \dots\ ,\\
  \delta_\xi h_{\mu,\nu}^{A} &= 2 \partial_{(\mu} \xi^A_{\nu)}  +2\bar{\partial}_{(\mu}^A \bar{\xi}_{\nu)} + \dots
 \end{align}
 \end{subequations}
 whereas the ones of the constrained fields $\chi_M{}^{\ta} = \partial_M X^\ta$ are defined by the variations of $X^\ta$ as
 \bea
  \delta_\xi X_\mu &=& 2 \partial_A \xi^A_\mu +2 \bar \partial^A_\mu  \bar \lambda_A - \varepsilon_{\mu\nu\sigma} \bar \partial^\nu \xi^\sigma  + \dots \ ,\CR
 \delta_\xi X_{\mu\nu}^A &=& - 2  \partial_{[\mu} \xi^A_{\nu]}  + 2 \bar \partial_{[\mu}^A \bar\xi_{\nu]} - \varepsilon_{\mu\nu\sigma}  ( \bar \partial^\rho \lambda^A - \bar \partial^A \xi^\rho)+ \dots\; , \CR
 \delta_\xi X_{\mu\nu\sigma}^{AB} &=&- 2 \varepsilon_{\mu\nu\sigma}  \bar \partial^{[A}  \lambda^{B]}  + \dots  \ .
 \eea

We define the linearised field strengths in the same way as  
\begin{subequations}
 \begin{align}
 \Fs{3}_{\mu\nu\sigma}^{AB} &= 3 \partial_{[\mu} B_{\nu\sigma]}^{AB} \; , \\
 \Fs{3}_{\mu\nu;\sigma}^{A} &= 2 \partial_{[\mu} h_{\nu],\sigma}^A + \partial_\sigma X_{\mu\nu}^A \; , \\
 \Fs{1}_{\mu\nu}{}^A&= 2 \partial_{[\mu} A_{\nu]}^{A} - \partial_B B_{\mu\nu}^{AB} - f^{AB}{}_{C} \partial_B X_{\mu\nu}^C \ ,\nn\\
 \Fs{1}_{\mu;\nu} &= \partial_\mu X_\nu + \partial_A X^A_{\mu\nu} - \partial_A h^{A}_{\mu,\nu}  \\
  \Fs{-1}_{\mu\nu}{}^\sigma &= 2 \partial_{[\mu} h_{\nu]}{}^\sigma +  \tfrac1{14} \acute{\partial}^{\sigma}_{AB} \acute{B}_{\mu\nu}^{AB} +4\partial^\sigma B_{\mu\nu} +2 \partial^\sigma_A  X_{\mu\nu}^{A}  +2\delta^\sigma_{[\mu} \bigl(\partial_A A_{\nu]}^A  +\tfrac17 \acute{\partial}^\rho_{AB} \acute{B}_{\nu]\rho}^{AB} + 8 \partial^\rho B_{\nu]\rho}   ) \,,\nn\\
 \Fs{-1}_{\mu A}  &=  \partial_\mu \Phi_A + f_{AB}{}^C \partial_C A_\mu^B + \partial_A X_\mu  \nn\\
 &\quad + \tfrac17 f_{AB}{}^C {\partial}^\nu_{CD} {B}^{BD}_{\mu\nu} - \partial^\nu_{AB} ( h^B_{\mu,\nu} + X^B_{\mu\nu}) - f_{AB}{}^C \partial_C^\nu (h^B_{\mu,\nu} - X^B_{\mu\nu})+  \dots \\
  \Fs{-3}_{AB} &= \bigl( 14 P^{CD}{}_{AB}  + \tfrac14 \kappa^{CD} \kappa_{AB} \bigr) \partial_C \Phi_D \nn = 2 \partial_{(A} \Phi_{B)} + f^{EC}{}_{A} f_{EB}{}^{D} \partial_{(C} \Phi_{D)}  \,,\\
\Fs{-3}_{\mu}{}^\nu_A  &=  \partial_\mu \bar{A}^{\nu}_A - \partial_A h_\mu{}^\nu - \partial^\nu_{AB} A_\mu^B - f_{AB}{}^C \partial_C^\nu A_\mu^A - \tfrac12 \delta_\mu^\nu f^{BC}{}_A \partial_B \Phi_C + \delta_\mu^\nu \partial^\sigma_{AB} A_\sigma^B  \ ,\\  
 \Fs{-5}_{\sigma AB}^{\hspace{1mm} \mu\nu} &= \partial_\sigma \bar{B}_{AB}^{\mu\nu} + 2 \partial_{AB}^{[\mu} h_\sigma{}^{\nu]}  - 2 \delta_\sigma^{[\mu} \bigl( ( 14 P^{CD}{}_{AB} + \tfrac14 \kappa^{CD} \kappa_{AB} ) \partial_C^{\nu]} \Phi_D + f^{CD}{}_{(A} \acute{\partial}_{B)C}^{\nu]} \Phi_D  \bigr) \; , \\
{}^{\scalebox{0.6}{$(\frac{-5}{2})$}}{}\hspace{-0.6mm} \grave{F}_{AB}^{\mu}  &= \bigl(14 P^{CD}{}_{AB} + \tfrac14 \kappa^{CD} \kappa_{AB} \bigr)  ( \partial_C \bar{A}_D^{\mu} + \partial_C^\mu \Phi_D )  -2 \partial_{AB}^\nu h_\nu{}^\mu  - f^{CD}{}_{(A} \acute{\partial}_{B)C}^{\mu} \Phi_D    \, ,\\
\Fs{-5}_{\sigma A}^{\hspace{0.2mm} \mu,\nu} &= \partial_\sigma  \bar{h}_{A}^{\mu,\nu}   - 2 \partial_A^{(\mu} h_\sigma{}^{\nu)} + \delta_\sigma^{(\mu} \bigl(  \partial_{AB}^{\nu)} \Phi^B + 4 \partial^{\nu)} \Phi_A + 3 f^{BC}{}_A \partial_B^{\nu)} \Phi_C \bigr)   \; , \\ 
{}^{\scalebox{0.6}{$(\frac{-5}{2})$}}{}\hspace{-0.6mm} \check{F}_{AB}^{\mu} &= 2\partial_{[A}  \bar{A}^{\mu}_{B]}  -   2\partial^\mu_{[A} \Phi_{B]}  +  f^{CD}{}_{[A} \acute{\partial}^\mu_{B]C} \Phi_D - \tfrac16  f_{AB}{}^C \acute{\partial}^\mu_{CD} \Phi^D  + \tfrac16  f_{AB}{}^C \partial^\mu \Phi_C \; ,
 \end{align}
 \end{subequations}
which determine the coefficients  $C^{IM}{}_{\widehat{\alpha}}$. One obtains the coefficient $C^{IM}{}_\Lambda$ up to a normalisation constant using the same construction as in Appendix \ref{app:MID}, by proving that there is a unique highest weight vector in the module $R(\Lambda_{10})$ and a unique highest weight vector in the module $R(2\Lambda_3)$ of the form $\overline{Y}_{\Lambda} = \Omega_{IJ} C^{I M}{}_\Lambda \partial_M F^J $ given together by 
\be \overline{Y}_{AB} = \partial_\mu \bigl(  \Fs{-5}_{\nu  AB}^{\hspace{1mm} \mu\nu}- {}^{\scalebox{0.6}{$(\frac{-5}{2})$}}{}\hspace{-0.6mm} \grave{F}_{AB}^{\mu}    \bigr) + \partial_C \bigl( ( 14 P^{CD}{}_{AB} + \tfrac14 \kappa^{CD} \kappa_{AB} ) \Fs{-3}_{\mu}{}^\mu_D- f^{CD}{}_{(A}  \Fs{-3}_{B)D} \bigr) + \dots  \label{Ext2L3} \ee
Moreover, it follows from \eqref{BRSTextended} that the coefficient $C^{IM}{}_\Lambda$ for the irreducible representation $R(2\Lambda_3)$ is non-zero. This allows us to extend the proof of \eqref{eq:2prove} to the submodule $R(\Lambda_{10})\oplus R(2\Lambda_3) \subset L(\Lambda_{10})$.

In this paper we also need the structure coefficients $C^{\tilde{I}}{}_{M \wa}$ that are conveniently combined on the auxiliary field strength in $L(\Lambda_3)$
\be G^\tI = C^{\tI}{}_{M \wa} \cM^{MN} \cJ_N{}^\wa  \sim C^{\tI}{}_{M \wa}  \bar{\partial}^M \Phi^{\wa} \; . \ee
The $R(\Lambda_3)$ component $G^{\MLthree}$ of $G^{\tI}$ can be defined in the linearised approximation as
\begin{subequations}
\label{eq:GM3}
\begin{align}
 G &= \bar \partial^\mu X_\mu  + \bar \partial^A \Phi_A - 2 \bar \partial_\mu^A \bar{A}^{\mu}_A + \dots  \; , \label{eq:GM3a} \\
G^{ A}_\mu &= \bar \partial^\nu ( X_{\mu\nu}^A - h_{\mu,\nu}^A) + \bar \partial^A X_\mu + f_{BC}{}^A \bar \partial^B A_\mu^C-2 \bar \partial_\nu^A h_\mu{}^\nu + \bar \partial^{AB}_\mu \Phi_B - f^{AB}{}_C \bar \partial_\mu^C \Phi_B + \dots \; . \\
G_{\mu\nu}^{A;B} &= 2 \bar \partial^{[A} X_{\mu\nu}^{B]} - f_{BC}{}^A \bar \partial^C B^{BD}_{\mu\nu} - 4 \bar \partial_{[\mu}^A A_{\nu]}^B - 2 f_{CE}{}^{(A} f^{B)E}{}_D \bar \partial_{[\mu}^{C} A_{\nu]}^D + 2 f_{CD}{}^{[A} \bar\partial^{B]C}_{[\mu} A_{\nu]}^D \nn\\
& \hspace{60mm} +\bar \partial_{[\mu}^{AB} X_{\nu]} - f^{AB}{}_C \bar \partial_{[\mu}^C X_{\nu]} \\
G_{\mu,\nu}^{A;B} &= 2 \bar \partial^{(A} h_{\mu,\nu}^{B)} + f_{CE}{}^A f^{BE}{}_D \bar \partial^C h_{\mu,\nu}^D-2 f_{CD}{}^{(A} \bar \partial^{B)C}_{(\mu} A_{\nu)}^D\, .
\end{align}
\end{subequations}
At level $\frac72$, $G^{\tilde{I}}$ also includes components in $R(\Lambda_1+\Lambda_{10})$ starting with $\acute{\tilde{G}}_{\mu\nu}^{AB}$, components in $R(\Lambda_1+2\Lambda_{3})$ starting with ${\tilde{G}}_{\mu\nu}$ and  components in  $R(\Lambda_1+\Lambda_{4})$ starting with $\tilde{G}_{\mu\nu}^{A}$ that are given by 
\bea 
\tilde{G}_{\mu\nu}^{AB} &=& 2 \bar \partial^{(A} X_{\mu\nu}^{B)}  + f_{CE}{}^{(A} f^{B)E}{}_D \bar \partial^C X_{\mu\nu}^D - f_{CD}{}^{(A} \bar \partial^C B^{B)D}_{\mu\nu} -2 f_{CD}{}^{(A} \bar \partial_{[\mu}^{B)C} A_{\nu]}^D \\
&& \hspace{60mm} -4 \bar\partial_{[\mu}^{(A} A_{\nu]}^{B)} - 2 f_{CE}{}^{(A} f^{B)E}{}_D \bar \partial^C_{[\mu}  A_{\nu]}^D  -2\bar \partial_{[\mu}^{AB} X_{\nu]}\; ,   \CR
\tilde{G}_{\mu\nu}^{A} &=& f_{BC}{}^A \bar \partial^B X^C_{\mu\nu} - \tfrac13 \bar \partial_B B_{\mu\nu}^{AB} - \tfrac43 \bar \partial^A B_{\mu\nu} - 4 \bar \partial_{[\mu}^A X_{\nu]} - \tfrac23 \bar\partial^{AB}_{[\mu} A_{\nu] B} - \tfrac83 \bar \partial_{[\mu} A_{\nu]}^A - 2 f_{BC}{}^A \bar \partial^B_{[\mu} A_{\nu]}^C\nonumber\; , 
\eea
where we have combined $\tilde{G}_{\mu\nu}^{AB} = \acute{\tilde{G}}_{\mu\nu}^{AB} +\kappa^{AB} {\tilde{G}}_{\mu\nu}$. 

\subsection{Bilinear forms}

One  defines the following invariant bilinear forms. The Killing--Cartan form expands as 
\begin{multline} 
\kappa^{\alpha\beta} \Phi^+_\alpha \Phi^+_\beta = h^+_\mu{}^\nu h^+_\nu{}^\mu - \tfrac12 h^+_\mu{}^\mu h^+_\nu{}^\nu + \kappa^{AB} \Phi^+_A \Phi^+_B + 2  A^{-\mu}_A A_\mu^{+A} \\ + \frac1{14} \acute{B}^{-\mu\nu}_{AB} \acute{B}_{\mu\nu}^{+AB} + 4 B^{-\mu\nu} B^+_{\mu\nu} + 2 h^{-\mu,\nu}_A h_{\mu,\nu}^{+A} + \dots 
\end{multline}
One can also check the $K(E_{11})$ invariant bilinear form on $R(\Lambda_1)$ and $\mathcal{T}_{-1}$ respectively expand as
\begin{multline} 
\eta^{MN} \partial_M \partial_N  =\eta^{\mu\nu} \partial_\mu \partial_\nu + \delta^{AB} \partial_A \partial_B  + \frac{1}{14} \delta^{AC} \delta^{BD} \eta_{\mu\nu} \acute{\partial}^\mu_{AB}  \acute{\partial}^\nu_{CD} \\ + 4 \eta_{\mu\nu}\partial^\mu \partial^\nu + 2 \delta^{AB} \eta_{\mu\nu} \partial^\mu_A \partial^\nu_B + \dots  
\end{multline}
and 
 \begin{multline} 
 \eta_{IJ} F^I F^J = \tfrac12  \eta^{\mu\rho} \eta^{\nu\lambda}  \eta_{\sigma\kappa} F_{\mu\nu}{}^\sigma F_{\rho\lambda}{}^\kappa -  \eta^{\mu\nu} F_{\mu\sigma}{}^\sigma F_{\nu\rho}{}^\rho + \eta^{\mu\nu} \delta^{AB} F_{\mu A} F_{\nu B} \\ + \tfrac12 \eta^{\mu\sigma} \eta^{\nu\rho} \delta_{AB} F_{\mu\nu}^A F_{\sigma\rho}^B + \eta^{\mu\sigma} \eta^{\nu\rho} F_{\mu;\nu} F_{\rho;\sigma}  \\+ \tfrac1{14} \delta^{AC} \delta^{BD} \acute{F}_{AB} \acute{F}_{CD} + 4 F^2 +  \delta^{AB} ( F_{\mu}{}^\nu_A F_{\nu}{}^\mu_B -F_{\mu}{}^\mu_A F_{\nu}{}^\nu_B ) + \dots  \ ,
 \end{multline}
 with $F_{AB} = \acute{F}_{AB}  + \kappa_{AB} F $ and $ \acute{F}_{AB}$ in the $\bf 3875$. 
 
 We have  the $E_{11}$ invariant symplectic form 
 \begin{multline}\label{SympE8} 
 \Omega_{IJ} F^I F^{\prime J} = \tfrac12 
 \varepsilon^{\mu\nu\sigma} \Bigl(- F_{\sigma A} F^{\prime A}_{\mu\nu} + F_{\mu\nu}{}^\rho F^\prime_{\sigma;\rho}  - 2 F_\mu{}_A^\rho  F_{\rho\nu;\sigma}^{\prime \hspace{1.8mm} A}    - \tfrac1{3\cdot 14} \acute{F}_{AB} \acute{F}_{\mu\nu\sigma}^{\prime \hspace{0.3mm} AB}  - \tfrac{4}{3} F F^\prime_{\mu\nu\sigma}  \\
 +  F_{\mu\nu}^A F^\prime_{\sigma A}  - F_{\sigma;\rho}F_{\mu\nu}^{\prime \hspace{2.2mm} \rho}+  2F_{\rho\nu;\sigma}^{\hspace{2mm} A} F^\prime_\mu{}_A^\rho    + \tfrac1{3\cdot 14}\acute{F}_{\mu\nu\sigma}^{ \hspace{0.9mm} AB}   \acute{F}^\prime_{AB}  + \tfrac{4}{3}  F_{\mu\nu\sigma} F^\prime + \dots  \Bigr) \ .
 \end{multline}
We have also 
\begin{multline}  
\eta_{\MLthree\NLthree} G^{\MLthree} G^{\NLthree} = G^2 + \eta^{\mu\nu} \delta_{AB} G^A_\mu G^B_\nu + \frac{1}{14} \acute{G}_{\mu,\nu}^{AB} \acute{\bar{G}}^{\mu,\nu}_{AB}  + 4 G_{\mu,\nu} \bar G^{\mu,\nu}  + 2G^A_{\mu,\nu} \bar G_A^{\mu,\nu}  \\ +  \frac{1}{14} \acute{G}_{\mu\nu}^{AB} \acute{\bar{G}}^{\mu\nu}_{AB}  + 4 G_{\mu\nu} \bar G^{\mu\nu} + \frac14 G^{[A;B]}_{\mu\nu} \bar G^{\mu\nu}_{[A;B]} + \frac1{272} f_{AB}{}^E f^{CD}{}_E  G^{[A;B]}_{\mu\nu} \bar G^{\mu\nu}_{[C;D]}   + \dots \; .
\end{multline}
Here, we have used the indices $\MLthree$, $\NLthree$ to denote the representation $R(\Lambda_3)\subset L(\Lambda_3)$. 
One checks that \eqref{eq:ID7} is satisfied to this level for the  bilinear form $\eta_{\tI\tJ}$
\begin{align}
 \label{GG=FFE8}  \eta_{\tilde{I}\tilde{J}} G^{\tilde{I}} G^{\tilde{J}} =  \eta_{\MLthree\NLthree} G^{\MLthree} G^{\NLthree}  -  \frac{1}{28} \acute{\tilde{G}}_{\mu\nu}^{AB} \acute{\bar{\tilde{G}}}^{\mu\nu}_{AB}  - 2 \tilde{G}_{\mu\nu} \bar{ \tilde{G}}^{\mu\nu}  + \frac{9}{34} \tilde{G}^A_{\mu\nu} \bar {\tilde{G}}^{\mu\nu}_A\; . 
\end{align}
As in \eqref{FFGG11D}, we see that the irreducible components of $\eta_{\tI\tJ}$ come with alternating sign, positive for $R(\Lambda_3)$, negative for $R(\Lambda_1+\Lambda_{10})$ and $R(\Lambda_1+2\Lambda_3)$ and positive for $R(\Lambda_1+\Lambda_4)$. One also consistently finds that both  $R(\Lambda_3)$ and $R(\Lambda_1+\Lambda_{10})$ appear with the canonical normalisation. 

\subsection{Consistency checks of the identities}
\label{LowLevelFatherE8}

In order to give a consistency check of various $E_{11}$ identities that we have introduced in this paper, it is useful to check the results of Section \ref{sec:E8} by doing an explicit level expansion of the various terms up to level 2 fields. 

{\allowdisplaybreaks
In particular, one can check identities \eqref{eq:ID4} and \eqref{eq:ID2} from the computation of the topological term. The explicit expansion up to level 2 fields gives 
\begin{align}
 \Pi_{\tilde{\alpha}}{}^{MN}  {\Theta}_{MN}{}^{\tilde\alpha} &= \varepsilon^{\mu\nu\sigma} \Bigl( 2 \partial_\mu \chi_{\nu;\sigma} - \cJ_{\mu;\nu}{}^\rho \chi_{\sigma;\rho} + M_{AB} g^{\rho\lambda} \cJ_{\mu;}{}_\lambda^B ( \chi_{\nu;\sigma\rho}^{\hspace{2mm} A} - \cJ_{\nu;\sigma,\rho}^{\hspace{2mm} A} )  + \cJ_{\mu;\nu}^{\hspace{2mm} A} \cJ_{\sigma;A}  \CR
& \hspace{10mm} + \partial_\mu \chi_{A;}{}_{\nu\sigma}^A - \cJ_{\mu;\nu}{}^\rho \chi_{A;}{}_{\sigma\rho}^{A} - \tfrac12 \cJ_{\mu;C} f^{CA}{}_B \chi_{A;}{}_{\nu\sigma}^B - \cJ_{\mu;\nu}^{\hspace{2mm} A} \chi_{A;\sigma} \CR
& \hspace{50mm}  + \tfrac12 f_{BC}{}^A \cJ_{\mu;\nu}^{\hspace{2mm} B} \cJ_{A;}{}_\sigma^C+ \tfrac12 \cJ_{\mu;\nu\sigma}^{\hspace{2mm} AB} \cJ_{A;B} + \cJ_{\mu;\nu,\rho}^{\hspace{2mm} A} \cJ_{A;}{}_{\sigma}{}^\rho \CR
& \hspace{10mm} - \partial_A \chi_{\mu;\nu\sigma}^{\hspace{2mm} A}- \cJ_{A;\mu}{}^\rho \chi_{\nu;\sigma\rho}^{\hspace{2mm} A} + \tfrac12 \cJ_{A;C} f^{CA}{}_B \chi_{\mu;\nu\sigma}^{\hspace{2mm} B} - \cJ_{A;}{}_\mu^A \chi_{\nu;\sigma} \CR
& \hspace{50mm} + \tfrac12 f_{BC}{}^A \cJ_{A;}{}_\mu^B \cJ_{\nu;\sigma}^{\hspace{2mm} C}  - \tfrac12 \cJ_{A;}{}_{\mu\nu}^{AB} \cJ_{\sigma;B} + \cJ_{A;}{}_{\rho,\mu}^A \cJ_{\nu;\sigma}{}^\rho \CR
& \hspace{10mm}  + \tfrac13 \partial_A \chi_{B;}{}_{\mu\nu\sigma}^{AB} -  \cJ_{A;}{}_\mu^{[A} \chi_{B;}{}_{\nu\sigma}^{B]}+ \tfrac16 f_{AB}{}^{[C} \cJ_{C;}{}_{\mu\nu}^{D]A} \cJ_{D;}{}_\sigma^B  + \dots \Bigr) \CR
&= \varepsilon^{\mu\nu\sigma} \Bigl( 2 \partial_\mu \chi_{\nu;\sigma} + \partial_\mu \chi_{A;}{}_{\nu\sigma}^A - \partial_A \chi_{\mu;\nu\sigma}^{\hspace{2mm} A} + \tfrac13 \partial_A \chi_{B;}{}_{\mu\nu\sigma}^{AB} \CR
& \hspace{10mm} - \tfrac12 ( 2 \cJ_{\mu;\nu}{}^\rho + 2 \delta^\rho_\mu \cJ_{A;}{}_\nu^A) ( \chi_{\sigma;\rho} + \chi_{B;}{}_{\sigma\rho}^B - \cJ_{B;}{}_{\sigma,\rho}^B ) \CR
& \hspace{10mm} + ( M_{AB} g^{\rho\lambda} \cJ_{\mu;\lambda}^{\hspace{2mm} B} - \cJ_{A;\mu}{}^\rho - \tfrac12 \delta_\mu^\rho f^{BC}{}_A \cJ_{B;C} ) ( \chi_{\nu;\sigma\rho}^{\hspace{2mm} A} - \cJ_{\nu;\sigma,\rho}^{\hspace{2mm} A} ) \CR
& \hspace{5mm}  +\tfrac12 ( 2 \cJ_{\mu;\nu}^{\hspace{2mm} A} - \cJ_{B;}{}_{\mu\nu}^{AB}) ( \cJ_{\sigma;A} - \chi_{A;\sigma}) - \tfrac12 f^{AB}{}_C \chi_{B;}{}_{\mu\nu}^C (  \cJ_{\sigma;A} + f_{AD}{}^E \cJ_{E;}{}_\sigma^D +\chi_{A;\sigma}) \CR
& \hspace{10mm}  + \tfrac12 \cJ_{\mu;\nu\sigma}^{\hspace{2mm} AB} \cJ_{A;B} - \tfrac16 f_{AB}{}^{D} \cJ_{C;}{}_{\mu\nu}^{CA} \cJ_{D;}{}_\sigma^B +\dots \Bigr) \CR
&= \varepsilon^{\mu\nu\sigma} \Bigl( 2 \partial_\mu \chi_{\nu;\sigma} + \partial_\mu \chi_{A;}{}_{\nu\sigma}^A - \partial_A \chi_{\mu;\nu\sigma}^{\hspace{2mm} A} +\tfrac13 \partial_A \chi_{B;}{}_{\mu\nu\sigma}^{AB} \CR
& \hspace{10mm} - \tfrac12 \cF_{\mu\nu}{}^\rho \cF_{\sigma;\rho} +\tfrac12 \cF_{\mu\nu}^A \cF_{\sigma;A} + \cF_\mu{}_A^\rho \cF_{\rho\nu;\sigma}^{\hspace{2mm} A}    + \tfrac1{6\cdot 14} \acute{\cF}_{\mu\nu\sigma}^{\hspace{2mm} AB} \acute{\cF}_{AB} + \tfrac{2}{3} \cF_{\mu\nu\sigma} \cF  + \dots   \CR
& \hspace{15mm} - \cF_{\mu\nu}^{\hspace{2mm} A} \chi_{A;\sigma} -f_{AB}{}^C \cJ_{\mu;\nu}^{\hspace{2mm} A} \cJ_{C;\sigma}^{\hspace{2mm} B}   +\tfrac13  f_{AB}{}^{D} \cJ_{C;}{}_{\mu\nu}^{CA} \cJ_{D;}{}_\sigma^B   \Bigr) 
\end{align}
which indeed matches the results in \eqref{eq:Ltope8} using \eqref{SympE8}. The explicit level expansion of the other terms in the pseudo-Lagrangian is more complicated and we won't display it in this paper. We have computed them and we obtain indeed the same resulting $E_8$ Lagrangian provided the $E_8$ tensor identity 
\bea &&\Bigl(  \frac{1}{2} \delta_A^B \delta_C^E \delta_D^F  + 7 \delta_A^E \delta_C^B \delta_D^F + f_{GA}{}^B f^{GE}{}_C \delta_D^F \Bigr) \acute{B}^{CD} \acute{\bar{B}}_{EF} \CR
&=& \Bigl(  - 2 f_{AC}{}^G f^{BE}{}_G \delta_D^F - 7 \kappa_{AC} \kappa^{BE} \delta_D^F + \frac32 f_{AC}{}^E f^{BF}{}_D  \Bigr) \acute{B}^{CD} \acute{\bar{B}}_{EF} \eea
for  $\acute{B}^{CD} $ and $\acute{\bar{B}}_{EF} $ in the ${\bf 3875}$ is satisfied. This identity must therefore be a consequence of \eqref{e12Equation}. 
}

We can also check explicitly the first level component of Identity \eqref{Father}. The first non-trivial component in the $E_8$ decomposition is when $\tilde{I}$ is the highest weight singlet. At first-order one can take all the indices to be in the adjoint of $E_8$, with $C^{\tilde{I}}{}_{QA} \rightarrow \kappa_{QA}$ and \eqref{eq:ID5} reduces to 
\be -  \kappa_{PB}  f^{BM}{}_Q \kappa^{NA} \partial_M \partial_N = \bigl( f^{AB}{}_C f^{CM}{}_P f^N{}_{QB}+ 2 \delta^M_{[P}  f^{AN}{}_{Q]} \bigr) \partial_M \partial_N \ee
which is indeed equivalent to Equation (A.1) of \cite{Hohm:2014fxa}. This identity is necessary for the invariance of the potential  and the closure of the algebra in \cite{Hohm:2014fxa}, so that it is satisfactory that it appears exactly as the simplest component of the generalised identities required for $E_{11}$.


\section{\texorpdfstring{$E_{10}$ exceptional field theory and the $E_{10}$ sigma model}{E10 exceptional field theory and the E10 sigma model}}
\label{app:E10} 

The  $GL(1)\times E_{10}$ level decomposition of $E_{11}$ exceptional field theory as put forward in this paper should correspond to $E_{10}$ exceptional field theory in the same way that the $GL(D)\times E_{11-D}$ level decompositions correspond to $E_{11-D}$ exceptional field theory as we have shown for $D=11$ (meaning usual $D=11$ supergravity) in Section~\ref{sec:GL11} and for $D=3$ in Section~\ref{sec:E8}. As such $E_{10}$ exceptional field theory is expected to be able to describe $D=11$ supergravity, but this requires the internal coordinates and extra constrained fields. By contrast, the one-dimensional $E_{10}$ sigma model, proposed by Damour, Henneaux and Nicolai and studied in~\cite{Damour:2002cu,Damour:2002et,Damour:2007dt}, has no internal coordinates and no extra constrained fields but is also conjectured to describe full $D=11$ supergravity~\cite{Damour:2002cu}. 
In this appendix, we develop some initial ideas on the relationship between the $GL(1)\times E_{10}$ decomposition of our model and the one-dimensional $E_{10}$ sigma model. For the comparison we have to make a number of assumptions that will be highlighted along the way.

\subsection{\texorpdfstring{$E_{10}$ exceptional field theory from $E_{11}$}{E10 exceptional field theory from E11}}

To obtain $E_{10}$ exceptional field theory from our model we must branch all the relevant $E_{11}$ representations under $GL(1) \times E_{10}$. This branching is more involved because $\mf{e}_{10}$ is an indefinite Kac--Moody algebra and generically all $GL(1)$ levels will be infinite direct sums of irreducible $\mf{e}_{10}$ modules. We shall write irreducible modules of $\mf{e}_{10}$ as $R_\ten(\lambda)$ and bounded weight modules $L_\ten(\lambda)$, similar to $\mf{e}_{11}$. The numbering conventions for the $\mf{e}_{10}$ fundamental weights are such that $\Lambda_{10}$ denotes the exceptional fundamental weight while $\Lambda_1,\ldots,\Lambda_9$ are the fundamental weights along the `gravity line' $\mf{gl}(10)\subset \mf{e}_{10}$. 

The branching of the adjoint of $\mf{e}_{11}$ under $\mf{gl}(1)\oplus \mf{e}_{10}$ gives~\cite{Kleinschmidt:2003pt}
\begin{align}
\label{eq:e11e10}
 \mf{e}_{11}= \dots \oplus  L_\ten(\Lambda_3)^\ord{-2} \oplus R_\ten(\Lambda_1)^\ord{-1}\oplus \bigl( \mf{gl}(1)\oplus \mf{e}_{10}\bigr)^\ord{0} \oplus \overline{R_\ten(\Lambda_1)}^\ord{1}\oplus \dots 
\end{align}
where superscripts are the $\mf{gl}(1)$-grading and $L_\ten(\Lambda_3)$ is given by
\begin{align}
\label{eq:ASC10}
L_\ten(\Lambda_3) = R_\ten(\Lambda_1) \wedge R_\ten(\Lambda_1) \ominus R_\ten(\Lambda_2) \; .
\end{align}
This module is the representation in which antisymmetric derivatives $\partial_{[M} A \partial_{N]} B$ must vanish according to the section constraint. We define similarly for the symmetric section constraint the module 
\begin{align}
\label{eq:SSC10}
 L_\ten(\Lambda_9) = R_\ten(\Lambda_1) \SYM R_\ten(\Lambda_1) \ominus R_\ten(2\Lambda_1)
\end{align}
such that $\partial_{(M} A \partial_{N)} B|_{L_\ten(\Lambda_9)} = 0$.
The module  $R(\Lambda_2) \subset L(\Lambda_2)$ of the first rung of constrained fields $\chi_M{}^\ta$ branches similarly as
\begin{align}
R(\Lambda_2) = R_\ten(\Lambda_1)^\ord{0} \oplus ( R_\ten (\Lambda_2) \oplus L_\ten (\Lambda_3)\oplus L_\ten (\Lambda_9))^\ord{-1}\oplus \dots 
\end{align}
and the constrained fields $\zeta_M{}^\Lambda$ and $\zeta_M{}^{\tL}$ branch as
\begin{align}
 L(\Lambda_{10}) = L_\ten(\Lambda_9)^\ord{-1}\oplus \dots \; , \qquad L(\Lambda_{4}) = L_\ten(\Lambda_3)^\ord{-1}\oplus \dots 
\end{align}
The  module for the coordinates branches as
\begin{align}
R(\Lambda_1)= {\bf 1}^\ord{\frac12} \oplus R_\ten(\Lambda_1)^\ord{-\frac12} \oplus \bigl(L_\ten(\Lambda_9)\oplus L_\ten(\Lambda_3) \bigr) ^\ord{-\frac32} \dots
\end{align}
The interpretation of this branching is that the singlet corresponds to the external time direction $t$ (that will be identified with that of the one-dimensional $E_{10}$ model below) while $R_\ten(\Lambda_1)$ is the first set of internal coordinates subject to the section constraints that two derivatives vanish in the representations~\eqref{eq:ASC10} and~\eqref{eq:SSC10}. 

The branching of the field strength module $\cT_{-1}(\mf{e}_{11})$ under $\mf{e}_{10}$ is more complicated to obtain. Checking the representations that appear under the $GL(10)$ and $GL(2)\times E_8$ decompositions shows that 
\begin{align}
\cT_{-1}(\mf{e}_{11}) \supset  \bigl( R_\ten(\Lambda_1) \oleft \mf{e}_{10} \oleft R_\ten(\Lambda_1) \bigr)^\ord{-\frac12} \oplus  \bigl( \overline{R_\ten(\Lambda_1)} \oleft \mf{e}_{10}^* \oleft \overline{R_\ten(\Lambda_1)} \bigr)^\ord{\frac12} \  .  
\end{align}
This is consistent with the property that the weight $\frac{1}{2}$ component must include the dual of $\cT_0(\mf{e}_{10})\supset \mf{e}_{10} \oleft R_\ten(\Lambda_1)  $ as $\cT(\mf{e}_{10}) \subset \cT(\mf{e}_{11})$. It is conjectured in~\cite{Cederwall:2021ymp} that  $\cT_0(\mf{e}_{10})= \mf{e}_{10} \oleft R_\ten(\Lambda_1)  $.  In analogy with \eqref{DoubleExt}, we use the following notation for this doubly indecomposable module 
 \begin{align}
   \adjhhat_\ten = R_\ten(\Lambda_1) \oleft \mf{e}_{10} \oleft R_\ten(\Lambda_1) \; . 
 \end{align}
  We do not have a proof that these representations exhaust the field strength representations at levels $-\frac12$ and $\frac12$, but we will assume in this section that if any other bounded weight representation $L_\ten(\lambda_{\rm D})$  appears,  it does as a direct sum, such that
\begin{align}
\cT_{-1}(\mf{e}_{11}) = \dots \oplus \bigl(  \adjhhat_\ten \oplus L_\ten(\lambda_{\rm D}) \bigr)^\ord{-\frac12} \oplus \bigl(  \adjhhat_\ten^* \oplus \overline{L_\ten(\lambda_{\rm D})} \bigr)^\ord{\frac12} \oplus \dots  \ .  
\end{align}
It follows that the duality equation for the hypothetical field strengths in the modules $L_\ten(\lambda_{\rm D})$ and $\overline{L_\ten(\lambda_{\rm D})}$ do not mix with the equations for $\adjhhat$ and its conjugate. We assume that the duality equation for $L_\ten(\lambda_{\rm D})$ does not impose additional constraints on the $E_{10}$ fields, and we shall therefore ignore the corresponding  hypothetical field strengths. 
Using the results of Appendix \ref{app:e8tha} one obtains that $(\Lambda_2,\lambda_{\rm D})<-3$ and so $(\Lambda_{10},\lambda_{\rm D})<-6$.\footnote{This excludes the module $R_\ten(2\Lambda_1)$ and $R_\ten(3\Lambda_1)$ that define indecomposable extensions of   $ \mf{e}_{10} \oleft R_\ten(\Lambda_1) $.}

We write the level $\frac12$ derivative $\partial_t$ with a $0$ index when convenient, and the weight $-\frac12$ derivative $\partial_M$ where $M$ is now understood as an $E_{10}$ index in $\overline{R_\ten(\Lambda_1)}$.  The $E_{10}$ section constraint can be written as  
\be
 T^{\alpha P}{}_M T_\alpha{}^{Q}{}_N \partial_P \otimes \partial_Q = \partial_N \otimes \partial_M - \partial_M \otimes \partial_N \; .   
 \ee
We will ignore all the derivatives appearing at lower levels. The part of the level $\frac12$ field strength  in $ \adjhhat_\ten$ decomposes as
\begin{align}
\label{Fonehalf}
F^\ord{\frac12} = (  \acute{F}^M  \;  , \; F^\alpha \; ,\;F^M ) \; ,
\end{align}
where $\alpha$ is an adjoint $\mf{e}_{10}$ index. This field strength transforms indecomposably under an $\mf{e}_{10}$ transformation with parameter $\Lambda_\alpha$ as
\begin{align}
\delta_\Lambda \acute{F}^M &= \Lambda_\alpha   T^{\alpha M}{}_N  \acute{F}^N \; , \CR
\delta_\Lambda  F^\alpha &= \Lambda_\gamma  f^{\alpha \gamma}{}_\beta  F^\beta +\Lambda^\gamma    \overline{K}_{\gamma M}{}^{\alpha } \acute{F}^M \; ,  \CR
\delta_\Lambda {F}^M &=\Lambda_\alpha   T^{\alpha M}{}_N  {F}^N + \Lambda_\gamma    {K}^{\gamma M}{}_{\alpha}  F^\alpha  +   \Lambda_\alpha K^{\alpha M}{}_N \acute{F}^N \; . \label{E10Dind}
\end{align}
The indecomposable part $ {\widehat{{\rm adj}}}_\ten= \mf{e}_{10} \oleft R_\ten(\Lambda_1)$ of $\cT_0(\mf{e}_{10})$ is the quotient submodule that can be obtained by setting $\acute{F}^M=0$ in these transformations. The acute accent distinguishes $\acute{F}^M$ in the submodule $R_\ten(\Lambda_1)$ from $F^M$ in the quotient module $R_\ten(\Lambda_1) = \adjhhat_\ten/ {\widehat{{\rm adj}}}_\ten$.  The existence of the invariant bilinear form $\eta_{IJ}$ in $\cT_{-1}(\mf{e}_{11})$ implies that 
\begin{align}
\overline{K}_{\gamma M}{}^{\alpha } = \eta_{\gamma\delta} \eta^{\alpha\beta} \eta_{MN} K^{\delta N}{}_\beta \; . 
\end{align}
From $\cT(\mf{e}_{10})$ we also have the $\cT_0(\mf{e}_{10})$ representation matrices $\cT^{\wa P }{}_Q  = ( \cT^{\alpha  P}{}_Q , \cT^{M  P}{}_Q)$. We shall use $\wa = (\alpha , M)$ as an index for $ {\widehat{{\rm adj}}}_\ten$ and $\wwatext = ( M , \alpha , M) $ for $\adjhhat_\ten$, with $F^{\wwa}$ at $\mf{gl}(1)$ level $\frac12$ and $F_{\wwa}$ at level $-\frac12$. 

In order to derive the $E_{10}$ exceptional field theory duality equations and pseudo-Lagrangian, we use the semi-flat formulation as defined in Section \ref{sec:SF}. In this case the Levi subgroup $GL(1)\times E_{10}$ is also infinite-dimensional, so the expressions we write below are only defined for the minimal group  $E_{10}^{{\rm m}}$, and should be written in a specific level decomposition to be extended to the maximal group $E_{10}^{{\rm c}+}$, as discussed in Section \ref{sec:subtle}. We shall only briefly discuss this subtlety in the next section.

At level $-\frac12$ we have the field strengths 
\begin{align}
F^\ord{-\frac12} &=  (  \acute{F}_M  \;  , \; F_\alpha \; ,\;F_M ) 
\end{align}
that we can parametrise explicitly using~\eqref{eq:e11e10} with the help of an einbein $e$  (a.k.a. lapse) for the $GL(1)$ component, a generalised metric $M_{MN}$ for the adjoint of $E_{10}$ and a Kaluza--Klein vector $A^M$ in $R_\ten(\Lambda_1)$ for level $+1$. This leads to 
\begin{align}
 {F}_{\wwa} =   C_{\wwa}{}^{M}{}_{\wb} J_M{}^\wb  + \delta_{\wwa}^M \bigl( e^{2} M_{MN}  \partial_t A^N  - 2e^{-1} \partial_M e\bigr)
 \end{align}
with explicit components
\begin{align}
\label{eq:F.76}
F^\ord{-\frac12} 
&= 
( \acute{C}_{M}{}^{N}{}_{\wb} J_N{}^\wb\;  , \; C_{\alpha}{}^{N}{}_{\wb} J_N{}^\wb\; ,\; C_{M}{}^{N}{}_{\wb} J_N{}^\wb + e^{2} M_{MN} \partial_t A^N  - 2e^{-1} \partial_M e ) \; . 
\end{align}
As for $F^{\wwa}$, we distinguish the element $\acute{F}_M$ (respectively the tensor $\acute{C}_{M}{}^{N}{}_{\wb} $) that transforms in the submodule $\overline{R_\ten(\Lambda_1)}$ from the element $F_M$ that transforms indecomposably. The relative coefficient in the last term is determined by linearised gauge invariance with\footnote{The true gauge transformation is $\delta_\xi e = \partial_t  ( \xi^0 e)$ but the coefficient is halved when we do not consider the conjugate non-covariant term.}
\begin{align}
\delta_{\xi}^+ \bar A_M = \partial_M \xi^0 \; , \qquad \delta_\xi^+ e =\tfrac12  \partial_t  \xi^0 \; , 
\end{align}
with the notation of \eqref{Xiplus}. By representation theory it is consistent to include a vector of $R_\ten(\Lambda_1)$ as $ e^{2} M_{MN} \partial_t A^N  - 2e^{-1} \partial_M e $  in the definition of the field strength $F_M$. By contrast, trying to construct a linear combination of this field in $R_\ten(\Lambda_1)$ that transforms according to \eqref{E10Dind}, shows that there can be no such terms for $\acute{F}_M$ and $F_\alpha$.

We now study the constraints on the remaining components $C_{\wwa}{}^{M}{}_{\wb}$ appearing in~\eqref{eq:F.76} imposed by the doubly indecomposable structure. Firstly,  $\acute{F}_M$ is a tensor in the submodule $R_\ten(\Lambda_1)$, so $\acute{C}_{M}{}^{N}{}_{P}$ must be an invariant  tensor. But there is no such  invariant tensor since $R_\ten(\Lambda_1) \otimes R_\ten(\Lambda_1) \not\supset R_\ten(\Lambda_1)$, so we conclude that  $\acute{C}_{M}{}^{N}{}_{P}=  0 $. Then it follows that $\acute{C}_{M}{}^{N}{}_{\alpha}$ is an invariant tensor because $\Delta^\gamma \acute{C}_{M}{}^{N}{}_{\alpha} = - K^{\gamma P}{}_\alpha \acute{C}_{M}{}^{N}{}_{P}$, and using that there is a unique homomorphism from $R_\ten(\Lambda_1) \otimes \mf{e}_{10} \rightarrow R_\ten(\Lambda_1)$ (see footnote \ref{R1edE1}), one obtains that up to an overall coefficient 
\begin{align}
\label{eq:E.17}
\acute{C}_{M}{}^{N}{}_{\wb} J_N{}^\wb =   T_\beta{}^N{}_M J_N{}^\beta \; . 
\end{align}
For simplicity we assume that the coefficient is $1$.
The indecomposable structure is then consistent if one takes
\begin{align}
\label{eq:E.78}
C_{\alpha}{}^{N}{}_{\wb} J_N{}^\wb = 2 K_{(\alpha}{}^N{}_{\beta)} J_N{}^\beta + T_{\alpha}{}^M{}_N \chi_M{}^N\; . 
\end{align}
Indeed, as in \eqref{eq:Dcoc} and \eqref{eq:Dchi}
\begin{align}
\label{eq:Dcoc10}
\Delta^\alpha K^{\beta M}{}_\gamma =  - K^{\alpha M}{}_\delta f^{\beta\delta}{}_\gamma -T^{\beta M}{}_N K^{\alpha N}{}_\gamma \; , \qquad \Delta^\alpha  \chi_M{}^N = K^{\alpha N}{}_\beta J_M{}^\beta\; , 
\end{align}
so that 
\bea 
\Delta^\gamma \bigl( C_\alpha{}^M{}_\wb J_N{}^\wb \bigr)  &=& \Delta^\gamma \bigl( 2 K_{(\alpha}{}^N{}_{\beta)} J_N{}^\beta + T_{\alpha}{}^M{}_N \chi_M{}^N \bigr)  \CR &=& \bigl( 2 K^{\gamma M}{}_\delta f_{(\alpha\beta)}{}^{\delta} -2 T_{(\alpha}{}^{M}{}_N K^{\gamma N}{}_{\beta)} +  T_{\alpha}{}^M{}_N K^{\gamma N}{}_\beta \bigr) J_M{}^\beta \CR
&=& - K^{\gamma M }{}_\alpha  (  T_\beta{}^N{}_M J_N{}^\beta )= -K^{\gamma M }{}_\alpha  \bigl(  \acute{C}_M{}^N{}_\wb J_N{}^\wb \bigr)   \; . 
\eea
Using the same argument as in equation \eqref{XfromPhiPhi} and below one proves that there is no homomorphism from  $\mf{e}_{10}\otimes \mf{e}_{10}\to  R(\Lambda_1)$, so the structure coefficients  $C_{\alpha}{}^{N}{}_{\wb}$ are uniquely fixed to \eqref{eq:E.78}. The structure coefficients $C_{M}{}^{N}{}_{\wb}$ are  uniquely determined by the indecomposable representation as well, up to a term in $T_\beta{}^N{}_M $ as we saw in~\eqref{eq:E.17}. We will not  attempt to write these last coefficients in terms of other known coefficients as they will turn out to be irrelevant for the $E_{10}$ exceptional field theory.

In summary, we conclude that the level $-\frac12$ field strength~\eqref{eq:F.76} decomposes as
\be
F^\ord{-\frac12} = (  T_\beta{}^N{}_M J_N{}^\beta \;  , \; 2 K_{(\alpha}{}^N{}_{\beta)} J_N{}^\beta + T_{\alpha}{}^N{}_P \chi_N{}^P \; ,\; C_{M}{}^{N}{}_{\wb} J_N{}^\wb + e^{-2} M_{MN} \partial_t A^N - 2e^{-1} \partial_M e  ) \; . 
\ee

We shall now determine the level  $\frac{1}{2}$ field strength \eqref{Fonehalf}. Because $\acute{F}^M$ is a tensor in $R_\ten(\Lambda_1)$, it does not depend on $J_t{}^\alpha$ and $\partial_M A^N$ by representation theory. 
Therefore we conclude that 
\begin{align}
\acute{F}^M = \chi_N{}^{M;N}
\end{align}
for the constrained field $\chi_P{}^{M;N}$, whose symmetric component in $MN$ vanishes in $R_\ten(2\Lambda_1)$. In principle, we can write the same term for the $\zeta_P{}^{M;N}$ field, but we assume that the field $\chi_P{}^{M;N}$ has been redefined in order to absorb it. We can always do this because $ L_\ten(\Lambda_3)\oplus L_\ten(\Lambda_9)\subset R_\ten(\Lambda_2) \oplus L_\ten(\Lambda_3)\oplus L_\ten(\Lambda_9)$ trivially. We define the fields such that the free coefficients in this expression are all $1$, i.e. that there is no relative coefficient for each irreducible representation in $R_\ten(\Lambda_2) \oplus L_\ten(\Lambda_3)\oplus L_\ten(\Lambda_9)$.\footnote{Representation theory only gives that it is the sum over irreducible representations $\sum_\lambda c_\lambda \chi_N{}^{(M;N)_\lambda}$, but by a choice of normalisation we can set $c_\lambda = 1$ for all irreducible representations, unless some of them vanish.} Then the indecomposable structure implies that the remaining component of the weight $\frac12$ field strength is 
\begin{align}
 F^{\wa} = J_t{}^{\wa} + T^{\wa M}{}_N \partial_M A^N\; + C^{\wa P\! }{}_{M;N} \chi_P{}^{M;N} + \delta^{\wa}_M \zeta_{N}{}^{M;N} \; . 
\end{align}
The coefficients $C^{\wa P\! }{}_{M;N}$ are determined by the indecomposable structure such that 
\begin{align}
 \Delta^\gamma C^{\alpha  P\! }{}_{M;N} = \kappa^{\gamma \delta} \overline{K}_{\delta M}{}^\alpha \delta_N^P - \kappa^{\gamma \delta} \overline{K}_{\delta (M}{}^\alpha \delta_{N)_{2\Lambda_{\scalebox{0.4}{$1$}}}}^P \; , \label{FsCchi} 
\end{align}
where one removes the component $MN$ in $\overline{R(2\Lambda_1)}$ as indicated by the notation $(MN)_{2\Lambda_{\scalebox{0.4}{$1$}}}$. 

The duality equation $M_{\wwa\wwb} F^{\wwb}=F_{\wwa}$ relating $F^{\ord{\frac12}}$ and $F^{\ord{-\frac12}}$ is therefore
\begin{subequations} 
\label{E10Duality}
\begin{align}
\label{E10Dualitya}
  M_{MP} \chi_N{}^{P;N}   &=  e T_\beta{}^N{}_MJ_N{}^\beta  \; ,  \\
\label{E10Dualityb}  
e^{-1} \Bigl( J_{t \alpha} + T_{\alpha}{}^{M}{}_N M^{NP} M_{MQ}  \partial_P A^Q\;  \hspace{30mm} & 
\CR
+ M_{\alpha\beta} C^{\beta P}{}_{M;N} \chi_P{}^{M;N} + M_{\alpha M}  \chi_N{}^{M;N}\Bigr) &=2 K_{(\alpha}{}^N{}_{\beta)} J_N{}^\beta + T_{\alpha}{}^M{}_N \chi_M{}^N\; 
\end{align}
\vspace{-6mm}
\begin{align}
\label{chit} 
M_{M\wa} \Bigl(  J_t{}^{\wa} + T^{\wa P}{}_N \partial_P A^N + C^{\wa P}{}_{Q;N} \chi_P{}^{Q;N} \Bigr) + M_{MN}   \zeta_{P}{}^{N;P}  + \acute{M}_{MN}  \chi_{P}{}^{N;P} \nn\\
= e C_{M}{}^{N}{}_{\wb} J_N{}^\wb + e^{-1} M_{MN} \partial_t A^N - 2 \partial_M e\; .  
\end{align} 
\end{subequations}
This last equation determines $\chi_t{}^M =J_t{}^{\wa=M}$ in terms of the other fields. Note that $M_{MN}$ is identical to the same matrix for the module $R(\Lambda_1)$, and so is invertible, while $\acute{M}_{MN}$ only appears through the indecomposable structure
\begin{align}
\Omega_{\wwa}{}^{\wwg} M_{\wwg\wwb} = \left( \begin{array}{ccc} M_{MN} &  M_{M\beta} & \acute{M}_{MN} \\ 0 & M_{\alpha\beta} & M_{\alpha N} \\ 0 & 0 & M_{MN} \end{array}\right) \; , \qquad \Omega_{\wwa}{}^{\wwb}  = \left( \begin{array}{ccc} \, 0 \, & \, 0\,   & \, \delta_M^N \, \\ 0 & \delta_\alpha^\beta  & 0 \\ \delta_M^N  & 0 & 0 \end{array}\right) \; .
\end{align}
So we can indeed invert $M_{MN}$ in $M_{MN} \chi_t{}^N$ to solve this last equation \eqref{chit} by fixing $\chi_t{}^M$. Therefore we define the $E_{10}$ exceptional field theory to only depend on the fields $M_{MN}$, $A^M$, $\chi_M{}^M$ and $\chi_M{}^{N;P}$, which satisfy  the duality equations \eqref{E10Dualitya} and \eqref{E10Dualityb}.

We have checked that these equations in the $GL(10)$ level decomposition are indeed compatible with \eqref{F}.
\medskip

We shall now determine the $E_{10}$ exceptional field theory pseudo-Lagrangian. Using the same procedure as in Section \ref{sec:e8tame}, one computes that, up to a total derivative, the $E_{11}$ exceptional field theory pseudo-Lagrangian \eqref{eq:Lag} can be rewritten as a pseudo-Lagrangian for the $E_{10}$ exceptional field theory plus the infinite sum of the squares of the duality equations
\begin{align}
\cL = \sum_{k=0}^\infty \mathcal{O}_k  + \Pi_{\tilde{\gamma}}{}^{MN} \partial_M \chi_N{}^{\tilde{\gamma}} +  \cL_{E_{10}}
\end{align}
with
\begin{multline}
\label{eq:LagE10}
\cL_{E_{10}} =  \Pi_{\tilde{\gamma}_\dgr{0}}{}^{0N}  J_t{}^{\alpha_\dgr{0}}  K_{\alpha_\dgr{0}\!}{}^{\tilde{\gamma}_\dgr{0}}{}_{\beta_\dgr{0}} J_N{}^{\beta_\dgr{0}} + \Omega_{I_\dgr{-\frac12} J_\dgr{\frac12}} C^{I_\dgr{-\frac12} A }{}_{\tilde{\alpha}_\dgr{0}} \chi_A{}^{\tilde{\alpha}_\dgr{0}} F^{J_\dgr{\frac12}}   \\
- \tfrac14 \kappa_{\alpha_\dgr{0}\beta_\dgr{0}} m^{MN} J_M{}^{\alpha_\dgr{0}} J_N{}^{\beta_\dgr{0}} + \tfrac12 J_{M}{}^{\! \alpha_\dgr{0}} T_{\beta_\dgr{0} \!}{}^{M}{}_P  m^{PQ} T_{\alpha_\dgr{0} \!}{}^{N}{}_Q J_{N}{}^{\! \beta_\dgr{0}}\\
- \tfrac12 m_{\tilde{I}_\dgr{\frac12}\tilde{J}_\dgr{\frac12}} C^{\tilde{I}_\dgr{\frac12}}{}_{P\wa_\dgr{0}} C^{\tilde{J}_\dgr{\frac12}}{}_{Q \wb_\dgr{0}} m^{PN} m^{QM} J_M{}^{\wa_\dgr{0}} J_N{}^{\wb_\dgr{0}}\\
+ \tfrac14 \bigl( \kappa_{\alpha_\dgr{0}\beta_\dgr{0}}  -2  T_{\alpha_\dgr{0}}{}^0{}_0  T_{\beta_\dgr{0}}{}^0{}_0 \bigr) m^{00} J_t{}^{\alpha_\dgr{0}} J_t{}^{\beta_\dgr{0}}
- T_{\alpha_\dgr{0}}{}^0{}_0 m^{00}T_{\beta_\dgr{1}}{}^N{}_0   J_t{}^{\alpha_\dgr{0}} J_N{}^{\beta_\dgr{1}}\\ - m^{00} T_{\alpha_\dgr{1}\!}{}^{(M}{}_{0} T_{\beta_\dgr{1}\!}{}^{N)}{}_{0}   J_M{}^{\alpha_\dgr{1}} J_N{}^{\beta_\dgr{1}} \; , 
\end{multline} 
and where we use $m^{MN} = e M^{MN}$ and $m^{00} = -e^{-1}$ to distinguish the $GL(1)\times E_{10}$ matrices in $E_{11}$ from the $E_{10}$ matrix $M^{MN}$ and the lapse $e$. Note that $\alpha_\dgr{0}$ includes both the $\mf{e}_{10}$ adjoint index and the $\mf{gl}(1)$ index for the lapse. The pseudo-Lagrangian~\eqref{eq:LagE10} only depends on the fields $M_{MN}$, $A^M$ and $\chi_M{}^N$, but not on $\chi_M{}^{N;P}$, and can be written out in the explicit form
\begin{multline} \label{E10ExFTLagrange}
\cL_{E_{10}} =   J_t{}^{\alpha}  K_{\alpha}{}^{M}{}_{\beta} J_M{}^{\beta} + T_{\alpha}{}^M{}_P \chi_M{}^P \bigl( J_t{}^\alpha + T^{\alpha N}{}_Q \partial_N A^Q \bigr)   \\
- \tfrac14 e \kappa_{\alpha\beta} M^{MN} J_M{}^{\alpha} J_N{}^{\beta} + \tfrac12 e J_{M}{}^{\! \alpha} T_{\beta}{}^{M}{}_P  M^{PQ} T_{\alpha \!}{}^{N}{}_Q J_{N}{}^{\! \beta}\\
- \tfrac12 e M_{\tilde{I}\tilde{J}} C^{\tilde{I}\! }{}_{P\wa} C^{\tilde{J}\! }{}_{Q \wb} M^{PN} M^{QM} J_M{}^{\wa} J_N{}^{\wb}  +2 \partial_M e \partial_N M^{MN} \\
- \tfrac14 e^{-1} \kappa_{\alpha\beta}   J_t{}^{\alpha} J_t{}^{\beta} - \partial_t e^{-1} \partial_M A^M + e^{-1} \partial_M A^{(M} \partial_N A^{N)}  \; . 
\end{multline} 
Here, we have dropped the subscript $\frac12$ on the $\tilde{I}$ index, that is valued in $R_\ten(\Lambda_1)\otimes R_\ten(\Lambda_1)\ominus R_\ten(2\Lambda_1)$ with $C^{\tilde{I}}{}_{M\wa = P}=C^{\tilde{I}}{}_{M;P}$ the $E_{10}$ intertwiner and the bilinear form $\eta_{\tilde{I}\tilde{J}}$ determined such that
\be 
T^{\alpha M}{}_P T_\alpha{}^{N}{}_Q = C^{\tilde{I}}{}_{Q;P} \eta_{\tilde{I}\tilde{J}} \eta^{M\! R} \eta^{N\! S} C^{\tilde{J}}{}_{R;S} \; . 
\ee
The non-invariant coefficients $C^{\tilde{I}}{}_{M \alpha}$ are related to the field strength structure coefficients \eqref{FsCchi} through 
\be 
C^{\tilde{I}\! }{}_{M \alpha} = - \eta_{\alpha\beta} \eta_{M\! N} C^{\beta N\! }{}_{P;Q} \eta^{QR} \eta^{PS} C^{\tilde{I}\! }{}_{R;S}  \; . 
\ee 
One recognises the second and the third line in \eqref{E10ExFTLagrange} as the expected potential term for the internal current and the internal derivative of the lapse $e$. The first line is a topological term while the last line gives a kinetic term $- \tfrac14 e^{-1} \kappa_{\alpha\beta}   J_t{}^{\alpha} J_t{}^{\beta}$. Note, however, that it  has the opposite sign compared to a standard sigma model, but  this is due to the mixing with the topological term. The $K(E_{10}) / E_{10} $ coset fields are the only ones that are dynamical, whereas the Euler--Lagrange equations for the lapse $e$ and the generalised shift $A^M$ can  be  interpreted respectively as a Hamiltonian constraint and a generalised (spatial) momentum constraint. Although the latter is a total internal derivative 
\be \partial_M \bigl( \partial_t e^{-1} -  e^{-1} \partial_N A^N + \chi_N{}^N \bigr) - \partial_N \bigl( e^{-1} \partial_M A^N + \chi_M{}^N\bigr) = 0 \; ,  \label{GMC} \ee
the fact that it transforms in $R_\ten(\Lambda_1)$ is consistent with the proposed momentum constraint for the $E_{10}$ sigma model \cite{Damour:2007dt}.

Note that the duality equations~\eqref{E10Duality} do not follow from the Euler--Lagrange equations, and  the pseudo-Lagrangian $\cL_{E_{10}}$ is not a {\it bona fide} Lagrangian, unlike the case of exceptional field theories in odd dimensions greater than one. This difference can be understood as follows. The Euler--Lagrange equations of exceptional field theory Lagrangians never determine the equations for the non-propagating higher-form fields, so in this sense, could be considered as pseudo-Lagrangians even in odd dimensions. In higher dimensions, one separates the equations for the propagating and non-propagating fields and thus obtains a proper Lagrangian for the propagating fields. 
For $D=1$, however, the $K(E_{10}) / E_{10} $ coset fields already include non-propagating fields and there is no $E_{10}$-covariant distinction between propagating fields and non-propagating ones, so one cannot have a Lagrangian for the propagating fields alone.

\subsection{\texorpdfstring{Relation to the $E_{10}$ sigma model}{Relation to the E10 sigma model}}

The $E_{10}$ exceptional field theory gives an $E_{10}$-covariant formulation of maximal supergravity theories, so it is natural to ask how it may relate to the $E_{10}$ sigma model \cite{Damour:2002cu} that is conjectured to describe eleven-dimensional supergravity. Note that because of the section constraint, $E_{10}$ exceptional field theory does not provide an $E_{10}$-invariant formulation of eleven-dimensional supergravity, whereas the sigma model has $E_{10}$ symmetry. In order to recover the sigma model, it seems therefore necessary to remove the dependence on the internal coordinates for all the fields. In the spirit of to the `gradient representation' conjecture~\cite{Damour:2002cu}, it is natural to consider a gradient expansion of all fields in $E_{10}$ exceptional field theory.

The gradient expansion is valid in the limit in which the spatial gradients of the metric and the three-form are much smaller than their time derivative. So the corresponding small gradient approximation in $E_{10}$ exceptional field theory is to consider 
\be 
\label{E10GradientAppr} |  J_{t \alpha } |  \gg  |  J_{M \alpha } | \; , \qquad  |  J_{t \alpha } |  \gg  |  \partial_{M} A^N | \; , 
\ee
and to take fields that only depend on the time-coordinate at zeroth order in the gradient expansion. We remark that such a small gradient approximation is inconsistent with the duality equation in eleven dimensions, because 
\be
 F_{0ijk} \gg F_{ijkl} \quad \Leftrightarrow \quad F_{0i_1\dots i_6} \ll F_{i_1\dots i_7} \; , 
 \ee
and similarly for higher level duality equations.  Nevertheless, we shall assume in this section that the approximation \eqref{E10GradientAppr} makes sense prior to a choice of section thanks to the constrained fields. Looking at  \eqref{E10Dualityb} one finds that it may indeed be consistent to have \eqref{E10GradientAppr} as long as $| \chi_M{}^N|$ is of the same order as $ |  J_{t \alpha } |$. To prove the sigma model conjecture~\cite{Damour:2002cu} using $E_{10}$ exceptional field theory, one would need to prove that such a gradient expansion can systematically be solved iteratively as a formal power series such that the zeroth order fields $M_{MN}|_{y=0}$ determine the full dynamics unambiguously, where $y$ denotes the internal coordinates. The first order duality equation \eqref{E10Duality} may be solvable iteratively in this way, but we leave this question for future investigations. We shall only discuss here the zeroth order dynamics in the gradient expansion as an $E_{10}$ sigma model in which $M_{MN}$ is a function of time only. To define this sigma model we must determine at which order the constrained fields contribute, so in particular how $|\chi_{M}{}^N |$ and $|\chi_{M}{}^{N;P}|$ compare to $ |  J_{t \alpha } | $ and $ |  J_{M \alpha } |$ in the small gradient approximation.\footnote{The components $\chi_t{}^{N}$ are  fixed by duality from~\eqref{chit} and thus $\chi_M{}^{N}$ are the lowest components to consider.}

If we considered that $|\chi_{M}{}^N |$ and $|\chi_{M}{}^{N;P}|$ were both of the same order as the $E_{10}$ gradients $ |  J_{M \alpha } |$, the duality equation~\eqref{E10Dualityb} would give $e^{-1}  J_{t \alpha}|_{y=0}  = 0$ at the zeroth order in the gradient expansion, and the corresponding sigma model would be trivial. To get a non-empty sigma model we must therefore keep some constrained fields. To determine which ones, let us first consider the truncation in which one sets all internal derivatives to zero and takes fields that only depend on the time coordinate, while keeping both $\chi_{M}{}^N$ and  $\chi_{M}{}^{N;P}$ as time-dependent fields satisfying the section constraint.  With this definition, the duality equations truncate to 
\begin{align}
\label{E10DualityZero} 
 \chi_N{}^{M;N}   =  0  \; , \qquad e^{-1} \bigl( J_{t \alpha} + M_{\alpha\beta} C^{\beta P}{}_{M;N} \chi_P{}^{M;N}  \bigr) = T_{\alpha}{}^M{}_N \chi_M{}^N\; ,  
\end{align}
where the first equation comes from~\eqref{E10Dualitya} and the second one from~\eqref{E10Dualityb}, and we have set to zero all dependence of the $E_{11}$ coset fields on the internal coordinates, i.e. $J_N{}^\alpha=\partial_P A^Q = \partial_Ne = 0$.
The pseudo-Lagrangian~\eqref{eq:LagE10} plus the square of \eqref{E10DualityZero} simplifies to\footnote{By construction the kinetic term for $e$ vanishes with $h_0{}^0 h_0{}^0 - \frac12 h_0{}^0 h_0{}^0  - 2 ( h_0{}^0 - \tfrac12 h_0{}^0 )^2 = 0 $, and the equation of motion of $\chi_M{}^N$ being the duality equation, one gets that the term $ \frac14   e^{-1} \kappa^{\alpha\beta} J_{t \alpha} J_{t \beta} $ in the duality equation changes sign.}
\begin{align}
\label{LE10}
L^\ord{0}_{E_{10}} &=  \cL_{E_{10}}\big|_{\partial_M = 0}   \\
&  \hspace{5mm} + \tfrac12 e^{-1} \bigl(  J_{t \alpha}\! +\! M_{\alpha\gamma} C^{\gamma R\!}{}_{M;N} \chi_R{}^{M;N}  \!  -\! e T_\alpha{}^{ M}{}_N \chi_M{}^N  \bigr)  \bigl(  J_{t}{}^{\! \alpha} \! +\! C^{\alpha S\!}{}_{P;Q} \chi_S{}^{P;Q} \!  -\! e  M^{\alpha\beta} T_\beta{}^{ P}{}_Q \chi_P{}^Q  \bigr) \CR
&= \frac14   e^{-1} \kappa^{\alpha\beta} J_{t \alpha} J_{t \beta}  + e^{-1} J_{t \alpha}  C^{\alpha P\! }{}_{M;N} \chi_P{}^{\! M;N}   + \frac12 e^{-1}  M_{\alpha\beta} C^{\alpha R\! }{}_{M;N} \chi_R{}^{M;N}  C^{\beta S\!}{}_{P;Q} \chi_S{}^{P;Q} \; , \nn 
\end{align}
for fields that do not depend on the internal coordinates, but where we have kept the internal $M$ component of the constrained fields. Note that $ C^{\alpha P\! }{}_{M;N} \chi_P{}^{\! M;N}$ is an $E_{10}$ tensor on-shell using \eqref{FsCchi} since $\chi_N{}^{M;N}=0$ according to \eqref{E10DualityZero}. The corresponding quadratic term in~\eqref{E10Dualityb} would not be $E_{10}$ invariant without the truncation $J_{M}{}^\alpha = \partial_M A^N = 0$.

We now argue that this model cannot describe the full eleven-dimensional supergravity dynamics. After having shown this, we will  neglect $\chi_M{}^{N;P}$ at zeroth order and will  only keep $\chi_M{}^{N}$ in the sigma model. We begin by solving the section condition for $\chi_M{}^{N;P}$ and $\chi_M{}^{N}$ in the $GL(10)$ level decomposition of $E_{10}$. Then $\chi_M{}^{N;P}$ contributes to $F^\alpha = J_{t}{}^\alpha  +C^{\alpha P\! }{}_{M;N} \chi_P{}^{\! M;N}$ starting from level $k= 3$ and above. This means that we can rewrite~\eqref{LE10} as 
\bea 
L^\ord{0}_{E_{10}}  &=& \frac14   \sum_{k=-2}^2 e^{-1} \kappa^{\alpha_\dgr{k} \beta_\dgr{k} } J_{t \alpha_\dgr{k}} J_{t \beta_\dgr{k}} \CR
&& \qquad + \frac{1}{2} \sum_{k=3}^\infty    e^{-1}  M_{\alpha_\dgr{k}\beta_\dgr{k} } \bigl( J_t{}^{\alpha_\dgr{k}} +  C^{\alpha_\dgr{k} R\! }{}_{M;N} \chi_R{}^{M;N}  \bigr) \bigl( J_t{}^{\beta_\dgr{k}} +  C^{\beta_\dgr{k} S\!}{}_{P;Q} \chi_S{}^{P;Q} \bigr) \; , \qquad \label{L0E10Square} 
\eea
where we also used the Cartan involution to double the $M_{\alpha_\dgr{k}\beta_\dgr{k} }  J_t{}^{\alpha_\dgr{k}}J_t{}^{\beta_\dgr{k}} $ terms for $k\geq 3$. 
Because  $T_{\alpha_\dgr{k}  }{}^{M}{}_N \chi_M{}^N$ only contributes to negative levels $k \le 0 $ in \eqref{E10DualityZero}, the second line in~\eqref{L0E10Square} is quadratic in the duality equation \eqref{E10DualityZero} and can be consistently disregarded in the pseudo-Lagrangian. Moreover, one gets from \eqref{E10DualityZero} that $J_t{}^{\alpha_\dgr{1}} =J_t{}^{\alpha_\dgr{2}}=0$ and the only remaining field is the metric, with 
\be 
L^\ord{0}_{E_{10}} \approx  \frac14 e^{-1} \bigl(  g^{ik} g^{jl} \partial_t {g}_{ij}  \partial_t {g}_{kl} - g^{ij}  \partial_t {g}_{ij}g^{kl}  \partial_t {g}_{kl}\bigr) \; . 
\ee

We conclude that the constrained field $\chi_M{}^{N;P}$ has the effect of truncating the sigma model to the $GL(10)$ subgroup, so if we want the sigma model to describe the full eleven-dimensional dynamics we need to consider instead that  $|\chi_M{}^{N;P}| \ll  |  J_{t \alpha } | $ in the gradient approximation and to only keep $\chi_M{}^N$ non-zero at zeroth order in the gradient expansion. By construction this simply corresponds to set $\chi_M{}^{N;P}=0$ in the duality equation \eqref{E10DualityZero} and the pseudo-Lagrangian \eqref{LE10}. 

Doing so, one obtains the standard $E_{10}$ sigma model Lagrangian
\begin{align}
L =  \frac14   e^{-1} \kappa^{\alpha\beta} J_{t \alpha} J_{t \beta} \; , 
\end{align}
in one (time) dimension, together with the duality equation
\begin{align}
 \label{E10DualityZeroZero} 
  e^{-1} J_{t \alpha} =  T_\alpha{}^M{}_N \chi_M{}^N \; . 
\end{align}
This gives precisely the Lagrangian $L$ of the sigma model of~\cite{Damour:2002cu} with the additional constraint \eqref{E10DualityZeroZero}.  The sigma model Lagrangian and the duality equation \eqref{E10DualityZeroZero} determine the dynamics for the time-dependent fields $M_{MN}$ of $E_{10}$, the lapse $e$  and the constrained fields  $\chi_M{}^N$. Note that the section constraint for the constrained fields $\chi_M{}^N$ is now algebraic, and therefore the dynamics of this mechanical model is truly $E_{10}$ invariant. The quadratic constraint $\chi_M{}^N$ satisfies is an  algebraic constraint similar to Berkovits' pure spinor constraint in \cite{Berkovits:2000fe}. 

Solving explicitly the section constraint in the $GL(10)$ decomposition one obtains that \eqref{E10DualityZeroZero} implies that $e^{-1} J_{t \alpha} = Q_\alpha$ is in the $E_{10}$ orbit of an element in the positive Borel subalgebra ${\mf b}_+\subset \mf{e}_{10}$, i.e. there exists a time-independent $g\in E_{10}$ such that
\begin{align} 
\label{QGB}
g^{-1} Q g \in  {\mf b}_+ \; . 
\end{align}
This condition would be trivially satisfied for a finite-dimensional Lie algebra~\cite{CM}, but it is a priori a non-trivial constraint for a hyperbolic Kac--Moody algebra. 
 This problem depends on the precise definition of the Kac--Moody groups in which $\cV$ and $g$ are defined. Equation \eqref{QGB} is well-defined for $\cV$ and $g$ valued in the minimal group $E_{10}^{{\rm m}}$ and ${\mf b}_+$ the minimal module according to the terminology introduced in Section \ref{sec:subtle}, see~\cite{Kac:1983} for some results.  However, physically interesting solutions require a priori to consider $\cV$ in a non-trivial extension of the minimal group. For $\cV \in E_{10}^{{\rm c}+}$, the projection of the Maurer--Cartan form is valued in ${\mf e}_{10}^{{\rm c}+-}$ and the charge $Q$ is generally not defined. It seems therefore that one may need an intermediate completion of $E_{10}^{{\rm m}}$ that is yet to be discovered in order to properly define the model.

The analysis above made a number of assumptions whose validity and consistency would need to be investigated further. First, keeping only the constrained fields that are independent of the internal coordinates may be justifiable through a gauge-fixing of generalised diffeomorphisms (including ancillary transformations) which would not break $E_{10}$ symmetry. The consistency of the gradient expansion might require to introduce an additional momentum constraint as proposed in~\cite{Damour:2007dt}, that could be a consequence of the generalised momentum constraint~\eqref{GMC}. Second, the duality equation~\eqref{E10DualityZeroZero} restricts the $E_{10}$ current through the surviving field $\chi_M{}^N$. This could be related to the analysis of additional constraints to be imposed on the $E_{10}$ sigma model according to~\cite{Damour:2007dt} that are related to null (or potentially) imaginary roots of the $E_{10}$ root lattice. The first such constraint appears for the null root $\Lambda_1$ of $E_{10}$, see also~\eqref{GMC}.

\end{document}